\documentclass{aa}  

\usepackage{graphicx}
\usepackage{txfonts}

\usepackage{xspace}
\usepackage{booktabs}
\usepackage[flushleft]{threeparttable}
\usepackage{amsmath}
\usepackage{arydshln}
\usepackage{longtable}
\usepackage{threeparttablex} 
\usepackage{orcidlink}

\usepackage{enumerate}

\newcommand{\tess}{TESS\xspace}

\newcommand{\vsini}{$v\sin i_\star$\xspace}

\newcommand{\figref}[1]{Fig.~\ref{#1}}
\newcommand{\tabref}[1]{Table~\ref{#1}}
\newcommand{\sref}[1]{Section~\ref{#1}}
\newcommand{\eref}[1]{Eq.~(\ref{#1})}

\newcommand{\hdu}{$-74^{+6}_{-9}$}
\newcommand{\hdg}{$-23^{+25}_{-38}$}
\newcommand{\hdvsinig}{$4.9^{+0.4}_{-0.5}$}

\newcommand{\hdtwog}{$29^{+12}_{-13}$}
\newcommand{\hdtwos}{$14^{+7}_{-8}$}
\newcommand{\hdtwovsinig}{$6.3 \pm 0.4$}

\newcommand{\ksixoneu}{$29^{+35}_{-56}$}
\newcommand{\ksixoneg}{$32^{+48}_{-33}$}
\newcommand{\ksixonevsinig}{$1.3\pm0.4$}

\newcommand{\keightsevenu}{$21^{+14}_{-11}$}
\newcommand{\keightseveng}{$23^{+12}_{-13}$}
\newcommand{\keightsevens}{$37^{+8}_{-9}$}
\newcommand{\keightsevenvsinig}{$1.16\pm0.17$}

\newcommand{\keltthreeu}{$-5\pm4$}
\newcommand{\keltthreeg}{$-6\pm5$}
\newcommand{\keltthreevsinig}{$7.2\pm0.4$}

\newcommand{\keltfouru}{$77_{-23}^{+25}$}
\newcommand{\keltfourg}{$80_{-22}^{+25}$}
\newcommand{\keltfourvsinig}{$5.8_{-0.5}^{+0.4}$}

\newcommand{\lttg}{$22_{-83}^{+98}$}
\newcommand{\lttvsinig}{$1.4\pm0.5$}

\newcommand{\toifourfiveu}{$-27^{+64}_{-74}$}
\newcommand{\toifourfiveg}{$-23^{+37}_{-40}$}
\newcommand{\toifourfivevsinig}{$8.1\pm0.5$}

\newcommand{\toieightthirteenu}{$-53^{+22}_{-25}$}
\newcommand{\toieightthirteeng}{$-32\pm23$}
\newcommand{\toieightthirteens}{$-53^{+25}_{-28}$}
\newcommand{\toieightthirteenvsinig}{$8.4\pm0.5$}

\newcommand{\toieightnineu}{$10^{+39}_{-35}$}
\newcommand{\toieightnineg}{$9^{+30}_{-23}$}
\newcommand{\toieightnines}{$-12^{+18}_{-16}$}

\newcommand{\toieightninevsinis}{$6.5\pm0.5$}

\newcommand{\toithirtyu}{$4_{-6}^{+5}$}
\newcommand{\toithirtyg}{$4^{+5}_{-6}$}
\newcommand{\toithirtyvsinig}{$1.2\pm0.3$}
\newcommand{\toithirtys}{$6^{+10}_{-9}$}

\newcommand{\waspfiftyu}{$-2.9\pm1.2$}
\newcommand{\waspfiftyg}{$-2.9^{+1.2}_{-1.3}$}
\newcommand{\waspfiftys}{$-1.8\pm1.0$}
\newcommand{\waspfiftyvsinig}{$1.97^{+0.09}_{-0.10}$}
\newcommand{\waspfiftyvsiniu}{$1.97\pm0.10$}

\newcommand{\waspfiftynineu}{$16^{+18}_{-16}$}
\newcommand{\waspfiftynineg}{$16^{+17}_{-16}$}

\newcommand{\waspfiftyninevsiniu}{$0.58^{+0.18}_{-0.16}$}

\newcommand{\waspthirtyeightu}{$-57^{+9}_{-14}$}
\newcommand{\waspthirtyeightg}{$-38^{+16}_{-18}$}
\newcommand{\waspthirtyeightvsinig}{$12.2\pm0.5$}

\newcommand{\waspfortyeightu}{$-2\pm8$}
\newcommand{\waspfortyeightg}{$-2\pm8$}
\newcommand{\waspfortyeightvsinig}{$2.0^{+0.4}_{-0.3}$}

\newcommand{\waspseventwou}{$118\pm6$}
\newcommand{\waspseventwog}{$115\pm6$}
\newcommand{\waspseventwos}{$121^{+11}_{-13}$}
\newcommand{\waspseventwovsinig}{$13.0^{+0.4}_{-0.5}$}

\newcommand{\waspseventhreeu}{$16^{+38}_{-31}$}
\newcommand{\waspseventhreeg}{$11^{+32}_{-20}$}
\newcommand{\waspseventhrees}{$25_{-11}^{+12}$}
\newcommand{\waspseventhreevsinig}{$6.5^{+0.3}_{-0.4}$}
\newcommand{\waspseventhreevsinis}{$7.2^{+0.4}_{-0.4}$}

\newcommand{\waspeightsixu}{$-10^{+29}_{-32}$}
\newcommand{\waspeightsixg}{$-10^{+18}_{-16}$}
\newcommand{\waspeightsixvsinig}{$14.8\pm0.5$}
\newcommand{\waspeightsixvsiniu}{$11^{+4}_{-11}$}

\newcommand{\xosevenu}{$-4^{+27}_{-23}$}
\newcommand{\xoseveng}{$0^{+20}_{-14}$}
\newcommand{\xosevenvsinig}{$5.1\pm0.5$}

\usepackage[]{hyperref}
\usepackage{xcolor}
\hypersetup{
    colorlinks,
    linkcolor={red!50!black},
    citecolor={blue!50!black},
    urlcolor={blue!80!black}
}
\begin{document}

   \title{Obliquities of Exoplanet Host Stars}

   \subtitle{19 New and Updated Measurements,
   and Trends in the Sample of 205 Measurements }

    \author{
        E.~Knudstrup\inst{\ref{see},\ref{sac}}\orcidlink{0000-0001-7880-594X} \and
        S.~H.~Albrecht\inst{\ref{sac},\ref{prince}}\orcidlink{0000-0003-1762-8235} \and
        J.~N.~Winn\inst{\ref{prince}}\orcidlink{0000-0002-4265-047X} \and
        D.~Gandolfi\inst{\ref{torino}}\orcidlink{0000-0001-8627-9628} \and
        J.~J.~Zanazzi\inst{\ref{berk}}\orcidlink{0000-0002-9849-5886}\thanks{51 Pegasi b Fellow}
        \and
        C.~M.~Persson\inst{\ref{oso}} \and
        M.~Fridlund\inst{\ref{leiden},\ref{oso}}\orcidlink{0000-0002-0855-8426} \and
        M.~L.~Marcussen\inst{\ref{sac}}\orcidlink{0000-0003-2173-0689} \and
        A.~Chontos\inst{\ref{prince},\ref{hawaii}}\orcidlink{0000-0003-1125-2564} \and
        M.~A.~F.~Keniger\inst{\ref{warwick}}\orcidlink{0009-0005-2761-9190} \and
        N.~L.~Eisner\inst{\ref{flat}}\orcidlink{0000-0002-0786-7307} \and
        A.~Bieryla\inst{\ref{cfa}}\orcidlink{0000-0001-6637-5401} \and
        H.~Isaacson\inst{\ref{berk2}}\orcidlink{0000-0002-0531-1073} \and
        A.~W.~Howard\inst{\ref{caltech}}\orcidlink{0000-0001-8638-0320} \and
        L.~A.~Hirsch\inst{\ref{miss}}\orcidlink{0000-0001-8058-7443} \and
        F.~Murgas\inst{\ref{iac},\ref{ull}}\orcidlink{0000-0001-9087-1245} \and
        N.~Narita\inst{\ref{kom},\ref{ac},\ref{iac}}\orcidlink{0000-0001-8511-2981} \and
        E.~Palle\inst{\ref{iac},\ref{ull}}\orcidlink{0000-0003-0987-1593} \and
        Y.~Kawai\inst{\ref{multi}}\orcidlink{0000-0002-0488-6297} \and
        D.~Baker\inst{\ref{physics}}\orcidlink{0000-0002-2970-0532}
}
\authorrunning{Knudstrup et~al.}

   \institute{
    Department of Space, Earth and Environment, Chalmers University of Technology, 412 93, Gothenburg, Sweden\label{see} \email{emil.knudstrup@chalmers.se} \and
    Stellar Astrophysics Centre, Department of Physics and Astronomy, Aarhus University, Ny Munkegade 120, DK-8000 Aarhus C, Denmark\label{sac} \and
    Department of Astrophysical Sciences, Princeton University, 4 Ivy Lane, Princeton, NJ 08544, USA\label{prince} \and
    Dipartimento di Fisica, Universita degli Studi di Torino, via Pietro Giuria 1, I-10125, Torino, Italy\label{torino} \and
    Astronomy Department, Theoretical Astrophysics Center, and Center for Integrative Planetary Science, University of California, Berkeley, Berkeley, CA 94720, USA\label{berk} \and
    Chalmers University of Technology, Department of Space, Earth and Environment, Onsala Space Observatory, SE-439 92 Onsala, Sweden\label{oso} \and
    Leiden Observatory, University of Leiden, PO Box 9513, 2300 RA Leiden, The Netherlands\label{leiden} \and
    Institute for Astronomy, University of Hawai‘i, 2680 Woodlawn Drive, Honolulu, HI 96822, USA\label{hawaii} \and
    Department of Physics, University of Warwick, Gibbet Hill Road, Coventry CV4 7AL, UK\label{warwick} \and
    Center for Computational Astrophysics, Flatiron Institute, 162 Fifth Avenue, New York, NY 10010, USA\label{flat} \and
    Center for Astrophysics ${\rm \mid}$ Harvard {\rm \&} Smithsonian, 60 Garden Street, Cambridge, MA 02138, USA\label{cfa} \and
    501 Campbell Hall, University of California at Berkeley, Berkeley, CA 94720, USA\label{berk2} \and
    Department of Astronomy, California Institute of Technology, Pasadena, CA 91125, USA\label{caltech} \and
    Department of Chemical and Physical Sciences, University of Toronto Mississauga, Mississauga, ON, Canada\label{miss} \and
    Instituto de Astrofísica de Canarias (IAC), 38205 La Laguna, Tenerife, Spain\label{iac} \and
    Departamento de Astrofísica, Universidad de La Laguna (ULL), 38206 La Laguna, Tenerife, Spain\label{ull} \and
    Komaba Institute for Science, The University of Tokyo, 3-8-1 Komaba, Meguro, Tokyo 153-8902, Japan\label{kom} \and
    Astrobiology Center, 2-21-1 Osawa, Mitaka, Tokyo 181-8588, Japan\label{ac} \and
    Department of Multi-Disciplinary Sciences, Graduate School of Arts and Sciences, The University of Tokyo, 3-8-1 Komaba, Meguro, Tokyo 153-8902, Japan\label{multi} \and
    Physics Department, Austin College, Sherman, TX 75090, USA\label{physics}
   }



   \date{Received May 8, 2024; accepted August 16, 2024}


    \abstract{
        Measurements of the obliquities in exoplanet systems have revealed some remarkable architectures, some of which are very different from the Solar System. Nearly 200 obliquity measurements have been obtained through observations of the Rossiter-McLaughlin (RM) effect. 
   Here we report on observations of 19 planetary systems that led to 17 clear detections of the RM effect and 2 less secure detections. After adding the new measurements to the tally, we use the entire collection of RM measurements to investigate four issues
   that have arisen in the literature.
   i) Does the obliquity distribution show a peak at approximately
   90$^\circ$? We find tentative evidence that such a peak does exist 
   when restricting attention
   to the sample of
   sub-Saturn planets and hot Jupiters orbiting F stars. ii) Are high obliquities associated with high eccentricities? We find the association to be weaker than previously reported, and that a stronger association
   exists between obliquity and orbital
   separation, possibly due to tidal obliquity damping at small separations. iii) How low are the lowest known obliquities? 
   Among hot Jupiters around cool stars,
   we find the dispersion to be $1.4\pm0.7^\circ$, smaller than the 6$^\circ$ obliquity of the Sun, which serves as additional evidence for tidal damping. iv) What are the obliquities of stars with compact and flat systems of multiple planets? We find that they generally have obliquities lower than $10^\circ$, with several
   remarkable exceptions possibly caused by wide-orbiting stellar or planetary companions. 
    }


   \maketitle
%

\section{Introduction} \label{sec:intro}

Hot Jupiters, planets on extremely eccentric orbits, planets orbiting binary stars, planets orbiting white dwarfs and neutron stars --- these and other types of exotic planetary systems have been discovered over the last 30 years. A diversity of architectures has also been revealed by measurements of the host star's obliquity, i.e., the angle between the star's spin axis and a planet's orbital axis, or at least its sky projection.
Thanks to many research groups pursuing these measurements, usually by means of the Rossiter-McLaughlin (RM) effect, we now have a sample of nearly 200 measurements including 
prograde, polar, and retrograde orbits \citep[see, e.g.,][for a review]{Albrecht2022}. 

\begin{table*}
    \centering
    
\begin{threeparttable}
\caption{Systems observed and spectroscopic observations. }
\begin{tabular}{l c c c c c c c}
\toprule
Target & $V$ & Instrument & ID & Night & $N$ & $t$/$\Delta t$ & $T_{41}$ \\
& (mag) & & & (dd-mm-yyyy) & & (s) & (hr) \\
    \midrule
        HD~118203~b & 8.14 & FIES & 60-417 & 24-03-2020 & 12 & 900/1170 & 5.64\\
        HD~118203~b & 8.14 & FIES & 65-851 & 18-06-2022 & 21 & 600/768 & 5.64\\
        HD~148193~b & 9.77 & HARPS-N & A43TAC\_11 & 15-06-2021 & 23 & 1140/1168 & 6.61 \\
        K2-261~b & 10.61 & HARPS-N & A44TAC\_16 & 24-03-2022 & 40 & 600/625 & 5.11 \\
        K2-287~b & 11.41 & ESPRESSO & 105.20KD.003 & 05-07-2021 & 26 & 540/600 & 3.42 \\
         KELT-3~b & 9.87 & FIES & 65-851 & 08-04-2022 & 15 & 900/1080 & 3.15 \\
        KELT-4Ab & 9.5 & FIES & 62-506 & 27-01-2021 & 16 & 900/1090 & 3.46 \\
        LTT~1445Ab & 11.22 & HARPS-N & A41TAC\_19 & 05-09-2020 & 15 & 900/920 & 1.38 \\
        TOI-451Ab & 11.02 & ESPRESSO & 109.22Z4.002 & 24-09-2022 & 21 & 720/755 & 2.04 \\
        TOI-813~b & 10.32 & ESPRESSO & 109.22Z4.008 & 19-01-2023 & 29 & 800/832 & 13.10 \\
        TOI-892~b & 11.45 & HARPS-N & A44TAC\_16 & 01-01-2022 & 21 & 960/987 & 5.51 \\
        TOI-1130~c & 11.37 & ESPRESSO & 105.20KD.002 & 01-06-2021 & 23 & 660/700 & 2.02 \\
         WASP-50~b & 11.44 & ESPRESSO & 109.22Z4.007 & 23-07-2022 & 61 & 240/270 & 1.81 \\
         WASP-59~b & 12.78 & HIRES & N117Hr & 23-11-2016 & 20 & 787/835 &  2.45 \\
        WASP-136~b & 9.98 & FIES & 63-505 & 30-08-2021  & 20 & 1140/1320 & 5.57 \\
        WASP-148~b & 12.25 & HARPS-N & A42TAC\_22 & 21-03-2021 & 17 & 900/920 & 3.14 \\
        WASP-172~b & 11.0 & ESPRESSO & 109.22Z4.006 & 01-06-2022 & 31 & 900/940 & 5.47 \\
        WASP-173Ab & 11.3 & ESPRESSO & 109.22Z4.003 & 23-07-2022 & 34 & 555/590 & 2.33 \\
        WASP-186~b & 10.82 & FIES & 64-506 & 11-10-2021 & 15 & 900/1080 & 2.75 \\
        XO-7~b & 10.52 & FIES & 65-851 & 27-08-2022 & 12 & 840/1000 & 2.77 \\
    \bottomrule
    \end{tabular}
    \begin{tablenotes}
        \item In the columns we list the Johnson $V$ magnitudes, the instrument used, the program IDs, the dates of the observation nights, the number of exposures on a given night ($N$), the exposure duration ($t$), sampling time ($\Delta t$), and the total transit duration ($T_{41}$).
    \end{tablenotes}
    \end{threeparttable}
\label{tab:spec_obs} 
\end{table*}

Here we present observations of 19 transiting exoplanets aiming to detect the RM effect and measure the sky-projected stellar obliquities. Together with the sample drawn from the literature, we also wanted to investigate four  issues regarding the distribution of obliquities.

\begin{enumerate}[i)]
\item For most systems in the sample, the projected obliquity ($\lambda$) is known
but not the obliquity itself ($\psi$) because
of missing information about the inclination
of the stellar rotation axis with respect to the line of sight ($i_\star$).
Using the subset of 57 systems for which $\psi$ has been measured, \citet{Albrecht2021}
found evidence for a preponderance of perpendicular planets, i.e., a peak in the obliquity distribution near $90^\circ$.
Following up on this result, \citet{Siegel2023} and \citet{Dong2023} applied more advanced
statistical techniques that allowed the entire
sample to be used, and did not find
evidence for a peak.
With an enlarged sample, we wanted to revisit this issue and see whether there is a particular category of planets for which a peak does exist.

\item Three theories have been proposed to explain the existence of hot Jupiters: in-situ formation, disk-driven migration, and high-eccentricity or tidally-driven migration \citep[see][for a review]{Dawson2018}. The latter process would not only increase the orbit's eccentricity but might also raise its inclination relative to the star's equatorial plane.
One might therefore expect high eccentricities and obliquities to be statistically associated, and evidence for such an association
has been reported \citep{Rice2022}. After noticing some errors in this earlier study, we decided to revisit the issue with an enlarged sample.

\item Stars with $T_{\rm eff} \lesssim 6{,}250$~K that host hot Jupiters tend to have especially low obliquities. Indeed, the
most precise such measurements ($\sigma_\lambda<2^\circ$) show an obliquity dispersion less than one degree --- smaller
than 6.2$^\circ$ obliquity of the Sun relative
to the Solar System's invariable plane.
Highlighting this result, \citet{Albrecht2022} interpreted the very low obliquities of hot
Jupiters around cool stars as evidence 
for tidal obliquity damping. If this is true, then by continuing to perform precise measurements, we might gain a better understanding of the evolution of hot Jupiters and rates of tidal dissipation. With the enlarged sample, we wanted to see if this trend still holds and include the numerous but less precise obliquity measurements in the determination
of the dispersion.

\item The first five measurements of $\lambda$ for stars with compact multiple-transiting planetary systems (``multis'') were all consistent with low obliquities \citep{Albrecht2013}, following the blueprint of the Solar System. Since then, 
several exceptions have been discovered \citep{Hjorth2021,Huber2013,Chaplin2013}. The misalignments in these cases are suspected of being caused by the effects of
an outer massive companion. We wanted to
assess the obliquity
distribution of the multis, and
see whether all of the misaligned multis have outer companions.

\end{enumerate}

The paper is structured as follows. \sref{sec:obs} describes the general characteristics of the new data.
\sref{sec:analysis} presents our general
approach to analyzing the data, and \sref{sec:targets} discusses the particulars
for each of the 19
observed systems.
\sref{sec:Discussion} discusses
the issues and questions posed above,
and \sref{sec:conc} summarizes
our conclusions.


\section{Observations}\label{sec:obs}
To measure the projected obliquity of each
host star, we performed high-resolution optical spectroscopy over a time range (typically 4--8 hours) spanning a planetary transit and tried to detect the
RM effect.
\tabref{tab:spec_obs} gives the names of the 19 systems, the names of the telescopes and instruments employed, and some key observational
characteristics. The telescopes and instruments
are further described in \tabref{tab:spectro}. 
Tables~\ref{tab:planets} and \ref{tab:stars}
give some of the basic parameters of the host stars and their planets.
\tabref{tab:rvs} provides
all of the apparent radial velocity (RV) measurements.
These spectroscopic data were
supplemented by the best available
transit light curves, as described below.

\begin{table}[]
    \centering
    
    \begin{threeparttable}  
    \caption{Telescopes and spectrographs used in this study.}  
    \label{tab:spectro}
    \begin{tabular}{c c c c c}
    \toprule
    Telescope & Aperture & Instrument & $R$ & $\sigma_\mathrm{PSF}$ \\
     & (m) &  &  & (km~s$^{-1}$) \\ 
    \midrule    
    NOT & 2.56 & FIES & 67\,000 & 1.9 \\
    TNG & 3.6 & HARPS-N & 115\,000 & 1.1 \\
    VLT & 8.2 & ESPRESSO & 140\,000 & 0.9 \\
    Keck-1 & 10 & HIRES & 70\,000 & 1.8 \\
    \bottomrule
    \end{tabular}
    \begin{tablenotes}
    \item $R$ is the resolution and $\sigma_\mathrm{PSF}$ the width of the point spread function in velocity units.
    \end{tablenotes} 
    \end{threeparttable}
    
\end{table}


\subsection{Spectroscopy}

Seven systems were observed with
the Echelle SPectrograph for Rocky Exoplanets and Stable Spectroscopic Observations \citep[ESPRESSO;][]{Pepe2021} that can be fed by any of the four Very Large Telescopes at Paranal Observatory, Chile. 
Another seven systems were observed
with the FIber-fed Echelle Spectrograph \citep[FIES;][]{Frandsen1999,Telting2014} mounted on the Nordic Optical Telescope \citep[NOT;][]{Djupvik2010} located on the Roque de los Muchachos, La Palma, Spain.
For five
systems, we used the High Accuracy Radial velocity Planet Searcher for the Northern hemisphere \citep[HARPS-N;][]{Cosentino2014} mounted on the Telescopio Nazionale Galileo (TNG), also on the Roque de los Muchachos.
For one system, we used the High Resolution Echelle Spectrometer \citep[HIRES;][]{Vogt1994} mounted on the 10-m Keck-1 telescope on Mauna Kea, Hawai’i, USA.


The data from ESPRESSO and HARPS-N were reduced and radial velocities (RVs) were obtained using the standard data reduction software \citep[DRS;][]{Lovis2007,Pepe2021,Dumusque2021}.
The FIES data for all but two systems were reduced using FIEStool\footnote{\url{https://www.not.iac.es/instruments/fies/fiestool/}},
and RVs were extracted using FIESpipe\footnote{\url{https://fiespipe.readthedocs.io/en/}} following an approach similar to that described by \citet{Zechmeister2018}. The FIES observations of WASP-136 and WASP-186
were reduced and RVs extracted following the procedure
described by \citet{Gandolfi2015}. To account
for any instrumental RV drift of the FIES spectrograph, we observed the spectrum
of a Thorium-Argon (ThAr) lamp in between each science exposure. The HIRES spectra were obtained with the iodine cell in the light path, and the RV extraction followed the standard HIRES forward modelling pipeline of the California Planet Search \citep[CPS;][]{Howard2010} using the methodology of \citet{Butler1996}.

For some systems, in addition to the apparent RVs we modeled the shape
of the Cross Correlation Function (CCF)
of the absorption lines.
For the ESPRESSO and HARPS-N data, we used
the CCFs that are automatically produced
by the standard reduction pipelines.
For the FIES data, we derived CCFs using a template spectrum based on the most suitable
model atmosphere from the library of \citet{Castelli2003}, as implemented in \texttt{FIESpipe}.

 
\subsection{Photometry}

Light curves of all 19 of the systems are available from the databases of either the K2 mission \citep{Howell2014} or the Transiting Exoplanet Survey Satellite \citep[\tess;][]{Ricker2015}. For the K2 data, we used the light curves from the \texttt{EVEREST} pipeline \citep{Luger2016,Luger2018}.
The \tess data were downloaded and extracted using the \texttt{lightkurve} package \citep{lightkurve}, and the \texttt{RegressionCorrector} was applied to reduce the systematic variations due to scattered light.
Our parametric model for each
system was based on jointly fitting the photometric and spectroscopic data.
In some cases, we supplemented the space-based photometry with ground-based photometry to refine the transit ephemerides.

\begin{table*}[]
    \centering
    \begin{threeparttable}
    \caption{Key literature orbital and planetary parameters.}
    \label{tab:planets}
    \begin{tabular}{l c c c c c c c}
    \toprule
        Planet & $P$ & $e$ & $a/R_\star$ & $R_{\rm p}/R_\star$ & $b$ & $M_\mathrm{p}$ & $R_\mathrm{p}$ \\
         & (d) &   &  &  &   & (M$_\mathrm{J}$) & (R$_\mathrm{J}$) \\
\midrule
    HD~118203~b & $6.13499(3)$ & $0.314 \pm 0.017$ & $7.23^{+0.13}_{-0.14}$ & $0.05552^{+0.00017}_{-0.00019}$ & $0.111^{+0.076}_{-0.095}$ & $2.17^{+0.07}_{-0.08}$ & $1.14\pm0.03$  \\
    HD~148193~b & $20.38084(2)$ & $0.13_{-0.09}^{+0.12}$ & $20.2\pm0.5$ & $0.0462^{+0.0008}_{-0.0005}$ & $0.31^{+0.19}_{-0.21}$ & $0.092\pm0.015$ & $0.764^{+0.018}_{-0.017}$ \\
    K2-261~b & $11.63344(12)$ & $0.42 \pm 0.03$ & $13.38^{+0.16}_{-0.18}$ & $0.05178^{+0.0006}_{-0.0005}$ & $0.187$ & $0.22\pm0.03$ & $0.85^{+0.03}_{-0.02}$ \\
    K2-287~b & $14.89329(3)$ & $0.48\pm0.03$ & $23.9^{+0.3}_{-0.3}$ & $0.0801^{+0.0010}_{-0.0009}$ & $0.779$ & $0.32\pm0.03$ & $0.833\pm0.013$ \\
    KELT-3~b & $2.7033904(10)$ & $0$ & $6.0\pm0.2$ & $0.0939\pm0.0011$ & $0.606^{+0.037}_{-0.046}$ & $1.47\pm0.07$ & $1.35 \pm 0.07$ \\
    KELT-4Ab & $2.989594(5)$ & $0.03^{+0.03}_{-0.02}$ & $5.8\pm0.2$ & $0.1089\pm0.0005$ & $0.689^{+0.011}_{-0.012}$ & $0.90\pm0.06$ & $1.70\pm0.05$ \\
    LTT~1445Ab & $5.358766(4)$ & $<0.110$ & $30.9^{+1.1}_{-1.2}$ & $0.0451^{+0.0014}_{-0.0013}$ & $0.17^{+0.15}_{-0.12}$ & $0.0090\pm0.0008$ & $0.116^{+0.006}_{-0.005}$ \\
    TOI-451Ab & $1.85870(3)$ & $0$ & $6.93^{+0.11}_{-0.16}$ & $0.0199^{+0.0010}_{-0.0011}$ &$0.22^{+0.20}_{-0.15}$ & \ldots & $0.170\pm0.011$ \\
    TOI-813~b & $83.891\pm0.003$ & \ldots & $47.2^{+2.1}_{-3.8}$ & $0.03165^{+0.00072}_{-0.00061}$ & $0.30^{+0.18}_{-0.19}$ & \ldots & $0.60\pm0.03$ \\
    TOI-892~b & $10.62656(7)$ & $<0.125$ \tnote{a} & $14.2^{+0.8}_{-0.7}$ & $0.0790 \pm 0.0010$ & $0.43^{+0.09}_{-0.13}$ & $0.95\pm0.07$ & $1.07\pm0.02$ \\
    TOI-1130~c & $8.35038(3)$ & $0.047^{+0.040}_{-0.027}$ & $22.21^{+0.50}_{-0.43}$ & $0.218^{+0.037}_{-0.029}$ & $0.995^{+0.046}_{-0.043}$ & $0.97\pm0.04$ & $1.5^{+0.3}_{-0.2}$ \\
    WASP-50~b & $1.955096(5)$ & $0.009^{+0.011}_{-0.006}$ & $7.51^{+0.17}_{-0.15}$ & $0.1404\pm0.0013$ & $0.687^{+0.014}_{-0.016}$ & $1.47\pm0.09$ & $1.15\pm0.05$ \\
    WASP-59~b & $7.919585(10)$ & $0.100\pm0.042$ & $24\pm2$ & $0.130\pm0.006$ & $0.29\pm0.18$ & $0.863\pm0.045$ & $0.775\pm0.068$ \\
    WASP-136~b & $5.215357(6)$ & $0$ & $6.87\pm0.12$ & $0.064\pm0.012$ & $0.59^{+0.08}_{-0.14}$ & $1.51\pm0.08$ & $1.38 \pm 0.16$ \\
    WASP-148~b & $8.80381(4)$ & $0.220 \pm 0.063$ & $19.8 \pm 1.5$ & $0.0807 \pm 0.0007$ & $0.05 \pm 0.07$ & $0.29\pm0.03$ & $0.72\pm0.06$ \\
    WASP-172~b & $5.477433(7)$ & $0$ & $8.0\pm0.5$ & $0.085\pm0.014$ & $0.45\pm0.12$ & $0.47\pm0.10$ & $1.57\pm0.10$ \\
    WASP-173Ab & $1.3866532(3)$ & $0$ & $4.78\pm0.17$ & $0.111\pm0.014$ & $0.40\pm0.08$ & $3.69 \pm 0.18$ & $1.20 \pm 0.06$ \\
    WASP-186~b & $5.026799(13)$ & $0.33\pm0.01$ & $8.78\pm0.18$ & $0.078\pm0.017$ & $0.84^{+0.01}_{-0.02}$ & $4.22\pm0.18$ & $1.11\pm0.03$ \\
    XO-7~b & $2.864142(4)$ & $0.04 \pm 0.03$ & $6.43\pm0.14$ & $0.0953\pm0.0009$ & $0.71\pm0.02$ & $0.71\pm0.03$ & $1.37\pm0.03$ \\
    \bottomrule
    \end{tabular}
    \begin{tablenotes}
        \item In the period column, the
        number in parentheses is the
        uncertainty in the last digit of
        the period.
        \item Sources are given in \tabref{tab:stars} for the corresponding host.
    \item[(a)] With 98\% confidence.
   \end{tablenotes}
   \end{threeparttable}
\end{table*}


\section{Data analysis}\label{sec:analysis}


\subsection{Photometric data}\label{sec:phot_analysis}

The photometric data provide tight constraints on the orbital period ($P$), the
time of conjunction of a transit chosen
to be the reference epoch,
($T_0$), the planet-to-star radius ratio ($R_{\rm p}/R_\star$), the cosine of the orbital inclination ($\cos i$), and the
ratio of the semi-major axis and stellar
radius ($a/R_\star$). Uniform prior
probability distributions were adopted
for these parameters. 

To model the stellar limb darkening
function, we used a standard quadratic law and allowed
the sum of the two coefficients (the center-to-limb intensity ratio) to be an adjustable
parameter, while holding the difference
fixed at the value obtained from the tables of \citet{Claret2013,Claret2018} 
as appropriate for the photometric bandpass and the star's effective temperature ($T_{\rm eff}$), surface gravity ($\log g$), and metallicity ($[\rm Fe/H]$).
See \tabref{tab:LD} and \tabref{tab:wasp148_phot} for details. 
The sum of coefficients was subjected
to a Gaussian prior with a width of $0.1$
and a central value from the aforementioned tables.

In cases for which a nonzero orbital eccentricity has been reported (HD~118203~b, K2-261~b, K2-287~b, WASP-148~b, and WASP-186~b), we applied Gaussian priors on $e$ and the argument of periastron ($\omega$) using the values given in \tabref{tab:planets}. For the multis, we applied Gaussian priors on the parameters of all planets apart from the
transiting planet that was the target
of our observations.
The main effect of these planets
on our model is to change the
slope of the RV time series on the night
of the transit that was observed
spectroscopically. When two planets
are transiting simultaneously, the shape of the transit light curve is also
affected. Simultaneous transits occasionally occurred (for TOI-1130 and TOI-451A) within the long intervals
spanned by the photometric data,
but did not occur during any of our 19
spectroscopic observations.

\begin{table*}[]
    \centering
    \begin{threeparttable}        
    \caption{Key literature stellar parameters.}
    \label{tab:stars}
    \begin{tabular}{l c c c c c c c}
    \toprule
        Star & $T_{\rm eff}$ & $\log g$ & $[\rm Fe/H]$ & $M_\star$ & $R_\star$ & Age & Source \\
         & (K) & (cgs; dex) & (dex) & (M$_\odot$) & (R$_\odot$) & (Gyr) & \\
         \midrule
         HD 118203 & $5683 \pm 85$ & $3.889 \pm 0.018$ & $0.223\pm0.076$ & $1.13^{+0.05}_{-0.06}$ & $2.10\pm0.05$ & $5.32^{+0.96}_{-0.73}$ & 1 \\
         HD~148193 & $6198 \pm 100$ & $4.18 \pm 0.10$ & $-0.13 \pm 0.06$ & $1.23^{+0.02}_{-0.05}$ & $1.63^{+0.03}_{-0.02}$ & $3.5^{+1.3}_{-0.5}$ & 2 \\
         K2-261 & $5537\pm71$ & $4.21\pm0.11$ & $0.36\pm0.06$ & $1.105 \pm 0.019$ & $1.669 \pm 0.022$ & $8.8^{+0.4}_{-0.3}$ & 3 \\
         K2-287 & $5695 \pm 58$ & $4.398\pm0.015$ & $0.20\pm0.04$ & $1.06\pm0.02$ & $1.070\pm0.010$ & $4.5\pm1$ & 4 \\
         KELT-3 & $6306 \pm 50$ & $4.21\pm0.03$ & $0.04\pm0.08$ & $1.28\pm0.06$ & $1.47\pm0.07$ & $3\pm0.2$ & 5 \\
         KELT-4A & $6207\pm75$ & $4.11 \pm 0.03$ & $ 0.12 \pm 0.07$ & $1.20^{+0.07}_{-0.06}$ & $1.60\pm0.04$ & $4.44^{+0.78}_{-0.89}$ & 6 \\
         LTT~1445A & $3340\pm150$ & $5.0$ & $-0.34\pm0.09$ & $0.257\pm0.014$ & $0.265^{+0.011}_{-0.010}$ & \ldots & 7 \\ 
         TOI-451A & $5550\pm56$ & $4.53$ & $0.0$ & $0.95\pm0.02$ & $0.88\pm0.03$ & $0.125\pm0.008$ & 8 \\
         TOI-813 & $5907 \pm 150$ & $3.86\pm0.14$ & $0.10\pm0.10$ & $1.32\pm0.06$ & $1.94 \pm 0.10$ & $3.73\pm0.62$ & 9 \\
         TOI-892 & $6261\pm80$ & $4.26\pm0.02$ & $0.24 \pm 0.05$ & $1.28\pm0.03$ & $1.39\pm0.02$ & $2.2\pm0.5$ & 10 \\
         TOI-1130 & $4250\pm67$ & $4.60\pm0.02$ & $>0.2$ & $0.684^{+0.016}_{-0.017}$ & $0.687\pm0.015$ & $8.2^{+3.8}_{-4.9}$ & 11 \\
         WASP-50 & $5400 \pm 100$ & $4.5 \pm 0.1$ & $-0.12 \pm 0.08$ & $0.89^{+0.08}_{-0.07}$ & $0.84\pm0.03$ & $7\pm3.5$ & 12 \\
         WASP-59 & $4650 \pm 150$ & $4.55 \pm 0.15$ & $-0.15 \pm 0.11$ & $0.72\pm0.04$ & $0.61\pm0.04$ & $0.5^{+0.7}_{-0.4}$ & 13 \\
         WASP-136 & $6250 \pm 100$ & $3.9 \pm 0.1$ & $-0.18\pm0.10$ & $1.41\pm0.07$ & $2.2\pm0.2$ & $3.62\pm0.70$ & 14 \\
         WASP-148 & $5460\pm130$ & $4.40\pm0.15$ & $0.11\pm0.08$ & $1.00\pm0.08$ & $1.03\pm0.20$ & \ldots & 15 \\
        WASP-172 & $6900 \pm 150$ & $4.1 \pm 0.2$ & $-0.10 \pm 0.08$ & $1.49\pm0.07$ & $1.91\pm0.10$ & $1.79\pm0.28$ & 16 \\
        WASP-173A & $5700 \pm 150$ & $4.5 \pm 0.2$ & $0.16 \pm 0.14$ & $1.05 \pm 0.08$ & $1.11\pm0.05$ & $6.78\pm2.93$ & 17 \\
        WASP-186 & $6300\pm100$ & $4.1\pm0.2$ & $-0.08 \pm 0.14$ & $1.21^{+0.07}_{-0.08}$ & $1.46\pm0.02$ & $3.1^{+1.0}_{-0.8}$ & 18 \\
        XO-7 & $6250 \pm 100$ & $4.25 \pm 0.02$ & $0.43\pm0.06$ & $1.405 \pm 0.059$ & $1.480\pm0.022$ & $1.18^{+0.98}_{-0.71}$ & 19 \\
         \bottomrule
    \end{tabular}
    \begin{tablenotes}
        \item The sources refer to the following references: 1 \citet{Pepper2020}, 2 \citet{Chontos2024}, 3 \citet{Johnson2018}, 4 \citet{Jordan2019}, 5 \citet{Pepper2013}, 6 \citet{Eastman2016}, 7 \citet{Winters2022}, 8 \citet{Newton2021}, 9 \citet{Eisner2020}, 10 \citet{Brahm2020}, 11 \citet{Huang2020}, 12 \citet{Gillon2011}, 13 \citet{Hebrard2013}, 14 \citet{Lam2017}, 15 \citet{Hebrard2020}, 16 \citet{Hellier2019}, 17 \citet{Hellier2019}, 18 \citet{Schanche2020}, 19 \citet{Crouzet2020}.
    \end{tablenotes}
    \end{threeparttable}
\end{table*}

We modeled the light curves 
using the \texttt{batman} package \citep{Kreidberg2015}, which is
based on the equations of \citet{MandelAgol2002}.
In addition to accounting for the
transit and white noise in the measurements,
we included a Gaussian process (GP) using \texttt{celerite} \citep{celerite}. We generally employed a Matérn-3/2 kernel,
which is characterised by two hyperparameters: the amplitude ($A$) and the timescale ($\tau$). For the time series with 30-minute sampling, evenly spaced model light curves were created and integrated over to mimic a cadence of 2~min.

{\small\addtolength{\tabcolsep}{-1.5pt}
\begin{table}[]
    \centering
    \begin{threeparttable}        
    \caption{Limb-darkening coefficients and velocity fields.} 
    \label{tab:LD}
    \begin{tabular}{l c c c c c}
        \toprule
          & $q_1$/$q_2$ & $q_1$/$q_2$ & $q_1$/$q_2$ & $\zeta_\star$ & $\xi_\star$ \\
         & \tess & K2 & $V$ &  \multicolumn{2}{c}{(km~s$^{-1}$)}\\
         \midrule
         HD~118203 & 0.30/0.28 & \ldots & 0.31/0.20 & 4.11 & 1.00 \\
         HD~148193 & 0.23/0.31 & \ldots & 0.49/21 & 5.23 & 1.31 \\
         K2-261 & 0.41/0.22 & 0.54/0.15 & 0.58/0.16 & 2.20 & 1.30 \\
         K2-287 & 0.53/0.17\tnote{a} & 0.38/0.19 & 0.58/0.16 & 2.76 & 0.92 \\
         KELT-3 & 0.23/0.31 & \ldots & 0.47/0.23 & 5.45 & 1.39 \\
         KELT-4A & 0.23/0.31 & \ldots & 0.50/0.21 & 5.23 & 1.31 \\
         LTT~1445A & 0.17/0.44 & \ldots & 0.70/0.07 & 1 & 0.9 \\
        TOI-451A & 0.35/0.25 & \ldots & 0.57/0.16 & 2.60 & 0.95 \\
         TOI-813 & 0.26/0.30 & \ldots & 0.52/0.20 & 4.73 & 1.12 \\
         TOI-892 & 0.24/0.30 & \ldots & 0.44/0.24 & 5.17 & 1.35 \\ 
         TOI-1130 & 0.51/0.17 & \ldots & 0.49/0.25 & 3.59 & 0.90 \\
        WASP-50 & 0.35/0.25 & \ldots & 0.62/0.14 & 2.49 & 0.90 \\
        WASP-59 & 0.48/0.17 & 0.68/0.07\tnote{b} & 0.72/0.06 & 2.90 & 0.83 \\
         WASP-136 & 0.23/0.31 & \ldots & 0.45/0.21 & 5.89 & 1.35 \\
         WASP-148 & \ldots & \ldots & 0.62/0.14 & 2.75 & 0.92 \\
        WASP-172 & 0.18/0.33 & \ldots & 0.42/0.27 & 6.0 & 1.5 \\
        WASP-173A & 0.31/0.27 & \ldots & 0.54/0.18 & 3.28 & 1.04 \\ 
         WASP-186 & 0.23/0.31 & \ldots & 0.44/0.23 & 5.75 & 1.43 \\
         XO-7 & 0.25/0.32 & \ldots & 0.49/0.21 & 5.2 & 1.34 \\
         \bottomrule
     \end{tabular}
     \begin{tablenotes}
         \item The second, third, and fourth columns show the $q_1$ and $q_2$ quadratic limb darkening pairs we used as priors in the TESS, K2, and spectrograph bands. For the latter we simply assumed the $V$-filter. We obtained these using values for $T_{\rm eff}$, $\log g$, and $[\rm Fe/H]$ in tabulated in \tabref{tab:stars} and we queried the limb-darkening tables for the quadratic law by \citet[][for K2 and $V$]{Claret2013} and \citet[][for \tess]{Claret2018}. The macro- ($\zeta_\star$) and micro-turbulence ($\xi_\star$) priors were obtained from the relations in \citet{Doyle2014} and \citet{Bruntt2010}, respectively, again using the values from \tabref{tab:stars}, with the exception of WASP-172 and LTT~1445A (see \sref{sec:analysis}).
         \item[a] These values are for CHEOPS and not \tess. Queried from \citet{Claret2021}.
         \item[b] These values are for the $r^\prime$ passband and not K2. Queried from \citet{Claret2012}.
     \end{tablenotes}
    \end{threeparttable}
    
\end{table}
}

\subsection{Stellar rotation}

The analysis of the RM effect is aided by knowledge of the stellar rotation velocity or
period. We tried to use the available spectroscopic and photometric data to 
measure these quantities independently
of the transit data, as describe below.


\subsubsection{Projected rotation velocity}\label{sec:vsini}

The amplitude of the RM effect scales with $v\sin i_\star$, the product of the stellar rotation speed and the sine of the line-of-sight inclination of the stellar spin axis. 
Therefore, knowing \vsini helps to
plan observations and
analyze the results.
When the RM effect is detected with a high
signal-to-noise ratio (SNR), \vsini can
be determined precisely from the RM data.
Indeed, for stars with a high obliquity,
the effects of differential rotation
might be detectable 
\citep{GaudiWinn2007, Cegla2016}.

{\small\addtolength{\tabcolsep}{-3.1pt}
\begin{table}
    \centering
\begin{threeparttable}
\caption{Stellar rotation parameters.}
    \label{tab:rotation}
    \begin{tabular}{l c c c c}
       \toprule
      & \multicolumn{2}{c}{$v \sin i_\star$} & \multicolumn{2}{c}{$P_{\rm rot}$} \\
     & \multicolumn{2}{c}{(km~s$^{-1}$)} & \multicolumn{2}{c}{(d)} \\
    Star & Literature & This work & Literature & This work \\
     \midrule
     HD~118203 & $4.7$ & $4.9\pm0.4$ & \ldots & $6.1\pm0.4$ \\
     HD~148193 & $5.9\pm1.0$ & $6.27\pm0.02$ & \ldots & \ldots \\
     K2-261 & $3.70 \pm 0.17 $ & $1.59 \pm 0.17$ & \ldots & \ldots \\
     K2-287 & $3.2 \pm 0.2$ & $1.99\pm0.02$ & \ldots & \ldots \\
     KELT-3 & $10.2\pm0.5$ & $7.59\pm0.16$ & \ldots & \ldots \\
     KELT-4A & $6.2\pm1.2$ & $6.0\pm0.4$ & \ldots & \ldots \\
     LTT~1445A & $<3.4$ & $1.41\pm0.11$ & \ldots & \ldots \\
     TOI-451A & $7.9 \pm 0.5$ & $8.16\pm0.02$ & $5.1\pm0.1$ & $5.3\pm0.5$ \\
     TOI-813 & $8.2\pm0.9$ & $8.33\pm0.02$ & \ldots & \ldots \\ 
     TOI-892 & $7.69 \pm 0.5$ & $6.2\pm0.5$ & \ldots & \ldots \\ 
     TOI-1130 & $4.0 \pm 0.5$ & $1.28\pm0.05$ & \ldots & \ldots \\
     WASP-50 & $2.6\pm0.5$ & $1.40\pm0.12$ & $16.3\pm0.5$ & \ldots \\
     WASP-59 & $2.3\pm1.2$ & \ldots & \ldots & \ldots \\
     WASP-136 & $13.1\pm0.8$ & $12.06\pm0.04$ & \ldots & $5.4\pm0.5$ \\
     WASP-148 & $3 \pm 2 $ & $2.0\pm0.4$ & $26.2\pm1.3$ & $26 \pm 4$ \\
    WASP-172 & $13.7\pm1.0$ & $13.33\pm0.05$ & \ldots & $7.0 \pm 0.7$ \\
    WASP-173A & $6.1 \pm 0.3$ & $6.62\pm0.06$ & $7.9\pm0.1$ & $8.4\pm1.1$ \\
    WASP-186 & $15.6\pm0.9$ & $14.8\pm0.2$ & \ldots & $4.8\pm1.9$ \\
    XO-7 & $6\pm1$ & $5.2\pm0.3$ & \ldots & $3.62\pm0.12$ \\
     \bottomrule
    \end{tabular}
     \begin{tablenotes}
         \item Measurements of \vsini and $P_\mathrm{rot}$ from the literature (see \tabref{tab:stars} for citations) and those derived in this work. The uncertainties reported for \vsini generally refer to statistical
         uncertainties or the scatter in results of analyzing different spectra. Because they do not include systematic
         uncertainties, we imposed a minimum
         uncertainty of $0.5$~km~s$^{-1}$
         in our analysis.
     \end{tablenotes}
    \end{threeparttable}

\end{table}
}

In more typical cases in which the RM effect
is detected with a modest SNR,
it is useful to determine \vsini by modeling the spectral line broadening
outside of transits, and thereby gain the option of applying a prior constraint
on \vsini when modeling the RM effect.
For this purpose, we used the
broadening function (BF) method of \citet{Rucinski1999} as implemented in \texttt{FIESpipe}, which is based
on an appropriate model atmosphere from \citet{Castelli2003}. The BFs were created from high SNR orders from out-of-transit spectra, where we visually inspected the result to ensure that the BF created from a given order was well-behaved. This typically meant selecting central orders typically with wavelengths $\sim$5000-6000~\AA. Orders containing telluric lines were excluded. For a given epoch we stacked the BFs from the individual orders to create the highest possible SNR BF for that epoch. We applied
this method not only to the FIES
spectra but also the ESPRESSO and HARPS-N spectra. To extract \vsini we fitted the theoretical rotational BF from \citet[][Eq. (2)]{Kaluzny2006}. The model includes the
rotational broadening parameters along with a parameter representing non-rotational broadening (in most cases dominated by the finite
instrumental resolution).

\tabref{tab:rotation} gives the results.
The tabulated \vsini is the median
of the best-fit results of analyzing spectra from
different epochs, and the tabulated uncertainty
is the standard deviation between those epochs, which in most cases are unrealistically small. This is especially true for observations where we only have few out-of-transit observations. For systems where we only had two out-of-transit spectra, we used the first ingress as a third spectrum.
Our measurements are generally in agreement with previously published
results.
The most significant deviations are seen for the most slowly-rotating stars,
where the details of extracting line
profiles and modeling
non-rotational broadening are most
important.
When modeling the RM effect,
we imposed Gaussian priors
based on our \vsini determinations,
with a minimum width
of $\sigma=0.5$~km~s$^{-1}$ to account for possible systematic errors. 

To test if our choice of the width of the uncertainty interval in the \vsini prior does significantly affect our results for the projected obliquity, we selected four systems in our sample for which we expect the result of lambda to be most sensitive to \vsini (low impact parameter, incomplete transit coverage) and performed the same analysis as described below. However, now with an uncertainty in \vsini of 1.0~km~s$^{-1}$. In all four cases we find that the highest probability value for $\lambda$ changes only by a small fraction of its uncertainty interval, relative to the 0.5~km~s$^{-1}$ case.


\subsubsection{Photometric rotation period}\label{sec:prot}

We attempted to measure the rotation period ($P_\mathrm{rot}$) by seeking
periodicities in the 
K2 and \tess light curves.
The combination of the rotation period, \vsini, and $R_\star$
can be used to constrain
the stellar inclination \citep{Masuda2020}, which in turn can be used in combination 
with $i_{\rm orb}$ and $\lambda$
to determine the ``true'' or three-dimensional obliquity.

We used the method based on the auto-correlation function (ACF) described
by \citet{McQuillan2014}. After removing the data affected
by transits, we searched for peaks in the ACF as a function of lag. When multiple peaks
were seen, we fitted a linear function between peak number and lag,
the slope of which is an estimate
of the rotation period \citep[e.g.,][]{Hjorth2021}.

\tabref{tab:rotation} gives the results. Empty entries mark
the cases for which no period
could be determined.
For two systems, TOI-813 and TOI-892, peaks were seen in the ACF  (corresponding to periods of
$10.1\pm0.6$~d and $2.76\pm0.11$~d, respectively), but because the statistical significance was weak,
the results do not appear in
\tabref{tab:rotation}.
For WASP-50, a series of strong peaks
seemed to imply a period of
$7.2\pm1.8$~d, which contrasts
with the value
of $16.3\pm0.5$~d reported by \citet{Gillon2011}. They also
noted that the true rotation
period could be 32.6 days if
there are similar spot patterns
on opposite stellar hemispheres.
We folded the light curve using
trial periods of $16.3$~d or $32.6$~d but neither choice
seemed compelling.
Evidently, the variability
of WASP-50 is complex,
which is why \tabref{tab:rotation}
contains no entry for WASP-50.

We note that the estimated rotation periods of WASP-136 and WASP-186, $5.4\pm0.5$~d and $4.8\pm1.9$~d, are both close to
the orbital period of the transiting planets.
These systems might have been driven
into spin-orbit synchronization
by tidal interactions between the planet and star \citep[e.g.,][]{Albrecht2012,Brown2014,Penev2018}.

\begin{figure*}[h]
    \centering
    \includegraphics[width=\textwidth]{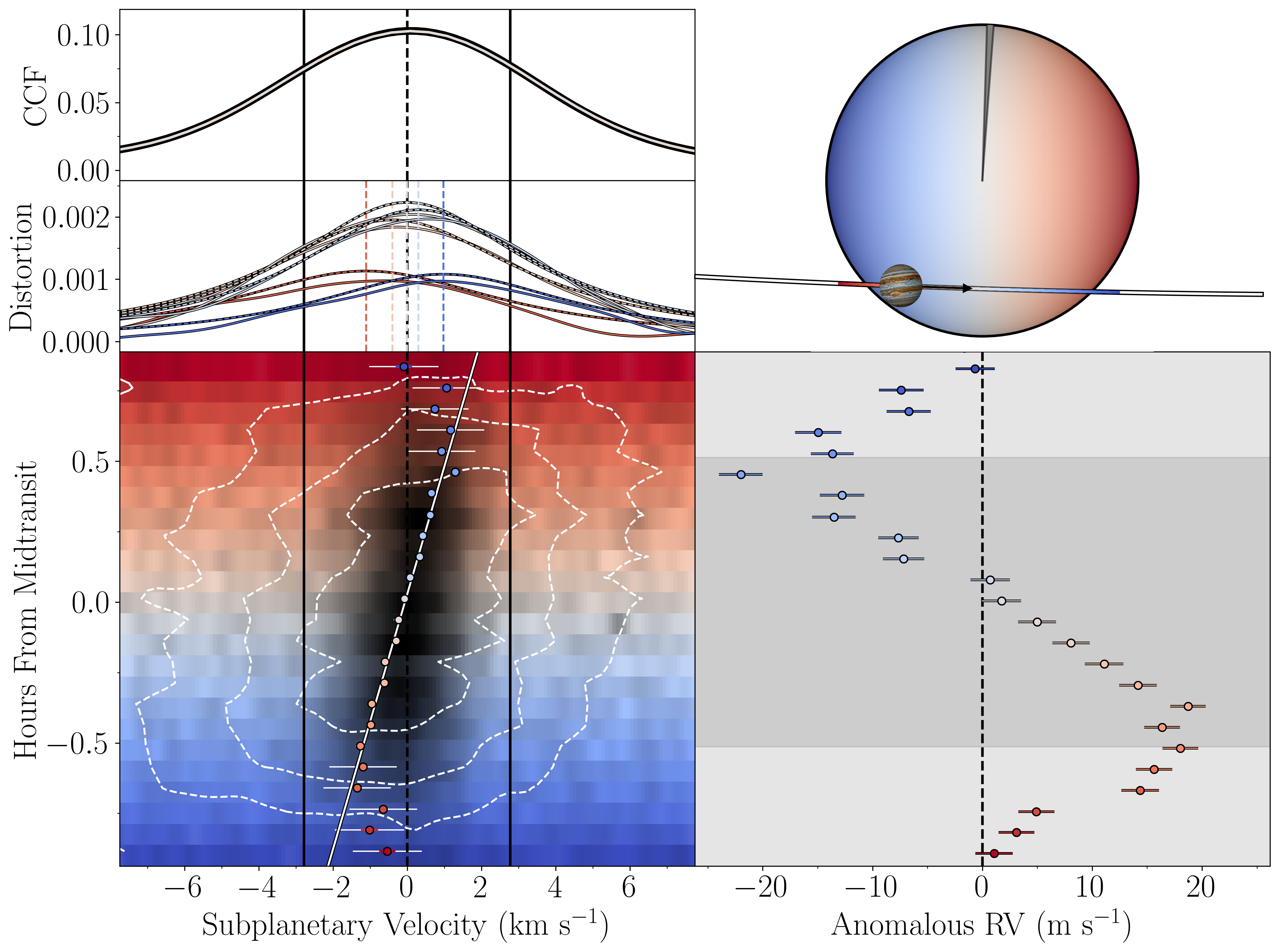}
    \caption{{\bf The Rossiter-McLaughlin effect}
    displayed in four ways, for the illustrative
    case of WASP-50. 
    {\it Top right:} Schematic of the planet transiting the stellar disk. The wedge represents the best-fit $\lambda$ and its uncertainty. The color scheme is based
    on the star's rotational Doppler shift;
    the same color scheme is used in the
    top left and lower left panels
    to convey the location of the planet
    on the star.
    {\it Top left: Spectral lines.}
    The thick black curve is the average out-of-transit cross-correlation function (CCF)
    and the superimposed white curve is
    a CCF observed near mid-transit.
    {\it Middle left: Spectral line
    distortions.} Shown are the results
    of subtracting five of the in-transit CCFs from the mean out-of-transit CCF.
    Solid curves show the data.
    Dashed curves are the best-fit
    Lorentzian profiles, with vertical
    lines marking their centers. The color coding is done according to the apparent velocity anomaly going from redshifted (in red) to blueshifted (in blue).
    {\it Bottom left: The Doppler shadow.}
    Each row shows the spectral
    line distortion at a particular time.
    The white contours are amplitude levels of the distortion.
    Each small circle marks the best-fit subplanetary velocity (the Doppler shift of
    the portion of the star directly behind
    the planet's center), with colors that convey
    the corresponding anomalous radial velocity
    shown in the lower right panel.
    Dark vertical lines mark
    the measured \vsini, and the diagonal white line is the best-fit trajectory.
    {\it Bottom right: Anomalous radial velocities.} The dark shaded area is the
    range of times when the planet's
    silhouette is completely contained
    within the stellar disk.}
    \label{fig:slope_shadow}
\end{figure*}


\subsection{The Rossiter-McLaughlin effect}\label{sec:rm}

The RM effect is the distortion of a star's
absorption lines that arises from the combination of stellar rotation and a transiting planet.
During a transit, the planet prevents
the light from a portion of the photosphere
from contributing to the disk-integrated spectral line. The missing spectral component causes the
velocity profile of the spectral lines to have a slightly reduced flux at velocities centered on
the rotational Doppler shift of the star at the
``subplanetary'' point, i.e., the point on the
the star directly behind the center of the planet.
Observations of the RM effect can therefore be used to determine the time series of
subplanetary velocities, or equivalently, the
trajectory of the transiting planet relative
to the stellar rotation axis.

Neglecting differential rotation,
the RM effect is mainly a function of
$\lambda$, \vsini, $R_{\rm p}/R_\star$,
and the transit
impact parameter $b \equiv d/R_\star \cos i$, where $d$ is the star-planet
distance at the time of conjunction.
For precise modeling it is also necessary
to take into account limb darkening
and non-rotational spectral line broadening.
To model non-rotational broadening due
to the star,
we used the \cite{Gray2005} model
with parameters for the velocity spread generated
by microturbulence ($\xi_\star$) and macroturbulence ($\zeta_\star$).
We included a separate term for
instrumental broadening, another source of non-rotational broadening.
We neglected the effects of differential
rotation and the convective blueshift,
which for solar-type stars
are not expected to produce detectable
effects given the quality of our data
(and indeed our modeling did not
uncover any evidence
for those effects, with one
possible exception; see \sref{sec:cb}).
We did not model the effects of
starspots, flares, or pulsations, since
there was no evidence for such phenomena
in our data.

The RM effect has been analyzed as a distortion of individual spectral lines \citep[e.g.,][]{Albrecht2007}, 
as distortion of the spectral
cross-correlation function
\citep[e.g.,][]{Cegla2016},
or as an ``anomalous RV''
obtained by an ordinary radial-velocity extraction
code to a spectrum affected by the RM effect \citep[e.g.,][]{Queloz2000}. 
We used a combination of these approaches, depending on the system, as described below.

We fitted the spectral CCFs with a
parameterized model that incorporates
the RM effect, using a code first described by \citet{Knudstrup2022a} and \citet{Knudstrup2023}. In this model, the stellar disk is discretized (with a radius of 100 pixels)
and a spectrum is assigned to each pixel
according to the model parameters
for limb darkening, stellar rotation,
and non-rotational broadening.
A disk-integrated spectral line is created
by summing the spectra of all pixels
that are not concealed by the planet,
and the disk-integrated line profile
is compared to the observed spectral CCF.

An alternative is to 
fit the time series of
subplanetary velocities derived
from the residuals between the out-of-transit CCF and the in-transit CCFs.
The subplanetary velocity is calculated as \begin{equation} \label{eq:v_p}
v_p(t)   = v\sin i_\star x(t) \, ,
\end{equation}
where $x(t)$ is the distance from
the projected rotation axis divided
by the stellar radius. Once the orbital
parameters are chosen, $x(t)$ can be calculated. It is also
instructive to calculate $x$ at
the times of ingress ($x_1$) and egress ($x_2$) \citep{Albrecht2011}:  
\begin{equation} \label{eq:slope}
\begin{split}
x_1  & = \sqrt{1-b^2} \cos \lambda - b \sin \lambda \, , \\ 
x_2  & = \sqrt{1-b^2} \cos \lambda + b \sin \lambda \, .
\end{split}
\end{equation}
Therefore, the very nearly linear
trajectory of the planet
can be calculated as a function
of time for given values
of $t$, $T_{\rm c}$, the transit
duration, $b$, and $\lambda$. 

The simplest and most common approach is to fit a time series of anomalous RVs. In the model, the anomalous RV is computed as a function of $v_p$ by taking into
account the size of the planet's shadow, the limb darkening function,
and the response of the
radial-velocity extraction code to
the RM distortion; for this
purpose we used the code
described by \citet{Hirano2011}. 

The different ways to sense the RM effect are illustrated in \figref{fig:slope_shadow} for the case of WASP-50.
For all 19 systems in our sample,
we fitted the time series
of anomalous RVs (the ``RV-RM'' method).
For some systems, we also analyzed the distortions of the CCF,
or fitted the time series
of subplanetary velocities. Multiple
methods were used to test for consistency in the results, and
because for some systems it seemed
plausible that the more sophisticated
models would give more information
about $\lambda$. In general, we expect
line-profile modeling to be advantageous when $(R_{\rm p}/R_\star)^2 v\sin i_\star$
is larger than the non-rotational
broadening \citep{Albrecht2022}. 

We determined the best-fit parameters
and their uncertainties via
Markov Chain Monte Carlo (MCMC) sampling of the posterior
probability distribution in parameter space. We used the \texttt{emcee} package \citep{emcee}. To ensure convergence we both inspected the chains visually and calculated the rank normalized R-hat\footnote{\url{https://python.arviz.org/en/latest/api/generated/arviz.rhat.html}}. As mentioned above, we always fitted the 
the transit light curve
alongside the spectroscopic data.

When fitting the RVs and the line distortion directly, we used Gaussian priors for the macroturbulence and microturbulence parameters, and we adopted a fixed value for the instrumental broadening parameter of  each spectrograph (see \tabref{tab:spectro}). The mean values for our Gaussian priors were obtained from the relationships presented by \citet{Doyle2014} for stars with $T_\mathrm{eff} \in \mathclose[5200~\mathrm{K},6400~\mathrm{K}\mathclose]$ and $\log g \in \mathclose[4.0,4.6\mathclose]$, and \citet{Bruntt2010} for stars with $T_\mathrm{eff} \in \mathclose[5000~\mathrm{K},6500~\mathrm{K}\mathclose]$ and $\log g>4.0$. The width of the prior in all cases was 1~km~s$^{-1}$. 

We note that the properties of WASP-136, TOI-813, and HD~118203 are on the borderline of the range
of properties for which the
aforementioned relationships involving turbulent broadening are applicable;
we applied them anyways.
On the other hand, the properties of WASP-172 and LTT~1445A are far outside the applicable ranges.
For WASP-172 ($T_\mathrm{eff}=6900\pm150$~K), we expect a high macroturbulent velocity.
We chose $6.0$~km~s$^{-1}$ and increased the width of the prior to $4.0$~km~s$^{-1}$ to reflect our uncertainty. For the microturbulent velocity, we adopted $\xi_\star=1.5\pm1.5$~km~s$^{-1}$. In the other end of the spectrum, we expect the turbulent velocities of the M dwarf LTT~1445A to be smaller.
We chose $\zeta=1.0\pm1.0$~km~s$^{-1}$ and $\xi_\star=1.0\pm1.0$~km~s$^{-1}$. All of the priors are summarized in \tabref{tab:LD}. 

For each system, we performed two
RV-RM analyses, one in which we adopted a uniform prior on \vsini, and one in which we applied a Gaussian prior on \vsini based on the observed
line broadening of the out-of-transit spectrum (\tabref{tab:rotation}) and
a width of $0.5$~km~s$^{-1}$. When analyzing the line profiles or Doppler shadow, we always applied the Gaussian 
prior on \vsini.

When fitting the time series of
subplanetary velocities, we did not simultaneously fit the light curve.
Instead, we applied Gaussian priors on $P$, $T_0$, $R_{\rm p}/R_\star$, $i$, and $a/R_\star$ based on
an external fit to the light curve,
and 
adopted uniform priors on $\lambda$ and \vsini.

\begin{figure*}[h]
    \centering
    \begin{tabular}{c c}
         \includegraphics[width=\columnwidth]{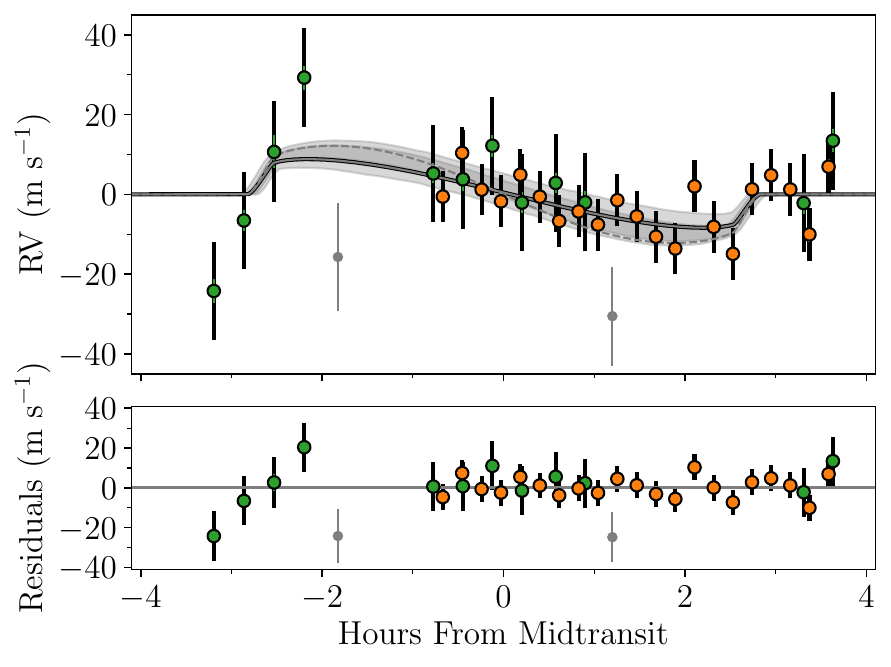} & \includegraphics[width=\columnwidth]{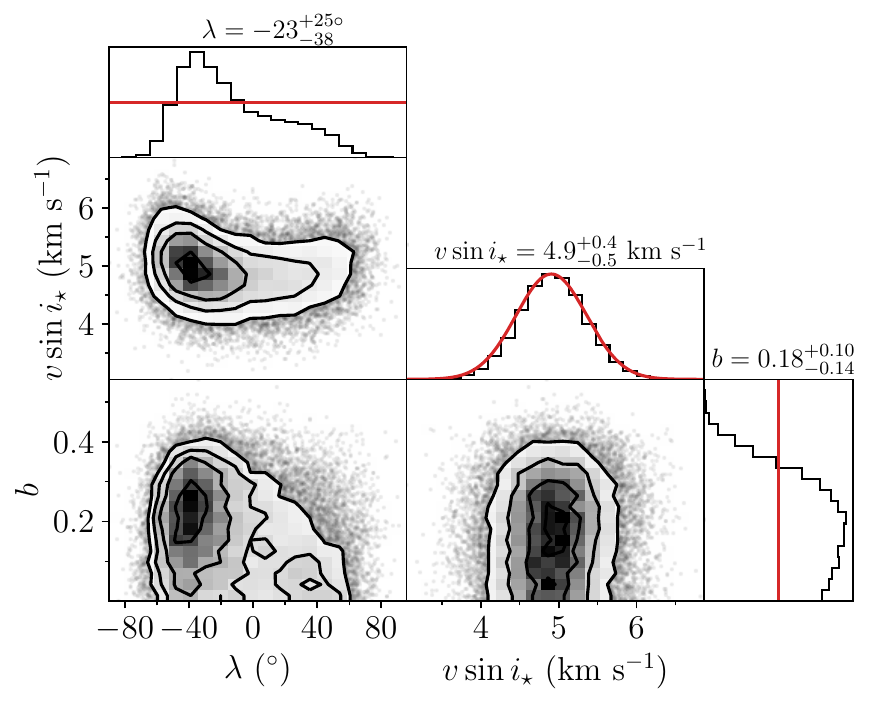} \\ 
    \end{tabular}
    \caption{{\bf The RM effect observed in HD~118203.} {\it Left:} Time series and residuals of the FIES+ RVs (orange) and FIES RVs (green). The black curve is the best-fit model
    and the gray shading shows the 1 and 2$\sigma$ confidence intervals created by randomly drawing values from the MCMC chains for $b$, \vsini, $R_{\rm p}/R_\star$, and $\lambda$.
    The dashed curve is
    the best-fit model assuming
    $\lambda=0^\circ$.
    Gray circles are the data that were obtained during poor
    atmospheric conditions
    and were excluded from the analysis. {\it Right:} Correlations between $\lambda$, \vsini, and $b$. The red lines illustrate the prior that was applied (either uniform or Gaussian).}
    \label{fig:rm_hd118203}
\end{figure*}

In some cases, to boost the SNR, we stacked the CCFs from different exposures in a way similar to \citet{Johnson2014}, but with a code
based more directly on the work of \citet{Hjorth2019}. In these cases, we extracted the Doppler shadow as shown in \figref{fig:slope_shadow}, and then transformed the velocities by correcting for the slope in \eref{eq:slope}. The white line/slope in \figref{fig:slope_shadow} is in this way used to bring each pixel directly beneath it in the Doppler shadow to be located at $0$~km~s$^{-1}$. We then collapse the slope corrected shadow by summing along the ordinate (time stamps) for a given geometry ($\lambda$,\vsini,$b$). When the model parameters match
reality, we expect to 
see a sharp peak at $0$~km~s$^{-1}$ in the summed spectrum, and we found the height of the peak by fitting a Gaussian to it. \citet{Hjorth2019} did this for a dense grid in the parameter space
of $b$, \vsini, and $\lambda$. For our purposes,
since $b$ was always tightly constrained
by the light curve,
we neglected the uncertainty in $b$
and simply adopted the best-fit value from the RV-RM method. We then searched the 2-d parameter space
of \vsini and $\lambda$ for the
strongest peak; in practice this was done by fitting a 2-d Gaussian
function. We did not run an MCMC when stacking the CCFs, nor did we fit the light curve simultaneously. The quoted uncertainties are taken as the widths ($\sigma$s) of the fitted 2D Gaussian and as such do not reflect the actual precision. The resulting values therefore only serves as an indication to whether the results in ($\lambda$,\vsini) from the RV-RM runs are in (qualitative) agreement with that seen in the CCFs.


\section{Analysis of the individual systems}\label{sec:targets}

Below, we describe the
particulars of each system, our analysis,
and the results.
The details for each system are
also summarized in
\tabref{tab:spec_obs},
\tabref{tab:stars}, and \tabref{tab:planets}.


\subsection{HD~118203}\label{obs:hd}

HD~118203~b was discovered by \citet{daSilva2006} through RV
monitoring, and was later found to be a transiting planet by \citet{Pepper2020} using \tess data. \tess data are available for HD~118203 from Sectors 15, 16, 22, and 49 with a 2-minute cadence. The planet is a hot Jupiter on an eccentric orbit around a slightly evolved G-type star \citep[][]{Pepper2020}. The \tess data centered on the mid-transit time are shown in \figref{fig:gridone}.

We observed two transits of HD~118203~b with FIES. The first
transit began on March 24, 2020 at around UT 21:00 and ended at 04:15. The observations were interrupted twice due to high humidity. The exposure time was 900~s
and the time between
exposures was 1171~s,
with the time in between spent
on ThAr calibration.
Two exposures on this night were heavily affected by weather
and the resulting RVs deviate significantly from the rest
(see the gray points in \figref{fig:rm_hd118203}). These two data points were omitted from the analysis.

Since the first transit had been interrupted by the weather, we
organized observations of a second transit on the night starting on June 18, 2022. Unfortunately, the first part of the transit was missed
because the telescope was being
used to observe a target-of-opportunity. Our transit observations started at around UT 22:30 and continued until around 03:00. As the FIES spectrograph was refurbished in the summer of 2021 resulting in an improved throughput, we opted for a shorter exposure time of 600~s for the second transit, which resulted in a sampling time of around 768~s. We note that the refurbishment introduced an offset in RV between observations. As such we allowed the systemic velocity
parameter to differ between the
two nights: $\gamma_{\rm FIES}$ applies
to the observations taken before summer 2021, and $\gamma_{\rm FIES+}$ applies to subsequent observations.

As mentioned above (\sref{sec:analysis}) we performed two MCMC analyses, one with a uniform prior on \vsini and one with a Gaussian prior. For this system,
the results differed substantially.
The uniform prior in \vsini yielded
$\lambda=$~\hdu$^\circ$ and the Gaussian prior yielded \hdg$^\circ$. The data obtained from the first transit observations are obviously very noisy. This paired with the second transit observations only covering the second half of the transit meant that the amplitude of the signal was allowed to vary significantly, which made \vsini wander off to large values when applying applying a uniform prior. In this case \vsini ended up being $26^{+13}_{-23}$~km~s$^{-1}$, which is both poorly constrained and at face value in disagreement with the value
we derived and the value
found in the literature. Therefore, our best estimate is \hdg$^\circ$ based on the analysis in which \vsini was subjected to a Gaussian prior.



\subsection{HD~148193}\label{obs:148193}

HD~148193 is also known as TOI-1836. The planet HD~148193~b (TOI-1836.01) was confirmed and characterized by \citet{Chontos2024}. It is a warm Saturn around a subgiant star in a system with a companion star. A second transiting planet candidate was announced later (TOI-1836.02), which is potentially a super-Earth-sized planet on a 1.7-day orbit. However, the signal-to-noise ratio of the
observed transits is modest
and the planet remains
unconfirmed.

\tess data are available with 30-min cadence from Sectors 16 and 25;
with 2-min cadence in Sectors 23, 24, 49, 50, 51, and 52; and with 20-sec cadence in Sector 56. The light curve is shown in \figref{fig:gridone}.

\begin{figure*}[h]
    \centering
    \begin{tabular}{c c}
         \includegraphics[width=\columnwidth]{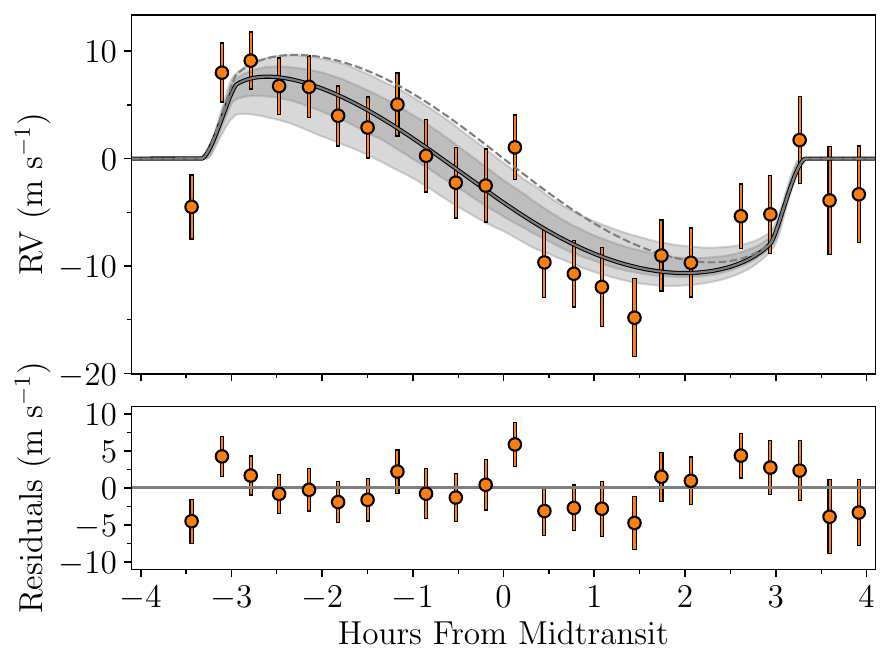} & \includegraphics[width=\columnwidth]{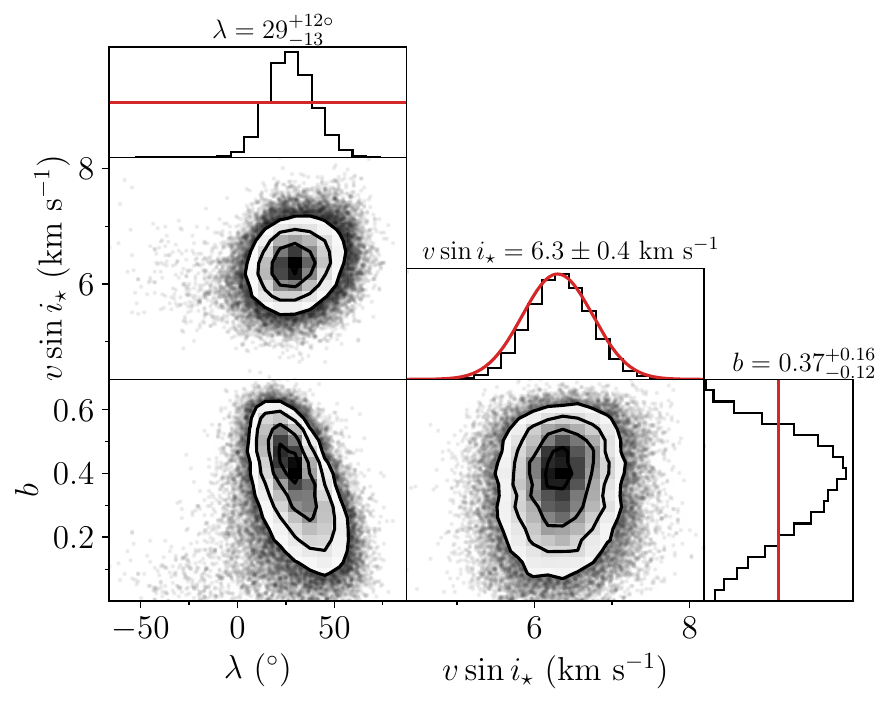} \\ 
    \end{tabular}
    \caption{{\bf The RV-RM effect for HD~148193.} Same as for \figref{fig:rm_hd118203}, but with RVs from HARPS-N.}
    \label{fig:rm_toi1836}
\end{figure*}

We used HARPS-N at the TNG to observe a transit of HD~148193~b. We started observations around UT 21:30 and continued until UT 05:45 on the night starting on June 15, 2021. The exposure time was 1140~s. After
accounting for overheads, the 
sampling time was 1168~s. The transit observations are shown in \figref{fig:rm_toi1836}.

\begin{figure*}[h]
    \centering
    \includegraphics[width=\textwidth]{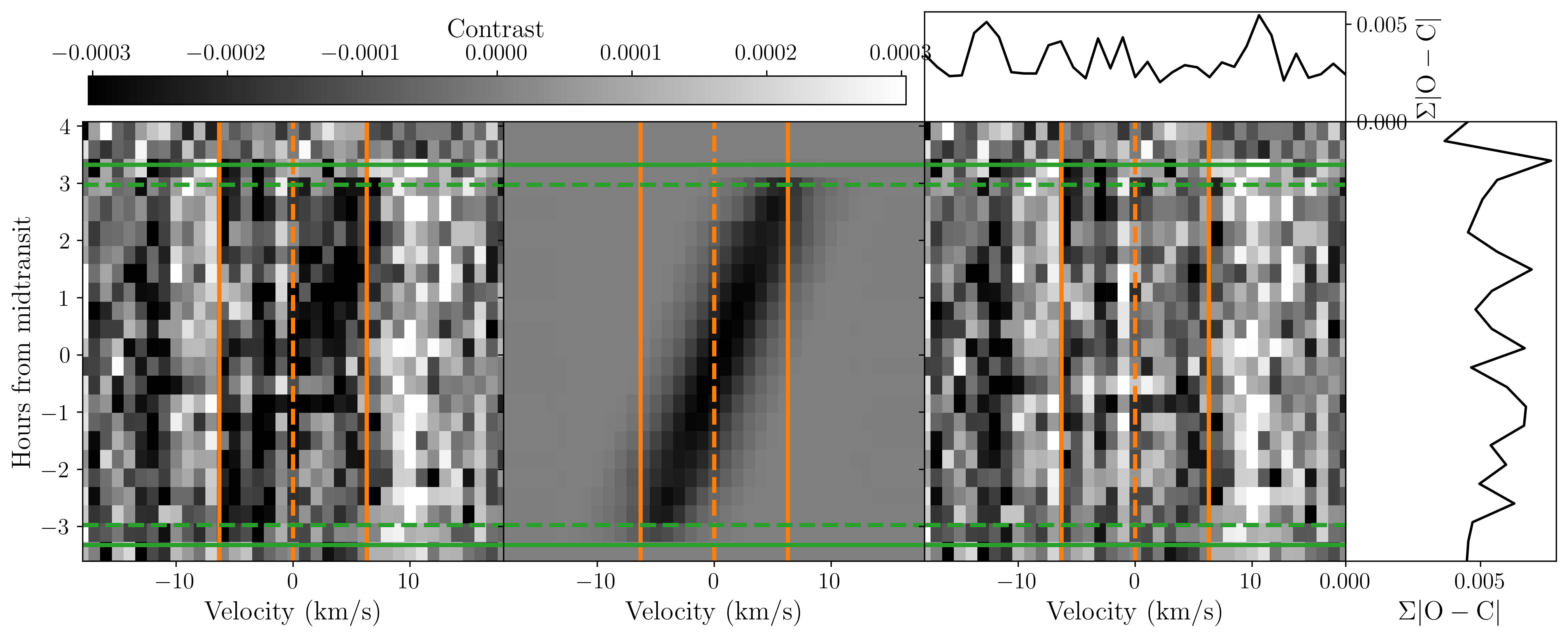}
    \caption{{\bf Doppler shadow for HD~148193.} Left: The observed Doppler shadow created by subtracting the in-transit CCFs from an average out-of-transit CCF. Middle: Model Doppler shadow. Right: Residuals from subtracting the model from the observed shadow. The summed residuals are given in the adjacent panels. The color bar demarcate the contrast. Green horizontal lines denote points of first (solid), second (dashed), third (dashed), and fourth (solid) contact. The solid, orange, vertical lines denote $\pm$\vsini.}
    \label{fig:shadow_toi1836}
\end{figure*}

The combined fit to the light curve and
RVs gave $\lambda = $\hdtwog$^\circ$.
However, we obtained only one pre-ingress RV and two post-egress RVs.
With such sparse out-of-transit data, any mismatch between the out-of-transit RV slope observed on that night and the
calculated slope based on the planet's orbital parameters
can lead to systematic errors. This
problem, combined with the reasons discussed in \sref{sec:rm} and the large projected rotation speed of this system ($\sim6$~km~s$^{-1}$), led us to analyze the line distortions and Doppler shadow. We found $\lambda=$~\hdtwos$^\circ$, with
results shown in \figref{fig:shadow_toi1836}. We adopted the value from the Doppler-shadow analysis as our best estimate for $\lambda$. 

\subsection{K2-261}\label{obs:261}

K2-261 was observed in Campaign 14 of the K2 mission with 30-minute cadence. The planet was confirmed and characterized independently by \citet{Johnson2018}, who named
the planet K2-261~b,
and \citet{Brahm2019}, who named
it K2-161~b. Here, we adopt the name K2-261~b, which is also the name chosen
by \citet{Ikwut2020}. The planet is a warm Saturn-sized planet on an eccentric orbit around a slightly evolved G-type star \citep[][]{Brahm2019}. K2-261 was observed by \tess in Sectors 9, 35, 45, and 46.
\tess data are available
with 2-min cadence in Sectors 9, 45, and 46, and with 20-sec cadence
in Sector 35. The light curve is shown in \figref{fig:gridone}.

\begin{figure*}[h]
    \centering
    \begin{tabular}{c c}
         \includegraphics[width=\columnwidth]{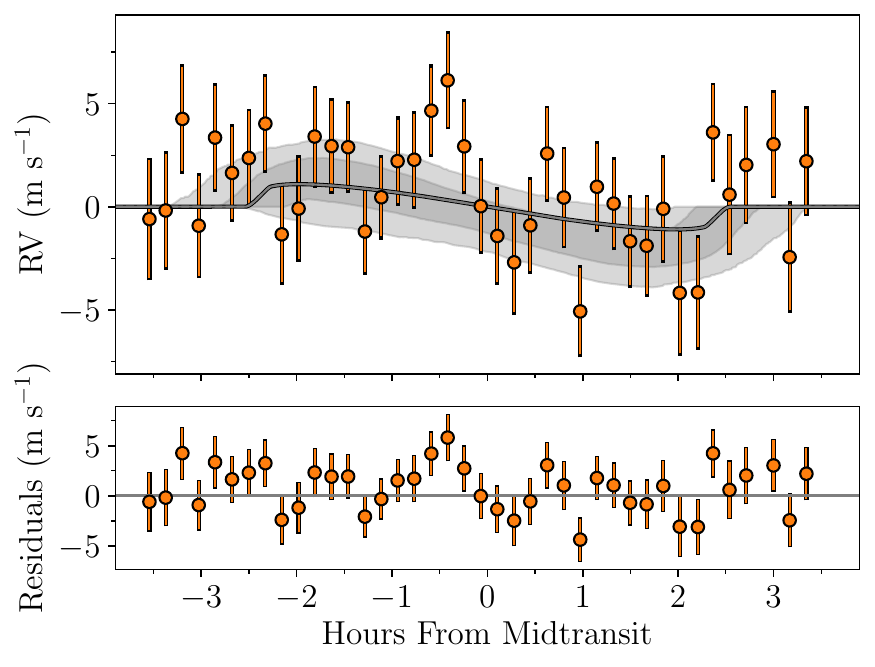} & \includegraphics[width=\columnwidth]{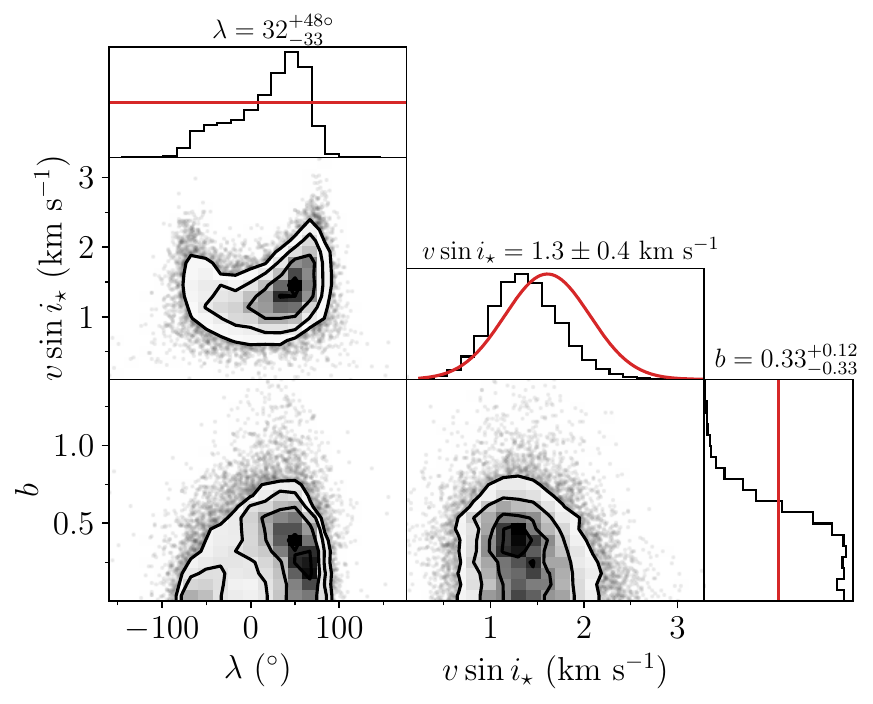} \\ 
    \end{tabular}
    \caption{{\bf The RV-RM effect for K2-261.} Same as for \figref{fig:rm_hd118203}, but with RVs from HARPS-N.}
    \label{fig:rm_k2261}
\end{figure*}

For K2-261~b we used HARPS-N at the TNG to observe a transit. The data are shown in \figref{fig:rm_k2261}. We carried out the observations on the night starting on March 24, 2022, from around UT 20:15 until 03:30. We used an exposure time of 600~s, resulting in a sampling time of around 625~s. 

We found a value for \vsini for K2-261 that is somewhat smaller than the value of $3.70\pm0.17$~km~s$^{-1}$ reported by \citet{Johnson2018}, both from our BF analysis ($1.6$~km~s$^{-1}$) and from our RM observations. The amplitude of the signal is not as strong as would be expected if \vsini
were really $3.70\pm0.17$~km~s$^{-1}$. A uniform prior on \vsini resulted in \vsini$=1.0^{+0.5}_{-0.6}$~km~s$^{-1}$ and $\lambda=$~\ksixoneu$^\circ$, and applying a Gaussian prior yielded a value for the projected obliquity of \ksixoneg$^\circ$. We adopt the latter as our best estimate for $\lambda$.



\subsection{K2-287}\label{obs:287}

K2-287 was observed in Campaign 15 of the K2 mission.
\citet{Jordan2019} discovered and characterized the system, finding the host to be a G-dwarf and the planet to be a warm Saturn-sized planet on an eccentric orbit. 
 
\begin{figure*}[h]
    \centering
    \begin{tabular}{c c}
         \includegraphics[width=\columnwidth]{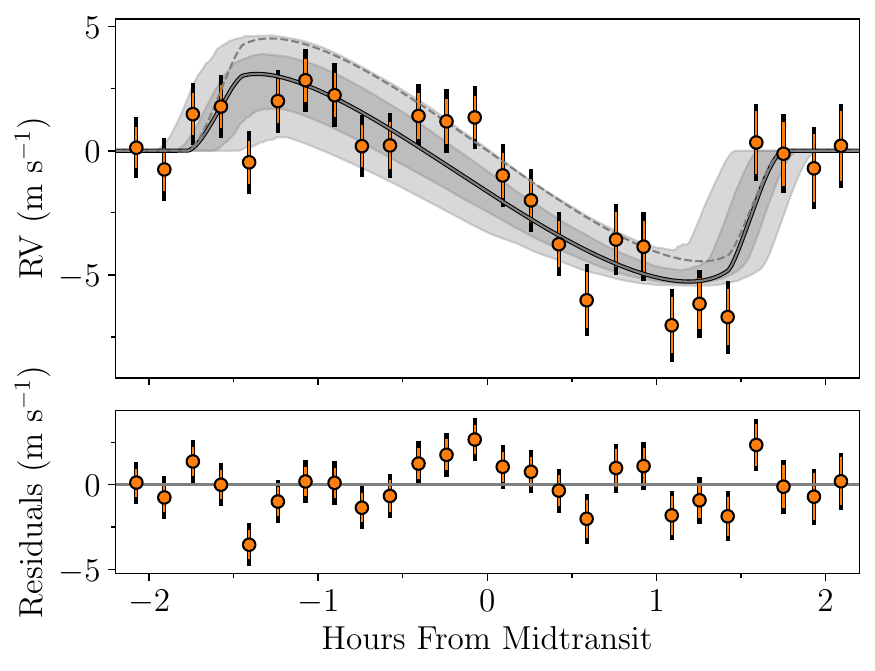} & \includegraphics[width=\columnwidth]{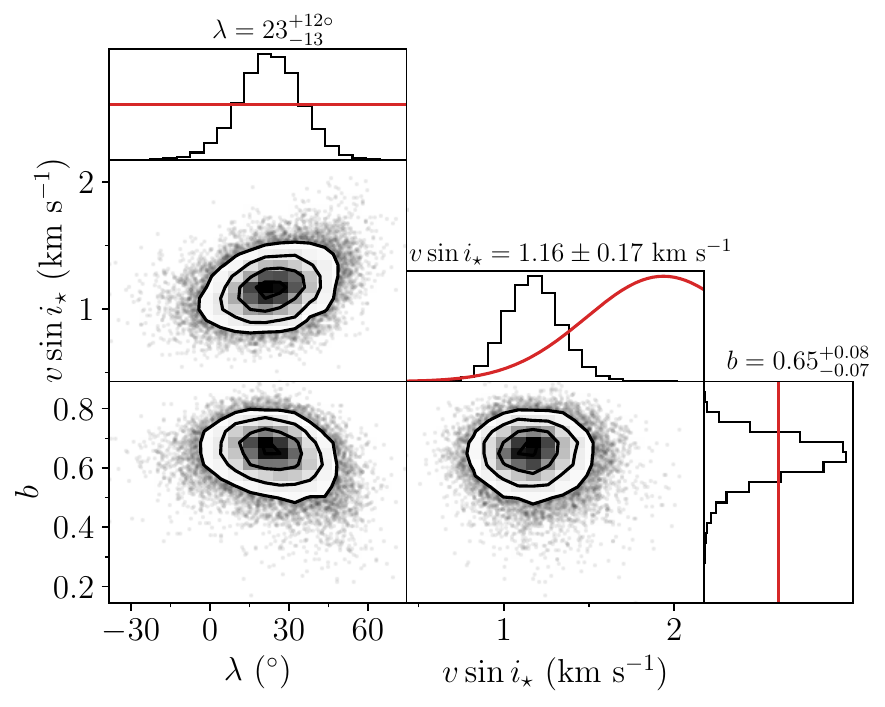} \\ 
    \end{tabular}
    \caption{{\bf The RV-RM effect for K2-287.} Same as for \figref{fig:rm_hd118203}, but with RVs from ESPRESSO.}
    \label{fig:rm_k2287}
\end{figure*}

We observed a transit of K2-287~b using ESPRESSO. Observations were carried out from UT 02:06 to 06:16 on the night July 6, 2021. The exposure time was 540~s, which including overhead comes out to a sampling time of roughly 600~s. The RVs derived from the spectra are shown in the left half of in \figref{fig:rm_k2287}.%

Given the 30-min time averaging of the K2 photometric time series, and the rather long orbital period of the planet, the K2 light curve is relatively poorly sampled. Furthermore, this star
has evaded \tess observations
to this point. As of the time
of writing, there are no scheduled
\tess observations that would
cover this star.

This situation, along with the lack of post-egress spectroscopoic observations, prompted us to search for a suitable archival ground-based light curve covering the egress.
The one we found was obtained
at the Adams Observatory at Austin College on the night starting June 30th, 2018\footnote{Available for download here: \url{https://exofop.ipac.caltech.edu/tess/target.php?id=73848324}.}. More recently, the system
was observed with the CHaracterising ExOPlanet Satellite \citep[CHEOPS;][]{Benz2021}. \citet{Borsato2021} observed three separate transits with a 1-min cadence, and the data from all
three provide good coverage
of the entiure transit.
In our analysis, we included the de-trended light curves presented by \citet{Borsato2021}. The K2, CHEOPS, and ground-based light curves are shown in \figref{fig:gridone}.

\begin{figure}[h!]
    \centering
    \includegraphics[width=\columnwidth]{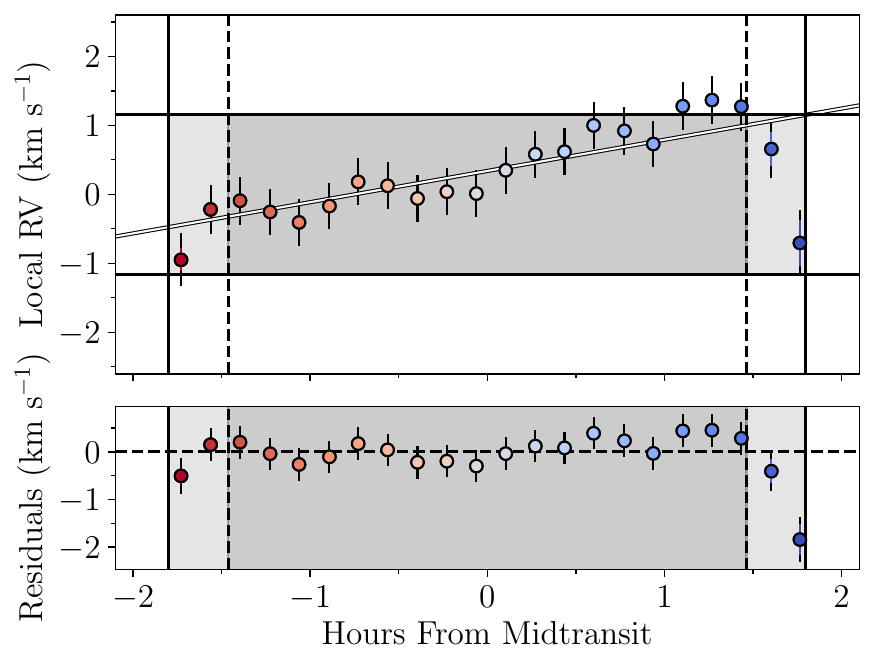}
    \caption{{\bf Subplanetary velocities for K2-287.} Subplanetary velocities for K2-287 with color-coding as described in \figref{fig:slope_shadow}. The dark shaded area corresponds to the time the planet is completely within the stellar disk and the light shaded area denote ingress and egress. The errorbars include a jitter term, where two separate terms have been added in quadrature; one for the epochs for the full transit and one for the epochs when the planet is partially outside of the limb of the star. The horizontal black lines in the top panel mark \vsini from the fit.}
    \label{fig:slope_k2287}
\end{figure}

The two RV-RM analyses yielded similar results for $\lambda$,
with \keightsevenu$^\circ$ for
the case of a uniform prior on \vsini and \keightseveng$^\circ$ for the case
of a Gaussian prior on \vsini.
We also investigated the in-transit subplanetary velocities, which are shown in \figref{fig:slope_k2287} along with the best-fitting slope. In addition to the errors obtained when extracting the location of the subplanetary velocities, we add two different jitter terms in quadrature; one for the epochs taken when the planet is completely within the stellar disk (dark gray area in \figref{fig:slope_k2287}), and another when the planet is at the limb (light gray). This is because there generally (see, e.g, WASP-50 in \sref{obs:50}) seems to be some additional scatter in the subplanetary velocities at the limb, which the errors from fitting a Gaussian to the residual peak do not fully capture. 

From this we found values of \vsini$=1.16\pm0.19$~km~s$^{-1}$ and $\lambda=$~\keightsevens$^\circ$, which are consistent with the values derived from the RV-RM effect. Somewhat
arbitrarily, we adopt $\lambda=$~\keightseveng$^\circ$ as our best estimate; this derives from the RV-RM fit with a Gaussian prior on \vsini. The posterior distributions are shown in \figref{fig:rm_k2287}.

\subsection{KELT-3}

\citet{Pepper2013} reported the discovery of the KELT-3 system, which consists of a hot Jupiter orbiting a late F star. \tess data with 2-min cadence are available from Sectors 21 and 48. The light curve
is shown in \figref{fig:gridone}.

\begin{figure*}[h]
    \centering
    \begin{tabular}{c c}
         \includegraphics[width=\columnwidth]{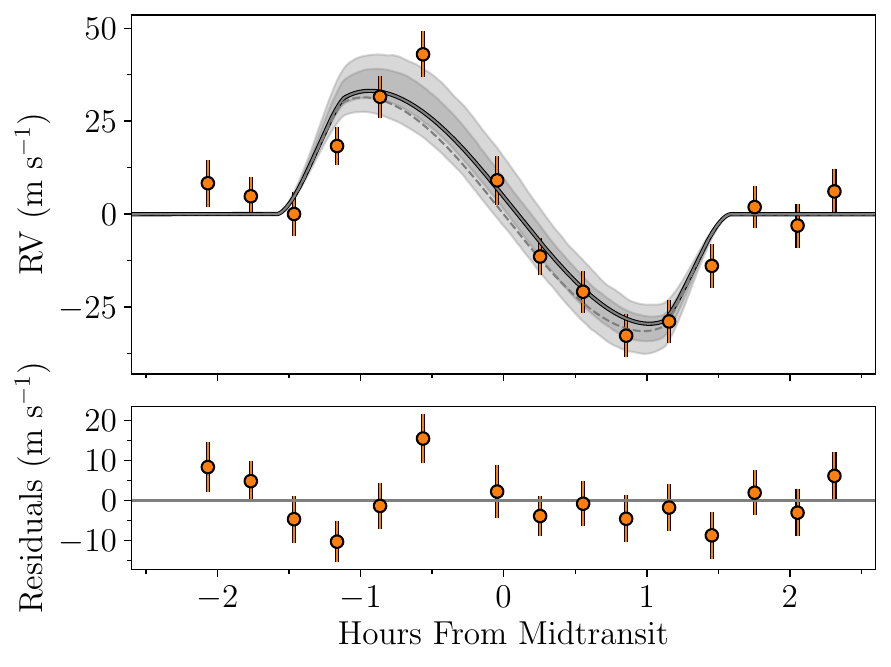} & \includegraphics[width=\columnwidth]{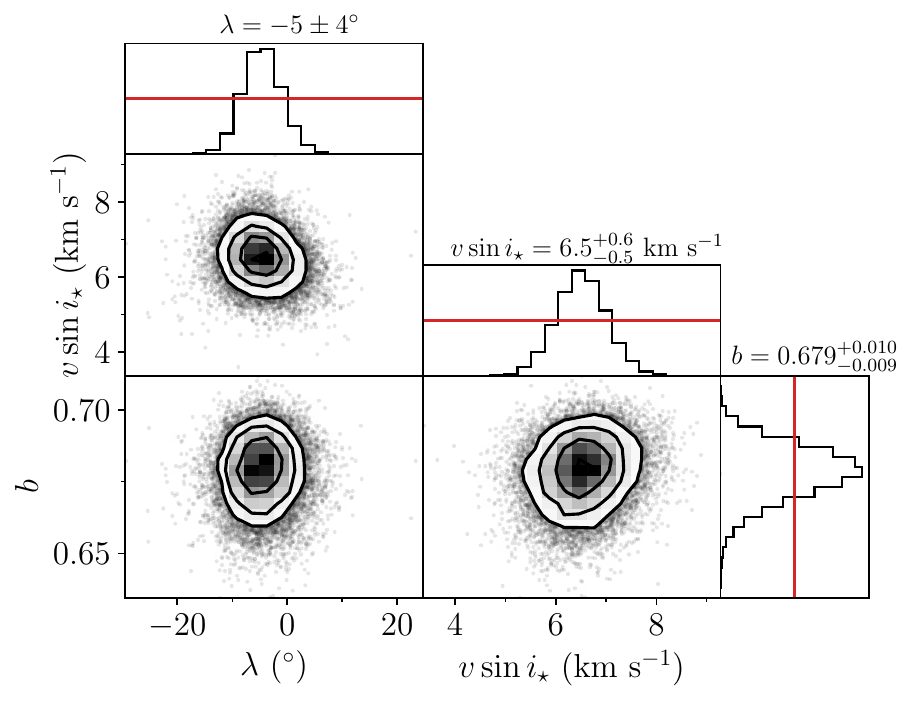} \\ 
    \end{tabular}
    \caption{{\bf KELT-3.} Same as for \figref{fig:rm_hd118203}. Here also with RVs from FIES.}
    \label{fig:rm_kelt3}
\end{figure*}

We used FIES to observe a transit on the night starting April 8, 2022, with observations starting around UT 20:50 and continuing until UT 01:30 April 9, 2022. The exposure time was 900~s, which including overhead resulted in a cadence of 1080~s. The RM-RV effect is shown on the left side of \figref{fig:rm_kelt3}.

As was the case for K2-261, the value for KELT-3's projected rotation velocity
drawn from the literature \citep[$10.2\pm0.5$~km~s$^{-1}$;][]{Pepper2013} is higher than what we obtained from the BF method
($7.6$~km~s$^{-1}$). The observed amplitude of the RM effect suggests that the true \vsini is even lower.
Our RV-RM analysis gave
$6.4\pm0.4$~km~s$^{-1}$ when using a uniform prior in \vsini, and
$\lambda$~=~\keltthreeu$^\circ$.
When we applied a Gaussian prior on \vsini, we obtained \keltthreeg$^\circ$. 

\begin{figure}[h]
    \centering
    \includegraphics[width=\columnwidth]{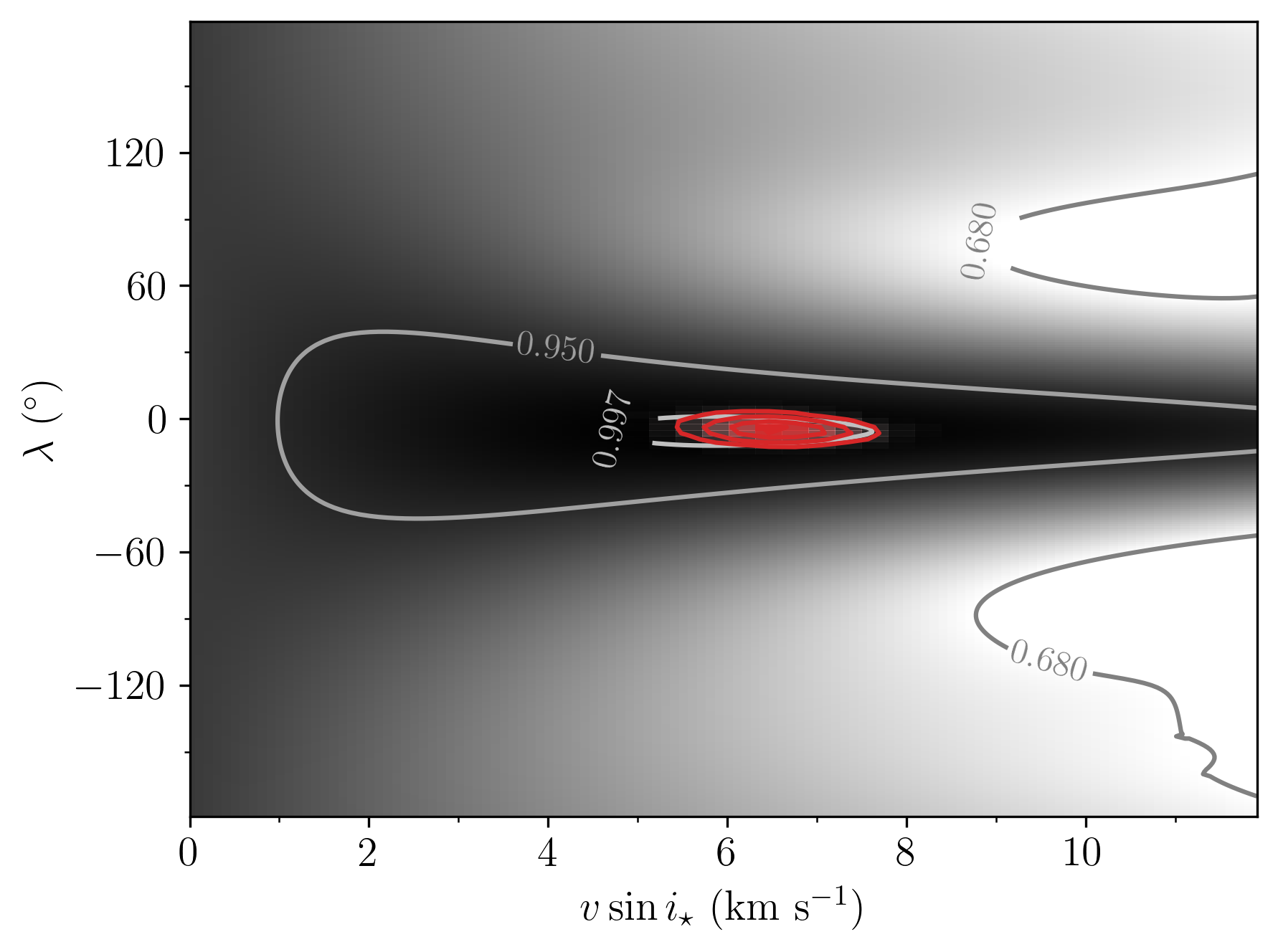}
    \caption{{\bf Peak of the stacked CCFs of KELT-3.} The peak of the stacked CCFs of KELT-3 created by assuming the best-fitting value of $b$ from the RV-RM fit. The gray contours show a peak around (\vsini,$\lambda$)=($6.0~\mathrm{km}~\mathrm{s}^{-1}$,$1.0^\circ$). The red contours are the same as created from the posterior shown to the right in \figref{fig:rm_kelt3}.}
    \label{fig:stack_kelt3}
\end{figure}

Given the discrepancy between the different measurements of \vsini, we decided to look at the stacked CCFs resulting from our FIES spectra. In \figref{fig:stack_kelt3} we show the contours of the peak values obtained from stacking the CCFS on the 2D grid in the space
of \vsini and $\lambda$, compared to our posterior from the RV-RM fit. From this analysis we found \vsini$=5.96\pm 0.03$~km~s$^{-1}$ and $\lambda=1.06\pm 0.16^\circ$, implying that the system
is well-aligned, and that \vsini is closer to 6~km~s$^{-1}$ than the
value of $10.2\pm0.5$~km~s$^{-1}$
reported by \citet{Pepper2013}. Although evidently
the uncertainties in some or all
of the \vsini determinations have
been underestimated, we believe
that the stacked-CCF method gives
the most reliable result for \vsini.

This lends further credence to the RV-RM fit with a uniform prior on \vsini, and as our observations of KELT-3 have a high SNR and cover the transit very well. We therefore adopt the result from this RV-RM fit,
$\lambda=$~\keltthreeu$^\circ$,
as our best estimate for this system.



\subsection{KELT-4A}\label{obs:kelt4}

KELT-4 is a hierarchical triple star system discovered by \citet{Eastman2016}. The primary star, KELT-4A, is an F star orbited by a hot Jupiter. The close binary system KELT-4BC is engaged in a wide
orbit with KELT-4A.
\tess data are available from Sector 48 with 2-min cadence. The light curve is shown in \figref{fig:gridone}. Since
the \tess cameras cannot resolve
KELT-4A from KELT-4BC,
we included a dilution factor in our analysis of $\delta M=3.5\pm0.1$ \citep[estimated from Table 3 in ][]{Eastman2016} to correct for the light from KELT-4BC.

On the night starting January 27, 2021, we observed a transit of KELT-4Ab, starting the observations UT 01:40 and ending them UT 06:45 January 28, 2021. The exposure time was 900~s, meaning the effective sampling was a spectrum every 1090~s.

\begin{figure*}[h]
    \centering
    \begin{tabular}{c c}
         \includegraphics[width=\columnwidth]{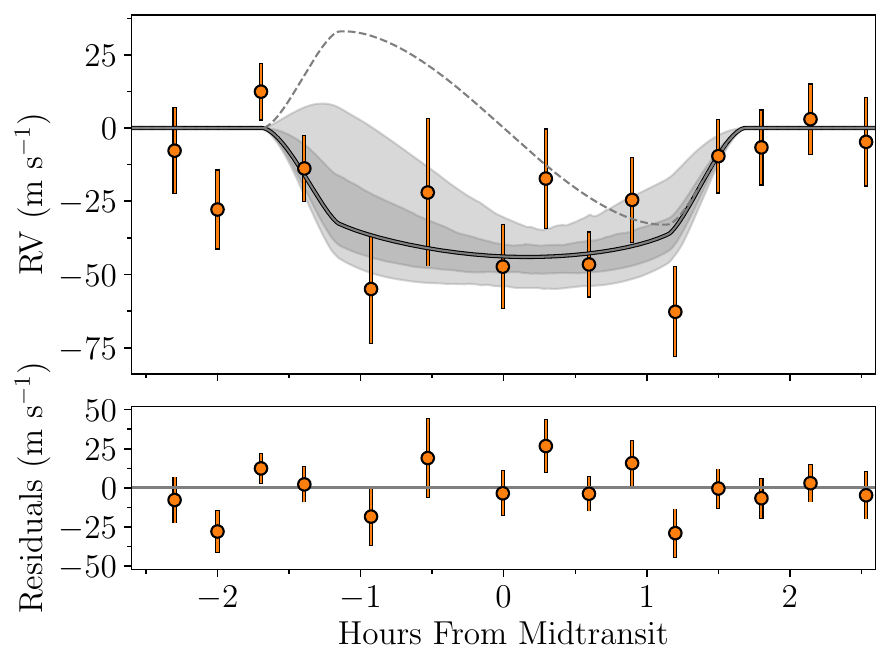} & \includegraphics[width=\columnwidth]{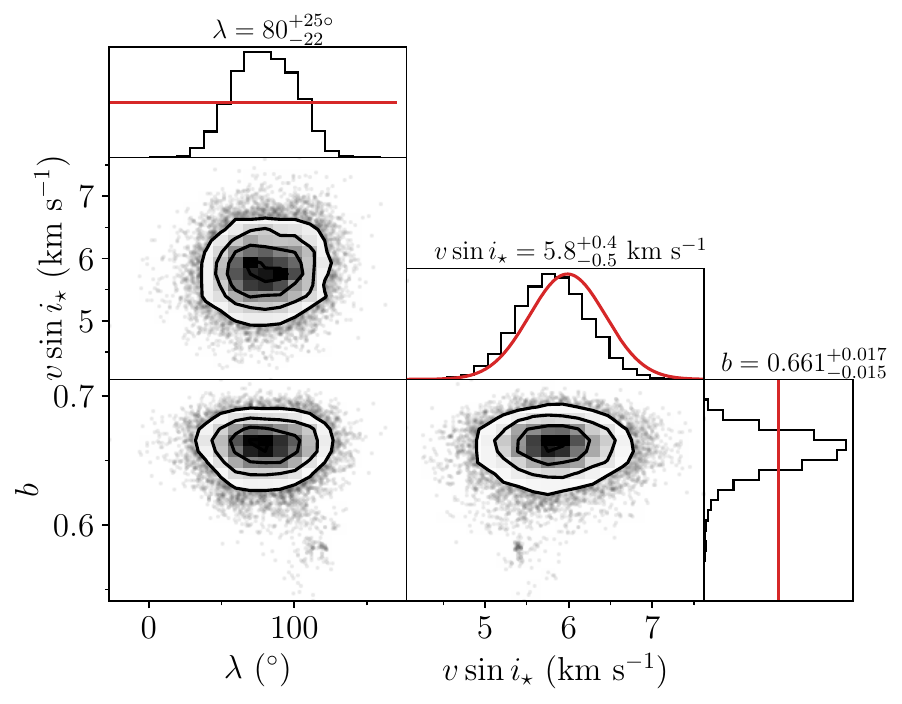} \\ 
    \end{tabular}
    \caption{{\bf KELT-4A.} Same as for \figref{fig:rm_hd118203}. Here also with RVs from FIES.}
    \label{fig:rm_kelt4}
\end{figure*}

The observations are shown in \figref{fig:rm_kelt4} with our results from a run invoking a Gaussian prior on \vsini. From this we got $\lambda=$~\keltfourg$^\circ$. Applying a uniform prior on \vsini we found $\lambda=$~\keltfouru$^\circ$. Evidently, applying a Gaussian or a uniform prior on \vsini makes little difference, despite the somewhat noisy data, most likely because of the rather large impact parameter and
large misalignment.

As for KELT-3, we investigated the stacked CCFs for this system.
The results coincided with the result from our RV-RM fit, and when fitting a 2D Gaussian to the grid of \vsini and
$\lambda$ we obtained values of \vsini$=6.0\pm0.4$~km~s$^{-1}$ and $\lambda=88.9\pm0.5^\circ$. The results are consistent with the polar configuration implied by the RV-RM fit. Our best estimate
for $\lambda$ for
this system, \keltfourg$^\circ$,
is based on the RV-RM fit with a Gaussian prior on \vsini.



\subsection{LTT~1445A}\label{obs:ltt1445}

LTT~1445 is a triple star system consisting of three M-dwarfs. The primary star hosts an Earth-sized planet on a 5-day orbit, LTT~1445Ab,
which was discovered by \citet{Winters2019}. A second transiting planet, LTT~1445Ac, was reported by \citet{Winters2022}, and an RV signal from a third planet was detected by \citet{Lavie2023}. The orbit of LTT~1445Ab is sandwiched between the orbits of the other two planets.

LTT~1445A was observed by \tess in Sectors 4 and 31, and the data are
available with 20-sec and 2-min cadence, respectively. The light curves show a photometric quasiperiodicity that probably arises from rotation of
either star B or star C \citep{Winters2019}. To account for
this rotational modulation, our model for the light curve including a GP
with a kernel consisting of two stochastically driven damped harmonic oscillators. As the amplitude of the rotational modulation seemed to change between the two sectors \citep{Winters2022}, we did not
require the hyperparameters to
be the same in both sectors.
The light curve is shown in \figref{fig:gridtwo}.

\begin{figure*}[h]
    \centering
    \begin{tabular}{c c}
         \includegraphics[width=\columnwidth]{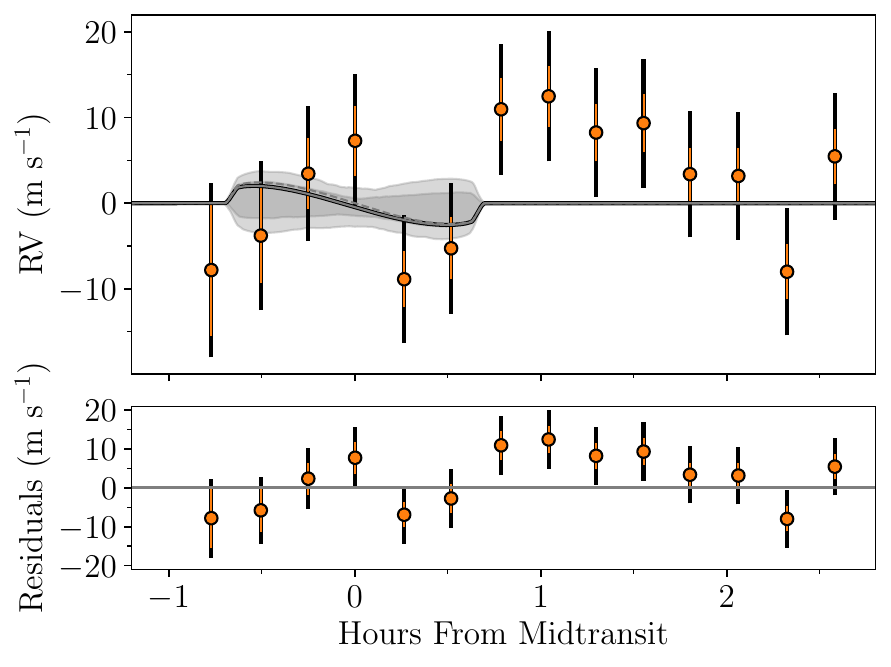} & \includegraphics[width=\columnwidth]{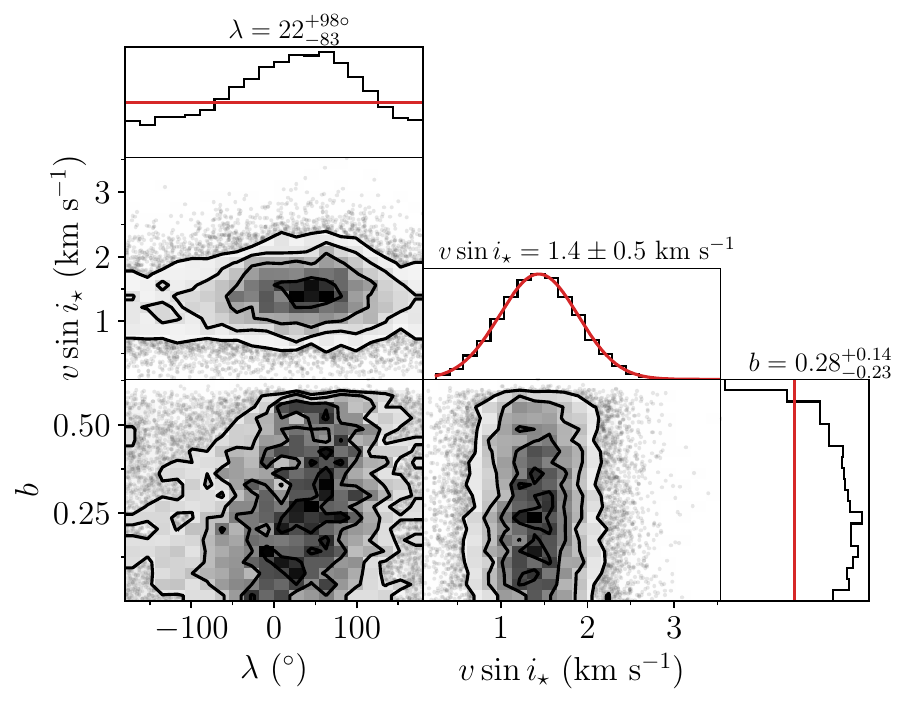} \\ 
    \end{tabular}
    \caption{{\bf LTT~1445A.} Same as for \figref{fig:rm_hd118203}, but with RVs from HARPS-N.}
    \label{fig:rm_ltt1445}
\end{figure*}

We observed a spectroscopic transit of LTT~1445Ab employing HARPS-N. The observations were carried out on the night starting September 5th, 2020. We used an exposure time of 900~s, yielding a sampling time of 920~s. The first exposure was not useful
because the telescope was erroneously aimed at LTT~1445BC. The remainder of the observations are shown in \figref{fig:rm_ltt1445}.

As can be seen in \figref{fig:rm_ltt1445} no RM effect is apparent in the RV data. Formally, we found $\lambda=$~\lttg$^\circ$,
implying a gentle
preference for a prograde orbit. However, as described by \cite{Albrecht2011}, such low SNR results are probably even more uncertain than the formal uncertainty suggests. 

Usually, if the RV-RM effect has
a detectable amplitude,
then the scatter of the RVs inside
the transit should be larger
than the scatter of the RVs
outside the transit.
For LTT~1445A, the in-transit
scatter is $\sigma=5.9$~m~s$^{-1}$ after subtracting the best-fit orbital solution. The out-of-transit scatter is $\sigma=7.1$~m~s$^{-1}$ (dropping to $\sigma=6.0$~m~s$^{-1}$ if the single pre-transit datum is excluded). The in-transit residuals relative
to the best-fit RM model have a scatter of $\sigma=5.5$~m~s$^{-1}$. In light of these results, we do not consider the RM effect to have been
detected at all.
Better results would probably require a bigger telescope.


\subsection{TOI-451A}
\label{obs:toi451}

TOI-451 is a young multiplanet system in the Pisces–Eridanus stream consisting of the G-dwarf planet
host, TOI-451A, and a wide-orbiting (likely) M-dwarf binary \citep[][]{Newton2021}. The planetary system
around TOI-451A consists of three  planets with radii 1.9, 3.1, and 4.1~R$_\oplus$ and periods of 1.9, 9.2, and 16~days. \tess observed the system in Sector 4, 5, and 31 with a cadence of 2~min. The \tess observations of TOI-451A are shown in \figref{fig:gridtwo}.

We observed a transit of TOI-451Ab, the innermost and smallest planet,
using ESPRESSO. The observations were carried out on the night starting on September 24, 2022, from around UT 04:15 until 08:30 September 25, 2022. The exposure time was set to 720~s, resulting in a spectrum every 755~s.

As reported by \citet{Newton2021}, the light curve of TOI-451 displays a rather large amplitude rotational modulation. We therefore adopted the same kernel as for LTT~1445 (described in the previous section) instead of the Matérn-3/2 kernel. We included all three planets in our transit modeling. We applied Gaussian priors with values drawn from \citet{Newton2021} for the transit parameters to the two, outer planets (c and d) for which we did not observe a spectroscopic transit. For planet b we followed our usual approach.

\begin{figure*}[h]
    \centering
    \begin{tabular}{c c}
         \includegraphics[width=\columnwidth]{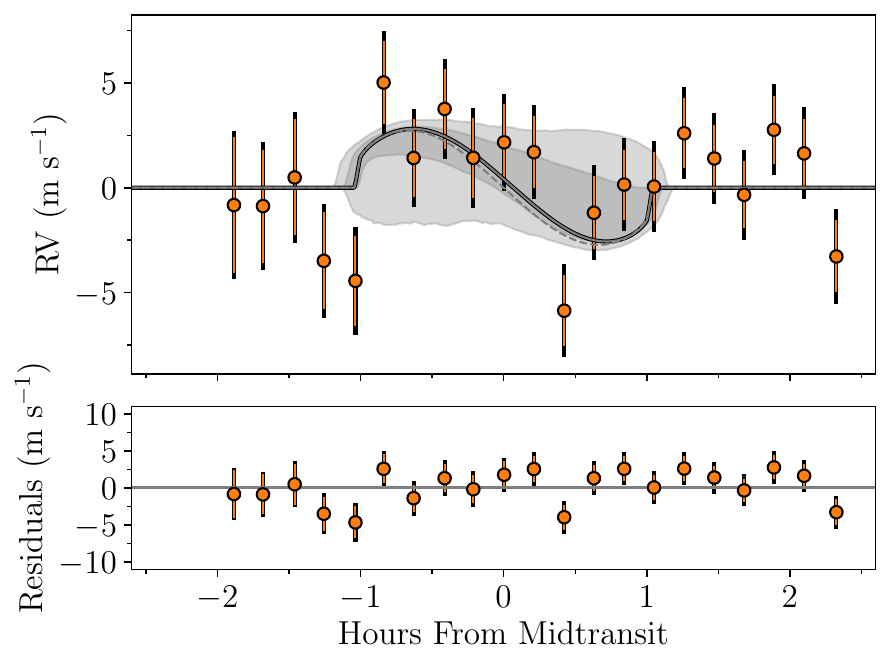} & \includegraphics[width=\columnwidth]{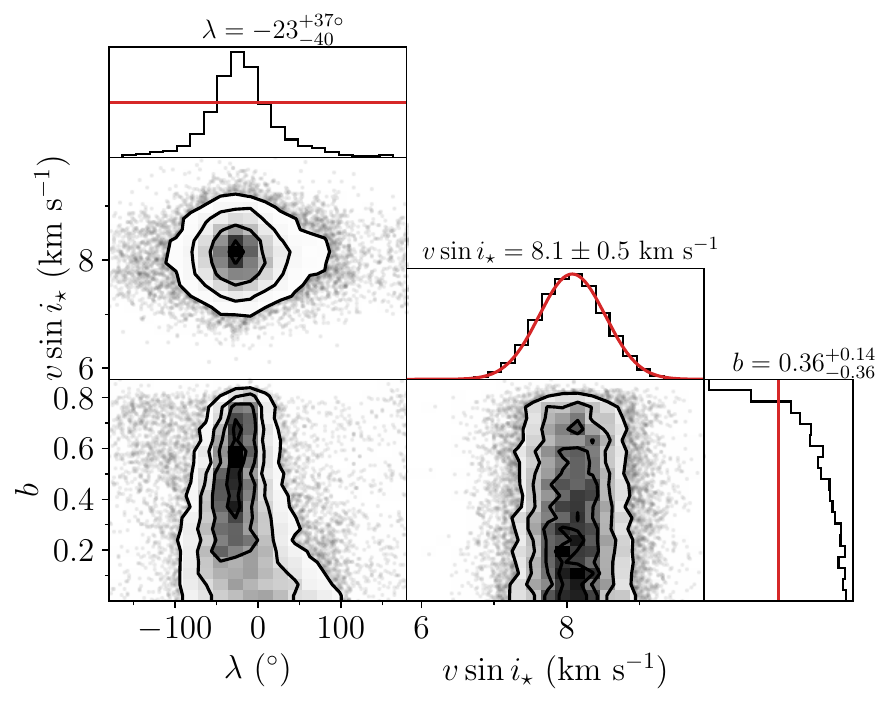} \\ 
    \end{tabular}
    \caption{{\bf The RV-RM effect for TOI-451.} Same as for \figref{fig:rm_hd118203}, but with RVs from ESPRESSO.}
    \label{fig:rm_toi451}
\end{figure*}

The RV-RM analysis with a uniform prior on \vsini gave
$\lambda=$~\toifourfiveu$^\circ$ 
and the analysis with a Gaussian prior gave \toifourfiveg$^\circ$. All
we can conclude is that the orbit is prograde.

We investigated the RV scatter in the same manner we did for LTT~1445.
We obtained
$\sigma=2.8$~m~s$^{-1}$ and $\sigma=2.4$~m~s$^{-1}$ for the in-transit and out-of-transit scatter, respectively, and when subtracting the RM effect from the in-transit data we found $\sigma=2.0$~m~s$^{-1}$. With these results we cannot be confident
that the RM effect was detected. For
better results, one might
observe a second transit of TOI-451Ab
or give up and choose to observe one of the larger planets (for which opportunities to observe transits are rarer).

We did not use any results
from LTT~1445A or TOI-451A
in the ensemble analyses
described in \sref{sec:elam}, \sref{sec:ppp}, and \sref{sec:precise}. However,
the results are displayed
in \sref{sec:multis}, which focuses
on multiplanet systems.



\subsection{TOI-813}\label{obs:toi813}
\citet{Eisner2020} used \tess data to discover a Saturn-sized planet on an 84-day orbit around this subgiant star.
\tess data are available with 30-min cadence in Sector 2;
2-min cadence in Sectors 
5, 8, 11, 30, 33, and 39;
and 20-sec cadence in Sectors
27 and 61.
Due to the planet's long orbital period, transits only occurred
during Sectors 2, 27, and 61.
\figref{fig:gridtwo} shows
the light curve.

\begin{figure*}[h]
    \centering
    \begin{tabular}{c c}
         \includegraphics[width=\columnwidth]{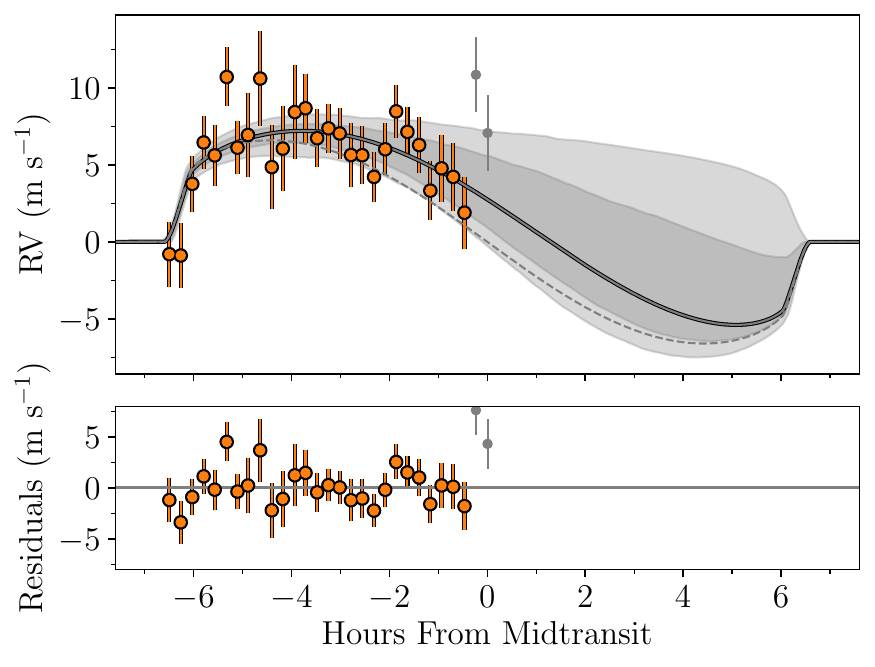} & \includegraphics[width=\columnwidth]{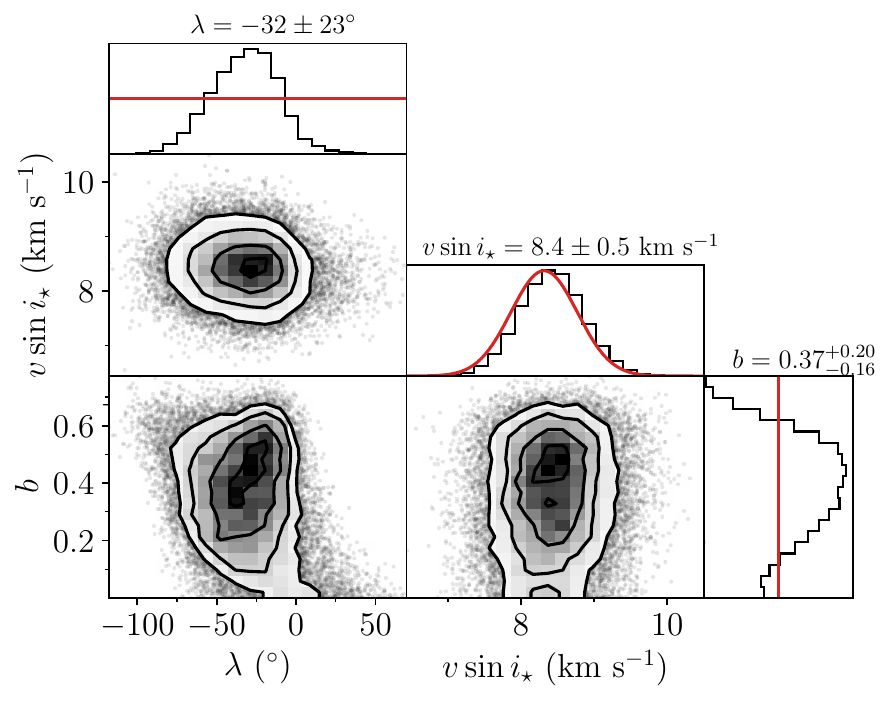} \\ 
    \end{tabular}
    \caption{{\bf The RV-RM effect for TOI-813.} Same as for \figref{fig:rm_hd118203}, but with RVs from ESPRESSO.}
    \label{fig:rm_toi813}
\end{figure*}

We observed a transit of TOI-813~b on the night January 20, 2023, using ESPRESSO. The exposure time was set to 800~s, yielding a sampling time of around 832~s. This transit was simultaneously observed with \tess. We observed the transit starting just before ingress and continued for for 6.5~hours until the target set. These observations cover roughly half of the 13.14~hr transit as shown in \figref{fig:rm_toi813}.

In an attempt to cover as much of the transit as possible, the two last exposures were acquired through
a large airmass ($>2.2$), and the
RVs based on those exposures
deviate from the trend established by the preceding RVs. Including the two high-airmass RVs in our fit resulted in $\lambda=-60^{+19}_{-21}$$^\circ$ suggesting that the system is significantly misaligned. Omitting these two exposures yielded \toieightthirteeng$^\circ$. In both cases described above, a Gaussian prior on \vsini was applied. Excluding the last two exposures while running an MCMC with a uniform prior in \vsini, we obtained \toieightthirteenu$^\circ$ for $\lambda$. 

We also analyzed the Doppler
shadow for this system and found a value of \toieightthirteens$^\circ$, which is more in-line with the results obtained using a uniform prior on \vsini or including the two data points at high airmass. However, the lack of observations outside of the transit left us without a spectral CCF
that is free of the RM effect.

Since only half of the transit was covered, we found it necessary to apply a Gaussian prior on \vsini to obtain reliable results. We also decided not to trust the two high-airmass RVs. We adopt the resulting value of \toieightthirteeng$^\circ$ as our best estimate
of $\lambda$
for TOI-813.
The RM effect has a high
enough amplitude to be detectable using smaller telescopes and less precise spectrographs. Another observation covering the second half of the transit is warranted, and would
allow us to be judge more definitively if the system is aligned or not.



\subsection{TOI-892}\label{obs:892}

TOI-892~b is a warm Jupiter on a circular orbit around an F-type dwarf \citep[][]{Brahm2020}. The system was first observed by \tess in Sector 6 with 30~min cadence, and then in Sector 33 with 2-min cadence. \figref{fig:gridtwo} shows the
light curve.

We used HARPS-N to observe a transit of TOI-892~b. We started observations around UT 19:15 and continued until 03:15 on the night starting on January 1, 2022. The exposure time was set to 960~s, which including overhead comes out to a sampling time of around 987~s. The observations were interrupted due to high humidity.

The results with and without the Gaussian prior on \vsini
are in agreement, giving $\lambda=$!~\toieightnineu$^\circ$ and \toieightnineg$^\circ$, respectively.
Since about half of the transit was
not observed, the run including a Gaussian prior on \vsini yielded a better constraint on $\lambda$, as one would expect. 

\begin{figure*}[h]
    \centering
    \begin{tabular}{c c}
         \includegraphics[width=\columnwidth]{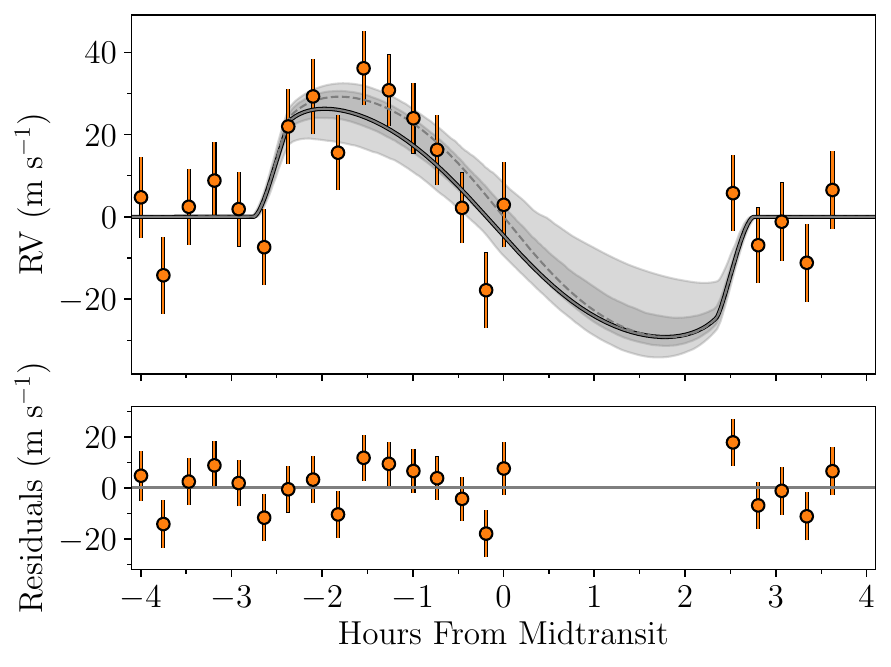} & \includegraphics[width=\columnwidth]{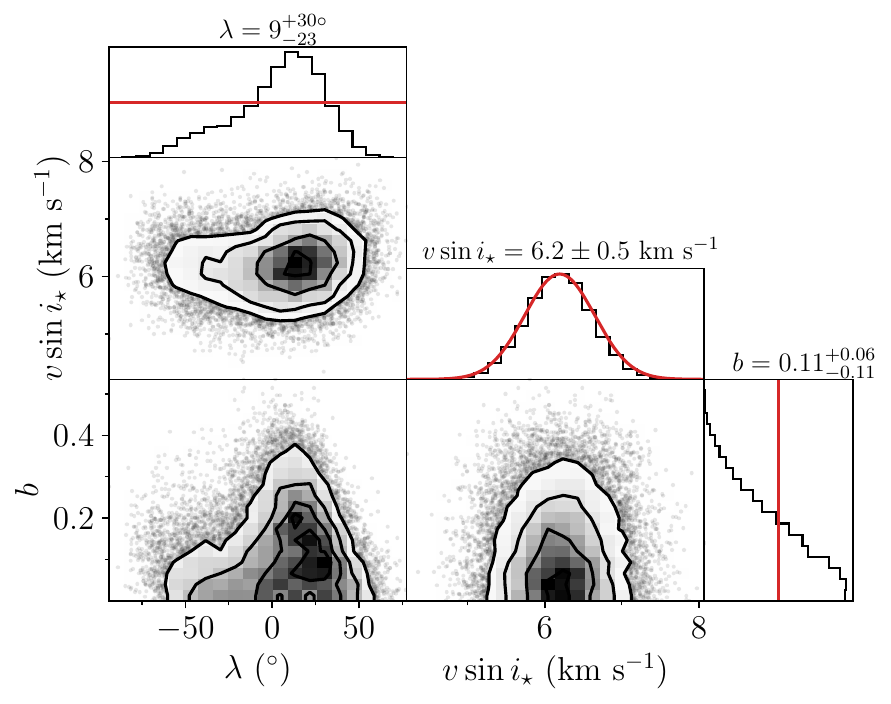} \\ 
    \end{tabular}
    \caption{{\bf The RV-RM effect for TOI-892.} Same as for \figref{fig:rm_hd118203}, but with RVs from HARPS-N.}
    \label{fig:rm_toi892}
\end{figure*}

\begin{figure*}[h]
    \centering
    \includegraphics[width=\textwidth]{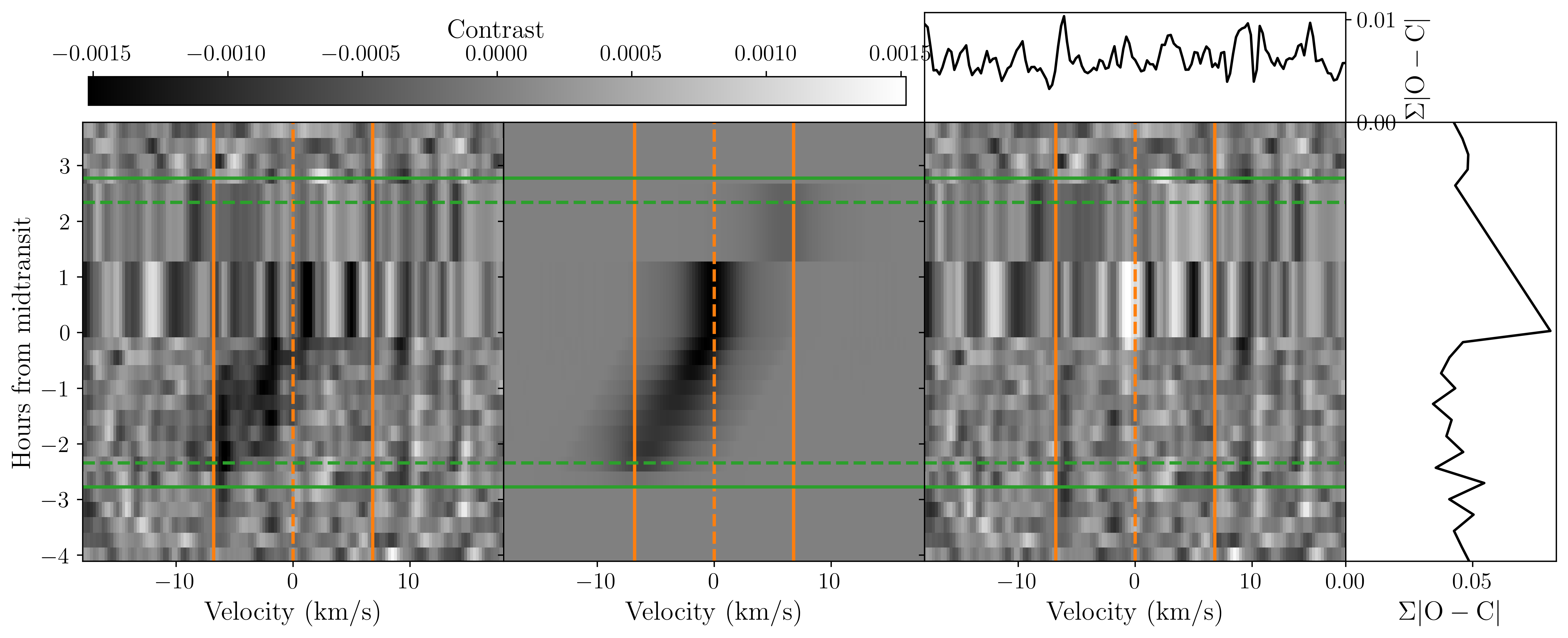}
    \caption{{\bf Doppler shadow for TOI-892.} Same as in \figref{fig:shadow_toi1836}, but for TOI-892.}
    \label{fig:shadow_toi892}
\end{figure*}

We analyzed the Doppler shadow and found $\lambda=$~\toieightnines$^\circ$, which differs from the RV-RM results
more than one would expect for
two analyses of the same data.
Since the the shadow shown in \figref{fig:shadow_toi892} looks convincing, we decided
to adopt the value of \toieightnines$^\circ$ as our best
estimate of $\lambda$ for this system.

\subsection{TOI-1130}
\label{obs:toi1130}

TOI-1130 is a K dwarf with two transiting planets, an inner Neptune and an outer Jupiter, discovered by \citet{Huang2020}. We used ESPRESSO to observe a transit of the larger planet, TOI-1130~c, on the night starting on June 1, 2021. The observations were carried out from around UT 01:30 to 06:00 on June 2, 2021. We opted for an exposure time of 660~s, yielding a sampling time of 701~s. TOI-1130 was observed by \tess in Sector 13 in 30-minute cadence and in Sectors 27 and 67 in 20-second cadence. The \tess light curve for planet c is shown in \figref{fig:gridtwo}.

Because the planets in TOI-1130 are in
or near a 2:1 mean-motion resonance, the planets show relatively
large transit timing variations (TTVs). The TTV amplitudes are on the order of a few hours for planet b and 15~min for planet c \citep{Korth2023}. 
Based on the comprehensive model of the system developed by \citet{Korth2023}, the mid-transit time of planet c
on June 2, 2021 should have
occurred at BJD $2459367.6304\pm0.0003$~d (J.~Korth, private communication). 
We used this result as a Gaussian prior on $T_0$ in our analysis.

The TTVs also affect the phase-folding process that is part of the construction of a high-SNR light curve. To deal with the TTVs,
we identified the transits of both planet b and c in the \tess data, and fitted a model
the \tess data alone,
holding fixed all of the parameters except for the individual mid-transit times. Then, in our RV-RM analysis,
we fitted the \tess data jointly
with the RV time series while applying Gaussian priors on the mid-transit times with a width of $0.005$~d. 
We also applied a Gaussian prior on the stellar mean density of $\rho_\star=2.98 \pm 0.18$~g~cm$^{-3}$ \citep{Korth2023}.
Most of the transits of planet b were irrelevant, but there was one
instance of overlapping transits
of planets b and c.

Our BF analysis gave a
lower projected rotation velocity for TOI-1130 ($1.3$~km~s$^{-1}$) than
we found in the literature \citep[$4.0\pm0.5$~km~s$^{-1}$,][]{Huang2020}. Our result is more in line with the upper limit of 3~km~s$^{-1}$ reported by \citet{Korth2023}. From the RV-RM analysis, we obtained
\vsini$=1.3\pm0.5$~km~s$^{-1}$
and $\lambda=$~\toithirtyu$^\circ$ 
when applying a uniform prior on \vsini. With a Gaussian prior on \vsini, we obtained
$\lambda=$~\toithirtyg$^\circ$, which we regard as the best estimate.
We need to stress that
the accuracy of the result does depend
on the prior on $T_0$ that was obtained
from the model of \citep{Korth2023}.

\begin{figure*}[h]
    \centering
    \begin{tabular}{c c}
         \includegraphics[width=\columnwidth]{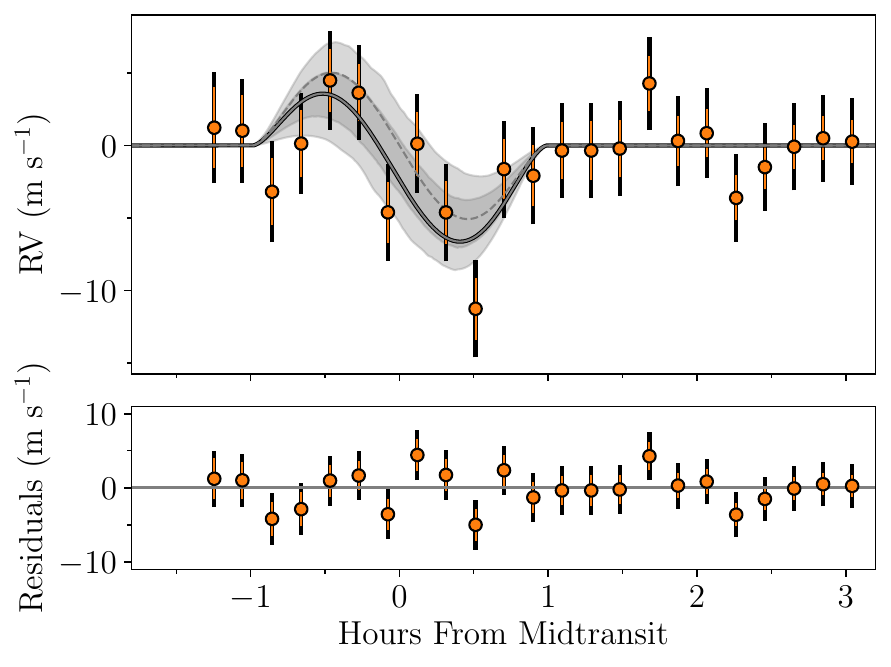} & \includegraphics[width=\columnwidth]{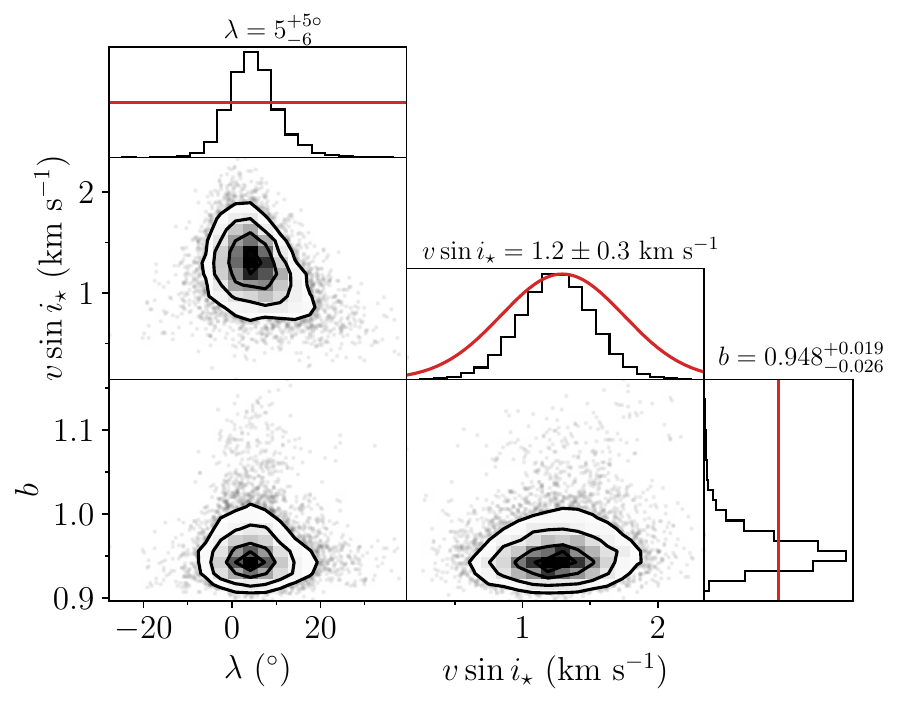} \\ 
    \end{tabular}
    \caption{{\bf The RV-RM effect for TOI-1130.} Same as for \figref{fig:rm_hd118203}, but with RVs from ESPRESSO.}
    \label{fig:rm_toi1130}
\end{figure*}

\begin{figure}[h]
    \centering
    \includegraphics[width=\columnwidth]{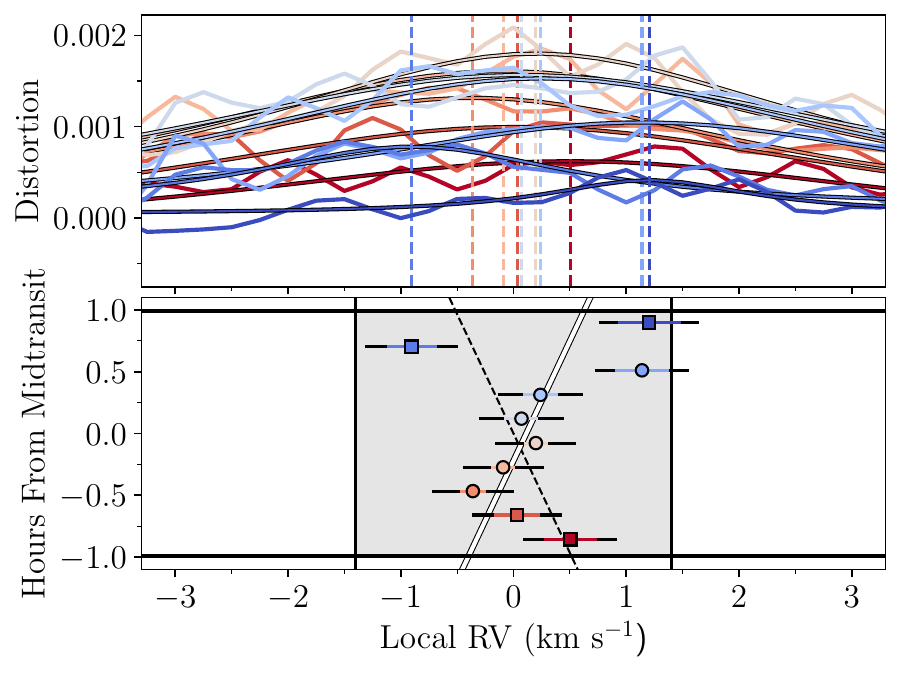}
    \caption{{\bf Subplanetary velocities for TOI-1130.} In the top panel the distortion from subtracting the in-transit CCF from the out-of-transit are shown as colored (wiggly) lines. Lines of corresponding colors with a black outline are (Gaussian) fits to the distortion in order to extract the position of the distortion denoted by the dashed vertical line. The bottom panel shows the location of the distortion plotted around the midtransit time. The squares denote epochs for which the decrease in flux from the transiting planet is less than 0.75\%. The solid white (black dashed) line shows a model for an (anti-)aligned configuration.}
    \label{fig:subp_toi1130}
\end{figure}

We investigated the subplanetary velocities for TOI-1130 to see if we could properly trace the deformation of the stellar lines for this grazing transit, where (unlike for K2-287) the planet only obscures part of the stellar limb. It proved more difficult to locate the distortion in the first and last few in-transit spectra, as shown in \figref{fig:subp_toi1130}.
We also found it necessary to adopt a Gaussian prior on \vsini. 

When fitting the subplanetary velocities using all of the data,
the posterior for $\lambda$ 
had a peak centered on $0^\circ$, and a secondary peak
near $\lambda=180^\circ$. We found the secondary peak to be a consequence
of including the noisiest data points
from the first and last few in-transit spectra. Out of concern over
systematic errors in our model for
limb-grazing transits,
we repeated the fit
after omitting the in-transit data that
were obtained when the loss of light was smaller than 0.75\% (the square data points in \figref{fig:subp_toi1130}). The result was $\lambda=$~\toithirtys$^\circ$. Although excluding data without a firm
justification is undesirable, we judge the evidence as indicating
that any misalignment is smaller
than about 20$^\circ$.



\subsection{WASP-50}\label{obs:50}
WASP-50 is a G9 dwarf hosting a hot Jupiter on a $1.9$~d orbit discovered by \citet{Gillon2011}. \tess data
are available from observations in Sector 4 and 31 with 2~min and 20~sec cadence, respectively.  \figref{fig:gridtwo} shows the \tess transit light curve.

On the night starting on August 24, 2022, we observed a transit of WASP-50~b with ESPRESSO. The observations started at around UT 05:10 and continued until UT 09:40 on August 25, 2022. An exposure time of 240~s ensured a cadence of 270~s. 

The time series is well-sampled and the SNR is very high, exactly the case where we do not expect the results
to be sensitive to the prior on \vsini. Indeed, for this system we found \waspfiftyu$^\circ$ when applying a uniform prior and \waspfiftyg$^\circ$ for a Gaussian prior.

\begin{figure*}[h]
    \centering
    \begin{tabular}{c c}
         \includegraphics[width=\columnwidth]{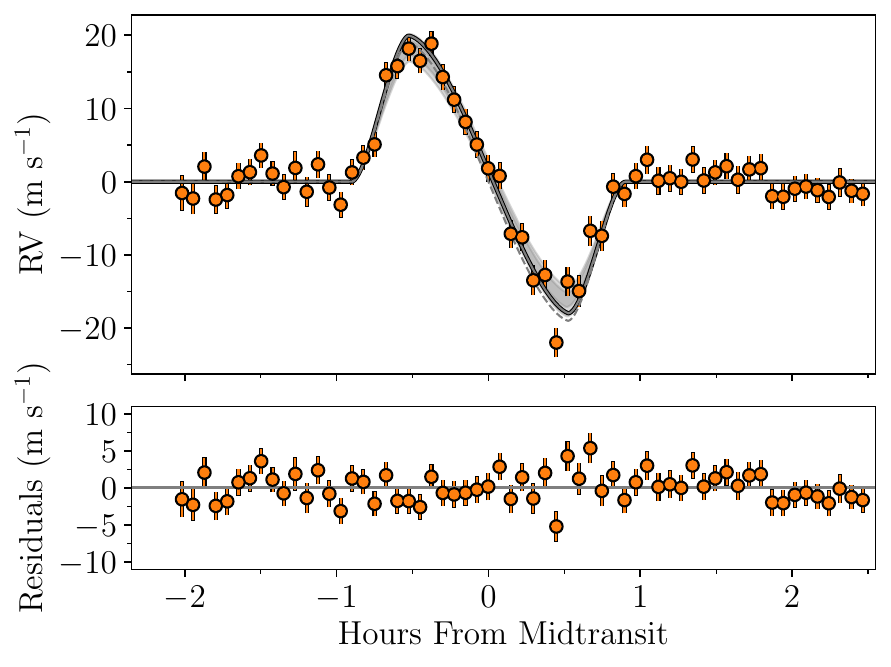} & \includegraphics[width=\columnwidth]{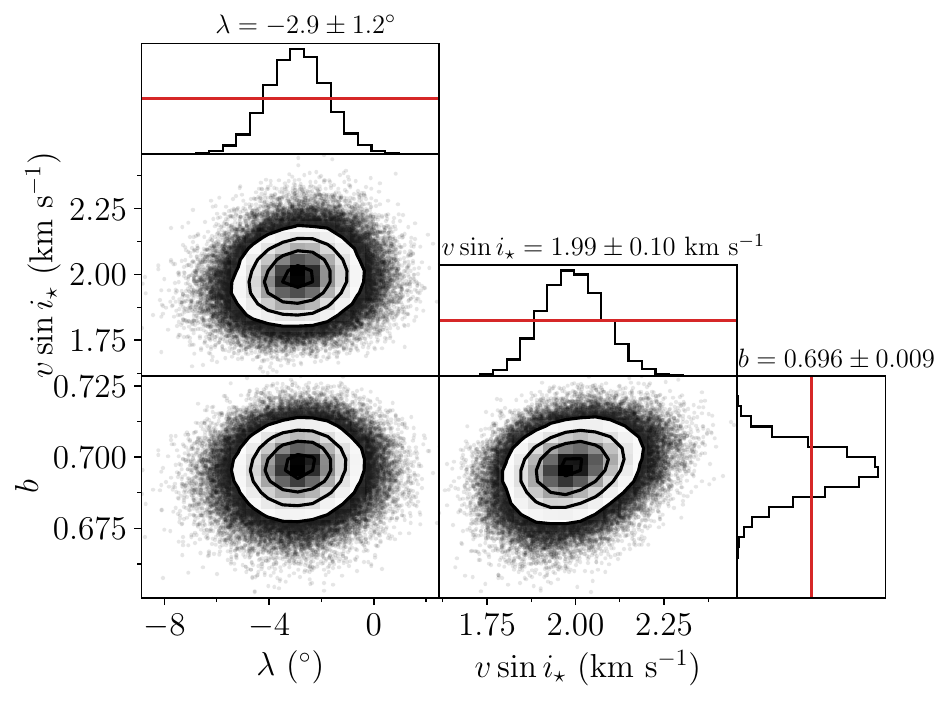} \\ 
    \end{tabular}
    \caption{{\bf The RM-RV effect for WASP-50.} Same as for \figref{fig:rm_hd118203}, but with RVs from ESPRESSO.}
    \label{fig:rm_wasp50}
\end{figure*}

\begin{figure}[h]
    \centering
    \includegraphics[width=\columnwidth]{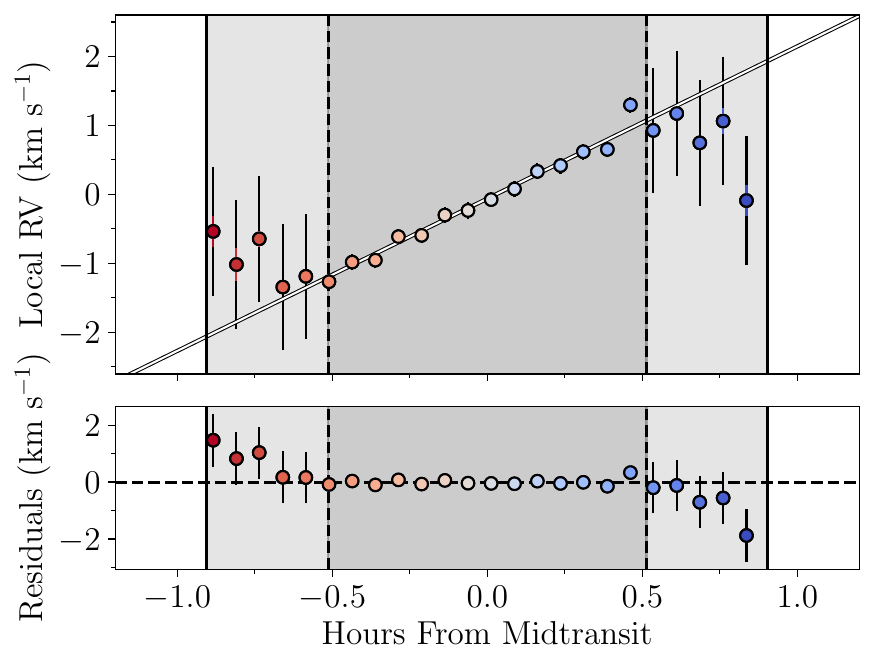}
    \caption{{\bf Subplanetary velocities for WASP-50.} Subplanetary velocities for WASP-50 with color-coding as described in \figref{fig:slope_shadow}. The dark shaded area correspond to the time the planet is completely within the stellar disk and the light shaded area denote ingress and egress. The errorbars include a jitter term, where two separate terms have been added in quadrature; one for the epochs for the full transit and one for the epochs when the planet is partially outside of the limb of the star. The horizontal black lines in the top panel mark \vsini from the fit.}
    \label{fig:slope_wasp50}
\end{figure}

The high-quality data prompted us to investigate the RM effect further.
We examined the subplanetary velocitiesm, which were shown in \figref{fig:slope_shadow}, but 
in \figref{fig:slope_wasp50} 
we display them in a slightly different manner (similar to the format
of \figref{fig:slope_k2287} for K2-287). As was the case for K2-287, we included two jitter terms, one for when the planet's silhouette is on or near the limb, and another when it is completely within the stellar disk. The result of our analysis using the subplanetary velocities was \vsini$=2.77\pm0.14$~km~s$^{-1}$ and $\lambda=$~\waspfiftys$^\circ$. 
The agreement with the RV-RM result is reassuring, but for our best estimate
we chose the result \waspfiftyu$^\circ$ from the RV-RM analysis.

\subsubsection{Convective blueshift}\label{sec:cb}

This term ``convective blueshift''
refers to the net blueshift of the disk-integrated stellar spectrum
caused by the imbalance in brightness between different components of convective cells. The upwelling gas is hotter, brighter, and blueshifted, while
the sinking gas is cooler, fainter,
and redshifted.
\citet{Shporer2011} showed that
this effect contributes to
the anomalous RVs with amplitudes of a few m~s$^{-1}$, and variability
on a timescale of
several minutes.
Within our sample, WASP-50 is the only system for which the quality of the data is high enough to
contemplate detecting and modeling the
convective blueshift.

We modelled the convective blueshift as described by \citet{Shporer2011}. In the model, the stellar disk is pixelated, and each pixel
is assigned an intensity according
to a limb-darkening law
and a net convective velocity
($V_{\rm CB}$) directed toward
the center of the star.
The radial component of each blueshifted surface element is therefore $V_{\rm CB}\cos \theta$
where $\theta$ is the angle between
the normal to the stellar surface
and our line of sight. The value of $V_{\rm CB}$ is expected to range from about $-200$~m~s$^{-1}$ for K-type stars to $-1000$~m~s$^{-1}$ for F-type stars \citep{Dravins1990a,Dravins1990b,Dravins1999}. The solar value
is about $-300$~m~s$^{-1}$ \citep{Dravins1987}. Since
WASP-50 is similar to the Sun,
with $T_{\rm eff}=5400\pm100$~K,
a reasonable expectation is $V_{\rm CB}\approx -300$~m~s$^{-1}$.

\begin{figure}[h]
    \centering
    \includegraphics[width=\columnwidth]{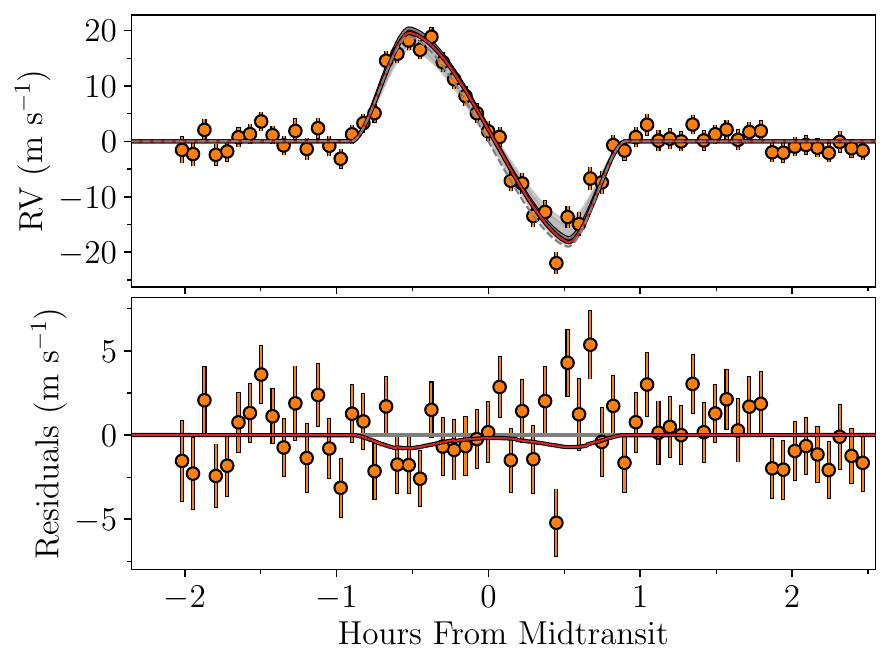}
    \caption{{\bf Convective blueshift in WASP-50.} Same as the left panel of \figref{fig:rm_wasp50}, but also showing a model as a red solid line that includes the effect of convective blueshift ($V_{\rm CB}=-300$~m~s$^{-1}$).}
    \label{fig:cb_w50}
\end{figure}

Since the anomalous RV caused by the convective  blueshift is smallest near the limb, and WASP-50~b has a high impact
parameter ($b\approx0.7$), the overall effect of the convective blueshift is expected to be rather weak, as shown in \figref{fig:cb_w50}. Nevertheless, we tested whether the inclusion of the convective blueshift in our RV-RM model leads to a substantial improvement in the quality of the fit
to the data,.
We created a grid of $\lambda$ values between $-45^\circ$ and $+45^\circ$ in steps of $0.25^\circ$. For each
choice of $\lambda$, we computed the Bayesian information criterion \citep[BIC;][]{Schwarz1978} for RV-RM models with and without
the convective blueshift. The standard model has a minimum BIC for $\lambda=-2.9^\circ$ (as expected given our result of \waspfiftyu$^\circ$), whereas the model with the convective blueshift has a minimum BIC for $\lambda=-3.9^\circ$. The difference in the optimized BIC values is 4.5, 
favoring the model without the convective blueshift.
We calculated the difference in BIC ($\Delta$BIC) between the two models at each $\lambda$. $\Delta$BIC changes sign around $\lambda=-4.9^\circ$ meaning that the net effect of including convective blueshift results in slightly lower (more misaligned) values for $\lambda$.

Even though we did not
find that including the convective
blueshift significantly improves the fit, we decided to
repeat our RV-RM analysis
after including the convective
blueshift. For convenience,
we did not jointly fit the \tess light curve; we simply held $b$, $R_{\rm p}/R_\star$, $P$, and $T_0$ fixed at the best-fit values from our previous run above. We applied a Gaussian prior on $V_{\rm CB}$ centered on $-400$~m~s$^{-1}$ with a width of $100$~m~s$^{-1}$. From this run we obtained $\lambda=-3.5\pm1.3^\circ$ again suggesting a slightly more misaligned system. A smaller absolute value for $V_{\rm CB}$ (probably more in-line with the expected value for WASP-50) would make the difference in $\lambda$ from the two runs even smaller as such we do not expect convective blueshift to significantly affect the results.
Thus, when attempting to measure
$\lambda$ with a precision at the level
of one degree,
the results depend on how much one
trusts the model for the convective blueshift and its assumed amplitude.



\subsection{WASP-59}

The WASP-59 system was discovered by \citet{Hebrard2013}. The system consists of a warm Jupiter, WASP-59~b, orbiting a K-dwarf. We observed a transit of WASP-59~b with HIRES. The observations were carried out on November 23, 2016, starting at UT 05:19 to 09:46 with an exposure time of $787$~s. Including overhead the sampling time came out to $835$~s. The series of 20 observations spanned a transit. 

\tess observed the system in Sector 56 with a cadence of with a cadence of 2~min. Given the long time interval that elapsed between
our HIRES observations and the \tess observations, one might be concerned
about the extra uncertainty in the transit ephemerides.
Fortunately, we also
arranged for ground-based photometric observations of a transit a few weeks after our spectroscopic observation.
We used KeplerCam, which is mounted on the 1.2 m telescope at the Fred L. Whipple Observatory on Mount Hopkins, Arizona \citep{Szentgyorgyi2005}. We observed with an $r^{\prime}$ filter on the night starting that began on the night of December 8, 2016. The light curves are shown in \figref{fig:gridthree}.

\begin{figure*}
    \centering
    \begin{tabular}{c c}
         \includegraphics[width=\columnwidth]{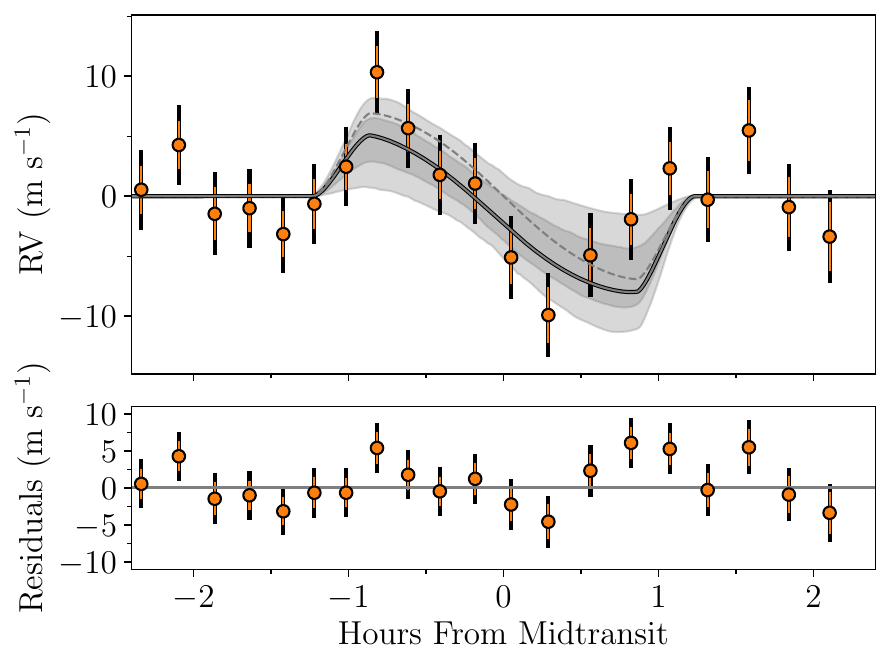} & \includegraphics[width=\columnwidth]{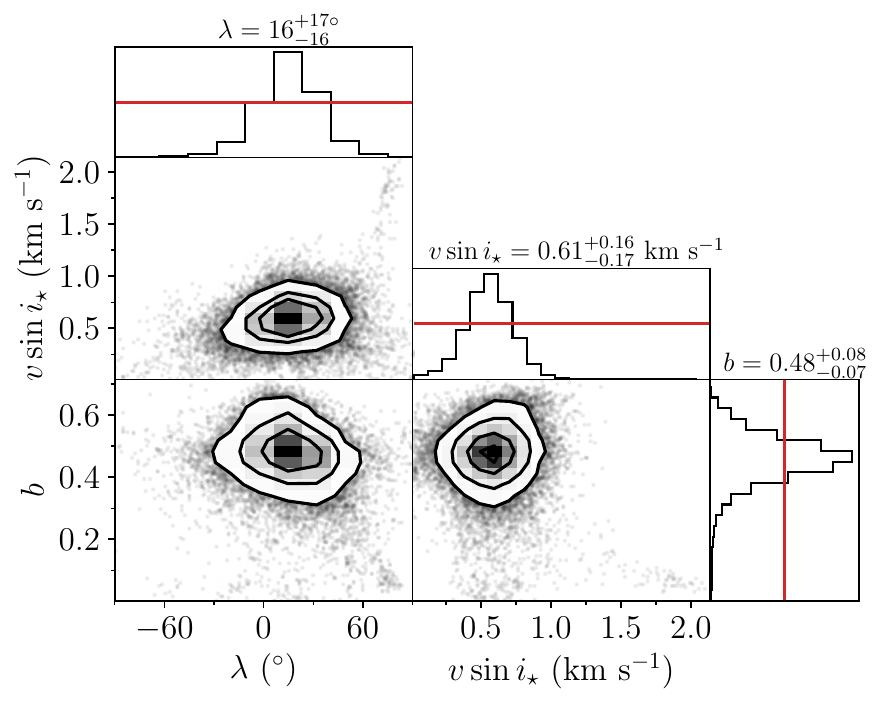} \\ 
    \end{tabular}
    \caption{{\bf The RM-RV effect for WASP-59.} Same as for \figref{fig:rm_hd118203}, but with RVs from HIRES.}
    \label{fig:rm_wasp59}
\end{figure*}

Because the HIRES spectra were
imprinted with iodine absorption lines,
we did not derive a value for \vsini with the BF method. Instead, we adopted the value of $2.3 \pm 1.2$~km~s$^{-1}$ reported by \citet{Hebrard2013}. When using
this result as a Gaussian prior,
we obtained $\lambda=$~\waspfiftynineg$^\circ$ and \vsini$=$\waspeightsixvsinig~km~s$^{-1}$, clearly suggesting that the Gaussian prior is not particularly informative. Indeed assuming a uniform prior on \vsini we got \vsini$=$\waspeightsixvsiniu~km~s$^{-1}$ and $\lambda=$~\waspfiftynineu$^\circ$, which we adopt as a out final value for $\lambda$. The resulting RM curve and posteriors are shown in \figref{fig:rm_wasp59}.



\subsection{WASP-136}

WASP-136~b is an inflated hot Jupiter orbiting an F9 subgiant star. The system was discovered by \citet{Lam2017}. \tess data are available with 2-min cadence
in Sector 29 and 20-sec cadence
in Sector 41. The light curve is shown in \figref{fig:gridthree}.

We observed a transit with FIES on the night starting August 30, 2021, with observations running from UT 23:00 to UT 05:50  August 31, 2021. We opted for an exposure time of 1140~s, yielding a sampling time of 1320~s. 

\begin{figure*}[h]
    \centering
    \begin{tabular}{c c}
         \includegraphics[width=\columnwidth]{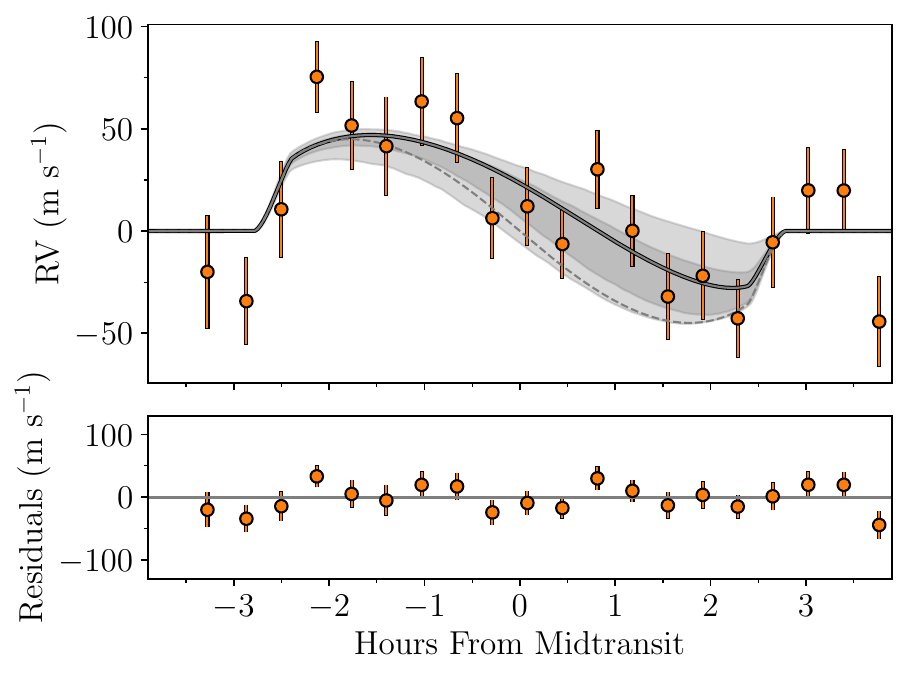} & \includegraphics[width=\columnwidth]{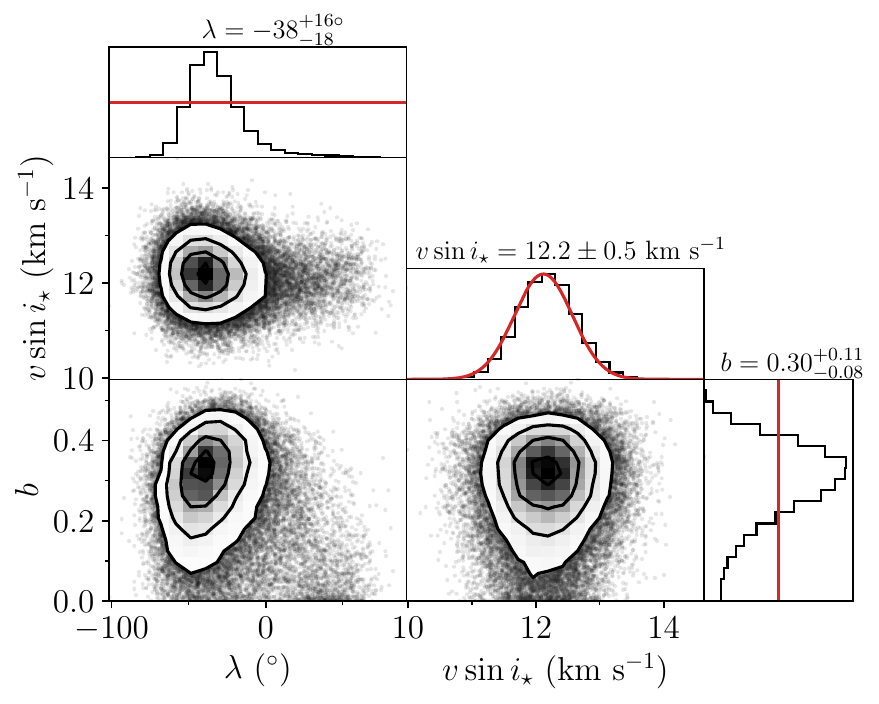} \\ 
    \end{tabular}
    \caption{{\bf The RV-RM effect for WASP-136.} Same as for \figref{fig:rm_hd118203}. Here also with RVs from FIES.}
    \label{fig:rm_wasp136}
\end{figure*}

Both of the RV-RM fits resulted in a significant misalignment for the WASP-136 system. We obtained
$\lambda=$~\waspthirtyeightu$^\circ$ when applying a uniform prior on \vsini, and \waspthirtyeightg$^\circ$ when applying a Gaussian prior. We adopt the latter as our best estimate of the projected obliquity. 




\subsection{WASP-148}\label{obs:wasp148}

The WASP-148 system was discovered by \citet[][]{Hebrard2020}. The system consists of two giant planets orbiting a late-G dwarf with strong TTVs.
The inner planet, WASP-148~b, is
a transiting hot Jupiter with a period of 8.8 days. The star's projected obliquity with respect to the orbit of planet b was measured by \citet{Wang2022} to be $\lambda=-8^{+9}_{-10}$$^\circ$.

We observed a transit of WASP-148~b using HARPS-N on the night starting March 21, 2021, with observations starting around UT 01:45 to 06:15 March 22, 2021. The exposure time was set to 900~s, resulting in a sampling time of 920~s. Because the strong TTVs lead to extra uncertainty in the
transit ephemerides, we also arranged for simultaneous photometric
observations of the transit
with the Multicolor Simultaneous Camera for studying Atmospheres of Transiting exoplanets \citep[MuSCAT-2;][]{Narita2015} mounted on the Telescopio Carlos S\'{a}nches, Tenerife, Spain. MuSCAT-2 allows for simultaneous observations in four filters; we used a $griz$ filter set. The light curves, after detrending, are shown in \figref{fig:lc_wasp148}. When modeling these light curves, we did not use GP regression. The limb-darkening coefficients tabulated by \citet{Claret2013} for these filters
and for the appropriate stellar parameters are reproduced in \tabref{tab:wasp148_phot}.

\begin{figure*}[h]
    \centering
    \begin{tabular}{c c}
         \includegraphics[width=\columnwidth]{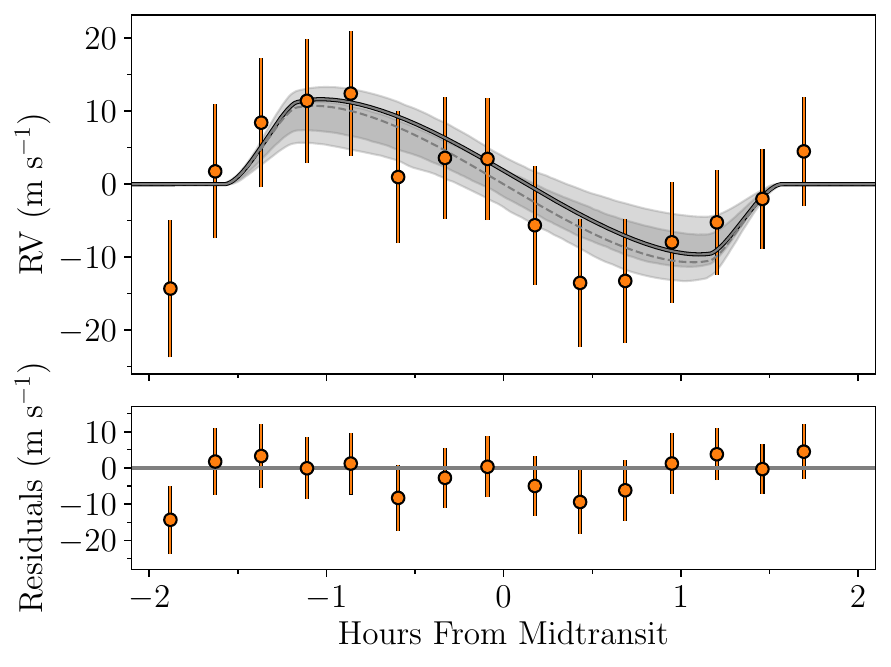} & \includegraphics[width=\columnwidth]{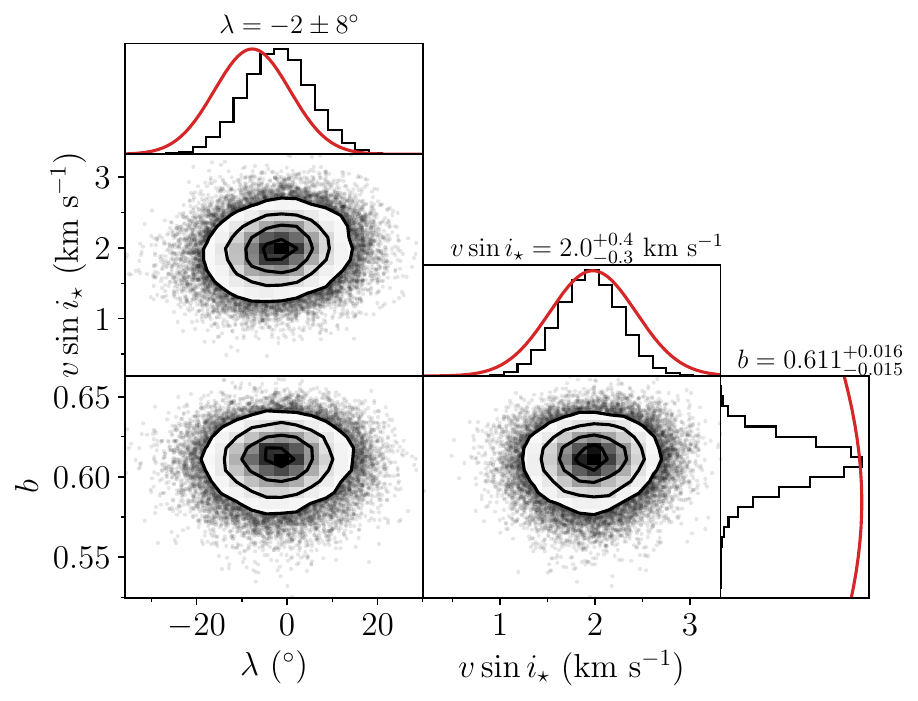} \\ 
    \end{tabular}
    \caption{{\bf WASP-148.} Same as for \figref{fig:rm_hd118203}, but with RVs from HARPS-N.}
    \label{fig:rm_wasp148}
\end{figure*}

At first, we fitted the RM effect for this system jointly with the
\tess and MuSCAT-2 light curves,
allowing the mid-transit time of each transit to be a freely adjustable parameter (as we did for TOI-1130, \sref{obs:toi1130}).
We also tried fitting only the MuSCAT-2 light curves and not the \tess light curve. In both cases, we found $\lambda=16\pm40^\circ$, which is consistent with, but less precise than, the measurement reported by \citet{Wang2022}. 

\citet{Almenara2022} performed a ``photodynamical'' analysis of the system, in which the light curve is fitted with a model in which all the mutual gravitational interactions between the star and both planets
are taken into account.
They used the data presented by \citet{Hebrard2020} along with
additional ground-based data and
\tess data. Given their comprehensive modeling effort, we decided to use the parameters from \citet{Almenara2022} as priors, and fitted only the HARPS-N RV measurements and the MuSCAT-2 light curve. The free parameters relevant to the light curve were the mid-transit time, $a/R_\star$, $\cos i$, $R_{\rm p}/R_\star$, and the sum-of-limb-darkening coefficients for each filter. The period was held fixed. In addition, we included a Gaussian prior on the projected obliquity using the value of $\lambda=-8^{+9}_{-10}$$^\circ$ as to incorporate all the information available for the system. Again, we performed two MCMCs obtaining a value of \waspfortyeightu$^\circ$ for both the run with a uniform and the one with a Gaussian prior on \vsini, respectively. 


{\small\addtolength{\tabcolsep}{-3pt}
\begin{table}[]
    \centering
    \begin{threeparttable}
    \caption{WASP-148, photometry only.}
    \label{tab:wasp148_phot}
    \begin{tabular}{l c c}
        \toprule
        Parameter & Value & Prior \\ 
        \midrule
        $T \rm _{0} \ (BJD)$ & $2459295.6598\pm0.0002$ & $\mathcal{U}(2459295,2459296)$ \\ 
        $a/R_\star$ & $19.7^{+0.4}_{-0.3}$ & $\mathcal{U}(10,30)$ \\ 
        $R_\mathrm{p}/R_\star$ & $0.0853^{+0.0008}_{-0.0007}$ & $\mathcal{U}(0,1)$ \\ 
        $q_1 + q_2: g$ & $0.97\pm0.05$ & $\mathcal{N}(0.781,0.1)$ \\ 
        $q_1 + q_2: i$ & $0.78 \pm 0.06$ & $\mathcal{N}(0.613,0.1)$ \\ 
        $q_1 + q_2: r$ & $0.83 \pm 0.05$ & $\mathcal{N}(0.698,0.1)$ \\ 
        $q_1 + q_2: z_s$ & $0.60 \pm 0.06$ & $\mathcal{N}(0.547,0.1)$ \\ 
        $\cos i $ & $0.011_{-0.011}^{+0.004}$ & $\mathcal{U}(0.0,0.5)$ \\ 
        \midrule
        Filter & $q_1$ & $q_2$ \\
        \midrule
            $g$ & 0.73 & 0.05 \\
            $r$ & 0.53 & 0.17 \\
            $i$ & 0.40 & 0.22 \\
            $z_s$ & 0.23 & 0.32 \\
        \bottomrule        
\end{tabular}
\begin{tablenotes}
    \item Top: Results from an MCMC using only the MuSCAT-2 photometry for the transit of WASP-148~b. Bottom: The limb-darkening coefficients for each filter using \citet{Claret2013}.
\end{tablenotes}
\end{threeparttable}    

\end{table}
}


As this is a dynamically interesting system, for which a large
collection of transit times over
a long interval
can help to constrain the parameters of the planetary system, we measured
the mid-transit time based on a fit solely to the MuSCAT-2 light curves.
The results are given in
\tabref{tab:wasp148_phot}.



\subsection{WASP-172}

WASP-172 is an F star with a
bloated hot Jupiter discovered by \citet{Hellier2019}. Data
from \tess are available with 30-min cadence from Sector 11 and 20-sec cadence from Sectors 37 and 38. The \tess light curve is shown in \figref{fig:gridthree}.

\begin{figure*}[h]
    \centering
    \begin{tabular}{c c}
         \includegraphics[width=\columnwidth]{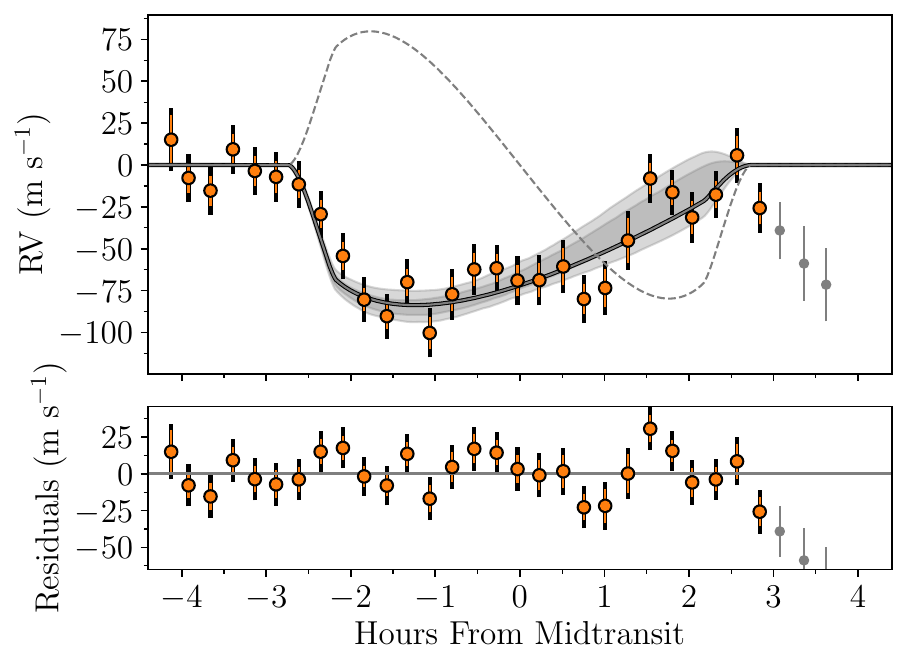} & \includegraphics[width=\columnwidth]{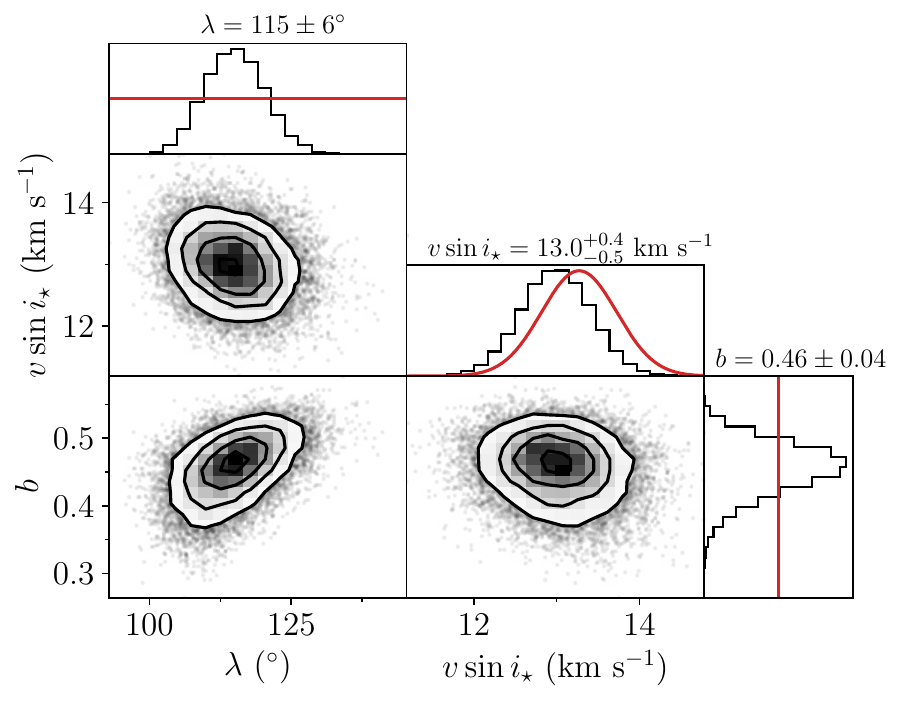} \\ 
    \end{tabular}
    \caption{{\bf WASP-172.} Same as for \figref{fig:rm_hd118203}, but with RVs from ESPRESSO. The gray data points were excluded in the fit due to high airmass.}
    \label{fig:rm_wasp172}
\end{figure*}

On the night starting June 1, 2022, we observed a transit using ESPRESSO with observations being carried out from UT 23:00 to UT 06:40. We opted for an exposure time of 900~s, resulting in a sampling time of 940~s. The RVs are shown in \figref{fig:rm_wasp172}, where three data points are marked with gray points as they were excluded from the analysis, since they were taken at high airmass.

\begin{figure*}[h]
    \centering
    \includegraphics[width=\textwidth]{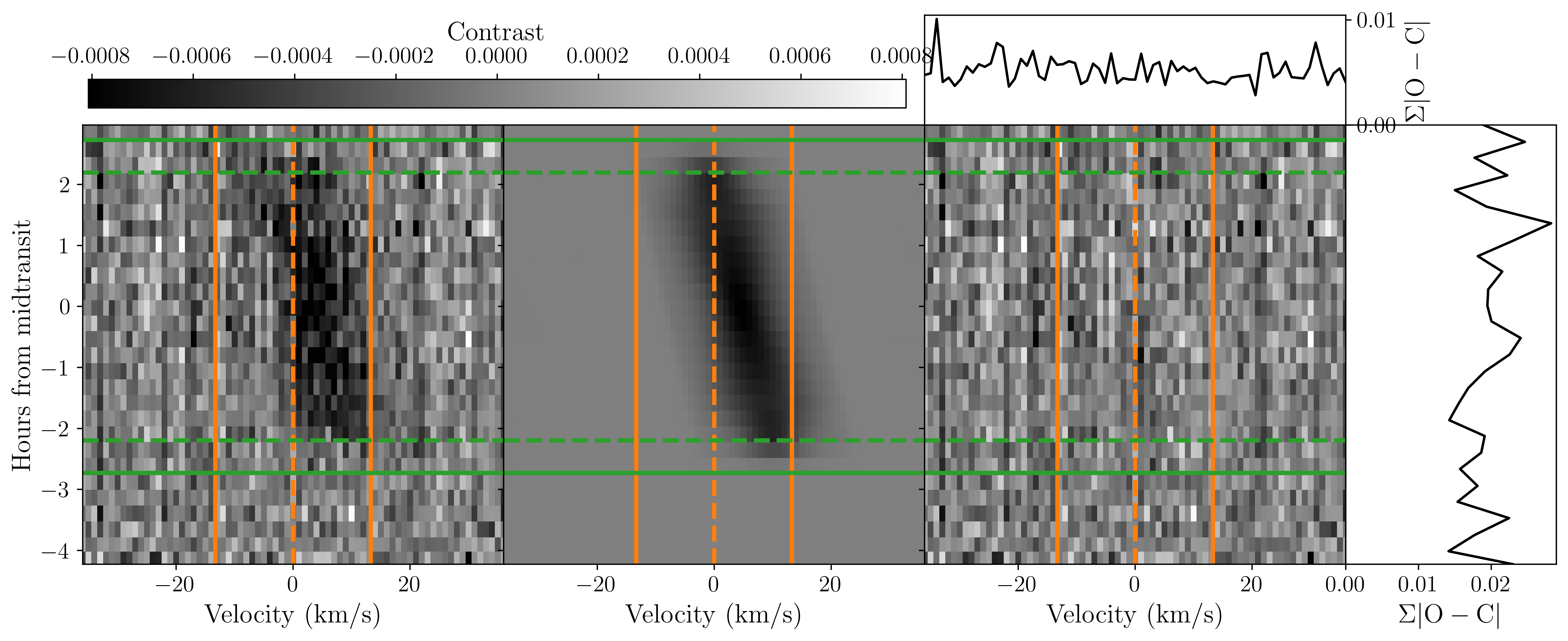}
    \caption{{\bf Doppler shadow for WASP-172.} Same as in \figref{fig:shadow_toi1836}, but for WASP-172.}
    \label{fig:shadow_wasp172}
\end{figure*}

By fitting the RV-RM effect and the
\tess light curve, we found
$\lambda=$~\waspseventwou$^\circ$ and \waspseventwog$^\circ$ for $\lambda$ when applying a uniform prior or a Gaussian prior on \vsini, respectively. Given the high quality of the data, and the relatively rapid
rotation of the star (\vsini~$=$\waspseventwovsinig~km~s$^{-1}$), we decided to analyze the Doppler shadow.
The results are displayed in
\figref{fig:shadow_wasp172}. The Doppler-shadow analysis gave $\lambda=$~\waspseventwos$^\circ$, consistent with the RV-RM results.
We adopted \waspseventwog$^\circ$ as our best estimate for $\lambda$,
based on the RV-RM fit with a Gaussian prior on \vsini.

The ESPRESSO transit spectroscopy presented here was also used 
by \citet{Seidel2023}
in an investigation of the planet's atmosphere. They reported the detection of sodium and hydrogen absorption, as well as a tentative detection of iron absorption. In their study, a value of $\lambda=121\pm13^\circ$ was adopted to correct for the RM effect, which was
taken from our Doppler-shadow analysis prior to the completion of this paper.



\subsection{WASP-173A}

WASP-173Ab is a hot Jupiter discovered by \citet{Hellier2019}. The host star is of spectral type G3V and has a wide-orbiting stellar companion. 
The planet was also dubbed
KELT-22Ab by \citet{Labadie2019},
who reported an independent discovery of the transits using data from the KELT data. Here, we use the name WASP-173Ab. 

\begin{figure*}[h]
    \centering
    \begin{tabular}{c c}
         \includegraphics[width=\columnwidth]{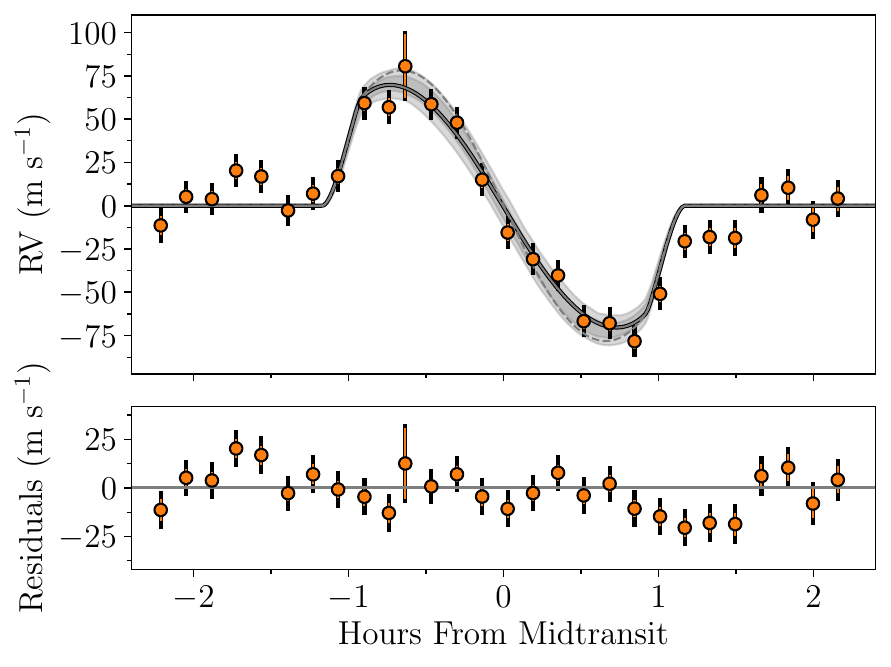} & \includegraphics[width=\columnwidth]{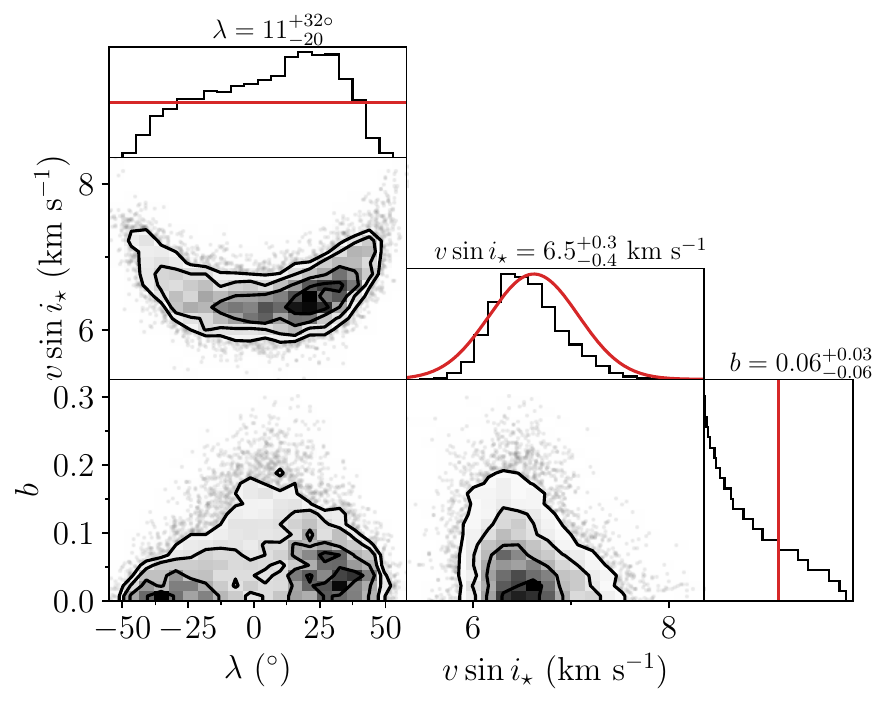} \\ 
    \end{tabular}
    \caption{{\bf The RV-RM effect for WASP-173A.} Same as for \figref{fig:rm_hd118203}, but with RVs from ESPRESSO.}
    \label{fig:rm_wasp173}
\end{figure*}

\tess observed the system twice.
Data with 2-min cadence are available from Sector 2, and with 20-sec cadence
from Sector 29. The \tess light curve is shown in \figref{fig:gridthree}. We observed a transit on the night starting July 23, 2022, with observations beginning at UT 05:40 until UT 10:00. The exposure time was set to 555~s, yielding a sampling time of 590~s. The RVs are shown to the left in \figref{fig:rm_wasp173}.

\begin{figure*}[h]
    \centering
    \includegraphics[width=\textwidth]{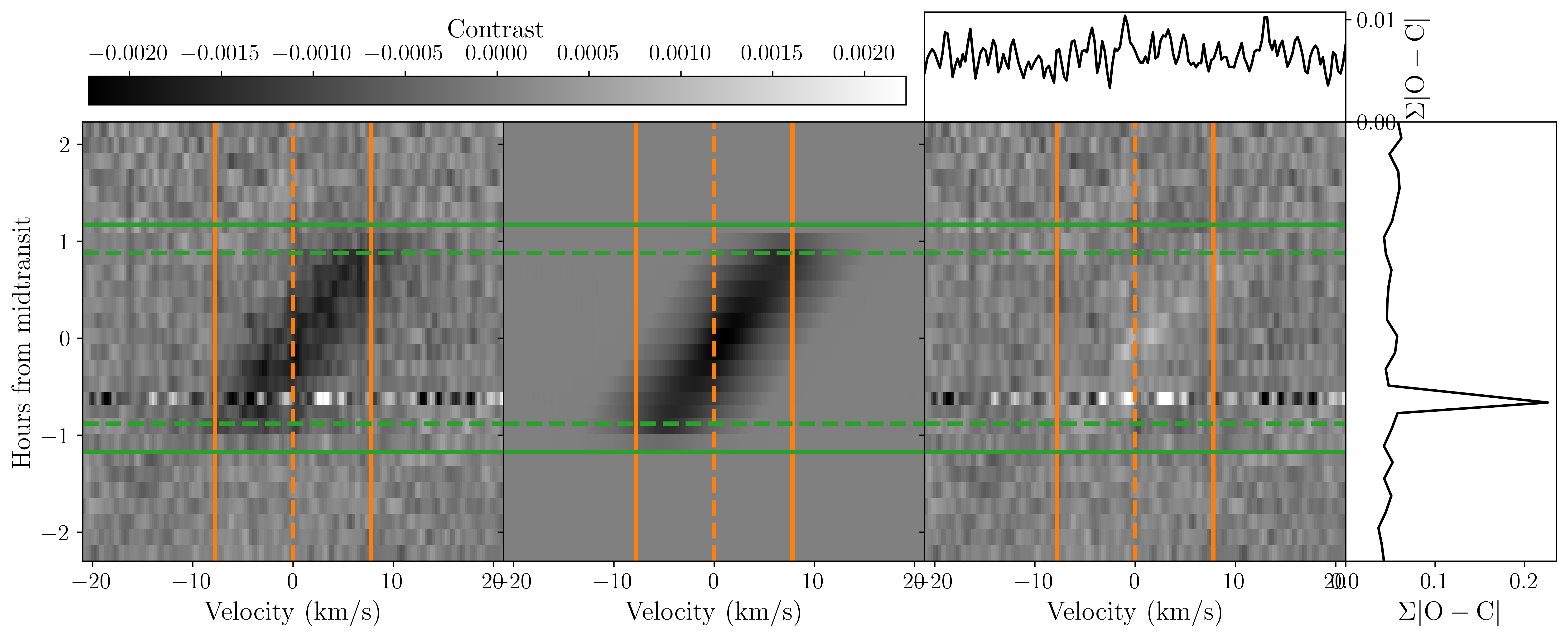}
    \caption{{\bf Doppler shadow for WASP-173A.} Same as in \figref{fig:shadow_toi1836}, but for WASP-173A. }
    \label{fig:shadow_wasp173}
\end{figure*}

Our analysis followed similar steps as for WASP-172 (see the previous section). We performed the usual RV-RM
analyses and also an analysis
of the Doppler shadow, for
which the results
are shown in \figref{fig:shadow_wasp173}. The RV-RM results were $\lambda=$~\waspseventhreeu$^\circ$ when applying a uniform prior on \vsini and \waspseventhreeg$^\circ$ for a Gaussian prior, while the shadow run resulted in a value for $\lambda$ of \waspseventhrees$^\circ$.

The \vsini derived from the Doppler-shadow analysis was \waspseventhreevsinis~km~s$^{-1}$, which is a bit higher than
the results obtained from the
RV-RM anlayses ($6.6_{-0.8}^{+0.6}$~km~s$^{-1}$) and a Gaussian prior (\waspseventhreevsinig~km~s$^{-1}$). The RV-RM-based values are closer to the value derived from the BF of 6.6~km~s$^{-1}$. As our best estimate
of $\lambda$, we therefore adopt the value of \waspseventhreeg$^\circ$ from our run applying a Gaussian prior on \vsini.



\subsection{WASP-186}\label{obs:186}

\citet{Schanche2020} discovered the WASP-186~b, a massive hot Jupiter with an eccentric orbit around a mid-F star. As noted by \citet{Schanche2020}, WASP-186 was observed by \tess in Sector 17 and produced data with a 30-min cadence. In addition,
\tess observed this system in Sectors 42 and 57 and the data are available
with 2-min and 20-sec cadence, respectively. The light curves are shown in \figref{fig:gridthree}.

\begin{figure*}[h]
    \centering
    \begin{tabular}{c c}
         \includegraphics[width=\columnwidth]{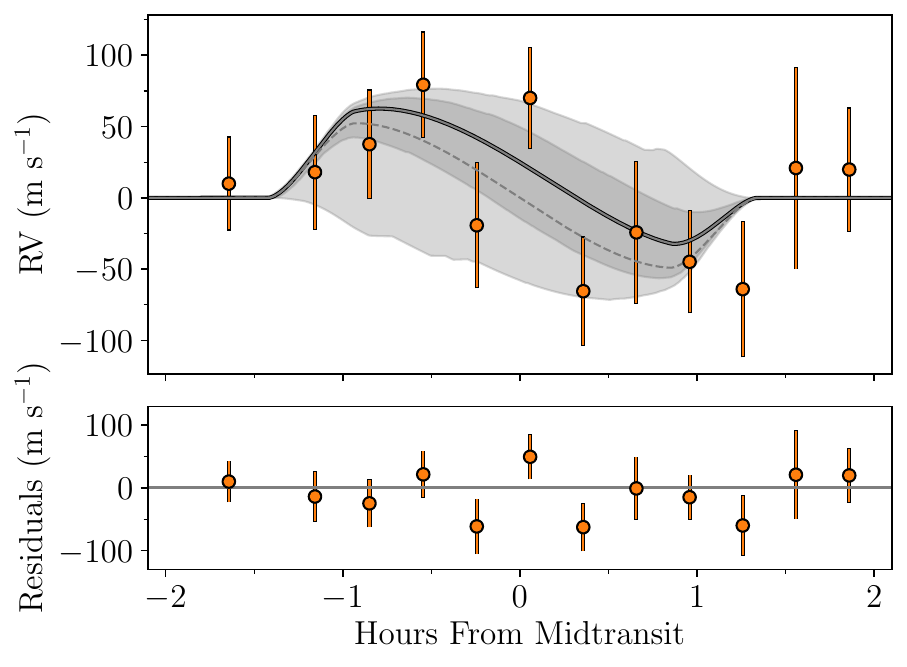} & \includegraphics[width=\columnwidth]{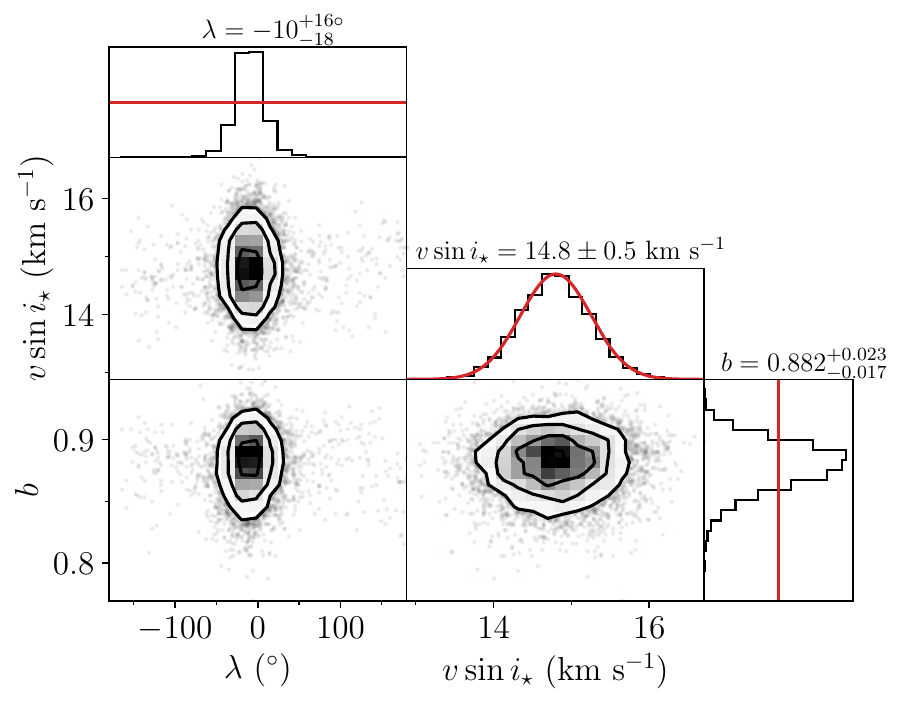} \\ 
    \end{tabular}
    \caption{{\bf The RV-RM effect for WASP-186.} Same as for \figref{fig:rm_hd118203}. Here also with RVs from FIES.}
    \label{fig:rm_wasp186}
\end{figure*}

Using FIES, we observed a transit on the night October 11, 2021, with observations starting UT 22:30 and continuing until UT 02:50 October 12, 2021. The exposure time was set to 900~s, and the time between mid-exposures was 1080~s. The transit observations are shown in \figref{fig:rm_wasp186}.

Our two RV-RM fits gave $\lambda=$~\waspeightsixu$^\circ$ and $\lambda=$~\waspeightsixg$^\circ$ for the run with a uniform and the one with a Gaussian prior on \vsini, respectively. The values are in  agreement and the prior on
\vsini leads to a more precise result for $\lambda$. Therefore, we chose $\lambda=$~\waspeightsixg$^\circ$ as our best estimate for WASP-186.



\subsection{XO-7}\label{obs:xo7}

XO-7~b was discovered by \citet{Crouzet2020} as part of the XO project \citep{McCullough2005}. The transiting planet is a hot Jupiter.
There is also a massive companion on an orbit with a period
of at least 2 years around the G2V host star \citep[][]{Crouzet2020}. \tess data are available with 30-min cadence in Sectors 18, 19, and 20;
with 2-min cadence in
Sectors 40, 47, 53, 53;
and with 20-sec cadence
in Sectors 59 and 60. The \tess light curves are shown in \figref{fig:gridthree}.

\begin{figure*}[h]
    \centering
    \begin{tabular}{c c}
         \includegraphics[width=\columnwidth]{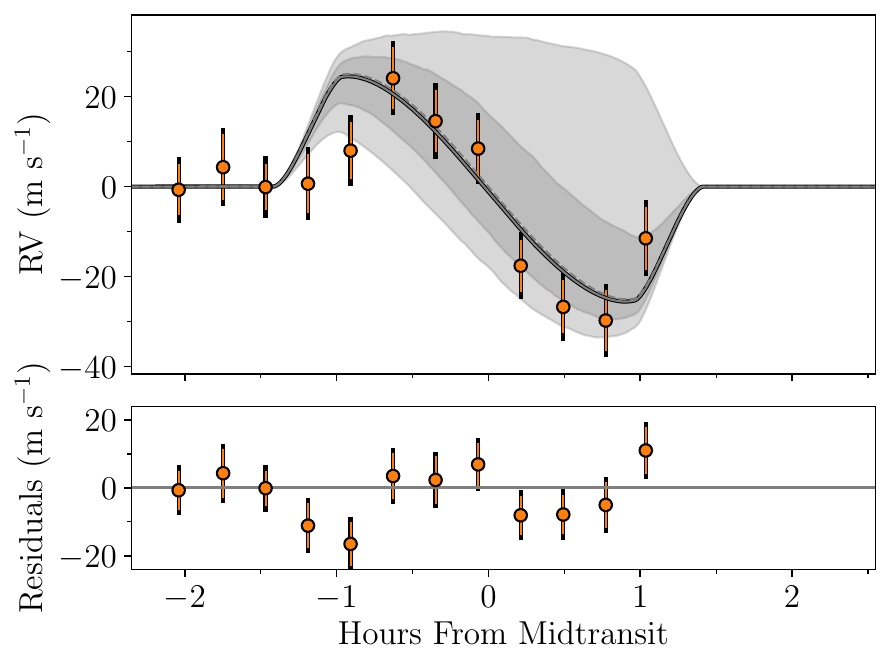} & \includegraphics[width=\columnwidth]{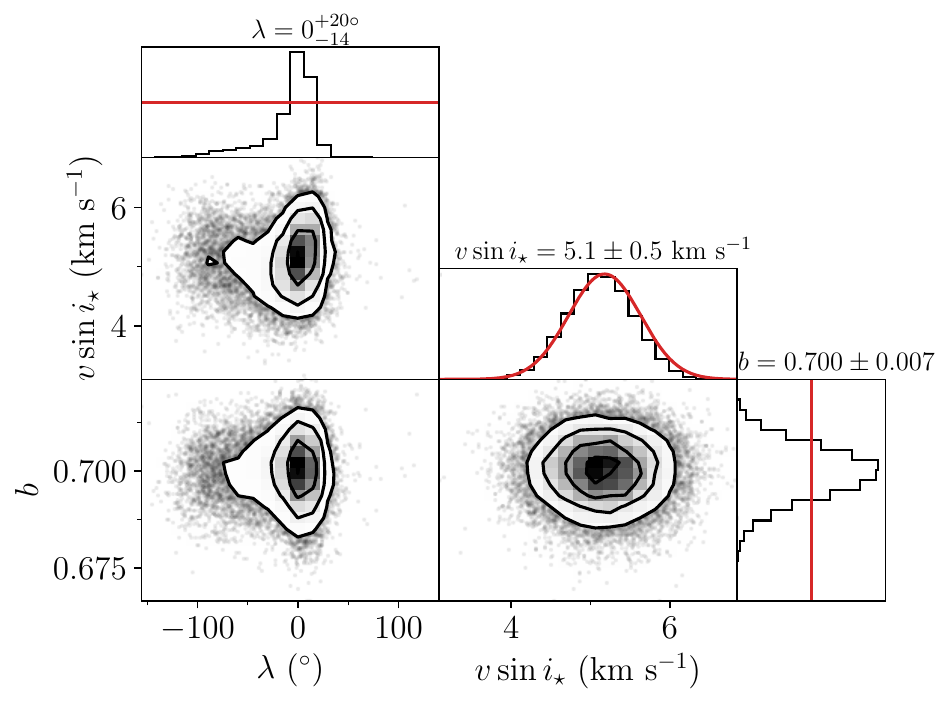} \\ 
    \end{tabular}
    \caption{{\bf The RV-RM effect for XO-7.} Same as for \figref{fig:rm_hd118203}. Here also with RVs from FIES.}
    \label{fig:rm_xo7}
\end{figure*}

We observed a transit of XO-7~b with FIES on the night starting on August 27, 2022, with observations starting around UT 00:15 and continuing until 04:00 August 28, 2022. We opted for an exposure time of 840~s, resulting in a sampling time of 1000~s. The RVs from the transit night are shown in \figref{fig:rm_xo7}. The egress was not completely covered, and no  observations were made after the egress.

The RV-RM results for $\lambda$ are \xosevenu$^\circ$ and \xoseveng$^\circ$ for the runs with a uniform and a Gaussian prior on \vsini, respectively, with the posterior distribution for the latter case shown in \figref{fig:rm_xo7}. Despite the incomplete coverage of the transit, the high impact parameter of $b\sim0.7$ enabled decent constraints to be placed upon $\lambda$.

The posterior for $\lambda$,
shown on the right side of
\figref{fig:rm_xo7}, has a tail on the left side extending toward $-90^\circ$, allowing for the possibility that system is strongly misaligned. This tail is probably caused by the lack of egress/post-egress observations and also shows up in the 2$\sigma$ contours in the left panel of \figref{fig:rm_xo7}. We wondered if this unusual feature is related to the
lack of egress and post-egress observations. We investigated the issue by examining the stacked spectral CCFs. The stacked CCFs showed contours that largely agree with the posteriors from our RV-RM. However, a peak is seen at around (\vsini,$\lambda$)=($4$~km~s$^{-1}$,$-60^\circ$) that is more consistent with the negative tail of the
posterior from the RV-RM analysis,  favoring a more misaligned system. A second and more complete transit observation seems warranted. For now, we adopted \xoseveng$^\circ$ as our best estimate, based on the run with a Gaussian prior on \vsini.



\subsection{WASP-26}\label{obs:w26}

\begin{figure*}[ht]
    \centering
    \begin{tabular}{c c}
         \includegraphics[width=\columnwidth]{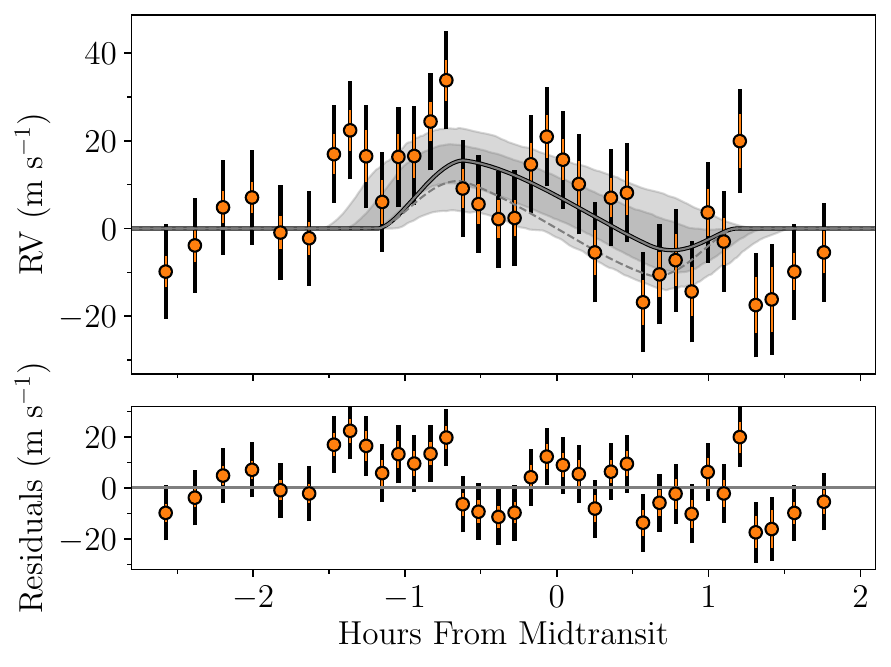} & \includegraphics[width=\columnwidth]{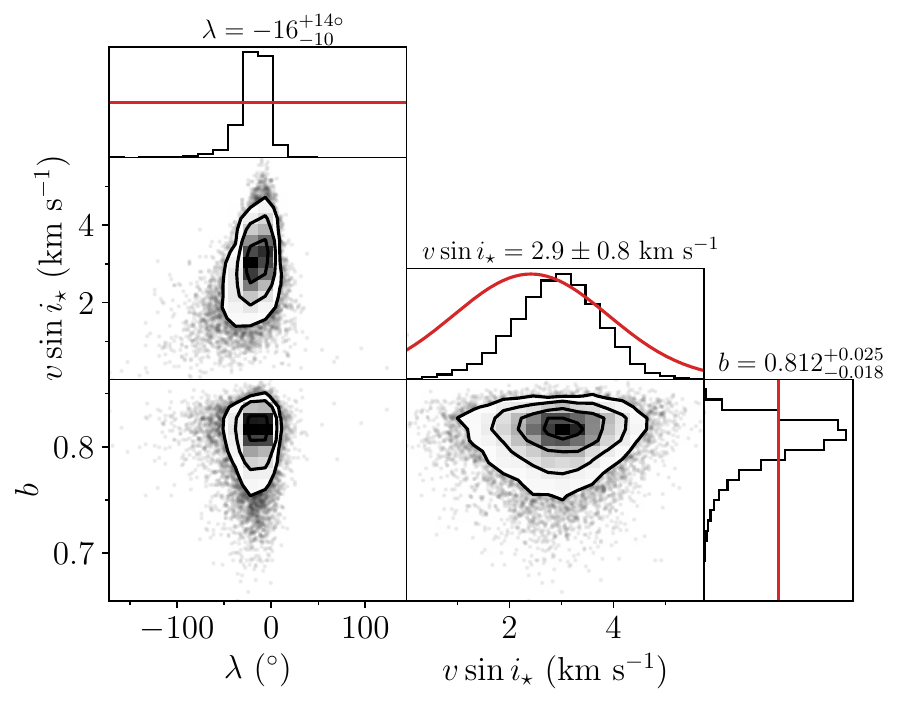} \\ 
    \end{tabular}
    \caption{{\bf The RV-RM effect for WASP-26.} Same as for \figref{fig:rm_hd118203}, but with RVs from PFS.}
    \label{fig:rm_wasp26}
\end{figure*}

WASP-26~b is a hot Jupiter orbiting an early G-type star, discovered by \citet{Smalley2010}. \citet{Albrecht2012} analyzed the RV-RM effect based on data obtained with the Planet Finder Spectrograph \citep[PFS;][]{Crane2010} mounted at the 6.5~m Magellan II telescope. The RV time series is shown in the left
half of \figref{fig:rm_wasp26}. 
They reported $\lambda=-34^{+36}_{-26}$$^\circ$.

This case is unlike all of other cases
presented in this paper in that we did not obtain any new data. Rather,
we wanted to take
advantage of \tess data that
did not exist at the time of the prior study, and correct an error that was
made in the prior study. The error
was the adoption of an incorrect
prior constraint on the transit
duration ($T_{12}$, in the terminology
of \citealt{Albrecht2012}),
which can be seen in their Table 3.
This error resulted in an incorrect
value of $b$ that affected
the results for $\lambda$.

Rather than placing a prior on $T_{21}$, we followed our usual
procedure of fitting the \tess light curve jointly with the RVs. We used
\tess 20-sec data from Sector 29, shown in \figref{fig:lc_wasp26}). 
As in \citet{Albrecht2012}, we applied a Gaussian prior on \vsini of $2.4\pm 1.5$~km~s$^{-1}$, and as usual,
we applied Gaussian priors on the macro- and microturbulent velocities of $4.0\pm1.0$~km~s$^{-1}$ and $1.1\pm1.0$~km~s$^{-1}$, respectively. We also applied a Gaussian prior on $T_0$ taken from \citet{Anderson2011},
as did \citet{Albrecht2012}. The reason for this choice is
that about 10 years elapsed
between the spectroscopic transit observations and the \tess observations; small systematic uncertainties in the \tess-based ephemeris might propagate into large uncertainties in the relevant
transit time. We found a projected obliquity of $-16^{+14}_{-10}$$^\circ$. \figref{fig:rm_wasp26} shows the RV-RM curve and posteriors. 




\subsection{Results}\label{sec:results}

\tabref{tab:lam_res} gives our best estimates of $\lambda$ and \vsini
for each system.
For cases in which the rotation
period could be measured, the
table also includes the stellar inclination, $i_\star$, calculated as per \citet{Masuda2020},
and the true obliquity.
Other key parameters
are given in
\tabref{tab:poster}.

\begin{table}
    \centering
    \begin{threeparttable}
    \caption{Final results for \vsini, $\lambda$, $i_\star$, and $\psi$.}
    \label{tab:lam_res}
    \begin{tabular}{l c c c c c}
      \toprule
     System  & \vsini\tnote{a} & $\lambda$ & $i_\star$ & $\psi$ \\
      & (km~s$^{-1}$) & ($^\circ$) & ($^\circ$) & ($^\circ$) \\
     \midrule
     HD~118203 & \hdvsinig & \hdg & $17\pm2$ & $75^{+3}_{-5}$ \\
     HD~148193 & \hdtwovsinig & \hdtwos & \ldots & \ldots \\
     K2-261 & \ksixonevsinig & \ksixoneg & \ldots & \ldots \\
     K2-287 &  \keightsevenvsinig & \keightseveng & \ldots & \ldots \\
     KELT-3 & \keltthreevsinig & \keltthreeu & \ldots & \ldots \\
     KELT-4A & \keltfourvsinig & \keltfourg & \ldots &  \ldots \\
     LTT~1445A\tnote{b} & \lttvsinig & \lttg & \ldots & \ldots \\
     TOI-451A\tnote{b} & \toifourfivevsinig & \toifourfiveg & $69^{+11}_{-8}$ & $40^{+22}_{-28}$  \\
     TOI-813 & \toieightthirteenvsinig & \toieightthirteeng & \ldots & \ldots \\
     TOI-892 & \toieightninevsinis & \toieightnines & \ldots & \ldots \\
     TOI-1130 & \toithirtyvsinig & \toithirtyg & \ldots & \ldots \\
     WASP-50 &  \waspfiftyvsiniu & \waspfiftyu & \ldots & \ldots \\
     WASP-59 &  \waspfiftyninevsiniu & \waspfiftynineu &  \ldots & \ldots \\
     WASP-136 & \waspthirtyeightvsinig & \waspthirtyeightg & $38^{+7}_{-6}$ & $60\pm8$ \\
     WASP-148 & \waspfortyeightvsinig & \waspfortyeightg & $68^{+8}_{-21}$ & $21^{+9}_{-17}$ \\
    WASP-172 & \waspseventwovsinig & \waspseventwog & $73^{+9}_{-14}$ & $112\pm6$ \\
    WASP-173A & \waspseventhreevsinig & \waspseventhreeg & $71^{+10}_{-13}$ & $30\pm14$ \\
    WASP-186 & \waspeightsixvsinig & \waspeightsixg & $75^{+7}_{-15}$ & $22^{+11}_{-14}$ \\
    XO-7 & \xosevenvsinig & \xoseveng & $14.5^{+1.6}_{-1.4}$ & $70\pm1.7$ \\ 
    WASP-26 & $2.9\pm0.8$ & $-16^{+14}_{-10}$ &  \ldots & \ldots \\
     \bottomrule
    \end{tabular}
    \begin{tablenotes}
        \item[a] In general the \vsini value is largely constrained by the Gaussian prior of $\sigma = 0.5$~km~s$^{-1}$ applied.
        \item[b]  We caution that we might not have detected the RM effect in these systems and these measurements are therefore dubious.
    \end{tablenotes}
    \end{threeparttable}

\end{table}




\section{Discussion}\label{sec:Discussion}

\begin{figure*}[ht]
    \centering
    \includegraphics[width=1.\textwidth]{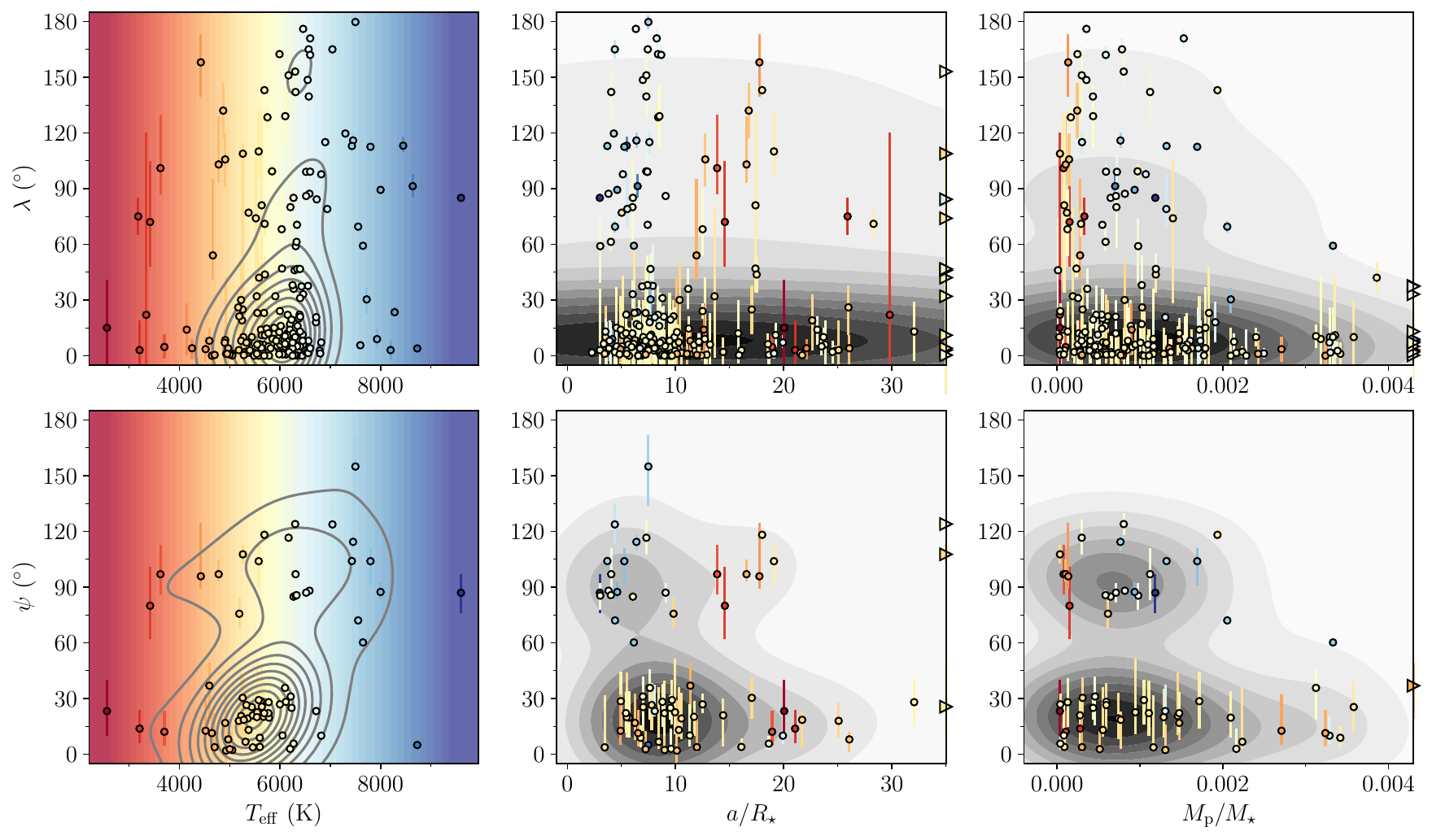}
    \caption{{\bf  Obliquities and projected obliquities} as a function of $T_{\rm eff}$ (left), $a/R_\star$ (center), and $M_{\rm p}/M_\star$ (right). Triangles indicate the a measurement whose value is outside of the plotted range. In all panels, the color of each data point conveys $T_{\rm eff}$ using the color scale shown in the lower left panel.
    The contours are KDEs illustrating the density of measurements in a given parameter space.}
    \label{fig:oblit}
\end{figure*}

We now return to the four questions posed in the introduction about the distribution of obliquities of
stars with transiting planets.
We combined the $\lambda$ measurements presented in this paper (and listed in \tabref{tab:lam_res}) with those measured previously.
Specifically, we used the catalog assembled by \citet{Albrecht2022} with the help of the TEPcat catalog \citep{Southworth2011}, and several other measurements
that have appeared in the more recent literature (up until February 26th, 2024).
\tabref{tab:lamlit} gives the key parameters for these additional systems, along with the systems presented in this study. \figref{fig:oblit} shows plots of $\lambda$ and $\psi$ as a function
of $T_{\rm eff}$, $a/R_\star$ and $M_{\rm p}/M_\star$.

As mentioned we excluded our results for LTT~1445Ab and TOI-451Ab in the ensemble analyses below. Furthermore as this study builds upon the sample presented in \citet{Albrecht2022} systems excluded in there are also excluded here (based on the criteria outlined in their Appendix A).

\subsection{A Preference for Polar Planets?}\label{sec:ppp}

\citet{Albrecht2021} investigated a sample of 57 systems for which $\psi$ has been determined by combining measurements of the projected obliquity ($\lambda$) and the stellar inclination angle ($i_\star$). Of the 19 values of $\psi$ that are inconsistent with $0^\circ$, 18 are between $80^\circ$ and $125^\circ$. They hypothesized that the misaligned systems show a preference for approximately polar orbits. \citet{Siegel2023} and \citet{Dong2023} followed up by replicating the preceding work and   analyzing the larger sample of systems for which only $\lambda$ has been measured but not $i_\star$. The larger sample did not show a significant preference  for polar orbits, suggesting that the sample analyzed by \citet{Albrecht2021} was somehow biased toward polar orbits.

We revisited the issue with a sample that is about 40\% larger. The current tally of systems with $\lambda$ measurements is 205, of which $\psi$ is known for 87. In the analysis that follows, when considering the $\psi$ measurements, we decided to omit the 6 systems for which the gravity-darkening technique was employed, because parameter degeneracies make it difficult to measure $\psi$ when it is near $0^\circ$, $90^\circ$, or $180^\circ$. When considering only the $\lambda$ measurements, we retained the 6 gravity darkening measurements.

First, we repeated the statistical tests performed by \cite{Albrecht2021} using the enlarged set of $\psi$ measurements. We selected the 27 systems for which $\cos\psi<0.75$ and computed two statistics that quantify clustering: the dispersion around $\cos\psi = 0$ and the standard deviation relative to the mean. We then used Monte Carlo simulations to find the probability of obtaining clustering statistics at least as large as those that were observed if the obliquities were drawn from an isotropic distribution (i.e., a uniform distribution in $\cos\psi$). A subtlety of these calculations is that for given values of the observables $|\cos i|$ (from the transit impact parameter), $|\sin i_\star|$ (from rotational broadening) and $\lambda$ (from the RM effect), there are two closely-spaced solutions for $\psi$:
\begin{equation}
\cos\psi = |\sin i_\star \sin i|\,\cos\lambda \pm |\cos i_\star \cos i|.
\end{equation}
Following \cite{Albrecht2021}, our Monte Carlo simulations take this discrete degeneracy into account by randomly choosing one of the two solutions in each iteration. The results were $p=9.4\times 10^{-4}$ for the dispersion and $p=3.8\times 10^{-4}$ for the standard deviation. 
Applying the same test to the sample of 14 systems that were available to \cite{Albrecht2021} and were not based on gravity darkening gives $p=1.4\times 10^{-2}$ and $2.1\times 10^{-3}$, respectively. Thus, the inclusion of 13 new data points has reduced the $p$-values by factors of 15 and 5.5, allowing a firmer rejection of the null hypothesis.

\begin{figure*}[ht]
    \centering
    \includegraphics[width=\textwidth]{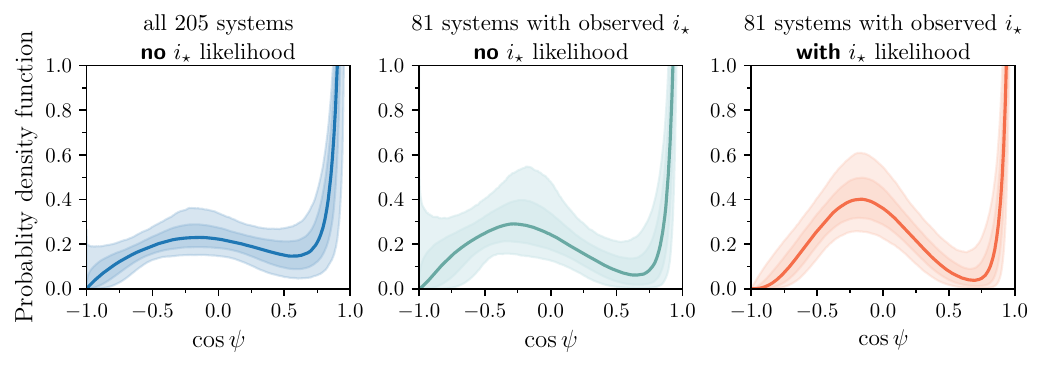}
    \caption{{\bf Hierarchical Bayesian inference of the obliquity distribution} using the code of \citet{Dong2023}. The left panel is for all systems for which $\lambda$
    was measured, the middle panel is for systems for which both $\lambda$ and $i_\star$ were measured but without
    using the $i_\star$ information,
    and the right panel is for
    systems for which both $\lambda$ and $i_\star$ were measured and making
    use of the $i_\star$ information.}
    \label{fig:jd_dfm}
\end{figure*}

Next, we used the ``DFM'' (Dong \& Foreman-Mackey) code\footnote{\url{https://github.com/jiayindong/obliquity}} provided by \citet{Dong2023} to fit a model to the obliquity data consisting of the sum of two Beta distributions, representing the aligned and misaligned populations.\footnote{We note that in the DFM code, the orbital inclination $i$ is generally assumed to be $90^\circ$ as \citet{Dong2023} found the small departures from 90$^\circ$ for transiting planets to produce negligible effects on the resulting inference
of the obliquity distribution. However, it is possible to include measurements of the orbital inclination following \url{https://github.com/jiayindong/obliquity/blob/main/obliquity_distribution_demos.ipynb}.}
Each Beta distribution is characterized by three hyperparameters: a weight $w$,
a mean $\mu$, and an inverse variance
$\kappa$. We adopted the same priors on the hyperparameters as outlined in \citet{Dong2023}. \tabref{tab:beta} gives
the results for the hyperparameters
when the DFM code is applied to different samples of the data, and \figref{fig:jd_dfm}
shows the posterior probability density
for $\cos\psi$.

The left panel of \figref{fig:jd_dfm}
shows the results of analyzing the complete sample of projected obliquity measurements.
There is a strong peak
at $\cos\psi=1$ representing the
aligned systems, and a much broader distribution representing the misaligned
systems.
The distribution falls off toward
retrograde systems ($\cos\psi<0.75$) but otherwise there is little or no evidence for a concentration near $\cos\psi=0$,
in agreement with the findings by \citet{Dong2023} and \citet{Siegel2023}.

The middle panel of \figref{fig:jd_dfm}
shows the results of analyzing
only those systems for
which constraints on $i_\star$
are available, but {\it without} using
the $i_\star$ information to
perform the inference. The inferred
probability density of $\cos\psi$ does
not look too different from the
case described above.

The right panel of \figref{fig:jd_dfm}
shows the results of analyzing 
only those systems for
which constraints on $i_\star$
are available, and {\it using}
the $i_\star$ information to
perform the inference.
This is the type of sample that led
\citet{Albrecht2021} to propose that
there is a peak near $\cos\psi=0$,
and indeed, we find a similar result.

The preceding results suggest that the obliquity distribution of the systems for which measurements of both $\lambda$ and $i_\star$ have been reported is
different from that of the systems
for which $\lambda$ was measured
and no constraints are available on $i_\star$.
This is true even though we
have excluded inclination measurements obtained via gravity darkening, thereby eliminating the bias discussed by \citet{Siegel2023}. 

It is probably relevant that the $\lambda$ \& $i_\star$
sample is enriched in the
types of systems that
are known to show frequent
misalignments: sub-Saturns ($0.025$~M$_{\rm J}<M_{\rm p}<0.2$~M$_{\rm J}$),
and hot Jupiters around hot stars
($a/R_\star<7$, $M_{\rm p}>0.3$~M$_{\rm J}$, $T_{\rm eff}\geq6250$~K).
Stellar inclination measurements
in those systems
almost all come from the $v\sin i_\star$ method -- the combination
of a photometric rotation period,
rotational Doppler broadening of the
spectral absorption lines, and 
the stellar radius.
In the entire sample, there are 17 sub-Saturns, of which inclination measurements are available for 11,
of which 5 have $\cos\psi < 0.75$.  
The entire sample also contains 41 hot Jupiters around hot stars, of which 15 have inclination measurements. After excluding the gravity-darkening measurements, we are left with 10 hot Jupiters, all of which have $\cos\psi < 0.75$. 

\begin{figure*}[ht]
    \centering
    \includegraphics[width=\textwidth]{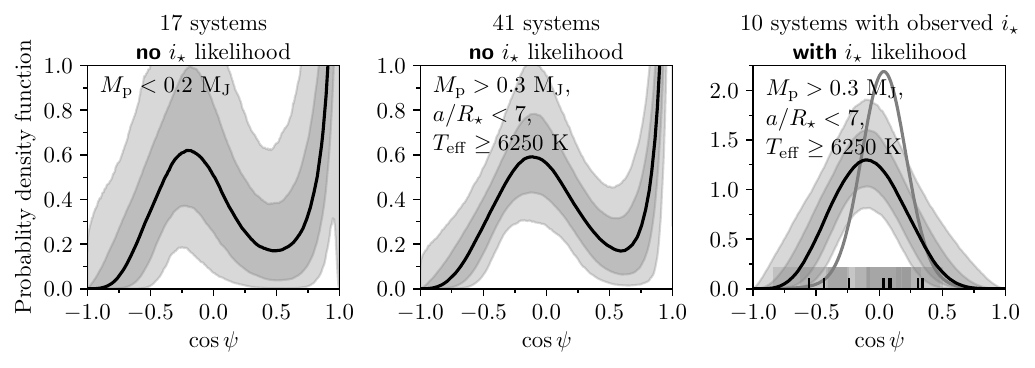}
    \caption{ {\bf Hierarchical Bayesian inference of the obliquity distribution} for subsets of
    the sample that are suspected
    of being especially prone to
    misalignment. The left panel
    is for all sub-Saturns with $\lambda$
    measurements, without using
    any $i_\star$ information even when
    it is available.
    The middle panel is for all hot Jupiters around hot stars with
    $\lambda$ measurements, without using
    any $i_\star$ information even when
    it is available.
    The right panel is for hot Jupiters around hot stars for which
    both $\lambda$ and $i_\star$ are known
    and utilized.
    In the right panel,
    the black upward tick marks on the
    horizontal axis are the
    individual values of $\cos \psi$,
    and the horizontal bands surrounding
    the tick marks convey the uncertainties.
    The solid gray curve is a KDE estimated from these measurements.}
    \label{fig:hotnnep}
\end{figure*}

By combining the 5 misaligned
hot Neptunes and the 10 misaligned hot Jupiters, we constructed a sample of 15 systems and subjected it to the same
Monte Carlo experiments described above
to test the null hypothesis that
they are drawn from an isotropic
distribution.
For the dispersion-around-zero and standard deviation, we found $p=3.1\times10^{-3}$ and $p=3.5\times10^{-3}$, respectively.
We also applied the DFM code to
a sample consisting only of sub-Saturns,
and a sample consisting only of hot Jupiters around hot stars,
with results
that are given in 
\tabref{tab:beta}
and displayed in \figref{fig:hotnnep},
in the same format as in \figref{fig:jd_dfm}.
For the sub-Saturns, a central peak appears to be favored by the data, although the uncertainty in the distribution is large due to the small number of systems (17). A central peak is even more pronounced in the sample of hot Jupiters around hot stars (41 systems including the measurements employing the gravity darkening method).
For the subset of systems with inclination measurements from this hot Jupiter sample,
since all of the systems are misaligned,
we modeled the distribution with a single Beta distribution rather than using
two components. There is a clear peak at $\cos \psi \approx -0.2$ or $\psi\approx 100^\circ$. 
Indeed, all of these 10 systems are consistent with $\psi=90^\circ$
with varying
degrees of uncertainty, and a simple
kernel-density estimate of the distribution shows a peak at $\cos\psi=0$.
Both distributions are shown in \figref{fig:hotnnep}. 

\begin{figure*}
    \centering
    \includegraphics[width=\textwidth]{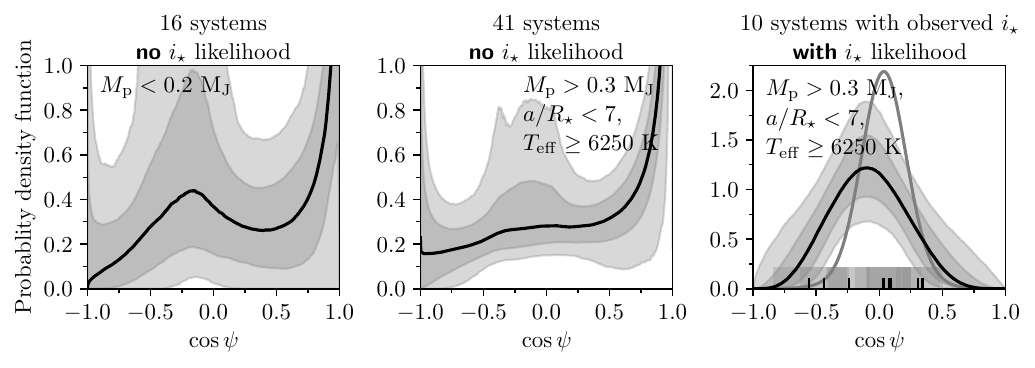}
    \caption{{\bf Hierarchical Bayesian inference of the obliquity distribution} for the same subsets as in \figref{fig:hotnnep}, but here a uniform prior, $\mathcal{U}(-4,10)$, was applied to $\log \kappa$ in the ``DFM'' code.}
    \label{fig:dfm_uni}
\end{figure*}

To investigate the importance of the priors assumed for the hyperparameters in the ``DFM'' code for these subpopulations, we tried adopting less informative priors on the variance, $\kappa$ or rather $\log \kappa$, hyperparameter. Here we tried a uniform prior instead of the Gaussian prior used in \citet[][i.e., $\mathcal{U}(-4,10)$ instead of $\mathcal{N}(0,3)$]{Dong2023}. The resulting hyperparameters are tabulated in \tabref{tab:beta} and distributions are displayed in \figref{fig:dfm_uni}. As the variance ($\kappa$) is now poorly constrained the distributions do look somewhat different, and the peaks for the misaligned populations, especially for the hot Jupiter sample, are not as sharp. At present there are probably too few systems or too little information to properly infer the distributions for these two subpopulations, without assuming some value for the variance to describe the morphology. 
When applying the information on $i_\star$ we have for the subset of 10 systems, the resulting distribution looks very similar to that in \figref{fig:hotnnep}.

Although the results are too prior-dependent to be sure, these results lead 
us to hypothesize that
the preponderance of polar planets
is a phenomenon specific to hot Jupiters
around hot stars and possibly also sub-Saturns.
The hot stars,
in particular, are more amenable to
the determination of $i_\star$ via
the $v\sin i$ method, given their
higher rotation velocities, possibly
explaining their over-representation
in the $\lambda$~\&~$i_\star$ sample.

If there truly is a preponderance of systems with nearly polar orbits, it would seem to be a clue about the history of these systems.
For example, for the sub-Saturn population,
\citet{Petrovich2020} proposed that nearly polar orbits
of Neptune-mass planets
can result from
a secular resonance between the nodal precession frequencies induced by the protoplanetary disk and by
an outer companion. 
One might test this theory by searching for distant massive companions to the polar-orbiting planets.
In any case, \cite{Petrovich2020} note that their theory is not
applicable to hot Jupiters.

Tidal evolution might lead to a pile-up of polar orbits
of hot Jupiters around
hot stars.
\citet{Lai2012} demonstrated that orbital orientations can linger near $90^\circ$ when damping is from the dissipation of inertial waves in the convective envelope of a star.
In general, this theory requires the planet's orbital angular momentum to be larger than the star's spin angular momentum
to avoid producing
more systems near 180$^\circ$ than are observed
\citep[e.g.][]{LiWinn2016}.



\subsection{Obliquity -- Eccentricity correlation?}\label{sec:elam}

Explaining the existence of giant planets on orbits well within the ``ice line'' of the protoplanetary disk is one of the oldest unsolved problems in exoplanetary science.
Disk-driven migration is one possibility,
and would generically
yield low eccentricities and obliquities \citep[e.g.,][]{Lin1996,Dawson2018}. In this scenario, the
observed cases of high eccentricity and obliquity arise from other processes.
An alternative idea is to combine eccentricity-raising dynamical interactions -- such as planet-planet scattering or Kozai-Lidov cycles -- and tidal dissipation \citep[e.g.,][]{Fabrycky2007,Chatterjee2008,Nagasawa2008}.
If the eccentricity reaches a high enough value, or more
to the point, if the periastron distance reaches a low enough value, the damping of tidal oscillations in either the planet or star can convert orbital energy into heat and thereby shrink the semi-major axis.

In many of the proposed scenarios for raising of the eccentricity, the orbital inclination relative to the initial plane is also increased, which could lead to a high stellar obliquity.
Therefore, if the dynamical scenarios are correct, one might expect to see a correlation between $e$ and $\lambda$ for the affected planets. 
An important caveat is that tidal damping tends to reduce both the eccentricity and stellar obliquity, but on timescales that might
differ by orders of magnitude. This is not only because eccentricity and obliquity
tides excite different types of perturbations, but also
because eccentricity damping can take place inside
either body while significant obliquity damping requires the energy to be dissipated inside the star \citep{Ogilvie2014}.
Thus, tides might confound the interpretation of any correlation.
For example, if eccentricity damping is much faster
than obliquity damping,
then any initial correlation might be diminished on astronomically
short timescales. Or, tidal dissipation might increase
the correlation by boosting the number of low-$e$, low-$\lambda$ systems even if there were little or no correlation before tidal evolution.

Regarding the latter point, $a/R_\star$ and $\lambda$ are known to be positively correlated for cool stars \citep{Albrecht2022}. This has been interpreted as evidence for tidal dissipation, since dissipation rates should decline sharply with $a/R_\star$. 
Furthermore, $a/R_\star$ is known to be positively correlated with $e$,
presumably for the same reason (although the causal variable might
be $a/R_{\rm p}$ in that case). Taken together, these two correlations 
might cause $e$ and $\lambda$ to be correlated
without invoking any new explanation.

Nevertheless, it is well worth
testing for a correlation between $e$ and $\lambda$ in
the current sample and \citet{Rice2022} investigated both of these parameters.
Specifically they found at the Kraft break, the  cumulative sum of $\lambda$ for systems with eccentric orbits differs by $6.5\sigma$ from the $\lambda$ distribution of systems with circular orbits. For systems with giant-planet they obtain an $8.7\sigma$ result.
Their procedure was as follows, as we understand it. They selected systems with periastron distances $\leq0.1$~AU for which a value of $\lambda$ was reported in the TEPCat obliquities table,
looked up $e$ in the
NASA Exoplanet Archive (NEA) Confirmed Planets list, 
and separated the systems into two samples: an eccentric sample ($e\geq0.1$), and a circular sample ($e=0$). They omitted borderline systems with $0<e< 0.1$.
They arranged the planets in order of the star's
effective temperature, giving each planet a position $i$ in a line of $N$ planets. They examined the cumulative sums of $|\lambda|$ for the eccentric sample, with the $j$th cumulative sum defined as the sum of all $|\lambda_i|$ values for $i\leq j$. Since the eccentric sample contained fewer
systems and had a different distribution of stellar types
than the circular sample, \citet{Rice2022} used a Monte Carlo procedure to make ``matching'' eccentric samples
for comparison. They created 5000 matching circular samples by drawing randomly (with replacement) from the circular sample the same number
of stars with $T_{\rm eff}<6100$\,K and the same
number of stars with $T_{\rm eff}>6100$\,K as in
the eccentric sample. They then compared
the $N$th cumulative sums of the eccentric sample
and the ensemble of matching circular samples,
and found them to differ by $6.5\sigma$, where $\sigma$
was determined from the spread amongst the 5000 trials.
For convenience we will denote the ``number-of-sigma'' by the symbol $\Delta$.

They also tried some variations on the samples. When performing the same steps but with the samples restricted to giant planets, defined as $M_{\rm p}>0.3$~M$_{\rm J}$, they found $\Delta = 8.7$.
To test the notion that $a/R_\star$ is a ``hidden variable'' responsible
for the $e/\lambda$ correlation,
they created samples that were not divided by eccentricity,
but rather on whether $a/R_\star$ is greater or smaller than 12.
In this case, they found $\Delta=4.9$, and since this is smaller
than 6.5 they concluded that the variation in $a/R_\star$ cannot
explain the entire effect they observed between $e$ and $\lambda$.

\begin{figure*}[ht]
    \centering
    \includegraphics[width=\textwidth]{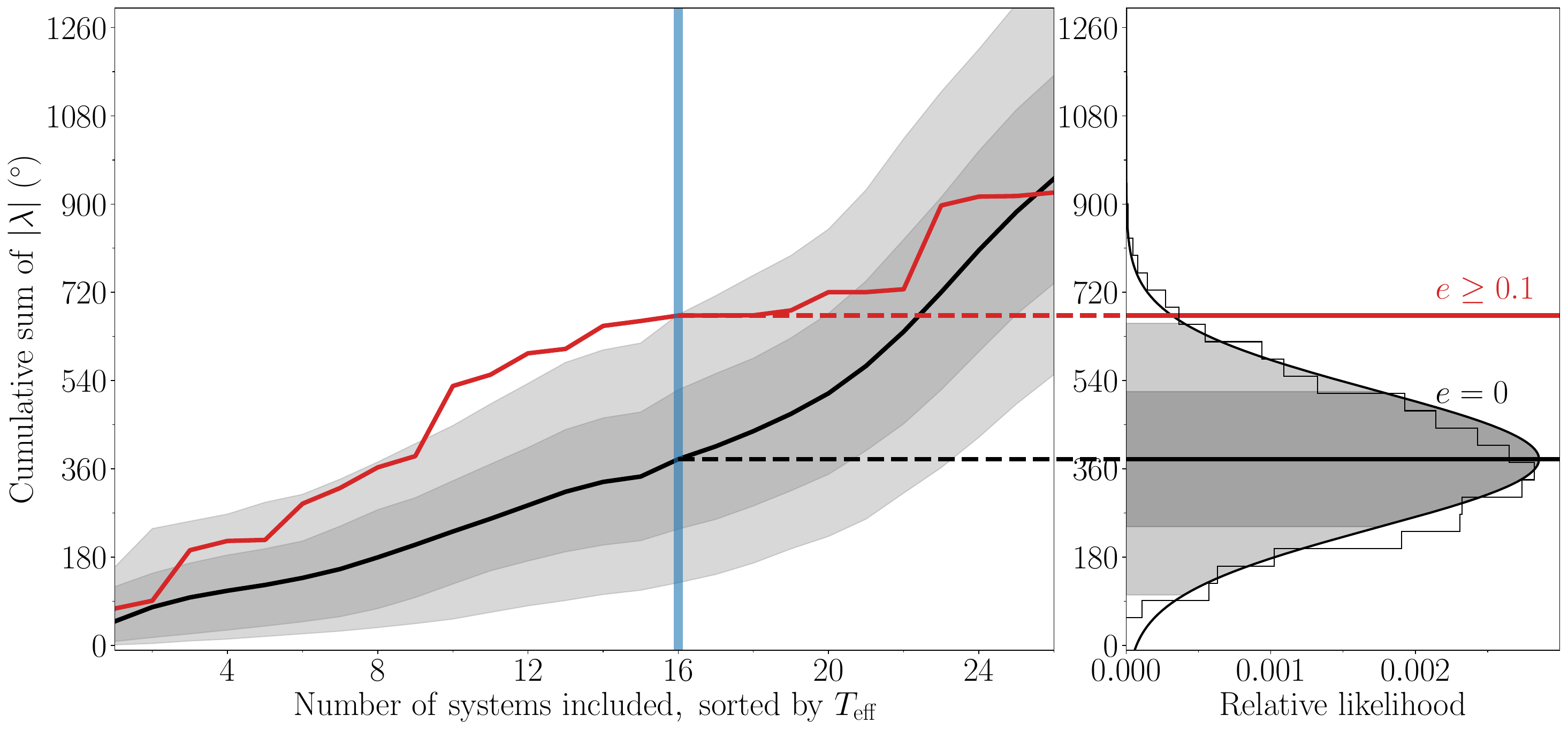}
    \caption{{\bf Obliquity distribution for stars hosting planets on eccentric and circular orbits.} {\it Left:} The cumulative sum of $|\lambda|$ for eccentric planets shown in red, sorted by $T_{\rm eff}$. The black line is the average of 5000 randomly sampled sets from the circular population with the gray shaded area denoting the 1$\sigma$ and 2$\sigma$ intervals. The vertical blue line denotes the Kraft break ($T_{\rm eff}=6100$~K). {\it Right:} Histogram showing the distribution of the circular population and the value of the eccentric population at the Kraft break. The Gaussian is a fit to the histogram. The circular and eccentric samples only differ by $2.1\sigma$, meaning the eccentric population does not appear to be significantly more misaligned. Adapted from \citet{Rice2022}.}
    \label{fig:elam}
\end{figure*}

With the enlarged sample, we decided to revisit these intriguing results. After implementing the
method of \citet{Rice2022} to the best of our ability, we found $\sigma = 2.1$ for the whole sample shown in \figref{fig:elam} and $\sigma = 4.1$ for planets with $M_{\rm p}>0.3$~M$_{\rm J}$. To check on our code, we tried reproducing their results using the systems known at that time and were unsuccessful.
Further investigation 
showed that the
main reason for the
discrepancy can be traced to 6 systems in the eccentric sample that should have been in
the circular sample.
These 6 systems (with details in
the footnotes) are
HAT-P-23~b and HAT-P-32~b\footnote{
\cite{Bonomo2017} reported the best available eccentricity and
implied both systems were ``circular.''
}
KELT-6~b\footnote{\citet[][Table 4]{Damasso2015} 
reported the best available eccentricity and implied
the system is ``circular''. Here, the eccentricities of planets
b and c might have been mixed up, as c travels on an eccentric orbit}.;
and HATS-14~b,
HATS-70~b,
\& Kepler-63~b\footnote{These three systems should have been
classified as ``circular'' but the reported upper limits on
eccentricity were mistaken for the eccentricity itself.
For HATS-70~b, see Table 6 of \citet{Zhou2019}.
For Kepler-63~b, see Table 2 of \citet[][]{Sanchis2013}.
HATS-14~b was excluded from our sample due to the precision in $\lambda$ (\sref{sec:Discussion}), but see Table 4 of \citet[][]{Mancini2015}.} 
Taken together, these errors caused $\sigma$ to appear
larger than in reality.

\begin{figure*}[ht]
    \centering
    \includegraphics[width=\textwidth]{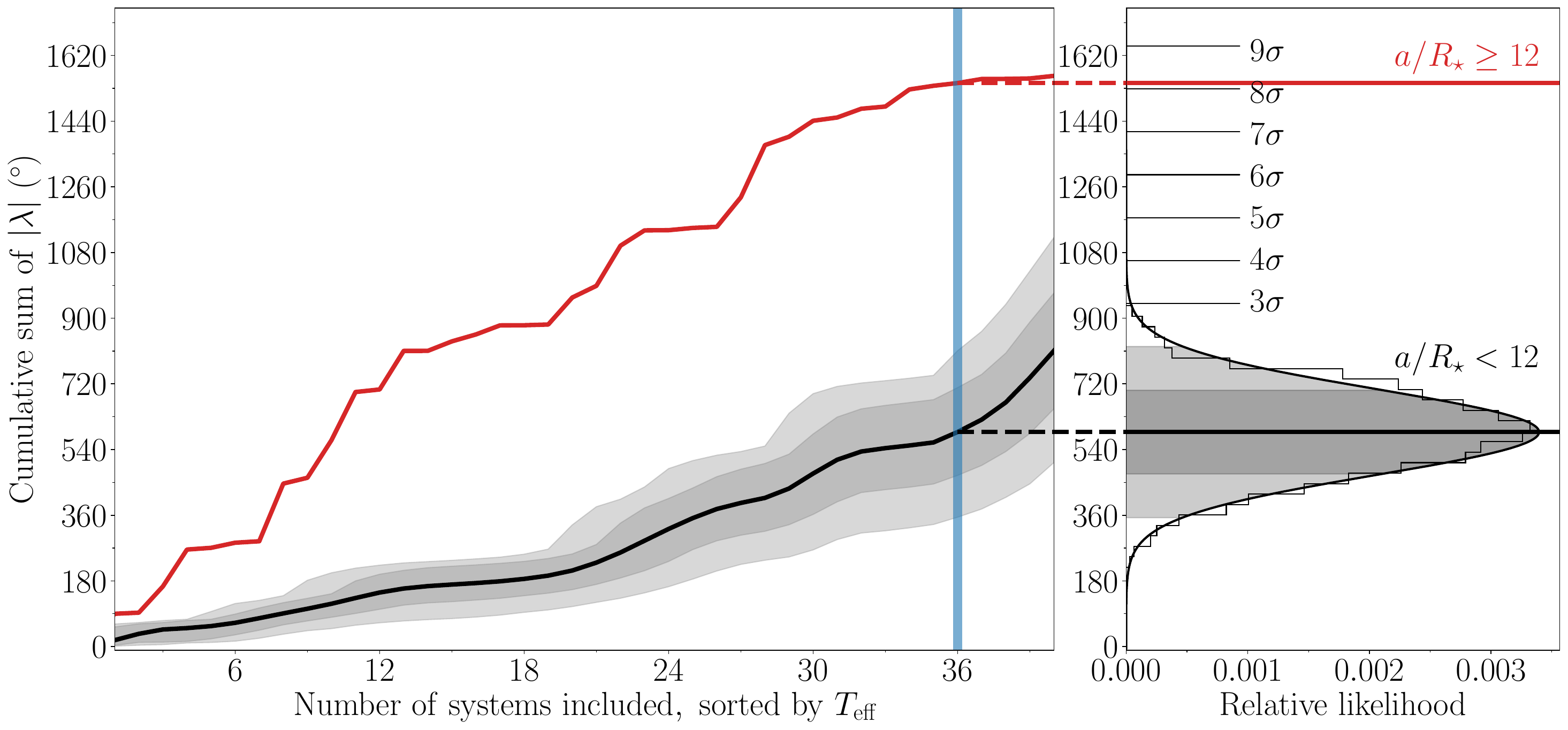}
    \caption{{\bf Obliquity distribution for stars hosting planets on close-in and wide orbits.} Same as \figref{fig:elam}, but for systems with planets on close-in or wide orbits with the division given for $a/R_\star=12$. The systems with planets on wider orbits appear to be significantly more misaligned compared to those with planets closer in. }
    \label{fig:arlam}
\end{figure*}

We tried some variations on the samples.
When including systems with $0<e<0.1$ in the circular group, instead
of omitting them, we found $\Delta = 1.8$.
When focusing on giant planets ($M_{\rm p}>0.3$~M$_{\rm J}$), we obtained $\Delta = 4.6$. We note that there are only 10 known eccentric giant planets around cool stars, making the results sensitive to an error for an individual system.
When we tried dividing the sample according to whether $a/R_\star$
is smaller or greater than 12, instead of by eccentricity,
we found $\Delta=8.1$ as shown in \figref{fig:arlam}. Thus, in our case, the $\sigma$ statistic 
is larger when testing the effect of $a/R_\star$ than when testing
for the effect of eccentricity.

\subsection{Tidal realignment}\label{sec:precise}
 
The main evidence that tidal obliquity damping occurs on astrophysically relevant timescales, despite many researchers' early expectations 
\citep[see, e.g.,][]{Queloz2000, Winn+2005}, is that the types of systems
where one might expect tidal dissipation rates to be fastest --- with massive, close-orbiting planets and host stars with thick convective envelopes --- do indeed have an obliquity distribution that is
concentrated around $0^\circ$. And conversely, systems for which
misalignment is proposed to be part of the formation process but for which tidal dissipation rates are expected to be much slower --- somewhat wider-orbiting giant planets, and stars with
radiative envelopes --- have a broad obliquity distribution.
For some details on these findings, see
\cite{Schlaufman2010, Winn2010}, and \citet{Hebrard2011}.
To demonstrate this point more explicitly,
\cite{Albrecht2012} constructed a crude ``tidal dissipation figure of merit'' for each system, meant to be a proxy for the tidal dissipation rate,
and showed that the obliquity distribution does indeed narrow
for the systems where the figure of merit predicts more rapid dissipation.

However, the precise mechanisms of tidal dissipation are not well understood and are probably quite complex, depending on forcing frequency,
stellar structure, rotation rate, etc. Another problem with tidal
damping is that the expected endpoint of tidal evolution for
most of the known close-orbiting giant planets
is tidal orbital decay and destruction \citep{Brown2011}.
Thus, an observed planet around a star that
experienced tidal damping must somehow have
avoided orbital decay
before the star was aligned to the observed degree. 

Several theories exist that might prevent or at least delay planet destruction for long enough to be consistent with the data.
The orbital angular momentum of some lucky planets may exceed the host's spin angular momentum, causing the star to align before its orbital decays \citep[see, e.g.,][]{Hansen2012,Valsecchi2014,Dawson2014}. If tidal damping is primarily driven by the dissipation of inertial waves in a star's convective zone, then alignment could be faster than orbital decay \citep{Lai2012,Lin2017,Damiani2018}. The damping of gravity waves in radiative cores might similarly cause alignment before tidal in-spiral \citep{Zanazzi+2024}. Another idea is
that if the efficiency of tidal dissipation drops sharply with increasing forcing frequency (shorter orbital periods), then alignment can be arranged to occur before tidal evolution slows down dramatically, allowing the orbit to survive \citep{Barker2010,Penev2018,Anderson2021,Ma2021}. Alternatively, and more speculatively, tides from the planet might only affect a star's outer convective zone, allowing the planet to donate a smaller amount of its
orbital angular momentum to spin up the star and save itself \citep{Winn2010}. With all of this theoretical uncertainty, is there a relatively model-independent way to seek corroborating evidence for tidal obliquity damping?

The Sun's obliquity is $7.155^\circ$ relative to the ecliptic
\citep{Beck2005} and $6.2^\circ$ relative to the Solar System's invariable plane \citep[see, e.g.,][]{Souami2012,Gomes2017}. If we assume this angle to be representative of the degree to which
stars and their planetary systems are initially aligned ---
admittedly, a big assumption --- then we would expect some systems that have experienced strong tidal damping to have obliquities much smaller than 6$^\circ$. This can be tested with high-precision RM observations of
the types of systems where strong tidal damping is expected,
and comparisons to systems where it is not expected.

\begin{figure}[h]
    \centering
    \includegraphics[width=\columnwidth]{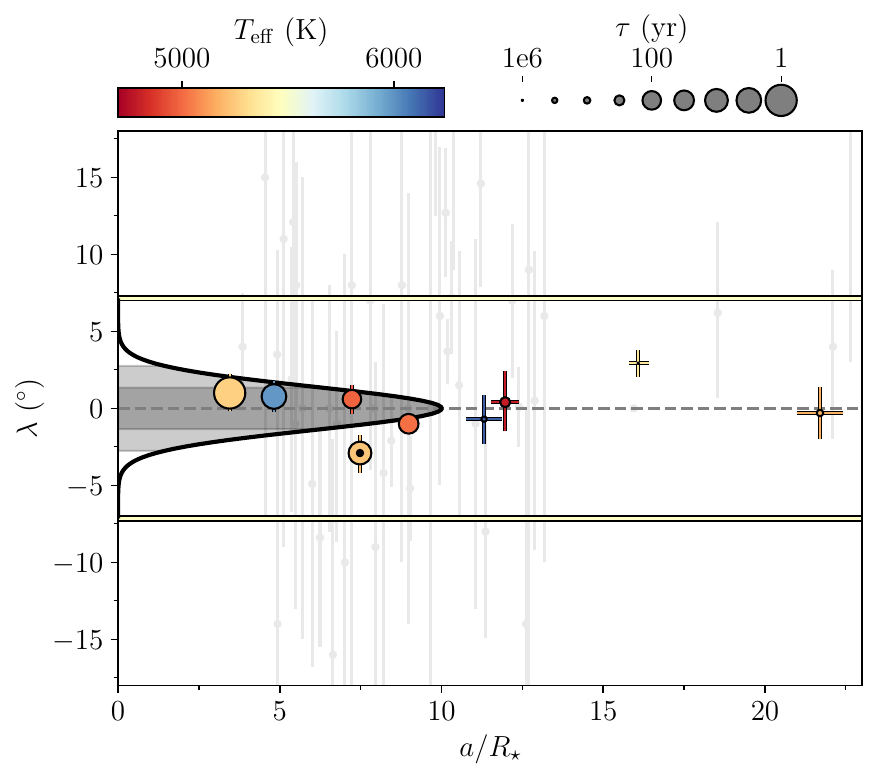}
    \caption{{\bf Precise projected obliquity measurements.} Here we highlight the most precise ($\sigma_\lambda \leq 2^\circ$) measurements of $\lambda$ for systems with cool stellar hosts ($T_{\rm eff}<6100$~K). WASP-50 is highlighted with a black center. The color-coding is done according to $T_\mathrm{eff}$ and the marker sizes scale with the tidal alignment timescale \citep[as given by][]{Zahn1977}. The gray shaded area shows the Gaussian distribution inferred from cool hosts ($T_{\rm eff}<6100$~K) with close-in ($a/R_\star<10$), massive ($M_{\rm p}<0.3$~M$_{\rm J}$) planets. The horizontal lines demarcate the Solar System obliquity of $7.155^\circ$ \citep{Beck2005}. The light gray error bars show the less precise measurements. Adapted from \citet[][Figure 10]{Albrecht2022}.}
    \label{fig:precise}
\end{figure}

Focusing on only the most precise measurements ($\sigma_\lambda<2^\circ$) in cool hot Jupiter systems, \citet[][Section 3.1.8]{Albrecht2022} found the dispersion of $\lambda$ measurements to be $0.91^\circ$. Given an average measurement uncertainty of $0.82^\circ$, this should probably be regarded as an upper limit on the true dispersion of a few degrees.
There are drawbacks to this simple approach. First, one must decide what constitutes a ``precise'' measurement. Somewhat arbitrarily, \cite{Albrecht2022} chose a threshold precision of $2^\circ$. Second, it ignores the information contained in the much more numerous and less precise measurements. 

We tried overcoming these problems with a Hierarchical Bayesian Model (HBM), following up on the work by \citet{Siegel2023}, who carried out similar tests to those described below. (To eliminate any suspense, we obtained similar results, too.)
The hyper-parameter of interest is the dispersion of the 
underlying $\lambda$ distribution around $0^\circ$, independent
of measurement uncertainties. To identify systems for which rapid tidal dissipation is expected, we required $T_{\rm eff}<6100$~K, $a/R_\star<10$, and $M_{\rm p}>0.3$~M$_{\rm J}$. We also required $\sigma_\lambda<50^\circ$.
This led to a sample of 60 systems, only one of which stood out:
TOI-858Bb \citep{Hagelberg2023}, which is located in a wide binary system
and has a retrograde orbit. Since tidal dissipation is evidently
not the only important obliquity-altering process in this system ---
perhaps due to the stellar companion --- 
we decided to omit it from consideration,
leaving a sample of 59 systems. 
\figref{fig:precise} shows the key characteristics of those 59 systems (as well as some systems with higher values of $a/R_\star$). The data points
with a precision in $\lambda$ better than $2^\circ$ are highlighted;
the measurement of WASP-50 of \waspfiftyu$^\circ$ presented in this paper
is included in this sample.

The result of the HBM was a dispersion in $\lambda$ of $1.4 \pm 0.7^\circ$, depicted in \figref{fig:precise} as a Gaussian function
with a width of $1.4^\circ$. 
For prograde systems the expected endpoint of the alignment process is $\lambda=0^{\circ}$, which is why our HBM did not include a hyperparameter
for the mean of the obliquity distribution. However, as an experiment to
check for surprises or systematic errors (such as the neglect of the convective blueshift) we repeated the analysis after allowing for the mean to be a second hyperparameter. The results were a a dispersion of $1.5 \pm 0.8^\circ$ and a mean of $-0.2 \pm 0.7^\circ$, giving no clear evidence for surprises or systematic errors.

In the light of these results, our measurement of WASP-50 (\waspfiftyu$^\circ$) is more interesting than we originally
anticipated. According to the rough criteria described above,
the tidal dissipation timescale for this system should be faster than in most of the other systems in the sample.
\citet{Gillon2011} reported an age of $7.0\pm3.5$~Gyr based on comparing the usual photometric and spectroscopic observables to
stellar-evolutionary models. However, the estimated age based on the chromospheric activity and rotational period was only $0.8 \pm 0.4$~Gyr \citep[which could also be a result of tidal spin-up, e.g.,][]{Tejada2021}. If the old age is correct, the system is more likely to have had enough time to achieve good alignment than if the young age is
correct. As discussed in \sref{sec:prot}, \citet{Gillon2011} reported a rotation period for WASP-50 of $16.3\pm0.5$~d or potentially twice this value. Using these values $\psi$ comes out to $34^{+12}_{-9}$ and $5.0^{+1.9}_{-3.0}$, respectively. The value of the rotation period has dramatic consequences for the inferred orientation of the system; the shorter period we identified in the \tess light curve results implies an even more misaligned system, which seems worth checking on, given that so many other similar systems are well-aligned. Future \tess observations might be able to resolve the issue.

Considering planets on wider orbits, with $10\leq a/R_\star <20$, the HBM yielded a dispersion of $8 \pm 2^\circ$. The spread in $\lambda$ is therefore seen to increase with orbital separation (as also discussed in \sref{sec:elam}). When restricted to planets with $20\leq a/R_\star <30$ (a sample of only 7 systems), the HBM yielded a dispersion of $4 \pm 3^\circ$, which is consistent with the $10\leq a/R_\star <20$ results.
It would be interesting to obtain more measurements for planets
with large separations to bolster these results and possibly even investigate the separation-dependence of tidal dissipation rates.



\subsection{Compact multi-transiting systems, with and without outer companions}\label{sec:multis}

At a time when only five $\lambda$ measurements had been made for
stars with more than one known transiting planets,
\citet{Albrecht2013} noted that they were all consistent with 
good alignment.
The number is now 22, including four systems we have added to the tally in this work. For WASP-148, $\lambda$ had already been measured by \citet{Wang2022}, and our result is consistent with theirs but with lower precision. For TOI-1130, we found $\lambda=$~\toithirtyg$^\circ$, while for two other systems, LTT~1445A and TOI-451A, our formal results were that the orbits are prograde, but given the low SNR of our
measurements we do not include them in the following discussion.
We do include an additional
4 systems for which there are good
constraints on the stellar inclination even though $\lambda$ is not known. Our discussion is therefore based on the properties of
24 systems. 

\begin{figure*}[ht]
    \centering
    \includegraphics[width=0.9\textwidth]{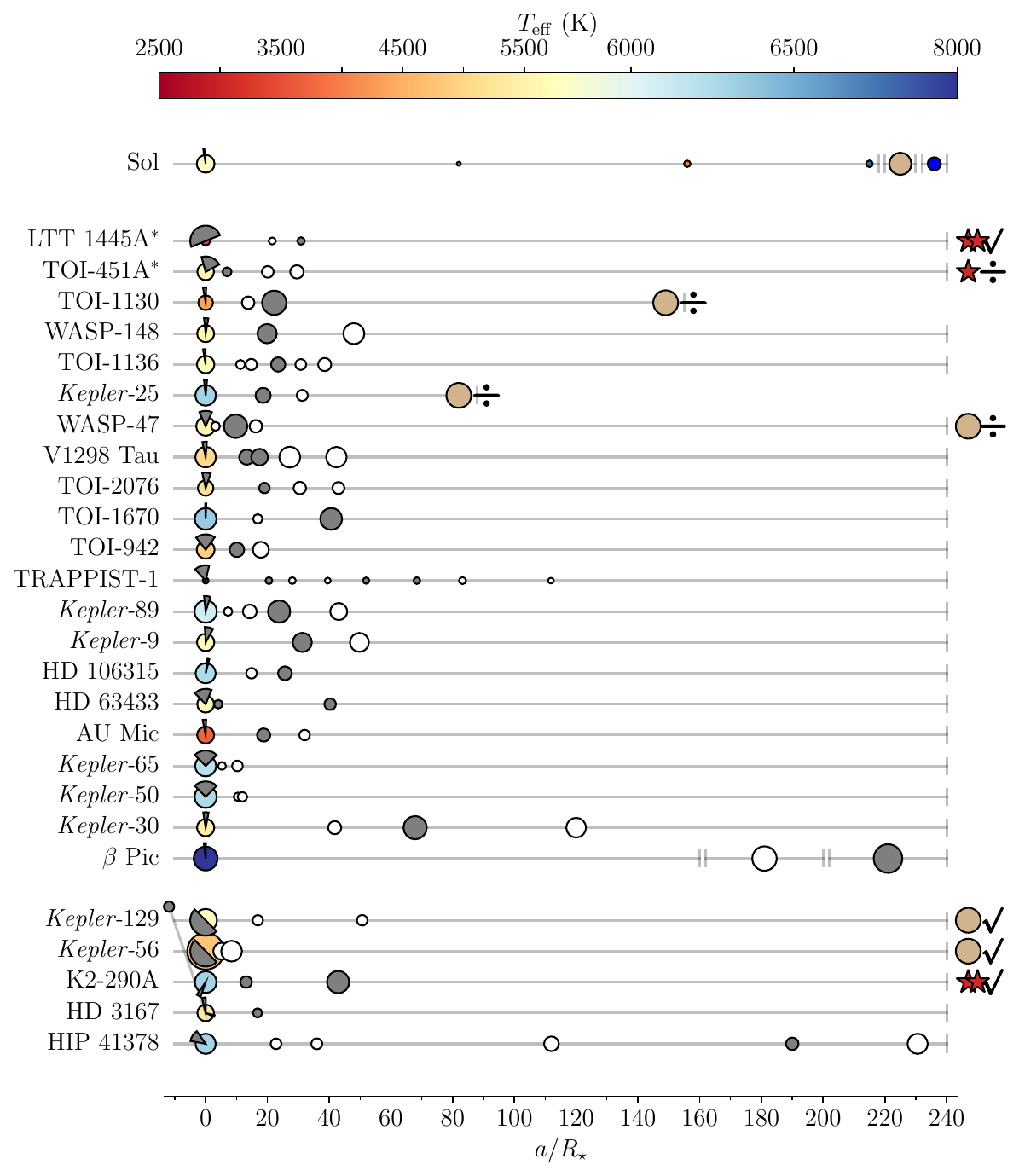}
    \caption{{\bf Obliquities of stars with multiple transiting planets,} as constrained by either the RM effect or  asteroseismology. Each horizontal line
    shows the $a/R_\star$ values of
    the planets in a given system, with the solar system on top (with gaps to
    allow Jupiter and Neptune to be shown). For the planets, the size of each circle conveys the planet's radius
    relative to Jupiter. For the stars,
    symbol size conveys the radius relative
    to the Sun, with a color conveying the effective temperature. 
    Gray circles are planets for which the RM effect has been measured -- except for $\beta$~Pic~b where spectro-interferometry was used. The wedge indicates the projected obliquity for a given system, based on the most precise measurement for any planet in the system, with the exception of HD~3167 for which two planets appear to have very different inclinations. The asteroseismic measurements constrain $i_\star$ and not $\lambda$; in those cases the wedge should therefore be read as aligned (\emph{Kepler}-50 and \emph{Kepler}-65) or misaligned (\emph{Kepler}-56, \emph{Kepler}-129).  Planetary (tan circle) or stellar companions (red star, M-dwarfs) to a given system are shown to the right, where $\surd$ ($\div$) denotes that the companion could (not) have influenced the obliquity of the inner planetary system. The asterisks for LTT~1445A and TOI-451A denote that these measurements should be taken with a grain of salt. Adapted from \citet[][Figure 3]{Wang2022}.}
    \label{fig:multis}
\end{figure*}

\figref{fig:multis} is a family portrait of these systems. The Solar System is shown at the top for scale and reference. 
There are 19 systems for which the data are consistent with zero obliquity. There are 5 systems with large obliquities relative to at least one transiting planet: HD~3167, \emph{Kepler}-56, HIP~41378, \emph{Kepler}-129, and K2-290. HD~3167 has a very unusual architecture in which the planets appear to be on perpendicular orbits \citep{Bourrier2021}. This case is illustrated in \figref{fig:multis} as the outer planet floating over \emph{Kepler}-129.
HIP~41378 is a five-planet system in which \citet{Grouffal2022} measured the projected obliquity with respect to planet d to be $57^{+26\circ}_{-18}$. However, due to the long duration ($\sim12.5$~hr and $P\sim278$~d), a full transit was not observed, preventing us
from having high confidence in the result.

In \figref{fig:multis} \emph{Kepler}-30 is grouped with the aligned systems, but there are conflicting indications about the obliquity.
By using two different methods \citet{Sanchis2012} found $\lambda=-1\pm10^\circ$ and $\lambda=4\pm10^\circ$, indicating a well-aligned system. However, \citet{Morgan2023} 
found $i_\star=43^{+15\circ}_{-9}$,
indicating a large misalignment,
based on the combination of
the rotation period, $v\sin i_\star$, and $R_\star$.
However, this result depends critically on the value \vsini$=2.0\pm0.2$~km/s
reported by \citet{Fabrycky2012},
and the measurement uncertainty of 0.2~km/s
is likely to have been underestimated.
Systematic errors of at least 0.5~km/s are typical,
due to effects such as turbulent and instrumental
broadening.
A value of $3.0\pm0.5$~km~s$^{-1}$ would allow for consistency with alignment ($i_\star\approx90^\circ$).

In contrast to hot Jupiter systems, where misalignments might have occurred in the tumultuous process of high-eccentricity migration after disk dispersal, the misalignments
in multiplanet systems might have occurred while the disk was present.  Gravitational interactions between a planet-forming disk, a binary companion, and a spinning host star can be enhanced when the disk looses mass, magnifying the host star's tilt through a secular resonance \citep[e.g.][]{BatyginAdams2013}.  For a binary companion to generate spin-orbit misalignments, the stellar spin must be gravitationally decoupled from the planet forming in the disk, otherwise the star's spin axis and planet's orbit normal precess in-unison (eq. 80 of \citealt{Zanazzi2018}). The spin is decoupled from the orbits of LTT~1445Ab, and K2-290Ab,c, but not for TOI-451Ab, when torqued by their binary companions.  
 
Similarly, sweeping secular resonances from a dispersing disk can enhance gravitational interactions between forming planets, magnifying their inclinations \citep[e.g.][]{Ward1981}. However, for planetary companions to generate misalignments, they must also supply the system with a large angular momentum deficit to misalign an inner planet. Based on the work by \cite{Petrovich2020, ZanazziChiang2014}, we assume a planetary companion (mass $m_2$, semi-major axis $a_2$) could secularly misalign an inner planet (mass $m_1$, semi-major axis $a_1$) if they are well separated ($a_2/a_1>2$), and they have a large angular momenta ratio ($(m_2 a_2^{1/2})/(m_1 a_1^{1/2}) > 10$, see e.g. eq. 22 of \citealt{ZanazziChiang2014}). The planet pairs \emph{Kepler}-56~c,d and \emph{Kepler}-129~c,d meet this criteria, while \emph{Kepler}-25~c,d, WASP 47 b,c, and TOI-1130~c,d do not, although we note that the companion TOI-1130~d is as of yet poorly constrained \citep{Korth2023}. LTT~144A might be aligned when a misalignment is predicted by \cite{Zanazzi2018}, while the other systems with distant, massive companions appear consistent with spin-orbit misalignments being generated (or not) while gas disk is present (\figref{fig:multis}).

Secular interactions from companions exciting an inner planet's inclination during the disk-hosting phase is not the only mechanism which can explain misalignments in multiplanet systems. For instance, \cite{GratiaFabrycky2017} postulate a dynamical instability between three outer planets could have  misaligned \emph{Kepler}-56~b and~c, while \cite{BestPetrovich2022} find secular chaos induced by the two binary companions of K2-290A is sufficient to misalign K2-290Ab and~c, with both mechanisms taking place after disk dispersal.  Additional stellar obliquity measurements in multiplanet systems with companions (e.g. LTT 1445A), and further constraints on the presence or absence of companions in inclined multiplanet systems (e.g. HD~3167, HIP~41378), could constrain which mechanisms are responsible for generating misalignments.

For the multiplanet systems which have obliquities consistent with alignment (again barring TOI-451A and LTT~1445A), we applied the HBM framework described in the previous section to infer the mean $\mu$ and dispersion $\sigma$
of the intrinsic distribution of obliquities.
The results were $\mu=0.0 \pm 1.7^\circ$ and $\sigma=2.0 \pm 1.5 ^\circ$. When we removed \emph{Kepler}-30~c from the sample, out of the concern raised above,
we found $\mu=0.0 \pm 1.8^\circ$ and $\sigma=2.1 \pm 1.6 ^\circ$.  Thus, the obliquity dispersion of the host stars
of multis is similar to that of 
cool stars with hot Jupiters.
However, tidal obliquity damping is expected to be much weaker
for the multis than for the hot Jupiters,
because of the planets' lower masses and relatively wider orbits. The low obliquities of the multis might
therefore be primordial, i.e., due to the flatness
of the protoplanetary disk and its strong coupling to the accreting young star. From this point of view,
the solar obliquity of 6$^\circ$ may be unusually high --- and if the high solar obliquity is simply a peculiarity
of the Solar System, then the evidence for tidal
obliquity damping presented in the previous section
would be undermined.

We finally note that an additional aligned system could potentially be the HD~148193 system ($\lambda=$~\hdtwos$^\circ$, \sref{obs:148193}), if the inner transiting candidate is confirmed as a bona fide planet at some point.




\section{Conclusions}\label{sec:conc}

We presented new observations of the RM effect for 19 stars
with transiting planets. We combined these measurements with results from the literature to arrive at a sample of 205 systems for which $\lambda$ has been determined with
reasonable confidence. With this sample, we revisited four  questions:

\begin{enumerate}[i)]

   \item Is there a preponderance of perpendicular planets? 
   First raised as a possibility by \citet{Albrecht2021},
   and then questioned \cite{Siegel2023} and \cite{Dong2023},
   our updated sample does not settle the matter.
   It remains the case that a ``polar peak'' is seen
   in the obliquity distribution of the sample of planets for
   which $\lambda$ and $i_\star$ have been measured,
   but no such peak is seen clearly when analyzing the
   larger sample of systems for which $\lambda$ has
   been measured but not necessarily $i_\star$.
   We showed that the two samples might differ from each other
   because certain kinds of systems --- namely, hot stars
   with hot Jupiters --- are more amenable
   to $i_\star$ measurements than others and also
   more likely to have near-polar orbits.
   The subsample of sub-Saturns around cooler stars also
   seems to have a preference for polar orbits, although
   the sample needs to be increased to be sure.
   Thus, the resolution of this
   issue still seems likely to be astrophysically interesting.
 
   \item Is there a correlation between orbital eccentricity and stellar obliquity for close-orbiting giant planets?
   We concluded that the correlation is weaker
   than had been previously reported by \cite{Rice2022}.
   Furthermore, we argued that the correlation exists
   because both eccentricity and obliquity are
   correlated with a third variable, orbital separation.
   The causal mechanism is likely to be tidal dissipation,
   which lowers both eccentricity and obliquity
   at short orbital separations.
    
    \item How low are the obliquities of stars suspected
    of having been subject to tidal obliquity damping?
    {\bf Measurements of $\lambda$} of cool stars with close-orbiting giant planets
    ($T_{\rm eff} < 6100$~K, $a/R_\star < 10$, $M_{\rm p} > 0.3$~M$_{\rm J}$) have an intrinsic dispersion
    of $1.4\pm0.7^\circ$ in $\lambda$, according
    to our Hierarchical Bayesian Model (HBM).
    Such fine alignment can be considered
    as supporting evidence for tidal obliquity
    damping, if the solar obliquity of 6$^\circ$
    is taken as representative of a typical un-damped
    ``primordial'' obliquity.
    More precise measurements of $\lambda$ in systems hosting massive planets around cool stars might even help to quantify the dependence
    of tidal dissipation rates 
    on orbital separation. 
    
    \item How well aligned are the central stars
    of compact systems of multiple transiting planets? Except for a few notable exceptions, the $\lambda$ measurements in compact multiplanet systems are consistent with low obliquities. For this sample, we found an intrinsic dispersion
    in $\lambda$ of $2.1\pm1.6^\circ$. In this sense, the
    multis are better aligned than the Solar System,
    indicating that compact multis are born unusually flat
    or that the Sun is unusually tilted, complicating the
    conclusions drawn above about tidal obliquity damping.
    The few examples of misaligned stars with compact systems
    of multiple transiting planets might be explained by interactions between the protoplanetary disk and a stellar companion, or by interactions with outer giant planets after the epoch of planet formation.
    
\end{enumerate}


\begin{acknowledgements}
The authors thank the anonymous referee for helpful and insightful comments that greatly improved the quality of the manuscript.
The authors would like to thank R.~I.~Dawson, J.~Siegel, J.~Dong, D.~Foreman-Mackey, M.~Rice, and S.~Wang for helpful comments and discussions.
Funding for the Stellar Astrophysics Centre is provided by The Danish National Research Foundation (Grant agreement no.: DNRF106). 
E.K. and S.H.A. acknowledge the support from the Danish Council for Independent Research through grant No.2032-00230B.
%
Parts of the numerical results presented in this work were obtained at the Centre for Scientific Computing, Aarhus, \url{https://phys.au.dk/forskning/faciliteter/cscaa/}.
This research has made use of data obtained from or tools provided by the portal \url{https://exoplanet.eu/home/} of The Extrasolar Planets Encyclopaedia.
This research has made use of NASA’s Astrophysics Data System.
This research has made use of the NASA Exoplanet Archive, which is operated by the California Institute of Technology, under contract with the National Aeronautics and Space Administration under the Exoplanet Exploration Program.
This work is based on observations made with the Very Large Telescope (VLT) at Cerro Paranal, Chile, the Telescopio Nazionale Galileo (TNG) at Roque de los Muchachos, La Palma, Spain, and the Nordic Optical Telescope (NOT) also at Roque de los Muchachos.
NOT is owned in collaboration by the University of Turku and Aarhus University, and operated jointly by Aarhus University, the University of Turku and the University of Oslo, representing Denmark, Finland and Norway, the University of Iceland and Stockholm University at the Observatorio del Roque de los Muchachos, La Palma, Spain, of the Instituto de Astrof\'{\i}sica de Canarias.
Some of the data presented herein were obtained at the W. M. Keck Observatory, which is operated as a scientific partnership among the California Institute of Technology, the University of California and the National Aeronautics and Space Administration. The Observatory was made possible by the generous financial support of the W. M. Keck Foundation. The authors wish to recognize and acknowledge the very significant cultural role and reverence that the summit of Maunakea has always had within the indigenous Hawaiian community. We are most fortunate to have the opportunity to conduct observations from this mountain.
This paper includes data collected by the \emph{Kepler} mission and obtained from the MAST data archive at the Space Telescope Science Institute (STScI). Funding for the \emph{Kepler} mission is provided by the NASA Science Mission Directorate. STScI is operated by the Association of Universities for Research in Astronomy, Inc., under NASA contract NAS 5–26555.
This paper includes data collected with the TESS mission, obtained from the MAST data archive at the Space Telescope Science Institute (STScI). Funding for the TESS mission is provided by the NASA Explorer Program. STScI is operated by the Association of Universities for Research in Astronomy, Inc., under NASA contract NAS 5–26555.
The K2 light curves were extracted through the EVEREST pipeline \citep{Luger2016,Luger2018} hosted on the Mikulski Archive for Space Telescopes (MAST).
The TESS light curves were extracted using Lightkurve \citep{lightkurve}.
EsoReflex \citep{Freudling2013} was used to extract CCFs from the ESPRESSO data.
This work made use of the following \texttt{Python} packages; NumPy \citep{numpy}, SciPy \citep{scipy}, Matplotlib \citep{matplotlib}, corner \citep{corner}, PyMC \citep{pymc}, and Astropy \citep{astropy:2013, astropy:2018, astropy:2022}.
\end{acknowledgements}


%
%

\bibliographystyle{aa} 
\bibliography{mybib} 

\begin{thebibliography}{229}
\expandafter\ifx\csname natexlab\endcsname\relax\def\natexlab#1{#1}\fi

\bibitem[{{Ahlers} {et~al.}(2020){Ahlers}, {Kruse}, {Col{\'o}n}, {Dorval},
  {Talens}, {Snellen}, {Albrecht}, {Otten}, {Ricker}, {Vanderspek}, {Latham},
  {Seager}, {Winn}, {Jenkins}, {Haworth}, {Cartwright}, {Morris}, {Rowden},
  {Tenenbaum}, \& {Ting}}]{2020ApJ...888...63A}
{Ahlers}, J.~P., {Kruse}, E., {Col{\'o}n}, K.~D., {et~al.} 2020, \apj, 888, 63

\bibitem[{{Albrecht} {et~al.}(2007){Albrecht}, {Reffert}, {Snellen},
  {Quirrenbach}, \& {Mitchell}}]{Albrecht2007}
{Albrecht}, S., {Reffert}, S., {Snellen}, I., {Quirrenbach}, A., \& {Mitchell},
  D.~S. 2007, \aap, 474, 565

\bibitem[{{Albrecht} {et~al.}(2012{\natexlab{a}}){Albrecht}, {Winn}, {Butler},
  {Crane}, {Shectman}, {Thompson}, {Hirano}, \&
  {Wittenmyer}}]{2012ApJ...744..189A}
{Albrecht}, S., {Winn}, J.~N., {Butler}, R.~P., {et~al.} 2012{\natexlab{a}},
  \apj, 744, 189

\bibitem[{{Albrecht} {et~al.}(2011){Albrecht}, {Winn}, {Johnson}, {Butler},
  {Crane}, {Shectman}, {Thompson}, {Narita}, {Sato}, {Hirano}, {Enya}, \&
  {Fischer}}]{Albrecht2011}
{Albrecht}, S., {Winn}, J.~N., {Johnson}, J.~A., {et~al.} 2011, \apj, 738, 50

\bibitem[{{Albrecht} {et~al.}(2012{\natexlab{b}}){Albrecht}, {Winn}, {Johnson},
  {Howard}, {Marcy}, {Butler}, {Arriagada}, {Crane}, {Shectman}, {Thompson},
  {Hirano}, {Bakos}, \& {Hartman}}]{Albrecht2012}
{Albrecht}, S., {Winn}, J.~N., {Johnson}, J.~A., {et~al.} 2012{\natexlab{b}},
  \apj, 757, 18

\bibitem[{{Albrecht} {et~al.}(2012{\natexlab{c}}){Albrecht}, {Winn}, {Johnson},
  {Howard}, {Marcy}, {Butler}, {Arriagada}, {Crane}, {Shectman}, {Thompson},
  {Hirano}, {Bakos}, \& {Hartman}}]{2012ApJ...757...18A}
{Albrecht}, S., {Winn}, J.~N., {Johnson}, J.~A., {et~al.} 2012{\natexlab{c}},
  \apj, 757, 18

\bibitem[{{Albrecht} {et~al.}(2013){Albrecht}, {Winn}, {Marcy}, {Howard},
  {Isaacson}, \& {Johnson}}]{Albrecht2013}
{Albrecht}, S., {Winn}, J.~N., {Marcy}, G.~W., {et~al.} 2013, \apj, 771, 11

\bibitem[{{Albrecht} {et~al.}(2022){Albrecht}, {Dawson}, \&
  {Winn}}]{Albrecht2022}
{Albrecht}, S.~H., {Dawson}, R.~I., \& {Winn}, J.~N. 2022, \pasp, 134, 082001

\bibitem[{{Albrecht} {et~al.}(2021){Albrecht}, {Marcussen}, {Winn}, {Dawson},
  \& {Knudstrup}}]{Albrecht2021}
{Albrecht}, S.~H., {Marcussen}, M.~L., {Winn}, J.~N., {Dawson}, R.~I., \&
  {Knudstrup}, E. 2021, \apjl, 916, L1

\bibitem[{{Almenara} {et~al.}(2022){Almenara}, {H{\'e}brard}, {D{\'\i}az},
  {Laskar}, {Correia}, {Anderson}, {Boisse}, {Bonfils}, {Brown}, {Casanova},
  {Cameron}, {Fern{\'a}ndez}, {Jenkins}, {Kiefer}, {des {\'E}tangs},
  {Lissauer}, {Maciejewski}, {McCormac}, {Osborn}, {Pollacco}, {Ricker},
  {S{\'a}nchez}, {Seager}, {Udry}, {Verilhac}, \& {Winn}}]{Almenara2022}
{Almenara}, J.~M., {H{\'e}brard}, G., {D{\'\i}az}, R.~F., {et~al.} 2022, \aap,
  663, A134

\bibitem[{{Alsubai} {et~al.}(2018){Alsubai}, {Tsvetanov}, {Latham}, {Bieryla},
  {Esquerdo}, {Mislis}, {Pyrzas}, {Foxell}, {McCormac}, {Baranec}, {Vilchez},
  {West}, {Esamdin}, {Dang}, {Dalee}, {Al-Rajihi}, \&
  {Al-Harbi}}]{2018AJ....155...52A}
{Alsubai}, K., {Tsvetanov}, Z.~I., {Latham}, D.~W., {et~al.} 2018, \aj, 155, 52

\bibitem[{{Anderson} {et~al.}(2011){Anderson}, {Collier Cameron}, {Gillon},
  {Hellier}, {Jehin}, {Lendl}, {Queloz}, {Smalley}, {Triaud}, \&
  {Vanhuysse}}]{Anderson2011}
{Anderson}, D.~R., {Collier Cameron}, A., {Gillon}, M., {et~al.} 2011, \aap,
  534, A16

\bibitem[{{Anderson} {et~al.}(2021){Anderson}, {Winn}, \&
  {Penev}}]{Anderson2021}
{Anderson}, K.~R., {Winn}, J.~N., \& {Penev}, K. 2021, \apj, 914, 56

\bibitem[{{Astropy Collaboration} {et~al.}(2022){Astropy Collaboration},
  {Price-Whelan}, {Lim}, {Earl}, {Starkman}, {Bradley}, {Shupe}, {Patil},
  {Corrales}, {Brasseur}, {N{\"o}the}, {Donath}, {Tollerud}, {Morris},
  {Ginsburg}, {Vaher}, {Weaver}, {Tocknell}, {Jamieson}, {van Kerkwijk},
  {Robitaille}, {Merry}, {Bachetti}, {G{\"u}nther}, {Aldcroft},
  {Alvarado-Montes}, {Archibald}, {B{\'o}di}, {Bapat}, {Barentsen},
  {Baz{\'a}n}, {Biswas}, {Boquien}, {Burke}, {Cara}, {Cara}, {Conroy},
  {Conseil}, {Craig}, {Cross}, {Cruz}, {D'Eugenio}, {Dencheva}, {Devillepoix},
  {Dietrich}, {Eigenbrot}, {Erben}, {Ferreira}, {Foreman-Mackey}, {Fox},
  {Freij}, {Garg}, {Geda}, {Glattly}, {Gondhalekar}, {Gordon}, {Grant},
  {Greenfield}, {Groener}, {Guest}, {Gurovich}, {Handberg}, {Hart},
  {Hatfield-Dodds}, {Homeier}, {Hosseinzadeh}, {Jenness}, {Jones}, {Joseph},
  {Kalmbach}, {Karamehmetoglu}, {Ka{\l}uszy{\'n}ski}, {Kelley}, {Kern},
  {Kerzendorf}, {Koch}, {Kulumani}, {Lee}, {Ly}, {Ma}, {MacBride}, {Maljaars},
  {Muna}, {Murphy}, {Norman}, {O'Steen}, {Oman}, {Pacifici}, {Pascual},
  {Pascual-Granado}, {Patil}, {Perren}, {Pickering}, {Rastogi}, {Roulston},
  {Ryan}, {Rykoff}, {Sabater}, {Sakurikar}, {Salgado}, {Sanghi}, {Saunders},
  {Savchenko}, {Schwardt}, {Seifert-Eckert}, {Shih}, {Jain}, {Shukla}, {Sick},
  {Simpson}, {Singanamalla}, {Singer}, {Singhal}, {Sinha}, {Sip{\H{o}}cz},
  {Spitler}, {Stansby}, {Streicher}, {{\v{S}}umak}, {Swinbank}, {Taranu},
  {Tewary}, {Tremblay}, {de Val-Borro}, {Van Kooten}, {Vasovi{\'c}}, {Verma},
  {de Miranda Cardoso}, {Williams}, {Wilson}, {Winkel}, {Wood-Vasey}, {Xue},
  {Yoachim}, {Zhang}, {Zonca}, \& {Astropy Project
  Contributors}}]{astropy:2022}
{Astropy Collaboration}, {Price-Whelan}, A.~M., {Lim}, P.~L., {et~al.} 2022,
  \apj, 935, 167

\bibitem[{{Astropy Collaboration} {et~al.}(2018){Astropy Collaboration},
  {Price-Whelan}, {Sip{\H{o}}cz}, {G{\"u}nther}, {Lim}, {Crawford}, {Conseil},
  {Shupe}, {Craig}, {Dencheva}, {Ginsburg}, {VanderPlas}, {Bradley},
  {P{\'e}rez-Su{\'a}rez}, {de Val-Borro}, {Aldcroft}, {Cruz}, {Robitaille},
  {Tollerud}, {Ardelean}, {Babej}, {Bach}, {Bachetti}, {Bakanov}, {Bamford},
  {Barentsen}, {Barmby}, {Baumbach}, {Berry}, {Biscani}, {Boquien}, {Bostroem},
  {Bouma}, {Brammer}, {Bray}, {Breytenbach}, {Buddelmeijer}, {Burke},
  {Calderone}, {Cano Rodr{\'\i}guez}, {Cara}, {Cardoso}, {Cheedella}, {Copin},
  {Corrales}, {Crichton}, {D'Avella}, {Deil}, {Depagne}, {Dietrich}, {Donath},
  {Droettboom}, {Earl}, {Erben}, {Fabbro}, {Ferreira}, {Finethy}, {Fox},
  {Garrison}, {Gibbons}, {Goldstein}, {Gommers}, {Greco}, {Greenfield},
  {Groener}, {Grollier}, {Hagen}, {Hirst}, {Homeier}, {Horton}, {Hosseinzadeh},
  {Hu}, {Hunkeler}, {Ivezi{\'c}}, {Jain}, {Jenness}, {Kanarek}, {Kendrew},
  {Kern}, {Kerzendorf}, {Khvalko}, {King}, {Kirkby}, {Kulkarni}, {Kumar},
  {Lee}, {Lenz}, {Littlefair}, {Ma}, {Macleod}, {Mastropietro}, {McCully},
  {Montagnac}, {Morris}, {Mueller}, {Mumford}, {Muna}, {Murphy}, {Nelson},
  {Nguyen}, {Ninan}, {N{\"o}the}, {Ogaz}, {Oh}, {Parejko}, {Parley}, {Pascual},
  {Patil}, {Patil}, {Plunkett}, {Prochaska}, {Rastogi}, {Reddy Janga},
  {Sabater}, {Sakurikar}, {Seifert}, {Sherbert}, {Sherwood-Taylor}, {Shih},
  {Sick}, {Silbiger}, {Singanamalla}, {Singer}, {Sladen}, {Sooley},
  {Sornarajah}, {Streicher}, {Teuben}, {Thomas}, {Tremblay}, {Turner},
  {Terr{\'o}n}, {van Kerkwijk}, {de la Vega}, {Watkins}, {Weaver}, {Whitmore},
  {Woillez}, {Zabalza}, \& {Astropy Contributors}}]{astropy:2018}
{Astropy Collaboration}, {Price-Whelan}, A.~M., {Sip{\H{o}}cz}, B.~M., {et~al.}
  2018, \aj, 156, 123

\bibitem[{{Astropy Collaboration} {et~al.}(2013){Astropy Collaboration},
  {Robitaille}, {Tollerud}, {Greenfield}, {Droettboom}, {Bray}, {Aldcroft},
  {Davis}, {Ginsburg}, {Price-Whelan}, {Kerzendorf}, {Conley}, {Crighton},
  {Barbary}, {Muna}, {Ferguson}, {Grollier}, {Parikh}, {Nair}, {Unther},
  {Deil}, {Woillez}, {Conseil}, {Kramer}, {Turner}, {Singer}, {Fox}, {Weaver},
  {Zabalza}, {Edwards}, {Azalee Bostroem}, {Burke}, {Casey}, {Crawford},
  {Dencheva}, {Ely}, {Jenness}, {Labrie}, {Lim}, {Pierfederici}, {Pontzen},
  {Ptak}, {Refsdal}, {Servillat}, \& {Streicher}}]{astropy:2013}
{Astropy Collaboration}, {Robitaille}, T.~P., {Tollerud}, E.~J., {et~al.} 2013,
  \aap, 558, A33

\bibitem[{{Bakos} {et~al.}(2010){Bakos}, {Torres}, {P{\'a}l}, {Hartman},
  {Kov{\'a}cs}, {Noyes}, {Latham}, {Sasselov}, {Sip{\H{o}}cz}, {Esquerdo},
  {Fischer}, {Johnson}, {Marcy}, {Butler}, {Isaacson}, {Howard}, {Vogt},
  {Kov{\'a}cs}, {Fernandez}, {Mo{\'o}r}, {Stefanik}, {L{\'a}z{\'a}r}, {Papp},
  \& {S{\'a}ri}}]{2010ApJ...710.1724B}
{Bakos}, G.~{\'A}., {Torres}, G., {P{\'a}l}, A., {et~al.} 2010, \apj, 710, 1724

\bibitem[{{Barker} \& {Ogilvie}(2010)}]{Barker2010}
{Barker}, A.~J. \& {Ogilvie}, G.~I. 2010, \mnras, 404, 1849

\bibitem[{{Batygin} \& {Adams}(2013)}]{BatyginAdams2013}
{Batygin}, K. \& {Adams}, F.~C. 2013, \apj, 778, 169

\bibitem[{{Beatty} {et~al.}(2017){Beatty}, {Stevens}, {Collins}, {Col{\'o}n},
  {James}, {Kreidberg}, {Pepper}, {Rodriguez}, {Siverd}, {Stassun}, \&
  {Kielkopf}}]{2017AJ....154...25B}
{Beatty}, T.~G., {Stevens}, D.~J., {Collins}, K.~A., {et~al.} 2017, \aj, 154,
  25

\bibitem[{{Beck} \& {Giles}(2005)}]{Beck2005}
{Beck}, J.~G. \& {Giles}, P. 2005, \apjl, 621, L153

\bibitem[{{Benz} {et~al.}(2021){Benz}, {Broeg}, {Fortier}, {Rando}, {Beck},
  {Beck}, {Queloz}, {Ehrenreich}, {Maxted}, {Isaak}, {Billot}, {Alibert},
  {Alonso}, {Ant{\'o}nio}, {Asquier}, {Bandy}, {B{\'a}rczy}, {Barrado},
  {Barros}, {Baumjohann}, {Bekkelien}, {Bergomi}, {Biondi}, {Bonfils},
  {Borsato}, {Brandeker}, {Busch}, {Cabrera}, {Cessa}, {Charnoz}, {Chazelas},
  {Collier Cameron}, {Corral Van Damme}, {Cortes}, {Davies}, {Deleuil},
  {Deline}, {Delrez}, {Demangeon}, {Demory}, {Erikson}, {Farinato}, {Fossati},
  {Fridlund}, {Futyan}, {Gandolfi}, {Garcia Munoz}, {Gillon}, {Guterman},
  {Gutierrez}, {Hasiba}, {Heng}, {Hernandez}, {Hoyer}, {Kiss}, {Kovacs},
  {Kuntzer}, {Laskar}, {Lecavelier des Etangs}, {Lendl}, {L{\'o}pez}, {Lora},
  {Lovis}, {L{\"u}ftinger}, {Magrin}, {Malvasio}, {Marafatto}, {Michaelis}, {de
  Miguel}, {Modrego}, {Munari}, {Nascimbeni}, {Olofsson}, {Ottacher},
  {Ottensamer}, {Pagano}, {Palacios}, {Pall{\'e}}, {Peter}, {Piazza}, {Piotto},
  {Pizarro}, {Pollaco}, {Ragazzoni}, {Ratti}, {Rauer}, {Ribas}, {Rieder},
  {Rohlfs}, {Safa}, {Salatti}, {Santos}, {Scandariato}, {S{\'e}gransan},
  {Simon}, {Smith}, {Sordet}, {Sousa}, {Steller}, {Szab{\'o}}, {Szoke},
  {Thomas}, {Tschentscher}, {Udry}, {Van Grootel}, {Viotto}, {Walter},
  {Walton}, {Wildi}, \& {Wolter}}]{Benz2021}
{Benz}, W., {Broeg}, C., {Fortier}, A., {et~al.} 2021, Experimental Astronomy,
  51, 109

\bibitem[{{Best} \& {Petrovich}(2022)}]{BestPetrovich2022}
{Best}, S. \& {Petrovich}, C. 2022, \apjl, 925, L5

\bibitem[{{Bieryla} {et~al.}(2014){Bieryla}, {Hartman}, {Bakos}, {Bhatti},
  {Kov{\'a}cs}, {Boisse}, {Latham}, {Buchhave}, {Csubry}, {Penev}, {de
  Val-Borro}, {B{\'e}ky}, {Falco}, {Torres}, {Noyes}, {Berlind}, {Calkins},
  {Esquerdo}, {L{\'a}z{\'a}r}, {Papp}, \& {S{\'a}ri}}]{2014AJ....147...84B}
{Bieryla}, A., {Hartman}, J.~D., {Bakos}, G.~{\'A}., {et~al.} 2014, \aj, 147,
  84

\bibitem[{{Bonomo} {et~al.}(2017{\natexlab{a}}){Bonomo}, {Desidera}, {Benatti},
  {Borsa}, {Crespi}, {Damasso}, {Lanza}, {Sozzetti}, {Lodato}, {Marzari},
  {Boccato}, {Claudi}, {Cosentino}, {Covino}, {Gratton}, {Maggio}, {Micela},
  {Molinari}, {Pagano}, {Piotto}, {Poretti}, {Smareglia}, {Affer}, {Biazzo},
  {Bignamini}, {Esposito}, {Giacobbe}, {H{\'e}brard}, {Malavolta}, {Maldonado},
  {Mancini}, {Martinez Fiorenzano}, {Masiero}, {Nascimbeni}, {Pedani},
  {Rainer}, \& {Scandariato}}]{Bonomo2017}
{Bonomo}, A.~S., {Desidera}, S., {Benatti}, S., {et~al.} 2017{\natexlab{a}},
  \aap, 602, A107

\bibitem[{{Bonomo} {et~al.}(2017{\natexlab{b}}){Bonomo}, {Desidera}, {Benatti},
  {Borsa}, {Crespi}, {Damasso}, {Lanza}, {Sozzetti}, {Lodato}, {Marzari},
  {Boccato}, {Claudi}, {Cosentino}, {Covino}, {Gratton}, {Maggio}, {Micela},
  {Molinari}, {Pagano}, {Piotto}, {Poretti}, {Smareglia}, {Affer}, {Biazzo},
  {Bignamini}, {Esposito}, {Giacobbe}, {H{\'e}brard}, {Malavolta}, {Maldonado},
  {Mancini}, {Martinez Fiorenzano}, {Masiero}, {Nascimbeni}, {Pedani},
  {Rainer}, \& {Scandariato}}]{2017A&A...602A.107B}
{Bonomo}, A.~S., {Desidera}, S., {Benatti}, S., {et~al.} 2017{\natexlab{b}},
  \aap, 602, A107

\bibitem[{{Borsato} {et~al.}(2021){Borsato}, {Piotto}, {Gandolfi},
  {Nascimbeni}, {Lacedelli}, {Marzari}, {Billot}, {Maxted}, {Sousa}, {Cameron},
  {Bonfanti}, {Wilson}, {Serrano}, {Garai}, {Alibert}, {Alonso}, {Asquier},
  {B{\'a}rczy}, {Bandy}, {Barrado}, {Barros}, {Baumjohann}, {Beck}, {Beck},
  {Benz}, {Bonfils}, {Brandeker}, {Broeg}, {Cabrera}, {Charnoz}, {Csizmadia},
  {Davies}, {Deleuil}, {Delrez}, {Demangeon}, {Demory}, {des Etangs},
  {Ehrenreich}, {Erikson}, {Escud{\'e}}, {Fortier}, {Fossati}, {Fridlund},
  {Gillon}, {Guedel}, {Hasiba}, {Heng}, {Hoyer}, {Isaak}, {Kiss}, {Kopp},
  {Laskar}, {Lendl}, {Lovis}, {Magrin}, {Munari}, {Olofsson}, {Ottensamer},
  {Pagano}, {Pall{\'e}}, {Peter}, {Pollacco}, {Queloz}, {Ragazzoni}, {Rando},
  {Rauer}, {Ribas}, {S{\'e}gransan}, {Santos}, {Scandariato}, {Simon}, {Smith},
  {Steller}, {Szab{\'o}}, {Thomas}, {Udry}, {Van Grootel}, \&
  {Walton}}]{Borsato2021}
{Borsato}, L., {Piotto}, G., {Gandolfi}, D., {et~al.} 2021, \mnras, 506, 3810

\bibitem[{{Bourrier} {et~al.}(2023){Bourrier}, {Attia}, {Mallonn}, {Marret},
  {Lendl}, {Konig}, {Krenn}, {Cretignier}, {Allart}, {Henry}, {Bryant},
  {Leleu}, {Nielsen}, {Hebrard}, {Hara}, {Ehrenreich}, {Seidel}, {dos Santos},
  {Lovis}, {Bayliss}, {Cegla}, {Dumusque}, {Boisse}, {Boucher}, {Bouchy},
  {Pepe}, {Lavie}, {Rey Cerda}, {S{\'e}gransan}, {Udry}, \&
  {Vrignaud}}]{2023A&A...669A..63B}
{Bourrier}, V., {Attia}, O., {Mallonn}, M., {et~al.} 2023, \aap, 669, A63

\bibitem[{{Bourrier} {et~al.}(2020){Bourrier}, {Ehrenreich}, {Lendl},
  {Cretignier}, {Allart}, {Dumusque}, {Cegla}, {Su{\'a}rez-Mascare{\~n}o},
  {Wyttenbach}, {Hoeijmakers}, {Melo}, {Kuntzer}, {Astudillo-Defru}, {Giles},
  {Heng}, {Kitzmann}, {Lavie}, {Lovis}, {Murgas}, {Nascimbeni}, {Pepe}, {Pino},
  {Segransan}, \& {Udry}}]{2020A&A...635A.205B}
{Bourrier}, V., {Ehrenreich}, D., {Lendl}, M., {et~al.} 2020, \aap, 635, A205

\bibitem[{{Bourrier} {et~al.}(2021{\natexlab{a}}){Bourrier}, {Lovis},
  {Cretignier}, {Allart}, {Dumusque}, {Delisle}, {Deline}, {Sousa},
  {Adibekyan}, {Alibert}, {Barros}, {Borsa}, {Cristiani}, {Demangeon},
  {Ehrenreich}, {Figueira}, {Gonz{\'a}lez Hern{\'a}ndez}, {Lendl}, {Lillo-Box},
  {Lo Curto}, {Di Marcantonio}, {Martins}, {M{\'e}gevand}, {Mehner}, {Micela},
  {Molaro}, {Oshagh}, {Palle}, {Pepe}, {Poretti}, {Rebolo}, {Santos},
  {Scandariato}, {Seidel}, {Sozzetti}, {Su{\'a}rez Mascare{\~n}o}, \& {Zapatero
  Osorio}}]{Bourrier2021}
{Bourrier}, V., {Lovis}, C., {Cretignier}, M., {et~al.} 2021{\natexlab{a}},
  \aap, 654, A152

\bibitem[{{Bourrier} {et~al.}(2021{\natexlab{b}}){Bourrier}, {Lovis},
  {Cretignier}, {Allart}, {Dumusque}, {Delisle}, {Deline}, {Sousa},
  {Adibekyan}, {Alibert}, {Barros}, {Borsa}, {Cristiani}, {Demangeon},
  {Ehrenreich}, {Figueira}, {Gonz{\'a}lez Hern{\'a}ndez}, {Lendl}, {Lillo-Box},
  {Lo Curto}, {Di Marcantonio}, {Martins}, {M{\'e}gevand}, {Mehner}, {Micela},
  {Molaro}, {Oshagh}, {Palle}, {Pepe}, {Poretti}, {Rebolo}, {Santos},
  {Scandariato}, {Seidel}, {Sozzetti}, {Su{\'a}rez Mascare{\~n}o}, \& {Zapatero
  Osorio}}]{2021A&A...654A.152B}
{Bourrier}, V., {Lovis}, C., {Cretignier}, M., {et~al.} 2021{\natexlab{b}},
  \aap, 654, A152

\bibitem[{{Brahm} {et~al.}(2019){Brahm}, {Espinoza}, {Rabus}, {Jord{\'a}n},
  {D{\'\i}az}, {Rojas}, {Vu{\v{c}}kovi{\'c}}, {Zapata}, {Cort{\'e}s}, {Drass},
  {Jenkins}, {Lachaume}, {Pantoja}, {Sarkis}, {Soto}, {Vasquez}, {Henning}, \&
  {Jones}}]{Brahm2019}
{Brahm}, R., {Espinoza}, N., {Rabus}, M., {et~al.} 2019, \mnras, 483, 1970

\bibitem[{{Brahm} {et~al.}(2020){Brahm}, {Nielsen}, {Wittenmyer}, {Wang},
  {Rodriguez}, {Espinoza}, {Jones}, {Jord{\'a}n}, {Henning}, {Hobson},
  {Kossakowski}, {Rojas}, {Sarkis}, {Schlecker}, {Trifonov}, {Shahaf},
  {Ricker}, {Vanderspek}, {Latham}, {Seager}, {Winn}, {Jenkins}, {Addison},
  {Bakos}, {Bhatti}, {Bayliss}, {Berlind}, {Bieryla}, {Bouchy}, {Bowler},
  {Brice{\~n}o}, {Brown}, {Bryant}, {Caldwell}, {Charbonneau}, {Collins},
  {Davis}, {Esquerdo}, {Fulton}, {Guerrero}, {Henze}, {Hogan}, {Horner},
  {Huang}, {Irwin}, {Kane}, {Kielkopf}, {Mann}, {Mazeh}, {McCormac}, {McCully},
  {Mengel}, {Mireles}, {Okumura}, {Plavchan}, {Quinn}, {Rabus}, {Saesen},
  {Schlieder}, {Segransan}, {Shiao}, {Shporer}, {Siverd}, {Stassun}, {Suc},
  {Tan}, {Torres}, {Tinney}, {Udry}, {Vanzi}, {Vezie}, {Vines}, {Vuckovic},
  {Wright}, {Yahalomi}, {Zapata}, {Zhang}, \& {Ziegler}}]{Brahm2020}
{Brahm}, R., {Nielsen}, L.~D., {Wittenmyer}, R.~A., {et~al.} 2020, \aj, 160,
  235

\bibitem[{{Brown}(2014)}]{Brown2014}
{Brown}, D.~J.~A. 2014, \mnras, 442, 1844

\bibitem[{{Brown} {et~al.}(2014){Brown}, {Anderson}, {Doyle}, {Maxted},
  {Smalley}, {McCormac}, {Almenera}, {Prieto-Arranz}, {Deleuil}, {Diaz},
  {Foxell}, {Hebrard}, {Lendl}, {Delrez}, {Gillon}, {Jehin}, {Lam}, {Triaud},
  {Turner}, {Armstrong}, {Bouchy}, {Collier Cameron}, {Pollacco}, {Faedi},
  {Gomez Maqueo Chew}, {Hebb}, {Hellier}, {Neveu-VanMalle}, {Palle}, {Queloz},
  {Segransan}, {Udry}, \& {West}}]{2014arXiv1412.7761B}
{Brown}, D.~J.~A., {Anderson}, D.~R., {Doyle}, A.~P., {et~al.} 2014, arXiv
  e-prints, arXiv:1412.7761

\bibitem[{{Brown} {et~al.}(2011){Brown}, {Collier Cameron}, {Hall}, {Hebb}, \&
  {Smalley}}]{Brown2011}
{Brown}, D.~J.~A., {Collier Cameron}, A., {Hall}, C., {Hebb}, L., \& {Smalley},
  B. 2011, \mnras, 415, 605

\bibitem[{{Brown} {et~al.}(2017){Brown}, {Triaud}, {Doyle}, {Gillon}, {Lendl},
  {Anderson}, {Collier Cameron}, {H{\'e}brard}, {Hellier}, {Lovis}, {Maxted},
  {Pepe}, {Pollacco}, {Queloz}, \& {Smalley}}]{2017MNRAS.464..810B}
{Brown}, D.~J.~A., {Triaud}, A.~H.~M.~J., {Doyle}, A.~P., {et~al.} 2017,
  \mnras, 464, 810

\bibitem[{{Bruntt} {et~al.}(2010){Bruntt}, {Bedding}, {Quirion}, {Lo Curto},
  {Carrier}, {Smalley}, {Dall}, {Arentoft}, {Bazot}, \& {Butler}}]{Bruntt2010}
{Bruntt}, H., {Bedding}, T.~R., {Quirion}, P.~O., {et~al.} 2010, \mnras, 405,
  1907

\bibitem[{{Butler} {et~al.}(1996){Butler}, {Marcy}, {Williams}, {McCarthy},
  {Dosanjh}, \& {Vogt}}]{Butler1996}
{Butler}, R.~P., {Marcy}, G.~W., {Williams}, E., {et~al.} 1996, \pasp, 108, 500

\bibitem[{{Castelli} \& {Kurucz}(2003)}]{Castelli2003}
{Castelli}, F. \& {Kurucz}, R.~L. 2003, in Modelling of Stellar Atmospheres,
  ed. N.~{Piskunov}, W.~W. {Weiss}, \& D.~F. {Gray}, Vol. 210, A20

\bibitem[{{Castro-Gonz{\'a}lez} {et~al.}(2022){Castro-Gonz{\'a}lez}, {D{\'\i}ez
  Alonso}, {Men{\'e}ndez Blanco}, {Livingston}, {de Leon}, {Lillo-Box},
  {Korth}, {Fern{\'a}ndez Men{\'e}ndez}, {Recio}, {Izquierdo-Ruiz}, {Coya
  Lozano}, {Garc{\'\i}a de la Cuesta}, {G{\'o}mez Hern{\'a}ndez}, {Vidal
  Blanco}, {Hevia D{\'\i}az}, {Pardo Silva}, {P{\'e}rez Acevedo}, {Polancos
  Ruiz}, {Padilla Tijer{\'\i}n}, {V{\'a}zquez Garc{\'\i}a}, {Su{\'a}rez
  G{\'o}mez}, {Garc{\'\i}a Riesgo}, {Gonz{\'a}lez Guti{\'e}rrez}, {Bonavera},
  {Gonz{\'a}lez-Nuevo}, {Rodr{\'\i}guez Pereira}, {S{\'a}nchez Lasheras},
  {S{\'a}nchez Rodr{\'\i}guez}, {Mu{\~n}iz}, {Santos Rodr{\'\i}guez}, \& {de
  Cos Juez}}]{2022MNRAS.509.1075C}
{Castro-Gonz{\'a}lez}, A., {D{\'\i}ez Alonso}, E., {Men{\'e}ndez Blanco}, J.,
  {et~al.} 2022, \mnras, 509, 1075

\bibitem[{{Cegla} {et~al.}(2016{\natexlab{a}}){Cegla}, {Lovis}, {Bourrier},
  {Beeck}, {Watson}, \& {Pepe}}]{Cegla2016}
{Cegla}, H.~M., {Lovis}, C., {Bourrier}, V., {et~al.} 2016{\natexlab{a}}, \aap,
  588, A127

\bibitem[{{Cegla} {et~al.}(2016{\natexlab{b}}){Cegla}, {Lovis}, {Bourrier},
  {Beeck}, {Watson}, \& {Pepe}}]{2016A&A...588A.127C}
{Cegla}, H.~M., {Lovis}, C., {Bourrier}, V., {et~al.} 2016{\natexlab{b}}, \aap,
  588, A127

\bibitem[{{Cegla} {et~al.}(2023){Cegla}, {Roguet-Kern}, {Lendl}, {Akinsanmi},
  {McCormac}, {Oshagh}, {Wheatley}, {Chen}, {Allart}, {Mortier}, {Bourrier},
  {Buchschacher}, {Lovis}, {Sosnowska}, {Sulis}, {Turner}, {Casasayas-Barris},
  {Palle}, {Yan}, {Burleigh}, {Casewell}, {Goad}, {Hawthorn}, \&
  {Wyttenbach}}]{2023A&A...674A.174C}
{Cegla}, H.~M., {Roguet-Kern}, N., {Lendl}, M., {et~al.} 2023, \aap, 674, A174

\bibitem[{{Chaplin} {et~al.}(2013){Chaplin}, {Sanchis-Ojeda}, {Campante},
  {Handberg}, {Stello}, {Winn}, {Basu}, {Christensen-Dalsgaard}, {Davies},
  {Metcalfe}, {Buchhave}, {Fischer}, {Bedding}, {Cochran}, {Elsworth},
  {Gilliland}, {Hekker}, {Huber}, {Isaacson}, {Karoff}, {Kawaler}, {Kjeldsen},
  {Latham}, {Lund}, {Lundkvist}, {Marcy}, {Miglio}, {Barclay}, \&
  {Lissauer}}]{Chaplin2013}
{Chaplin}, W.~J., {Sanchis-Ojeda}, R., {Campante}, T.~L., {et~al.} 2013, \apj,
  766, 101

\bibitem[{{Chatterjee} {et~al.}(2008){Chatterjee}, {Ford}, {Matsumura}, \&
  {Rasio}}]{Chatterjee2008}
{Chatterjee}, S., {Ford}, E.~B., {Matsumura}, S., \& {Rasio}, F.~A. 2008, \apj,
  686, 580

\bibitem[{{Chontos} {et~al.}(2024){Chontos}, {Huber}, {Grunblatt}, {Saunders},
  {Winn}, {McCormack}, {Knudstrup}, {Albrecht}, {Crossfield}, {Rodriguez},
  {Ciardi}, {Collins}, {Jenkins}, {Bieryla}, {Batalha}, {Beard}, {Dai},
  {Dalba}, {Fetherolf}, {Giacalone}, {Hill}, {Howard}, {Isaacson}, {Kane},
  {Lubin}, {MacDougall}, {Mo{\v{c}}nik}, {Akana Murphy}, {Petigura},
  {Pidhorodetska}, {Polanski}, {Robertson}, {Rubenzahl}, {Turtelboom}, {Weiss},
  {Van Zandt}, {Rocker}, {Vanderspek}, {Latham}, {Seager}, {Quinn}, {Shporer},
  {Eisner}, {Goeke}, {Levine}, {Ting}, {Howell}, {Schlieder}, {Benni}, {Boyle},
  {Gan}, {Girardin}, {Gonzalez}, {Gregorio}, {Horne}, {Livingston}, {Lund},
  {Mann}, {Massey}, {Matthews}, {McLeod}, {Palle}, {Popowicz}, {Relles},
  {Schwarz}, {Sefako}, {Srdoc}, {Tan}, {Wang}, \& {Ziegler}}]{Chontos2024}
{Chontos}, A., {Huber}, D., {Grunblatt}, S.~K., {et~al.} 2024, arXiv e-prints,
  arXiv:2402.07893

\bibitem[{{Christiansen} {et~al.}(2017){Christiansen}, {Vanderburg}, {Burt},
  {Fulton}, {Batygin}, {Benneke}, {Brewer}, {Charbonneau}, {Ciardi}, {Collier
  Cameron}, {Coughlin}, {Crossfield}, {Dressing}, {Greene}, {Howard}, {Latham},
  {Molinari}, {Mortier}, {Mullally}, {Pepe}, {Rice}, {Sinukoff}, {Sozzetti},
  {Thompson}, {Udry}, {Vogt}, {Barman}, {Batalha}, {Bouchy}, {Buchhave},
  {Butler}, {Cosentino}, {Dupuy}, {Ehrenreich}, {Fiorenzano}, {Hansen},
  {Henning}, {Hirsch}, {Holden}, {Isaacson}, {Johnson}, {Knutson}, {Kosiarek},
  {L{\'o}pez-Morales}, {Lovis}, {Malavolta}, {Mayor}, {Micela}, {Motalebi},
  {Petigura}, {Phillips}, {Piotto}, {Rogers}, {Sasselov}, {Schlieder},
  {S{\'e}gransan}, {Watson}, \& {Weiss}}]{2017AJ....154..122C}
{Christiansen}, J.~L., {Vanderburg}, A., {Burt}, J., {et~al.} 2017, \aj, 154,
  122

\bibitem[{{Claret}(2018)}]{Claret2018}
{Claret}, A. 2018, \aap, 618, A20

\bibitem[{{Claret}(2021)}]{Claret2021}
{Claret}, A. 2021, Research Notes of the American Astronomical Society, 5, 13

\bibitem[{{Claret} {et~al.}(2012){Claret}, {Hauschildt}, \&
  {Witte}}]{Claret2012}
{Claret}, A., {Hauschildt}, P.~H., \& {Witte}, S. 2012, \aap, 546, A14

\bibitem[{{Claret} {et~al.}(2013){Claret}, {Hauschildt}, \&
  {Witte}}]{Claret2013}
{Claret}, A., {Hauschildt}, P.~H., \& {Witte}, S. 2013, \aap, 552, A16

\bibitem[{{Cosentino} {et~al.}(2014){Cosentino}, {Lovis}, {Pepe}, {Collier
  Cameron}, {Latham}, {Molinari}, {Udry}, {Bezawada}, {Buchschacher},
  {Figueira}, {Fleury}, {Ghedina}, {Glenday}, {Gonzalez}, {Guerra}, {Henry},
  {Hughes}, {Maire}, {Motalebi}, \& {Phillips}}]{Cosentino2014}
{Cosentino}, R., {Lovis}, C., {Pepe}, F., {et~al.} 2014, in Society of
  Photo-Optical Instrumentation Engineers (SPIE) Conference Series, Vol. 9147,
  Ground-based and Airborne Instrumentation for Astronomy V, ed. S.~K.
  {Ramsay}, I.~S. {McLean}, \& H.~{Takami}, 91478C

\bibitem[{{Crane} {et~al.}(2010){Crane}, {Shectman}, {Butler}, {Thompson},
  {Birk}, {Jones}, \& {Burley}}]{Crane2010}
{Crane}, J.~D., {Shectman}, S.~A., {Butler}, R.~P., {et~al.} 2010, in Society
  of Photo-Optical Instrumentation Engineers (SPIE) Conference Series, Vol.
  7735, Ground-based and Airborne Instrumentation for Astronomy III, ed. I.~S.
  {McLean}, S.~K. {Ramsay}, \& H.~{Takami}, 773553

\bibitem[{{Cristo} {et~al.}(2024){Cristo}, {Esparza Borges}, {Santos},
  {Demangeon}, {Palle}, {Psaridi}, {Bourrier}, {Faria}, {Allart}, {Azevedo
  Silva}, {Borsa}, {Alibert}, {Figueira}, {Gonz{\'a}lez Hern{\'a}ndez},
  {Lendl}, {Lillo-Box}, {Lo Curto}, {Di Marcantonio}, {Martins}, {Nunes},
  {Pepe}, {Seidel}, {Sousa}, {Sozzetti}, {Stangret}, {Su{\'a}rez
  Mascare{\~n}o}, {Tabernero}, \& {Zapatero Osorio}}]{2024A&A...682A..28C}
{Cristo}, E., {Esparza Borges}, E., {Santos}, N.~C., {et~al.} 2024, \aap, 682,
  A28

\bibitem[{{Crossfield} {et~al.}(2017){Crossfield}, {Ciardi}, {Isaacson},
  {Howard}, {Petigura}, {Weiss}, {Fulton}, {Sinukoff}, {Schlieder}, {Mawet},
  {Ruane}, {de Pater}, {de Kleer}, {Davies}, {Christiansen}, {Dressing},
  {Hirsch}, {Benneke}, {Crepp}, {Kosiarek}, {Livingston}, {Gonzales},
  {Beichman}, \& {Knutson}}]{2017AJ....153..255C}
{Crossfield}, I. J.~M., {Ciardi}, D.~R., {Isaacson}, H., {et~al.} 2017, \aj,
  153, 255

\bibitem[{{Crouzet} {et~al.}(2020){Crouzet}, {Healy}, {H{\'e}brard},
  {McCullough}, {Long}, {Monta{\~n}{\'e}s-Rodr{\'\i}guez}, {Ribas},
  {Vilardell}, {Herrero}, {Garcia-Melendo}, {Conjat}, {Foote}, {Garlitz}, {Vo},
  {Santos}, {de Bruijne}, {Osborn}, {Dalal}, \& {Nielsen}}]{Crouzet2020}
{Crouzet}, N., {Healy}, B.~F., {H{\'e}brard}, G., {et~al.} 2020, \aj, 159, 44

\bibitem[{{Crouzet} {et~al.}(2017){Crouzet}, {McCullough}, {Long}, {Montanes
  Rodriguez}, {Lecavelier des Etangs}, {Ribas}, {Bourrier}, {H{\'e}brard},
  {Vilardell}, {Deleuil}, {Herrero}, {Garcia-Melendo}, {Akhenak}, {Foote},
  {Gary}, {Benni}, {Guillot}, {Conjat}, {M{\'e}karnia}, {Garlitz}, {Burke},
  {Courcol}, \& {Demangeon}}]{2017AJ....153...94C}
{Crouzet}, N., {McCullough}, P.~R., {Long}, D., {et~al.} 2017, \aj, 153, 94

\bibitem[{{Czesla} {et~al.}(2023){Czesla}, {Schneider}, \&
  {Hatzes}}]{2023A&A...677L..12C}
{Czesla}, S., {Schneider}, P.~C., \& {Hatzes}, A. 2023, \aap, 677, L12

\bibitem[{{da Silva} {et~al.}(2006){da Silva}, {Udry}, {Bouchy}, {Mayor},
  {Moutou}, {Pont}, {Queloz}, {Santos}, {S{\'e}gransan}, \&
  {Zucker}}]{daSilva2006}
{da Silva}, R., {Udry}, S., {Bouchy}, F., {et~al.} 2006, \aap, 446, 717

\bibitem[{{Dai} {et~al.}(2023){Dai}, {Masuda}, {Beard}, {Robertson},
  {Goldberg}, {Batygin}, {Bouma}, {Lissauer}, {Knudstrup}, {Albrecht},
  {Howard}, {Knutson}, {Petigura}, {Weiss}, {Isaacson}, {Kristiansen},
  {Osborn}, {Wang}, {Wang}, {Behmard}, {Greklek-McKeon}, {Vissapragada},
  {Batalha}, {Brinkman}, {Chontos}, {Crossfield}, {Dressing}, {Fetherolf},
  {Fulton}, {Hill}, {Huber}, {Kane}, {Lubin}, {MacDougall}, {Mayo},
  {Mo{\v{c}}nik}, {Akana Murphy}, {Rubenzahl}, {Scarsdale}, {Tyler}, {Zandt},
  {Polanski}, {Schwengeler}, {Terentev}, {Benni}, {Bieryla}, {Ciardi}, {Falk},
  {Furlan}, {Girardin}, {Guerra}, {Hesse}, {Howell}, {Lillo-Box}, {Matthews},
  {Twicken}, {Villase{\~n}or}, {Latham}, {Jenkins}, {Ricker}, {Seager},
  {Vanderspek}, \& {Winn}}]{2023AJ....165...33D}
{Dai}, F., {Masuda}, K., {Beard}, C., {et~al.} 2023, \aj, 165, 33

\bibitem[{{Dalal} {et~al.}(2019){Dalal}, {H{\'e}brard}, {Lecavelier des
  {\'E}tangs}, {Petit}, {Bourrier}, {Laskar}, {K{\"o}nig}, \&
  {Correia}}]{2019A&A...631A..28D}
{Dalal}, S., {H{\'e}brard}, G., {Lecavelier des {\'E}tangs}, A., {et~al.} 2019,
  \aap, 631, A28

\bibitem[{{Damasso} {et~al.}(2015){Damasso}, {Esposito}, {Nascimbeni},
  {Desidera}, {Bonomo}, {Bieryla}, {Malavolta}, {Biazzo}, {Sozzetti}, {Covino},
  {Latham}, {Gandolfi}, {Rainer}, {Petrovich}, {Collins}, {Boccato}, {Claudi},
  {Cosentino}, {Gratton}, {Lanza}, {Maggio}, {Micela}, {Molinari}, {Pagano},
  {Piotto}, {Poretti}, {Smareglia}, {Di Fabrizio}, {Giacobbe}, {Gomez-Jimenez},
  {Murabito}, {Molinaro}, {Affer}, {Barbieri}, {Bedin}, {Benatti}, {Borsa},
  {Maldonado}, {Mancini}, {Scandariato}, {Southworth}, \& {Zanmar
  Sanchez}}]{Damasso2015}
{Damasso}, M., {Esposito}, M., {Nascimbeni}, V., {et~al.} 2015, \aap, 581, L6

\bibitem[{{Damiani} \& {Mathis}(2018)}]{Damiani2018}
{Damiani}, C. \& {Mathis}, S. 2018, \aap, 618, A90

\bibitem[{{Dawson}(2014)}]{Dawson2014}
{Dawson}, R.~I. 2014, \apjl, 790, L31

\bibitem[{{Dawson} \& {Johnson}(2018)}]{Dawson2018}
{Dawson}, R.~I. \& {Johnson}, J.~A. 2018, \araa, 56, 175

\bibitem[{{Demangeon} {et~al.}(2018){Demangeon}, {Faedi}, {H{\'e}brard},
  {Brown}, {Barros}, {Doyle}, {Maxted}, {Collier Cameron}, {Hay}, {Alikakos},
  {Anderson}, {Armstrong}, {Boumis}, {Bonomo}, {Bouchy}, {Delrez}, {Gillon},
  {Haswell}, {Hellier}, {Jehin}, {Kiefer}, {Lam}, {Lendl}, {Mancini},
  {McCormac}, {Norton}, {Osborn}, {Palle}, {Pepe}, {Pollacco}, {Prieto-Arranz},
  {Queloz}, {S{\'e}gransan}, {Smalley}, {Triaud}, {Udry}, {West}, \&
  {Wheatley}}]{2018A&A...610A..63D}
{Demangeon}, O.~D.~S., {Faedi}, F., {H{\'e}brard}, G., {et~al.} 2018, \aap,
  610, A63

\bibitem[{{Djupvik} \& {Andersen}(2010)}]{Djupvik2010}
{Djupvik}, A.~A. \& {Andersen}, J. 2010, in Astrophysics and Space Science
  Proceedings, Vol.~14, Highlights of Spanish Astrophysics V, 211

\bibitem[{{Dong} \& {Foreman-Mackey}(2023)}]{Dong2023}
{Dong}, J. \& {Foreman-Mackey}, D. 2023, \aj, 166, 112

\bibitem[{{Dorval} {et~al.}(2020){Dorval}, {Talens}, {Otten}, {Brahm},
  {Jord{\'a}n}, {Torres}, {Vanzi}, {Zapata}, {Henry}, {Paredes}, {Jao},
  {James}, {Hinojosa}, {Bakos}, {Csubry}, {Bhatti}, {Suc}, {Osip}, {Mamajek},
  {Mellon}, {Wyttenbach}, {Stuik}, {Kenworthy}, {Bailey}, {Ireland},
  {Crawford}, {Lomberg}, {Kuhn}, \& {Snellen}}]{2020A&A...635A..60D}
{Dorval}, P., {Talens}, G.~J.~J., {Otten}, G.~P.~P.~L., {et~al.} 2020, \aap,
  635, A60

\bibitem[{{Doyle} {et~al.}(2014){Doyle}, {Davies}, {Smalley}, {Chaplin}, \&
  {Elsworth}}]{Doyle2014}
{Doyle}, A.~P., {Davies}, G.~R., {Smalley}, B., {Chaplin}, W.~J., \&
  {Elsworth}, Y. 2014, \mnras, 444, 3592

\bibitem[{{Doyle} {et~al.}(2023){Doyle}, {Cegla}, {Anderson}, {Lendl},
  {Bourrier}, {Bryant}, {Vines}, {Allart}, {Bayliss}, {Burleigh},
  {Buchschacher}, {Casewell}, {Hawthorn}, {Jenkins}, {Lafarga}, {Moyano},
  {Psaridi}, {Roguet-Kern}, {Sosnowska}, \& {Wheatley}}]{2023MNRAS.522.4499D}
{Doyle}, L., {Cegla}, H.~M., {Anderson}, D.~R., {et~al.} 2023, \mnras, 522,
  4499

\bibitem[{{Dravins}(1987)}]{Dravins1987}
{Dravins}, D. 1987, \aap, 172, 200

\bibitem[{{Dravins}(1999)}]{Dravins1999}
{Dravins}, D. 1999, in Astronomical Society of the Pacific Conference Series,
  Vol. 185, IAU Colloq. 170: Precise Stellar Radial Velocities, ed. J.~B.
  {Hearnshaw} \& C.~D. {Scarfe}, 268

\bibitem[{{Dravins} \& {Nordlund}(1990{\natexlab{a}})}]{Dravins1990a}
{Dravins}, D. \& {Nordlund}, A. 1990{\natexlab{a}}, \aap, 228, 184

\bibitem[{{Dravins} \& {Nordlund}(1990{\natexlab{b}})}]{Dravins1990b}
{Dravins}, D. \& {Nordlund}, A. 1990{\natexlab{b}}, \aap, 228, 203

\bibitem[{{Dumusque} {et~al.}(2021){Dumusque}, {Cretignier}, {Sosnowska},
  {Buchschacher}, {Lovis}, {Phillips}, {Pepe}, {Alesina}, {Buchhave},
  {Burnier}, {Cecconi}, {Cegla}, {Cloutier}, {Collier Cameron}, {Cosentino},
  {Ghedina}, {Gonz{\'a}lez}, {Haywood}, {Latham}, {Lodi}, {L{\'o}pez-Morales},
  {Maldonado}, {Malavolta}, {Micela}, {Molinari}, {Mortier}, {P{\'e}rez
  Ventura}, {Pinamonti}, {Poretti}, {Rice}, {Riverol}, {Riverol}, {San Juan},
  {S{\'e}gransan}, {Sozzetti}, {Thompson}, {Udry}, \& {Wilson}}]{Dumusque2021}
{Dumusque}, X., {Cretignier}, M., {Sosnowska}, D., {et~al.} 2021, \aap, 648,
  A103

\bibitem[{{Eastman} {et~al.}(2016){Eastman}, {Beatty}, {Siverd}, {Antognini},
  {Penny}, {Gonzales}, {Crepp}, {Howard}, {Avril}, {Bieryla}, {Collins},
  {Fulton}, {Ge}, {Gregorio}, {Ma}, {Mellon}, {Oberst}, {Wang}, {Gaudi},
  {Pepper}, {Stassun}, {Buchhave}, {Jensen}, {Latham}, {Berlind}, {Calkins},
  {Cargile}, {Col{\'o}n}, {Dhital}, {Esquerdo}, {Johnson}, {Kielkopf},
  {Manner}, {Mao}, {McLeod}, {Penev}, {Stefanik}, {Street}, {Zambelli},
  {DePoy}, {Gould}, {Marshall}, {Pogge}, {Trueblood}, \&
  {Trueblood}}]{Eastman2016}
{Eastman}, J.~D., {Beatty}, T.~G., {Siverd}, R.~J., {et~al.} 2016, \aj, 151, 45

\bibitem[{{Ehrenreich} {et~al.}(2020){Ehrenreich}, {Lovis}, {Allart}, {Zapatero
  Osorio}, {Pepe}, {Cristiani}, {Rebolo}, {Santos}, {Borsa}, {Demangeon},
  {Dumusque}, {Gonz{\'a}lez Hern{\'a}ndez}, {Casasayas-Barris},
  {S{\'e}gransan}, {Sousa}, {Abreu}, {Adibekyan}, {Affolter}, {Allende Prieto},
  {Alibert}, {Aliverti}, {Alves}, {Amate}, {Avila}, {Baldini}, {Bandy}, {Benz},
  {Bianco}, {Bolmont}, {Bouchy}, {Bourrier}, {Broeg}, {Cabral}, {Calderone},
  {Pall{\'e}}, {Cegla}, {Cirami}, {Coelho}, {Conconi}, {Coretti}, {Cumani},
  {Cupani}, {Dekker}, {Delabre}, {Deiries}, {D'Odorico}, {Di Marcantonio},
  {Figueira}, {Fragoso}, {Genolet}, {Genoni}, {G{\'e}nova Santos}, {Hara},
  {Hughes}, {Iwert}, {Kerber}, {Knudstrup}, {Landoni}, {Lavie}, {Lizon},
  {Lendl}, {Lo Curto}, {Maire}, {Manescau}, {Martins}, {M{\'e}gevand},
  {Mehner}, {Micela}, {Modigliani}, {Molaro}, {Monteiro}, {Monteiro},
  {Moschetti}, {M{\"u}ller}, {Nunes}, {Oggioni}, {Oliveira}, {Pariani},
  {Pasquini}, {Poretti}, {Rasilla}, {Redaelli}, {Riva}, {Santana Tschudi},
  {Santin}, {Santos}, {Segovia Milla}, {Seidel}, {Sosnowska}, {Sozzetti},
  {Span{\`o}}, {Su{\'a}rez Mascare{\~n}o}, {Tabernero}, {Tenegi}, {Udry},
  {Zanutta}, \& {Zerbi}}]{2020Natur.580..597E}
{Ehrenreich}, D., {Lovis}, C., {Allart}, R., {et~al.} 2020, \nat, 580, 597

\bibitem[{{Eisner} {et~al.}(2020){Eisner}, {Barrag{\'a}n}, {Aigrain},
  {Lintott}, {Miller}, {Zicher}, {Boyajian}, {Brice{\~n}o}, {Bryant},
  {Christiansen}, {Feinstein}, {Flor-Torres}, {Fridlund}, {Gandolfi},
  {Gilbert}, {Guerrero}, {Jenkins}, {Jones}, {Kristiansen}, {Vanderburg},
  {Law}, {L{\'o}pez-S{\'a}nchez}, {Mann}, {Safron}, {Schwamb}, {Stassun},
  {Osborn}, {Wang}, {Zic}, {Ziegler}, {Barnet}, {Bean}, {Bundy}, {Chetnik},
  {Dawson}, {Garstone}, {Stenner}, {Huten}, {Larish}, {Melanson}, {Mitchell},
  {Moore}, {Peltsch}, {Rogers}, {Schuster}, {Smith}, {Simister}, {Tanner},
  {Terentev}, \& {Tsymbal}}]{Eisner2020}
{Eisner}, N.~L., {Barrag{\'a}n}, O., {Aigrain}, S., {et~al.} 2020, \mnras, 494,
  750

\bibitem[{{Fabrycky} \& {Tremaine}(2007)}]{Fabrycky2007}
{Fabrycky}, D. \& {Tremaine}, S. 2007, \apj, 669, 1298

\bibitem[{{Fabrycky} {et~al.}(2012){Fabrycky}, {Ford}, {Steffen}, {Rowe},
  {Carter}, {Moorhead}, {Batalha}, {Borucki}, {Bryson}, {Buchhave},
  {Christiansen}, {Ciardi}, {Cochran}, {Endl}, {Fanelli}, {Fischer}, {Fressin},
  {Geary}, {Haas}, {Hall}, {Holman}, {Jenkins}, {Koch}, {Latham}, {Li},
  {Lissauer}, {Lucas}, {Marcy}, {Mazeh}, {McCauliff}, {Quinn}, {Ragozzine},
  {Sasselov}, \& {Shporer}}]{Fabrycky2012}
{Fabrycky}, D.~C., {Ford}, E.~B., {Steffen}, J.~H., {et~al.} 2012, \apj, 750,
  114

\bibitem[{Foreman-Mackey(2016)}]{corner}
Foreman-Mackey, D. 2016, The Journal of Open Source Software, 1, 24

\bibitem[{{Foreman-Mackey} {et~al.}(2017){Foreman-Mackey}, {Agol},
  {Ambikasaran}, \& {Angus}}]{celerite}
{Foreman-Mackey}, D., {Agol}, E., {Ambikasaran}, S., \& {Angus}, R. 2017, \aj,
  154, 220

\bibitem[{{Foreman-Mackey} {et~al.}(2013){Foreman-Mackey}, {Hogg}, {Lang}, \&
  {Goodman}}]{emcee}
{Foreman-Mackey}, D., {Hogg}, D.~W., {Lang}, D., \& {Goodman}, J. 2013, \pasp,
  125, 306

\bibitem[{{Frandsen} \& {Lindberg}(1999)}]{Frandsen1999}
{Frandsen}, S. \& {Lindberg}, B. 1999, in Astrophysics with the NOT, ed.
  H.~{Karttunen} \& V.~{Piirola}, 71

\bibitem[{{Frazier} {et~al.}(2023){Frazier}, {Stef{\'a}nsson}, {Mahadevan},
  {Yee}, {Ca{\~n}as}, {Winn}, {Luhn}, {Dai}, {Doyle}, {Cegla}, {Kanodia},
  {Robertson}, {Wisniewski}, {Bender}, {Dong}, {Gupta}, {Halverson}, {Hawley},
  {Hebb}, {Holcomb}, {Kowalski}, {Libby-Roberts}, {Lin}, {McElwain}, {Ninan},
  {Petrovich}, {Roy}, {Schwab}, {Terrien}, \& {Wright}}]{2023ApJ...944L..41F}
{Frazier}, R.~C., {Stef{\'a}nsson}, G., {Mahadevan}, S., {et~al.} 2023, \apjl,
  944, L41

\bibitem[{{Freudling} {et~al.}(2013){Freudling}, {Romaniello}, {Bramich},
  {Ballester}, {Forchi}, {Garc{\'{\i}}a-Dabl{\'o}}, {Moehler}, \&
  {Neeser}}]{Freudling2013}
{Freudling}, W., {Romaniello}, M., {Bramich}, D.~M., {et~al.} 2013, \aap, 559,
  A96

\bibitem[{{Gandolfi} {et~al.}(2015){Gandolfi}, {Parviainen}, {Deeg}, {Lanza},
  {Fridlund}, {Prada Moroni}, {Alonso}, {Augusteijn}, {Cabrera}, {Evans},
  {Geier}, {Hatzes}, {Holczer}, {Hoyer}, {Kangas}, {Mazeh}, {Pagano}, {Tal-Or},
  \& {Tingley}}]{Gandolfi2015}
{Gandolfi}, D., {Parviainen}, H., {Deeg}, H.~J., {et~al.} 2015, \aap, 576, A11

\bibitem[{{Garhart} {et~al.}(2020){Garhart}, {Deming}, {Mandell}, {Knutson},
  {Wallack}, {Burrows}, {Fortney}, {Hood}, {Seay}, {Sing}, {Benneke}, {Fraine},
  {Kataria}, {Lewis}, {Madhusudhan}, {McCullough}, {Stevenson}, \&
  {Wakeford}}]{2020AJ....159..137G}
{Garhart}, E., {Deming}, D., {Mandell}, A., {et~al.} 2020, \aj, 159, 137

\bibitem[{{Gaudi} \& {Winn}(2007)}]{GaudiWinn2007}
{Gaudi}, B.~S. \& {Winn}, J.~N. 2007, \apj, 655, 550

\bibitem[{{Gillon} {et~al.}(2011){Gillon}, {Doyle}, {Lendl}, {Maxted},
  {Triaud}, {Anderson}, {Barros}, {Bento}, {Collier-Cameron}, {Enoch}, {Faedi},
  {Hellier}, {Jehin}, {Magain}, {Montalb{\'a}n}, {Pepe}, {Pollacco}, {Queloz},
  {Smalley}, {Segransan}, {Smith}, {Southworth}, {Udry}, {West}, \&
  {Wheatley}}]{Gillon2011}
{Gillon}, M., {Doyle}, A.~P., {Lendl}, M., {et~al.} 2011, \aap, 533, A88

\bibitem[{{Gomes} {et~al.}(2017){Gomes}, {Deienno}, \&
  {Morbidelli}}]{Gomes2017}
{Gomes}, R., {Deienno}, R., \& {Morbidelli}, A. 2017, \aj, 153, 27

\bibitem[{{Gratia} \& {Fabrycky}(2017)}]{GratiaFabrycky2017}
{Gratia}, P. \& {Fabrycky}, D. 2017, \mnras, 464, 1709

\bibitem[{{Gray}(2005)}]{Gray2005}
{Gray}, D.~F. 2005, {The Observation and Analysis of Stellar Photospheres}

\bibitem[{{Grouffal} {et~al.}(2022{\natexlab{a}}){Grouffal}, {Santerne},
  {Bourrier}, {Dumusque}, {Triaud}, {Malavolta}, {Kunovac}, {Armstrong},
  {Attia}, {Barros}, {Boisse}, {Deleuil}, {Demangeon}, {Dressing}, {Figueira},
  {Lillo-Box}, {Mortier}, {Nardiello}, {Santos}, \& {Sousa}}]{Grouffal2022}
{Grouffal}, S., {Santerne}, A., {Bourrier}, V., {et~al.} 2022{\natexlab{a}},
  \aap, 668, A172

\bibitem[{{Grouffal} {et~al.}(2022{\natexlab{b}}){Grouffal}, {Santerne},
  {Bourrier}, {Dumusque}, {Triaud}, {Malavolta}, {Kunovac}, {Armstrong},
  {Attia}, {Barros}, {Boisse}, {Deleuil}, {Demangeon}, {Dressing}, {Figueira},
  {Lillo-Box}, {Mortier}, {Nardiello}, {Santos}, \&
  {Sousa}}]{2022A&A...668A.172G}
{Grouffal}, S., {Santerne}, A., {Bourrier}, V., {et~al.} 2022{\natexlab{b}},
  \aap, 668, A172

\bibitem[{{Hagelberg} {et~al.}(2023{\natexlab{a}}){Hagelberg}, {Nielsen},
  {Attia}, {Bourrier}, {Pearce}, {Venturini}, {Winn}, {Bouchy}, {Bouma},
  {Brice{\~n}o}, {Collins}, {Davis}, {Eastman}, {Evans}, {Falk}, {Grieves},
  {Guerrero}, {Hellier}, {Jones}, {Latham}, {Law}, {Mann}, {Marmier}, {Ottoni},
  {Radford}, {Restori}, {Rudat}, {Dos Santos}, {Seager}, {Stassun},
  {Stockdale}, {Udry}, {Wang}, \& {Ziegler}}]{Hagelberg2023}
{Hagelberg}, J., {Nielsen}, L.~D., {Attia}, O., {et~al.} 2023{\natexlab{a}},
  \aap, 679, A70

\bibitem[{{Hagelberg} {et~al.}(2023{\natexlab{b}}){Hagelberg}, {Nielsen},
  {Attia}, {Bourrier}, {Pearce}, {Venturini}, {Winn}, {Bouchy}, {Bouma},
  {Brice{\~n}o}, {Collins}, {Davis}, {Eastman}, {Evans}, {Falk}, {Grieves},
  {Guerrero}, {Hellier}, {Jones}, {Latham}, {Law}, {Mann}, {Marmier}, {Ottoni},
  {Radford}, {Restori}, {Rudat}, {Dos Santos}, {Seager}, {Stassun},
  {Stockdale}, {Udry}, {Wang}, \& {Ziegler}}]{2023A&A...679A..70H}
{Hagelberg}, J., {Nielsen}, L.~D., {Attia}, O., {et~al.} 2023{\natexlab{b}},
  \aap, 679, A70

\bibitem[{{Hansen}(2012)}]{Hansen2012}
{Hansen}, B. M.~S. 2012, \apj, 757, 6

\bibitem[{Harris {et~al.}(2020)Harris, Millman, van~der Walt, Gommers,
  Virtanen, Cournapeau, Wieser, Taylor, Berg, Smith, Kern, Picus, Hoyer, van
  Kerkwijk, Brett, Haldane, del R{\'{i}}o, Wiebe, Peterson,
  G{\'{e}}rard-Marchant, Sheppard, Reddy, Weckesser, Abbasi, Gohlke, \&
  Oliphant}]{numpy}
Harris, C.~R., Millman, K.~J., van~der Walt, S.~J., {et~al.} 2020, Nature, 585,
  357

\bibitem[{{H{\'e}brard} {et~al.}(2013){H{\'e}brard}, {Collier Cameron},
  {Brown}, {D{\'\i}az}, {Faedi}, {Smalley}, {Anderson}, {Armstrong}, {Barros},
  {Bento}, {Bouchy}, {Doyle}, {Enoch}, {G{\'o}mez Maqueo Chew}, {H{\'e}brard},
  {Hellier}, {Lendl}, {Lister}, {Maxted}, {McCormac}, {Moutou}, {Pollacco},
  {Queloz}, {Santerne}, {Skillen}, {Southworth}, {Tregloan-Reed}, {Triaud},
  {Udry}, {Vanhuysse}, {Watson}, {West}, \& {Wheatley}}]{Hebrard2013}
{H{\'e}brard}, G., {Collier Cameron}, A., {Brown}, D.~J.~A., {et~al.} 2013,
  \aap, 549, A134

\bibitem[{{H{\'e}brard} {et~al.}(2020){H{\'e}brard}, {D{\'\i}az}, {Correia},
  {Collier Cameron}, {Laskar}, {Pollacco}, {Almenara}, {Anderson}, {Barros},
  {Boisse}, {Bonomo}, {Bouchy}, {Bou{\'e}}, {Boumis}, {Brown}, {Dalal},
  {Deleuil}, {Demangeon}, {Doyle}, {Haswell}, {Hellier}, {Osborn}, {Kiefer},
  {Kolb}, {Lam}, {Lecavelier des {\'E}tangs}, {Lopez}, {Martin-Lagarde},
  {Maxted}, {McCormac}, {Nielsen}, {Pall{\'e}}, {Prieto-Arranz}, {Queloz},
  {Santerne}, {Smalley}, {Turner}, {Udry}, {Verilhac}, {West}, {Wheatley}, \&
  {Wilson}}]{Hebrard2020}
{H{\'e}brard}, G., {D{\'\i}az}, R.~F., {Correia}, A.~C.~M., {et~al.} 2020,
  \aap, 640, A32

\bibitem[{{H{\'e}brard} {et~al.}(2011){H{\'e}brard}, {Ehrenreich}, {Bouchy},
  {Delfosse}, {Moutou}, {Arnold}, {Boisse}, {Bonfils}, {D{\'\i}az},
  {Eggenberger}, {Forveille}, {Lagrange}, {Lovis}, {Pepe}, {Perrier}, {Queloz},
  {Santerne}, {Santos}, {S{\'e}gransan}, {Udry}, \&
  {Vidal-Madjar}}]{Hebrard2011}
{H{\'e}brard}, G., {Ehrenreich}, D., {Bouchy}, F., {et~al.} 2011, \aap, 527,
  L11

\bibitem[{{Hellier} {et~al.}(2019{\natexlab{a}}){Hellier}, {Anderson},
  {Bouchy}, {Burdanov}, {Collier Cameron}, {Delrez}, {Gillon}, {Jehin},
  {Lendl}, {Nielsen}, {Maxted}, {Pepe}, {Pollacco}, {Queloz}, {S{\'e}gransan},
  {Smalley}, {Triaud}, {Udry}, \& {West}}]{Hellier2019}
{Hellier}, C., {Anderson}, D.~R., {Bouchy}, F., {et~al.} 2019{\natexlab{a}},
  \mnras, 482, 1379

\bibitem[{{Hellier} {et~al.}(2019{\natexlab{b}}){Hellier}, {Anderson},
  {Triaud}, {Bouchy}, {Burdanov}, {Collier Cameron}, {Delrez}, {Ehrenreich},
  {Gillon}, {Jehin}, {Lendl}, {Linder}, {Nielsen}, {Maxted}, {Pepe},
  {Pollacco}, {Queloz}, {S{\'e}gransan}, {Smalley}, {Spake}, {Temple}, {Udry},
  {West}, \& {Wyttenbach}}]{2019MNRAS.488.3067H}
{Hellier}, C., {Anderson}, D.~R., {Triaud}, A.~H.~M.~J., {et~al.}
  2019{\natexlab{b}}, \mnras, 488, 3067

\bibitem[{{Hirano} {et~al.}(2011){Hirano}, {Suto}, {Winn}, {Taruya}, {Narita},
  {Albrecht}, \& {Sato}}]{Hirano2011}
{Hirano}, T., {Suto}, Y., {Winn}, J.~N., {et~al.} 2011, \apj, 742, 69

\bibitem[{{Hixenbaugh} {et~al.}(2023){Hixenbaugh}, {Wang}, {Rice}, \&
  {Wang}}]{2023ApJ...949L..35H}
{Hixenbaugh}, K., {Wang}, X.-Y., {Rice}, M., \& {Wang}, S. 2023, \apjl, 949,
  L35

\bibitem[{{Hjorth} {et~al.}(2021){Hjorth}, {Albrecht}, {Hirano}, {Winn},
  {Dawson}, {Zanazzi}, {Knudstrup}, \& {Sato}}]{Hjorth2021}
{Hjorth}, M., {Albrecht}, S., {Hirano}, T., {et~al.} 2021, Proceedings of the
  National Academy of Science, 118, e2017418118

\bibitem[{{Hjorth} {et~al.}(2019){Hjorth}, {Albrecht}, {Talens}, {Grundahl},
  {Justesen}, {Otten}, {Antoci}, {Dorval}, {Foxell}, {Fredslund Andersen},
  {Murgas}, {Palle}, {Stuik}, {Snellen}, \& {Van Eylen}}]{Hjorth2019}
{Hjorth}, M., {Albrecht}, S., {Talens}, G.~J.~J., {et~al.} 2019, \aap, 631, A76

\bibitem[{{Howard} {et~al.}(2010){Howard}, {Johnson}, {Marcy}, {Fischer},
  {Wright}, {Bernat}, {Henry}, {Peek}, {Isaacson}, {Apps}, {Endl}, {Cochran},
  {Valenti}, {Anderson}, \& {Piskunov}}]{Howard2010}
{Howard}, A.~W., {Johnson}, J.~A., {Marcy}, G.~W., {et~al.} 2010, \apj, 721,
  1467

\bibitem[{{Howell} {et~al.}(2014){Howell}, {Sobeck}, {Haas}, {Still},
  {Barclay}, {Mullally}, {Troeltzsch}, {Aigrain}, {Bryson}, {Caldwell},
  {Chaplin}, {Cochran}, {Huber}, {Marcy}, {Miglio}, {Najita}, {Smith},
  {Twicken}, \& {Fortney}}]{Howell2014}
{Howell}, S.~B., {Sobeck}, C., {Haas}, M., {et~al.} 2014, \pasp, 126, 398

\bibitem[{{Huang} {et~al.}(2020){Huang}, {Quinn}, {Vanderburg}, {Becker},
  {Rodriguez}, {Pozuelos}, {Gandolfi}, {Zhou}, {Mann}, {Collins}, {Crossfield},
  {Barkaoui}, {Collins}, {Fridlund}, {Gillon}, {Gonzales}, {G{\"u}nther},
  {Henry}, {Howell}, {James}, {Jao}, {Jehin}, {Jensen}, {Kane}, {Lissauer},
  {Matthews}, {Matson}, {Paredes}, {Schlieder}, {Stassun}, {Shporer}, {Sha},
  {Tan}, {Georgieva}, {Mathur}, {Palle}, {Persson}, {Van Eylen}, {Ricker},
  {Vanderspek}, {Latham}, {Winn}, {Seager}, {Jenkins}, {Burke}, {Goeke},
  {Rinehart}, {Rose}, {Ting}, {Torres}, \& {Wong}}]{Huang2020}
{Huang}, C.~X., {Quinn}, S.~N., {Vanderburg}, A., {et~al.} 2020, \apjl, 892, L7

\bibitem[{{Huber} {et~al.}(2013){Huber}, {Carter}, {Barbieri}, {Miglio},
  {Deck}, {Fabrycky}, {Montet}, {Buchhave}, {Chaplin}, {Hekker},
  {Montalb{\'a}n}, {Sanchis-Ojeda}, {Basu}, {Bedding}, {Campante},
  {Christensen-Dalsgaard}, {Elsworth}, {Stello}, {Arentoft}, {Ford},
  {Gilliland}, {Handberg}, {Howard}, {Isaacson}, {Johnson}, {Karoff},
  {Kawaler}, {Kjeldsen}, {Latham}, {Lund}, {Lundkvist}, {Marcy}, {Metcalfe},
  {Silva Aguirre}, \& {Winn}}]{Huber2013}
{Huber}, D., {Carter}, J.~A., {Barbieri}, M., {et~al.} 2013, Science, 342, 331

\bibitem[{Hunter(2007)}]{matplotlib}
Hunter, J.~D. 2007, Computing in Science \& Engineering, 9, 90

\bibitem[{{Ikwut-Ukwa} {et~al.}(2020){Ikwut-Ukwa}, {Rodriguez}, {Bieryla},
  {Vanderburg}, {Mocnik}, {Kane}, {Quinn}, {Col{\'o}n}, {Zhou}, {Eastman},
  {Huang}, {Latham}, {Dotson}, {Jenkins}, {Ricker}, {Seager}, {Vanderspek},
  {Winn}, {Barclay}, {Barentsen}, {Berta-Thompson}, {Charbonneau}, {Dragomir},
  {Daylan}, {G{\"u}nther}, {Hedges}, {Henze}, {McDermott}, {Schlieder},
  {Quintana}, {Smith}, {Twicken}, \& {Yahalomi}}]{Ikwut2020}
{Ikwut-Ukwa}, M., {Rodriguez}, J.~E., {Bieryla}, A., {et~al.} 2020, \aj, 160,
  209

\bibitem[{{Johnson} {et~al.}(2011){Johnson}, {Winn}, {Bakos}, {Hartman},
  {Morton}, {Torres}, {Kov{\'a}cs}, {Latham}, {Noyes}, {Sato}, {Esquerdo},
  {Fischer}, {Marcy}, {Howard}, {Buchhave}, {F{\H{u}}r{\'e}sz}, {Quinn},
  {B{\'e}ky}, {Sasselov}, {Stefanik}, {L{\'a}z{\'a}r}, {Papp}, \&
  {S{\'a}ri}}]{2011ApJ...735...24J}
{Johnson}, J.~A., {Winn}, J.~N., {Bakos}, G.~{\'A}., {et~al.} 2011, \apj, 735,
  24

\bibitem[{{Johnson} {et~al.}(2014){Johnson}, {Cochran}, {Albrecht},
  {Dodson-Robinson}, {Winn}, \& {Gullikson}}]{Johnson2014}
{Johnson}, M.~C., {Cochran}, W.~D., {Albrecht}, S., {et~al.} 2014, \apj, 790,
  30

\bibitem[{{Johnson} {et~al.}(2015){Johnson}, {Cochran}, {Collier Cameron}, \&
  {Bayliss}}]{2015ApJ...810L..23J}
{Johnson}, M.~C., {Cochran}, W.~D., {Collier Cameron}, A., \& {Bayliss}, D.
  2015, \apjl, 810, L23

\bibitem[{{Johnson} {et~al.}(2018){Johnson}, {Dai}, {Justesen}, {Gandolfi},
  {Hatzes}, {Nowak}, {Endl}, {Cochran}, {Hidalgo}, {Watanabe}, {Parviainen},
  {Hirano}, {Villanueva}, {Prieto-Arranz}, {Narita}, {Palle}, {Guenther},
  {Barrag{\'a}n}, {Trifonov}, {Niraula}, {MacQueen}, {Cabrera}, {Csizmadia},
  {Eigm{\"u}ller}, {Grziwa}, {Korth}, {P{\"a}tzold}, {Smith}, {Albrecht},
  {Alonso}, {Deeg}, {Erikson}, {Esposito}, {Fridlund}, {Fukui}, {Kusakabe},
  {Kuzuhara}, {Livingston}, {Monta{\~n}es Rodriguez}, {Nespral}, {Persson},
  {Purismo}, {Raimundo}, {Rauer}, {Ribas}, {Tamura}, {Van Eylen}, \&
  {Winn}}]{Johnson2018}
{Johnson}, M.~C., {Dai}, F., {Justesen}, A.~B., {et~al.} 2018, \mnras, 481, 596

\bibitem[{{Jord{\'a}n} {et~al.}(2019){Jord{\'a}n}, {Brahm}, {Espinoza},
  {Cort{\'e}s}, {D{\'\i}az}, {Drass}, {Henning}, {Jenkins}, {Jones}, {Rabus},
  {Rojas}, {Sarkis}, {Vu{\v{c}}kovi{\'c}}, {Zapata}, {Soto}, {Bakos},
  {Bayliss}, {Bhatti}, {Csubry}, {Lachaume}, {Moraga}, {Pantoja}, {Osip},
  {Shporer}, {Suc}, \& {V{\'a}squez}}]{Jordan2019}
{Jord{\'a}n}, A., {Brahm}, R., {Espinoza}, N., {et~al.} 2019, \aj, 157, 100

\bibitem[{{Kaluzny} {et~al.}(2006){Kaluzny}, {Pych}, {Rucinski}, \&
  {Thompson}}]{Kaluzny2006}
{Kaluzny}, J., {Pych}, W., {Rucinski}, S.~M., \& {Thompson}, I.~B. 2006,
  \actaa, 56, 237

\bibitem[{{Kawai} {et~al.}(2024){Kawai}, {Narita}, {Fukui}, {Watanabe}, \&
  {Inaba}}]{2024MNRAS.528..270K}
{Kawai}, Y., {Narita}, N., {Fukui}, A., {Watanabe}, N., \& {Inaba}, S. 2024,
  \mnras, 528, 270

\bibitem[{{Knudstrup} \& {Albrecht}(2022)}]{Knudstrup2022a}
{Knudstrup}, E. \& {Albrecht}, S.~H. 2022, \aap, 660, A99

\bibitem[{{Knudstrup} {et~al.}(2023{\natexlab{a}}){Knudstrup}, {Albrecht},
  {Gandolfi}, {Marcussen}, {Goffo}, {Serrano}, {Dai}, {Redfield}, {Hirano},
  {Csizmadia}, {Cochran}, {Deeg}, {Fridlund}, {Lam}, {Livingston}, {Luque},
  {Narita}, {Palle}, {Persson}, \& {Van Eylen}}]{Knudstrup2023}
{Knudstrup}, E., {Albrecht}, S.~H., {Gandolfi}, D., {et~al.}
  2023{\natexlab{a}}, \aap, 671, A164

\bibitem[{{Knudstrup} {et~al.}(2023{\natexlab{b}}){Knudstrup}, {Albrecht},
  {Gandolfi}, {Marcussen}, {Goffo}, {Serrano}, {Dai}, {Redfield}, {Hirano},
  {Csizmadia}, {Cochran}, {Deeg}, {Fridlund}, {Lam}, {Livingston}, {Luque},
  {Narita}, {Palle}, {Persson}, \& {Van Eylen}}]{2023A&A...671A.164K}
{Knudstrup}, E., {Albrecht}, S.~H., {Gandolfi}, D., {et~al.}
  2023{\natexlab{b}}, \aap, 671, A164

\bibitem[{{Knudstrup} {et~al.}(2022){Knudstrup}, {Serrano}, {Gandolfi},
  {Albrecht}, {Cochran}, {Endl}, {MacQueen}, {Tronsgaard}, {Bieryla},
  {Buchhave}, {Stassun}, {Collins}, {Nowak}, {Deeg}, {Barkaoui}, {Safonov},
  {Strakhov}, {Belinski}, {Twicken}, {Jenkins}, {Howard}, {Isaacson}, {Winn},
  {Collins}, {Conti}, {Furesz}, {Gan}, {Kielkopf}, {Massey}, {Murgas},
  {Murphy}, {Palle}, {Quinn}, {Reed}, {Ricker}, {Seager}, {Shiao}, {Schwarz},
  {Srdoc}, \& {Watanabe}}]{2022A&A...667A..22K}
{Knudstrup}, E., {Serrano}, L.~M., {Gandolfi}, D., {et~al.} 2022, \aap, 667,
  A22

\bibitem[{{Korth} {et~al.}(2023){Korth}, {Gandolfi}, {{\v{S}}ubjak}, {Howard},
  {Ataiee}, {Collins}, {Quinn}, {Mustill}, {Guillot}, {Lodieu}, {Smith},
  {Esposito}, {Rodler}, {Muresan}, {Abe}, {Albrecht}, {Alqasim}, {Barkaoui},
  {Beck}, {Burke}, {Butler}, {Conti}, {Collins}, {Crane}, {Dai}, {Deeg},
  {Evans}, {Grziwa}, {Hatzes}, {Hirano}, {Horne}, {Huang}, {Jenkins},
  {Kab{\'a}th}, {Kielkopf}, {Knudstrup}, {Latham}, {Livingston}, {Luque},
  {Mathur}, {Murgas}, {Osborne}, {Palle}, {Persson}, {Rodriguez}, {Rose},
  {Rowden}, {Schwarz}, {Seager}, {Serrano}, {Sha}, {Shectman}, {Shporer},
  {Srdoc}, {Stockdale}, {Tan}, {Teske}, {Van Eylen}, {Vanderburg},
  {Vanderspek}, {Wang}, \& {Winn}}]{Korth2023}
{Korth}, J., {Gandolfi}, D., {{\v{S}}ubjak}, J., {et~al.} 2023, \aap, 675, A115

\bibitem[{{Kreidberg}(2015)}]{Kreidberg2015}
{Kreidberg}, L. 2015, \pasp, 127, 1161

\bibitem[{{Kuhn} {et~al.}(2016){Kuhn}, {Rodriguez}, {Collins}, {Lund},
  {Siverd}, {Col{\'o}n}, {Pepper}, {Stassun}, {Cargile}, {James}, {Penev},
  {Zhou}, {Bayliss}, {Tan}, {Curtis}, {Udry}, {Segransan}, {Mawet}, {Dhital},
  {Soutter}, {Hart}, {Carter}, {Gaudi}, {Myers}, {Beatty}, {Eastman},
  {Reichart}, {Haislip}, {Kielkopf}, {Bieryla}, {Latham}, {Jensen}, {Oberst},
  \& {Stevens}}]{2016MNRAS.459.4281K}
{Kuhn}, R.~B., {Rodriguez}, J.~E., {Collins}, K.~A., {et~al.} 2016, \mnras,
  459, 4281

\bibitem[{{Labadie-Bartz} {et~al.}(2019){Labadie-Bartz}, {Rodriguez},
  {Stassun}, {Ciardi}, {Penev}, {Johnson}, {Gaudi}, {Col{\'o}n}, {Bieryla},
  {Latham}, {Pepper}, {Collins}, {Evans}, {Relles}, {Siverd}, {Bento}, {Yao},
  {Stockdale}, {Tan}, {Zhou}, {Eastman}, {Albrow}, {Bayliss}, {Beatty},
  {Berlind}, {Bozza}, {Calkins}, {Cohen}, {Curtis}, {Esquerdo}, {Feliz},
  {Fulton}, {Gregorio}, {James}, {Jensen}, {Johnson}, {Johnson}, {Joner},
  {Kasper}, {Kielkopf}, {Kuhn}, {Lund}, {Malpas}, {Manner}, {McCrady},
  {McLeod}, {Oberst}, {Penny}, {Reed}, {Sliski}, {Stephens}, {Stevens},
  {Villanueva}, {Wittenmyer}, {Wright}, \& {Zambelli}}]{Labadie2019}
{Labadie-Bartz}, J., {Rodriguez}, J.~E., {Stassun}, K.~G., {et~al.} 2019,
  \apjs, 240, 13

\bibitem[{{Lai}(2012)}]{Lai2012}
{Lai}, D. 2012, \mnras, 423, 486

\bibitem[{{Lam} {et~al.}(2017){Lam}, {Faedi}, {Brown}, {Anderson}, {Delrez},
  {Gillon}, {H{\'e}brard}, {Lendl}, {Mancini}, {Southworth}, {Smalley},
  {Triaud}, {Turner}, {Hay}, {Armstrong}, {Barros}, {Bonomo}, {Bouchy},
  {Boumis}, {Collier Cameron}, {Doyle}, {Hellier}, {Henning}, {Jehin}, {King},
  {Kirk}, {Louden}, {Maxted}, {McCormac}, {Osborn}, {Palle}, {Pepe},
  {Pollacco}, {Prieto-Arranz}, {Queloz}, {Rey}, {S{\'e}gransan}, {Udry},
  {Walker}, {West}, \& {Wheatley}}]{Lam2017}
{Lam}, K.~W.~F., {Faedi}, F., {Brown}, D.~J.~A., {et~al.} 2017, \aap, 599, A3

\bibitem[{{Lavie} {et~al.}(2023){Lavie}, {Bouchy}, {Lovis}, {Zapatero Osorio},
  {Deline}, {Barros}, {Figueira}, {Sozzetti}, {Gonz{\'a}lez Hern{\'a}ndez},
  {Lillo-Box}, {Rodrigues}, {Mehner}, {Damasso}, {Adibekyan}, {Alibert},
  {Allende Prieto}, {Cristiani}, {D'Odorico}, {Di Marcantonio}, {Ehrenreich},
  {G{\'e}nova Santos}, {Lo Curto}, {Martins}, {Micela}, {Molaro}, {Nunes},
  {Palle}, {Pepe}, {Poretti}, {Rebolo}, {Santos}, {Sousa}, {Su{\'a}rez
  Mascare{\~n}o}, {Tabrenero}, \& {Udry}}]{Lavie2023}
{Lavie}, B., {Bouchy}, F., {Lovis}, C., {et~al.} 2023, \aap, 673, A69

\bibitem[{{Li} \& {Winn}(2016)}]{LiWinn2016}
{Li}, G. \& {Winn}, J.~N. 2016, \apj, 818, 5

\bibitem[{{Libby-Roberts} {et~al.}(2023){Libby-Roberts}, {Schutte}, {Hebb},
  {Kanodia}, {Ca{\~n}as}, {Stef{\'a}nsson}, {Lin}, {Mahadevan}, {Parts},
  {Powers}, {Wisniewski}, {Bender}, {Cochran}, {Diddams}, {Everett}, {Gupta},
  {Halverson}, {Kobulnicky}, {Kowalski}, {Larsen}, {Monson}, {Ninan}, {Parker},
  {Ramsey}, {Robertson}, {Schwab}, {Swaby}, \& {Terrien}}]{2023AJ....165..249L}
{Libby-Roberts}, J.~E., {Schutte}, M., {Hebb}, L., {et~al.} 2023, \aj, 165, 249

\bibitem[{{Lightkurve Collaboration} {et~al.}(2018){Lightkurve Collaboration},
  {Cardoso}, {Hedges}, {Gully-Santiago}, {Saunders}, {Cody}, {Barclay}, {Hall},
  {Sagear}, {Turtelboom}, {Zhang}, {Tzanidakis}, {Mighell}, {Coughlin}, {Bell},
  {Berta-Thompson}, {Williams}, {Dotson}, \& {Barentsen}}]{lightkurve}
{Lightkurve Collaboration}, {Cardoso}, J.~V.~d.~M., {Hedges}, C., {et~al.}
  2018, {Lightkurve: Kepler and TESS time series analysis in Python},
  Astrophysics Source Code Library

\bibitem[{{Lin} {et~al.}(1996){Lin}, {Bodenheimer}, \& {Richardson}}]{Lin1996}
{Lin}, D.~N.~C., {Bodenheimer}, P., \& {Richardson}, D.~C. 1996, \nat, 380, 606

\bibitem[{{Lin} \& {Ogilvie}(2017)}]{Lin2017}
{Lin}, Y. \& {Ogilvie}, G.~I. 2017, \mnras, 468, 1387

\bibitem[{{Lovis} \& {Pepe}(2007)}]{Lovis2007}
{Lovis}, C. \& {Pepe}, F. 2007, \aap, 468, 1115

\bibitem[{{Lubin} {et~al.}(2023){Lubin}, {Wang}, {Rice}, {Dong}, {Wang},
  {Radzom}, {Robertson}, {Stefansson}, {Alvarado-Montes}, {Beard}, {Bender},
  {Gupta}, {Halverson}, {Kanodia}, {Li}, {Lin}, {Logsdon}, {Lubar},
  {Mahadevan}, {Ninan}, {Rajagopal}, {Roy}, {Schwab}, \&
  {Wright}}]{2023ApJ...959L...5L}
{Lubin}, J., {Wang}, X.-Y., {Rice}, M., {et~al.} 2023, \apjl, 959, L5

\bibitem[{{Luger} {et~al.}(2016){Luger}, {Agol}, {Kruse}, {Barnes}, {Becker},
  {Foreman-Mackey}, \& {Deming}}]{Luger2016}
{Luger}, R., {Agol}, E., {Kruse}, E., {et~al.} 2016, \aj, 152, 100

\bibitem[{{Luger} {et~al.}(2018){Luger}, {Kruse}, {Foreman-Mackey}, {Agol}, \&
  {Saunders}}]{Luger2018}
{Luger}, R., {Kruse}, E., {Foreman-Mackey}, D., {Agol}, E., \& {Saunders}, N.
  2018, \aj, 156, 99

\bibitem[{{Lund} {et~al.}(2017){Lund}, {Rodriguez}, {Zhou}, {Gaudi}, {Stassun},
  {Johnson}, {Bieryla}, {Oelkers}, {Stevens}, {Collins}, {Penev}, {Quinn},
  {Latham}, {Villanueva}, {Eastman}, {Kielkopf}, {Oberst}, {Jensen}, {Cohen},
  {Joner}, {Stephens}, {Relles}, {Corfini}, {Gregorio}, {Zambelli}, {Esquerdo},
  {Calkins}, {Berlind}, {Ciardi}, {Dressing}, {Patel}, {Gagnon}, {Gonzales},
  {Beatty}, {Siverd}, {Labadie-Bartz}, {Kuhn}, {Col{\'o}n}, {James}, {Pepper},
  {Fulton}, {McLeod}, {Stockdale}, {Calchi Novati}, {DePoy}, {Gould},
  {Marshall}, {Trueblood}, {Trueblood}, {Johnson}, {Wright}, {McCrady},
  {Wittenmyer}, {Johnson}, {Sergi}, {Wilson}, \&
  {Sliski}}]{2017AJ....154..194L}
{Lund}, M.~B., {Rodriguez}, J.~E., {Zhou}, G., {et~al.} 2017, \aj, 154, 194

\bibitem[{{Lund} {et~al.}(2019){Lund}, {Knudstrup}, {Silva Aguirre}, {Basu},
  {Chontos}, {Von Essen}, {Chaplin}, {Bieryla}, {Casagrande}, {Vanderburg},
  {Huber}, {Kane}, {Albrecht}, {Latham}, {Davies}, {Becker}, \&
  {Rodriguez}}]{2019AJ....158..248L}
{Lund}, M.~N., {Knudstrup}, E., {Silva Aguirre}, V., {et~al.} 2019, \aj, 158,
  248

\bibitem[{{Ma} \& {Fuller}(2021)}]{Ma2021}
{Ma}, L. \& {Fuller}, J. 2021, \apj, 918, 16

\bibitem[{{Mancini} {et~al.}(2018){Mancini}, {Esposito}, {Covino},
  {Southworth}, {Biazzo}, {Bruni}, {Ciceri}, {Evans}, {Lanza}, {Poretti},
  {Sarkis}, {Smith}, {Brogi}, {Affer}, {Benatti}, {Bignamini}, {Boccato},
  {Bonomo}, {Borsa}, {Carleo}, {Claudi}, {Cosentino}, {Damasso}, {Desidera},
  {Giacobbe}, {Gonz{\'a}lez-{\'A}lvarez}, {Gratton}, {Harutyunyan}, {Leto},
  {Maggio}, {Malavolta}, {Maldonado}, {Martinez-Fiorenzano}, {Masiero},
  {Micela}, {Molinari}, {Nascimbeni}, {Pagano}, {Pedani}, {Piotto}, {Rainer},
  {Scandariato}, {Smareglia}, {Sozzetti}, {Andreuzzi}, \&
  {Henning}}]{2018A&A...613A..41M}
{Mancini}, L., {Esposito}, M., {Covino}, E., {et~al.} 2018, \aap, 613, A41

\bibitem[{{Mancini} {et~al.}(2015){Mancini}, {Hartman}, {Penev}, {Bakos},
  {Brahm}, {Ciceri}, {Henning}, {Csubry}, {Bayliss}, {Zhou}, {Rabus}, {de
  Val-Borro}, {Espinoza}, {Jord{\'a}n}, {Suc}, {Bhatti}, {Schmidt}, {Sato},
  {Tan}, {Wright}, {Tinney}, {Addison}, {Noyes}, {L{\'a}z{\'a}r}, {Papp}, \&
  {S{\'a}ri}}]{Mancini2015}
{Mancini}, L., {Hartman}, J.~D., {Penev}, K., {et~al.} 2015, \aap, 580, A63

\bibitem[{{Mandel} \& {Agol}(2002)}]{MandelAgol2002}
{Mandel}, K. \& {Agol}, E. 2002, \apjl, 580, L171

\bibitem[{{Masuda} \& {Winn}(2020)}]{Masuda2020}
{Masuda}, K. \& {Winn}, J.~N. 2020, \aj, 159, 81

\bibitem[{{McCullough} {et~al.}(2005){McCullough}, {Stys}, {Valenti},
  {Fleming}, {Janes}, \& {Heasley}}]{McCullough2005}
{McCullough}, P.~R., {Stys}, J.~E., {Valenti}, J.~A., {et~al.} 2005, \pasp,
  117, 783

\bibitem[{{McQuillan} {et~al.}(2014){McQuillan}, {Mazeh}, \&
  {Aigrain}}]{McQuillan2014}
{McQuillan}, A., {Mazeh}, T., \& {Aigrain}, S. 2014, \apjs, 211, 24

\bibitem[{{Morgan} {et~al.}(2023){Morgan}, {Bowler}, {Tran}, {Petigura},
  {Nagpal}, \& {Blunt}}]{Morgan2023}
{Morgan}, M., {Bowler}, B.~P., {Tran}, Q.~H., {et~al.} 2023, arXiv e-prints,
  arXiv:2310.18445

\bibitem[{{Mounzer} {et~al.}(2022){Mounzer}, {Lovis}, {Seidel}, {Attia},
  {Allart}, {Bourrier}, {Ehrenreich}, {Wyttenbach}, {Astudillo-Defru},
  {Beatty}, {Cegla}, {Heng}, {Lavie}, {Lendl}, {Melo}, {Pepe}, {Pepper},
  {Rodriguez}, {S{\'e}gransan}, {Udry}, {Linder}, \&
  {Sousa}}]{2022A&A...668A...1M}
{Mounzer}, D., {Lovis}, C., {Seidel}, J.~V., {et~al.} 2022, \aap, 668, A1

\bibitem[{{Mo{\v{c}}nik} {et~al.}(2016){Mo{\v{c}}nik}, {Clark}, {Anderson},
  {Hellier}, \& {Brown}}]{2016AJ....151..150M}
{Mo{\v{c}}nik}, T., {Clark}, B.~J.~M., {Anderson}, D.~R., {Hellier}, C., \&
  {Brown}, D.~J.~A. 2016, \aj, 151, 150

\bibitem[{{Moya} {et~al.}(2011){Moya}, {Bouy}, {Marchis}, {Vicente}, \&
  {Barrado}}]{2011A&A...535A.110M}
{Moya}, A., {Bouy}, H., {Marchis}, F., {Vicente}, B., \& {Barrado}, D. 2011,
  \aap, 535, A110

\bibitem[{{Nagasawa} {et~al.}(2008){Nagasawa}, {Ida}, \&
  {Bessho}}]{Nagasawa2008}
{Nagasawa}, M., {Ida}, S., \& {Bessho}, T. 2008, \apj, 678, 498

\bibitem[{{Narita} {et~al.}(2015){Narita}, {Fukui}, {Kusakabe}, {Onitsuka},
  {Ryu}, {Yanagisawa}, {Izumiura}, {Tamura}, \& {Yamamuro}}]{Narita2015}
{Narita}, N., {Fukui}, A., {Kusakabe}, N., {et~al.} 2015, Journal of
  Astronomical Telescopes, Instruments, and Systems, 1, 045001

\bibitem[{{Narita} {et~al.}(2017){Narita}, {Hirano}, {Fukui}, {Hori}, {Dai},
  {Yu}, {Livingston}, {Ryu}, {Nowak}, {Kuzuhara}, {Sato}, {Takeda}, {Albrecht},
  {Kudo}, {Kusakabe}, {Palle}, {Ribas}, {Tamura}, {Van Eylen}, \&
  {Winn}}]{2017PASJ...69...29N}
{Narita}, N., {Hirano}, T., {Fukui}, A., {et~al.} 2017, \pasj, 69, 29

\bibitem[{{Neveu-VanMalle} {et~al.}(2014){Neveu-VanMalle}, {Queloz},
  {Anderson}, {Charbonnel}, {Collier Cameron}, {Delrez}, {Gillon}, {Hellier},
  {Jehin}, {Lendl}, {Maxted}, {Pepe}, {Pollacco}, {S{\'e}gransan}, {Smalley},
  {Smith}, {Southworth}, {Triaud}, {Udry}, \& {West}}]{2014A&A...572A..49N}
{Neveu-VanMalle}, M., {Queloz}, D., {Anderson}, D.~R., {et~al.} 2014, \aap,
  572, A49

\bibitem[{{Newton} {et~al.}(2021){Newton}, {Mann}, {Kraus}, {Livingston},
  {Vanderburg}, {Curtis}, {Thao}, {Hawkins}, {Wood}, {Rizzuto}, {Soubkiou},
  {Tofflemire}, {Zhou}, {Crossfield}, {Pearce}, {Collins}, {Conti}, {Tan},
  {Villeneuva}, {Spencer}, {Dragomir}, {Quinn}, {Jensen}, {Collins},
  {Stockdale}, {Cloutier}, {Hellier}, {Benkhaldoun}, {Ziegler}, {Brice{\~n}o},
  {Law}, {Benneke}, {Christiansen}, {Gorjian}, {Kane}, {Kreidberg}, {Morales},
  {Werner}, {Twicken}, {Levine}, {Ciardi}, {Guerrero}, {Hesse}, {Quintana},
  {Shiao}, {Smith}, {Torres}, {Ricker}, {Vanderspek}, {Seager}, {Winn},
  {Jenkins}, \& {Latham}}]{Newton2021}
{Newton}, E.~R., {Mann}, A.~W., {Kraus}, A.~L., {et~al.} 2021, \aj, 161, 65

\bibitem[{{Ogilvie}(2014)}]{Ogilvie2014}
{Ogilvie}, G.~I. 2014, \araa, 52, 171

\bibitem[{{Osborn} {et~al.}(2022){Osborn}, {Bonfanti}, {Gandolfi}, {Hedges},
  {Leleu}, {Fortier}, {Futyan}, {Gutermann}, {Maxted}, {Borsato}, {Collins},
  {Gomes da Silva}, {G{\'o}mez Maqueo Chew}, {Hooton}, {Lendl}, {Parviainen},
  {Salmon}, {Schanche}, {Serrano}, {Sousa}, {Tuson}, {Ulmer-Moll}, {Van
  Grootel}, {Wells}, {Wilson}, {Alibert}, {Alonso}, {Anglada}, {Asquier},
  {Barrado y Navascues}, {Baumjohann}, {Beck}, {Benz}, {Biondi}, {Bonfils},
  {Bouchy}, {Brandeker}, {Broeg}, {B{\'a}rczy}, {Barros}, {Cabrera}, {Charnoz},
  {Collier Cameron}, {Csizmadia}, {Davies}, {Deleuil}, {Delrez}, {Demory},
  {Ehrenreich}, {Erikson}, {Fossati}, {Fridlund}, {Gillon}, {G{\"o}mez-Munoz},
  {G{\"u}del}, {Heng}, {Hoyer}, {Isaak}, {Kiss}, {Laskar}, {Lecavelier des
  Etangs}, {Lovis}, {Magrin}, {Malavolta}, {McCormac}, {Nascimbeni},
  {Olofsson}, {Ottensamer}, {Pagano}, {Pall{\'e}}, {Peter}, {Piazza}, {Piotto},
  {Pollacco}, {Queloz}, {Ragazzoni}, {Rando}, {Rauer}, {Reimers}, {Ribas},
  {Demangeon}, {Smith}, {Sabin}, {Santos}, {Scandariato}, {Schroffenegger},
  {Schwarz}, {Shporer}, {Simon}, {Steller}, {Szab{\'o}}, {S{\'e}gransan},
  {Thomas}, {Udry}, {Walter}, \& {Walton}}]{2022A&A...664A.156O}
{Osborn}, H.~P., {Bonfanti}, A., {Gandolfi}, D., {et~al.} 2022, \aap, 664, A156

\bibitem[{{Penev} {et~al.}(2018){Penev}, {Bouma}, {Winn}, \&
  {Hartman}}]{Penev2018}
{Penev}, K., {Bouma}, L.~G., {Winn}, J.~N., \& {Hartman}, J.~D. 2018, \aj, 155,
  165

\bibitem[{{Pepe} {et~al.}(2021){Pepe}, {Cristiani}, {Rebolo}, {Santos},
  {Dekker}, {Cabral}, {Di Marcantonio}, {Figueira}, {Lo Curto}, {Lovis},
  {Mayor}, {M{\'e}gevand}, {Molaro}, {Riva}, {Zapatero Osorio}, {Amate},
  {Manescau}, {Pasquini}, {Zerbi}, {Adibekyan}, {Abreu}, {Affolter}, {Alibert},
  {Aliverti}, {Allart}, {Allende Prieto}, {{\'A}lvarez}, {Alves}, {Avila},
  {Baldini}, {Bandy}, {Barros}, {Benz}, {Bianco}, {Borsa}, {Bourrier},
  {Bouchy}, {Broeg}, {Calderone}, {Cirami}, {Coelho}, {Conconi}, {Coretti},
  {Cumani}, {Cupani}, {D'Odorico}, {Damasso}, {Deiries}, {Delabre},
  {Demangeon}, {Dumusque}, {Ehrenreich}, {Faria}, {Fragoso}, {Genolet},
  {Genoni}, {G{\'e}nova Santos}, {Gonz{\'a}lez Hern{\'a}ndez}, {Hughes},
  {Iwert}, {Kerber}, {Knudstrup}, {Landoni}, {Lavie}, {Lillo-Box}, {Lizon},
  {Maire}, {Martins}, {Mehner}, {Micela}, {Modigliani}, {Monteiro}, {Monteiro},
  {Moschetti}, {Murphy}, {Nunes}, {Oggioni}, {Oliveira}, {Oshagh}, {Pall{\'e}},
  {Pariani}, {Poretti}, {Rasilla}, {Rebord{\~a}o}, {Redaelli}, {Santana
  Tschudi}, {Santin}, {Santos}, {S{\'e}gransan}, {Schmidt}, {Segovia},
  {Sosnowska}, {Sozzetti}, {Sousa}, {Span{\`o}}, {Su{\'a}rez Mascare{\~n}o},
  {Tabernero}, {Tenegi}, {Udry}, \& {Zanutta}}]{Pepe2021}
{Pepe}, F., {Cristiani}, S., {Rebolo}, R., {et~al.} 2021, \aap, 645, A96

\bibitem[{{Pepper} {et~al.}(2020){Pepper}, {Kane}, {Rodriguez}, {Hinkel},
  {Eastman}, {Daylan}, {Mocnik}, {Dalba}, {Gaudi}, {Fetherolf}, {Stassun},
  {Campante}, {Vanderburg}, {Huber}, {Bossini}, {Crossfield}, {Howell},
  {Stephens}, {Furlan}, {Ricker}, {Vanderspek}, {Latham}, {Seager}, {Winn},
  {Jenkins}, {Twicken}, {Rose}, {Smith}, {Glidden}, {Levine}, {Rinehart},
  {Collins}, {Mann}, {Burt}, {James}, {Siverd}, \& {G{\"u}nther}}]{Pepper2020}
{Pepper}, J., {Kane}, S.~R., {Rodriguez}, J.~E., {et~al.} 2020, \aj, 159, 243

\bibitem[{{Pepper} {et~al.}(2013){Pepper}, {Siverd}, {Beatty}, {Gaudi},
  {Stassun}, {Eastman}, {Collins}, {Latham}, {Bieryla}, {Buchhave}, {Jensen},
  {Manner}, {Penev}, {Crepp}, {Cargile}, {Dhital}, {Calkins}, {Esquerdo},
  {Berlind}, {Fulton}, {Street}, {Ma}, {Ge}, {Wang}, {Mao}, {Richert}, {Gould},
  {DePoy}, {Kielkopf}, {Marshall}, {Pogge}, {Stefanik}, {Trueblood}, \&
  {Trueblood}}]{Pepper2013}
{Pepper}, J., {Siverd}, R.~J., {Beatty}, T.~G., {et~al.} 2013, \apj, 773, 64

\bibitem[{{Petrovich} {et~al.}(2020){Petrovich}, {Mu{\~n}oz}, {Kratter}, \&
  {Malhotra}}]{Petrovich2020}
{Petrovich}, C., {Mu{\~n}oz}, D.~J., {Kratter}, K.~M., \& {Malhotra}, R. 2020,
  \apjl, 902, L5

\bibitem[{{Piaulet} {et~al.}(2021){Piaulet}, {Benneke}, {Rubenzahl}, {Howard},
  {Lee}, {Thorngren}, {Angus}, {Peterson}, {Schlieder}, {Werner}, {Kreidberg},
  {Jaouni}, {Crossfield}, {Ciardi}, {Petigura}, {Livingston}, {Dressing},
  {Fulton}, {Beichman}, {Christiansen}, {Gorjian}, {Hardegree-Ullman}, {Krick},
  \& {Sinukoff}}]{2021AJ....161...70P}
{Piaulet}, C., {Benneke}, B., {Rubenzahl}, R.~A., {et~al.} 2021, \aj, 161, 70

\bibitem[{{Queloz} {et~al.}(2000){Queloz}, {Eggenberger}, {Mayor}, {Perrier},
  {Beuzit}, {Naef}, {Sivan}, \& {Udry}}]{Queloz2000}
{Queloz}, D., {Eggenberger}, A., {Mayor}, M., {et~al.} 2000, \aap, 359, L13

\bibitem[{{Rice} {et~al.}(2023{\natexlab{a}}){Rice}, {Wang}, {Gerbig}, {Wang},
  {Dai}, {Tyler}, {Isaacson}, \& {Howard}}]{2023AJ....165...65R}
{Rice}, M., {Wang}, S., {Gerbig}, K., {et~al.} 2023{\natexlab{a}}, \aj, 165, 65

\bibitem[{{Rice} {et~al.}(2022{\natexlab{a}}){Rice}, {Wang}, \&
  {Laughlin}}]{Rice2022}
{Rice}, M., {Wang}, S., \& {Laughlin}, G. 2022{\natexlab{a}}, \apjl, 926, L17

\bibitem[{{Rice} {et~al.}(2022{\natexlab{b}}){Rice}, {Wang}, {Wang},
  {Stef{\'a}nsson}, {Isaacson}, {Howard}, {Logsdon}, {Schweiker}, {Dai},
  {Brinkman}, {Giacalone}, \& {Holcomb}}]{2022AJ....164..104R}
{Rice}, M., {Wang}, S., {Wang}, X.-Y., {et~al.} 2022{\natexlab{b}}, \aj, 164,
  104

\bibitem[{{Rice} {et~al.}(2023{\natexlab{b}}){Rice}, {Wang}, {Wang}, {Shporer},
  {Barkaoui}, {Brahm}, {Collins}, {Jord{\'a}n}, {Lowson}, {Butler}, {Crane},
  {Shectman}, {Teske}, {Osip}, {Collins}, {Murgas}, {Boyle}, {Pozuelos},
  {Timmermans}, {Jehin}, \& {Gillon}}]{2023AJ....166..266R}
{Rice}, M., {Wang}, X.-Y., {Wang}, S., {et~al.} 2023{\natexlab{b}}, \aj, 166,
  266

\bibitem[{{Ricker} {et~al.}(2015){Ricker}, {Winn}, {Vanderspek}, {Latham},
  {Bakos}, {Bean}, {Berta-Thompson}, {Brown}, {Buchhave}, {Butler}, {Butler},
  {Chaplin}, {Charbonneau}, {Christensen-Dalsgaard}, {Clampin}, {Deming},
  {Doty}, {De Lee}, {Dressing}, {Dunham}, {Endl}, {Fressin}, {Ge}, {Henning},
  {Holman}, {Howard}, {Ida}, {Jenkins}, {Jernigan}, {Johnson}, {Kaltenegger},
  {Kawai}, {Kjeldsen}, {Laughlin}, {Levine}, {Lin}, {Lissauer}, {MacQueen},
  {Marcy}, {McCullough}, {Morton}, {Narita}, {Paegert}, {Palle}, {Pepe},
  {Pepper}, {Quirrenbach}, {Rinehart}, {Sasselov}, {Sato}, {Seager},
  {Sozzetti}, {Stassun}, {Sullivan}, {Szentgyorgyi}, {Torres}, {Udry}, \&
  {Villasenor}}]{Ricker2015}
{Ricker}, G.~R., {Winn}, J.~N., {Vanderspek}, R., {et~al.} 2015, Journal of
  Astronomical Telescopes, Instruments, and Systems, 1, 014003

\bibitem[{{Rodriguez} {et~al.}(2021){Rodriguez}, {Quinn}, {Zhou}, {Vanderburg},
  {Nielsen}, {Wittenmyer}, {Brahm}, {Reed}, {Huang}, {Vach}, {Ciardi},
  {Oelkers}, {Stassun}, {Hellier}, {Gaudi}, {Eastman}, {Collins}, {Bieryla},
  {Christian}, {Latham}, {Carleo}, {Wright}, {Matthews}, {Gonzales}, {Ziegler},
  {Dressing}, {Howell}, {Tan}, {Wittrock}, {Plavchan}, {McLeod}, {Baker},
  {Wang}, {Radford}, {Schwarz}, {Esposito}, {Ricker}, {Vanderspek}, {Seager},
  {Winn}, {Jenkins}, {Addison}, {Anderson}, {Barclay}, {Beatty}, {Berlind},
  {Bouchy}, {Bowen}, {Bowler}, {Brasseur}, {Brice{\~n}o}, {Caldwell},
  {Calkins}, {Cartwright}, {Chaturvedi}, {Chaverot}, {Chimaladinne},
  {Christiansen}, {Collins}, {Crossfield}, {Eastridge}, {Espinoza}, {Esquerdo},
  {Feliz}, {Fenske}, {Fong}, {Gan}, {Giacalone}, {Gill}, {Gordon}, {Granados},
  {Grieves}, {Guenther}, {Guerrero}, {Henning}, {Henze}, {Hesse}, {Hobson},
  {Horner}, {James}, {Jensen}, {Jimenez}, {Jord{\'a}n}, {Kane}, {Kielkopf},
  {Kim}, {Kuhn}, {Latouf}, {Law}, {Levine}, {Lund}, {Mann}, {Mao}, {Matson},
  {Mengel}, {Mink}, {Newman}, {O'Dwyer}, {Okumura}, {Palle}, {Pepper},
  {Quintana}, {Sarkis}, {Savel}, {Schlieder}, {Schnaible}, {Shporer}, {Sefako},
  {Seidel}, {Siverd}, {Skinner}, {Stalport}, {Stevens}, {Stibbards}, {Tinney},
  {West}, {Yahalomi}, \& {Zhang}}]{2021AJ....161..194R}
{Rodriguez}, J.~E., {Quinn}, S.~N., {Zhou}, G., {et~al.} 2021, \aj, 161, 194

\bibitem[{{Rosenthal} {et~al.}(2021){Rosenthal}, {Fulton}, {Hirsch},
  {Isaacson}, {Howard}, {Dedrick}, {Sherstyuk}, {Blunt}, {Petigura}, {Knutson},
  {Behmard}, {Chontos}, {Crepp}, {Crossfield}, {Dalba}, {Fischer}, {Henry},
  {Kane}, {Kosiarek}, {Marcy}, {Rubenzahl}, {Weiss}, \&
  {Wright}}]{2021ApJS..255....8R}
{Rosenthal}, L.~J., {Fulton}, B.~J., {Hirsch}, L.~A., {et~al.} 2021, \apjs,
  255, 8

\bibitem[{{Rucinski}(1999)}]{Rucinski1999}
{Rucinski}, S. 1999, in Astronomical Society of the Pacific Conference Series,
  Vol. 185, IAU Colloq. 170: Precise Stellar Radial Velocities, ed. J.~B.
  {Hearnshaw} \& C.~D. {Scarfe}, 82

\bibitem[{{Saha} {et~al.}(2021){Saha}, {Chakrabarty}, \&
  {Sengupta}}]{2021AJ....162...18S}
{Saha}, S., {Chakrabarty}, A., \& {Sengupta}, S. 2021, \aj, 162, 18

\bibitem[{{Sanchis-Ojeda} {et~al.}(2012){Sanchis-Ojeda}, {Fabrycky}, {Winn},
  {Barclay}, {Clarke}, {Ford}, {Fortney}, {Geary}, {Holman}, {Howard},
  {Jenkins}, {Koch}, {Lissauer}, {Marcy}, {Mullally}, {Ragozzine}, {Seader},
  {Still}, \& {Thompson}}]{Sanchis2012}
{Sanchis-Ojeda}, R., {Fabrycky}, D.~C., {Winn}, J.~N., {et~al.} 2012, \nat,
  487, 449

\bibitem[{{Sanchis-Ojeda} \& {Winn}(2011)}]{2011ApJ...743...61S}
{Sanchis-Ojeda}, R. \& {Winn}, J.~N. 2011, \apj, 743, 61

\bibitem[{{Sanchis-Ojeda} {et~al.}(2015){Sanchis-Ojeda}, {Winn}, {Dai},
  {Howard}, {Isaacson}, {Marcy}, {Petigura}, {Sinukoff}, {Weiss}, {Albrecht},
  {Hirano}, \& {Rogers}}]{2015ApJ...812L..11S}
{Sanchis-Ojeda}, R., {Winn}, J.~N., {Dai}, F., {et~al.} 2015, \apjl, 812, L11

\bibitem[{{Sanchis-Ojeda} {et~al.}(2013{\natexlab{a}}){Sanchis-Ojeda}, {Winn},
  {Marcy}, {Howard}, {Isaacson}, {Johnson}, {Torres}, {Albrecht}, {Campante},
  {Chaplin}, {Davies}, {Lund}, {Carter}, {Dawson}, {Buchhave}, {Everett},
  {Fischer}, {Geary}, {Gilliland}, {Horch}, {Howell}, \&
  {Latham}}]{Sanchis2013}
{Sanchis-Ojeda}, R., {Winn}, J.~N., {Marcy}, G.~W., {et~al.}
  2013{\natexlab{a}}, \apj, 775, 54

\bibitem[{{Sanchis-Ojeda} {et~al.}(2013{\natexlab{b}}){Sanchis-Ojeda}, {Winn},
  {Marcy}, {Howard}, {Isaacson}, {Johnson}, {Torres}, {Albrecht}, {Campante},
  {Chaplin}, {Davies}, {Lund}, {Carter}, {Dawson}, {Buchhave}, {Everett},
  {Fischer}, {Geary}, {Gilliland}, {Horch}, {Howell}, \&
  {Latham}}]{2013ApJ...775...54S}
{Sanchis-Ojeda}, R., {Winn}, J.~N., {Marcy}, G.~W., {et~al.}
  2013{\natexlab{b}}, \apj, 775, 54

\bibitem[{{Santerne} {et~al.}(2016){Santerne}, {H{\'e}brard}, {Lillo-Box},
  {Armstrong}, {Barros}, {Demangeon}, {Barrado}, {Debackere}, {Deleuil},
  {Delgado Mena}, {Montalto}, {Pollacco}, {Osborn}, {Sousa}, {Abe},
  {Adibekyan}, {Almenara}, {Andr{\'e}}, {Arlic}, {Barthe}, {Bendjoya},
  {Behrend}, {Boisse}, {Bouchy}, {Boussier}, {Bretton}, {Brown}, {Carry},
  {Cailleau}, {Conseil}, {Coulon}, {Courcol}, {Dauchet}, {Dalouzy}, {Deldem},
  {Desormi{\`e}res}, {Dubreuil}, {Fehrenbach}, {Ferratfiat}, {Girelli},
  {Gregorio}, {Jaecques}, {Kugel}, {Kirk}, {Labrevoir}, {Lachuri{\'e}}, {Lam},
  {Le Guen}, {Martinez}, {Maurin}, {McCormac}, {Pioppa}, {Quadri},
  {Rajpurohit}, {Rey}, {Rivet}, {Roy}, {Santos}, {Signoret}, {Strabla},
  {Suarez}, {Toublanc}, {Tsantaki}, {Vienney}, {Wilson}, {Bachschmidt},
  {Colas}, {Gerteis}, {Louis}, {Mario}, {Marlot}, {Montier}, {Perroud}, {Pic},
  {Romeuf}, {Ubaud}, \& {Verilhac}}]{2016ApJ...824...55S}
{Santerne}, A., {H{\'e}brard}, G., {Lillo-Box}, J., {et~al.} 2016, \apj, 824,
  55

\bibitem[{{Santerne} {et~al.}(2019){Santerne}, {Malavolta}, {Kosiarek}, {Dai},
  {Dressing}, {Dumusque}, {Hara}, {Lopez}, {Mortier}, {Vanderburg},
  {Adibekyan}, {Armstrong}, {Barrado}, {Barros}, {Bayliss}, {Berardo},
  {Boisse}, {Bonomo}, {Bouchy}, {Brown}, {Buchhave}, {Butler}, {Collier
  Cameron}, {Cosentino}, {Crane}, {Crossfield}, {Damasso}, {Deleuil}, {Delgado
  Mena}, {Demangeon}, {D{\'\i}az}, {Donati}, {Figueira}, {Fulton}, {Ghedina},
  {Harutyunyan}, {H{\'e}brard}, {Hirsch}, {Hojjatpanah}, {Howard}, {Isaacson},
  {Latham}, {Lillo-Box}, {L{\'o}pez-Morales}, {Lovis}, {Martinez Fiorenzano},
  {Molinari}, {Mousis}, {Moutou}, {Nava}, {Nielsen}, {Osborn}, {Petigura},
  {Phillips}, {Pollacco}, {Poretti}, {Rice}, {Santos}, {S{\'e}gransan},
  {Shectman}, {Sinukoff}, {Sousa}, {Sozzetti}, {Teske}, {Udry}, {Vigan},
  {Wang}, {Watson}, {Weiss}, {Wheatley}, \& {Winn}}]{2019arXiv191107355S}
{Santerne}, A., {Malavolta}, L., {Kosiarek}, M.~R., {et~al.} 2019, arXiv
  e-prints, arXiv:1911.07355

\bibitem[{{Schanche} {et~al.}(2020){Schanche}, {H{\'e}brard}, {Collier
  Cameron}, {Dalal}, {Smalley}, {Wilson}, {Boisse}, {Bouchy}, {Brown},
  {Demangeon}, {Haswell}, {Hellier}, {Kolb}, {Lopez}, {Maxted}, {Pollacco},
  {West}, \& {Wheatley}}]{Schanche2020}
{Schanche}, N., {H{\'e}brard}, G., {Collier Cameron}, A., {et~al.} 2020,
  \mnras, 499, 428

\bibitem[{{Schlaufman}(2010)}]{Schlaufman2010}
{Schlaufman}, K.~C. 2010, \apj, 719, 602

\bibitem[{Schwarz(1978)}]{Schwarz1978}
Schwarz, G. 1978, The Annals of Statistics, 6, 461

\bibitem[{{Sebastian} {et~al.}(2022){Sebastian}, {Guenther}, {Deleuil},
  {Dorsch}, {Heber}, {Heuser}, {Gandolfi}, {Grziwa}, {Deeg}, {Alonso},
  {Bouchy}, {Csizmadia}, {Cusano}, {Fridlund}, {Geier}, {Irrgang}, {Korth},
  {Nespral}, {Rauer}, {Tal-Or}, \& {CoRoT-team}}]{2022MNRAS.516..636S}
{Sebastian}, D., {Guenther}, E.~W., {Deleuil}, M., {et~al.} 2022, \mnras, 516,
  636

\bibitem[{{Sedaghati} {et~al.}(2023){Sedaghati}, {Jord{\'a}n}, {Brahm},
  {Mu{\~n}oz}, {Petrovich}, \& {Hobson}}]{2023AJ....166..130S}
{Sedaghati}, E., {Jord{\'a}n}, A., {Brahm}, R., {et~al.} 2023, \aj, 166, 130

\bibitem[{{Seidel} {et~al.}(2023){Seidel}, {Prinoth}, {Knudstrup},
  {Hoeijmakers}, {Zanazzi}, \& {Albrecht}}]{Seidel2023}
{Seidel}, J.~V., {Prinoth}, B., {Knudstrup}, E., {et~al.} 2023, \aap, 678, A150

\bibitem[{{Shporer} \& {Brown}(2011)}]{Shporer2011}
{Shporer}, A. \& {Brown}, T. 2011, \apj, 733, 30

\bibitem[{{Siegel} {et~al.}(2023){Siegel}, {Winn}, \& {Albrecht}}]{Siegel2023}
{Siegel}, J.~C., {Winn}, J.~N., \& {Albrecht}, S.~H. 2023, \apjl, 950, L2

\bibitem[{{Singh} {et~al.}(2023){Singh}, {Scandariato}, {Smith}, {Cubillos},
  {Lendl}, {Billot}, {Fortier}, {Queloz}, {Sousa}, {Csizmadia}, {Brandeker},
  {Carone}, {Wilson}, {Akinsanmi}, {Patel}, {Krenn}, {Demangeon}, {Bruno},
  {Pagano}, {Hooton}, {Cabrera}, {Santos}, {Alibert}, {Alonso}, {Asquier},
  {B{\'a}rczy}, {Barrado Navascues}, {Barros}, {Baumjohann}, {Beck}, {Beck},
  {Benz}, {Bergomi}, {Bonfanti}, {Bonfils}, {Borsato}, {Broeg}, {Charnoz},
  {Collier Cameron}, {Davies}, {Deleuil}, {Deline}, {Delrez}, {Demory},
  {Ehrenreich}, {Erikson}, {Fossati}, {Fridlund}, {Gandolfi}, {Gillon},
  {G{\"u}del}, {G{\"u}nther}, {Harre}, {Heitzmann}, {Helling}, {Hoyer},
  {Isaak}, {Kiss}, {Lam}, {Laskar}, {Lecavelier des Etangs}, {Magrin},
  {Maxted}, {Mischler}, {Mordasini}, {Nascimbeni}, {Olofsson}, {Ottensamer},
  {Pall{\'e}}, {Peter}, {Piotto}, {Pollacco}, {Ragazzoni}, {Rando}, {Rauer},
  {Ribas}, {Salmon}, {S{\'e}gransan}, {Simon}, {Stalport}, {Steinberger},
  {Szab{\'o}}, {Thomas}, {Udry}, {Ulmer}, {Van Grootel}, {Venturini},
  {Villaver}, {Walton}, \& {Zingales}}]{arXiv:2311.03264}
{Singh}, V., {Scandariato}, G., {Smith}, A.~M.~S., {et~al.} 2023, arXiv
  e-prints, arXiv:2311.03264

\bibitem[{{Siverd} {et~al.}(2018){Siverd}, {Collins}, {Zhou}, {Quinn}, {Gaudi},
  {Stassun}, {Johnson}, {Bieryla}, {Latham}, {Ciardi}, {Rodriguez}, {Penev},
  {Pinsonneault}, {Pepper}, {Eastman}, {Relles}, {Kielkopf}, {Gregorio},
  {Oberst}, {Aldi}, {Esquerdo}, {Calkins}, {Berlind}, {Dressing}, {Patel},
  {Stevens}, {Beatty}, {Lund}, {Labadie-Bartz}, {Kuhn}, {Col{\'o}n}, {James},
  {Yao}, {Johnson}, {Wright}, {McCrady}, {Wittenmyer}, {Johnson}, {Sliski},
  {Jensen}, {Cohen}, {McLeod}, {Penny}, {Joner}, {Stephens}, {Villanueva},
  {Zambelli}, {Stockdale}, {Evans}, {Tan}, {Curtis}, {Reed}, {Trueblood}, \&
  {Trueblood}}]{2018AJ....155...35S}
{Siverd}, R.~J., {Collins}, K.~A., {Zhou}, G., {et~al.} 2018, \aj, 155, 35

\bibitem[{{Smalley} {et~al.}(2010){Smalley}, {Anderson}, {Collier Cameron},
  {Gillon}, {Hellier}, {Lister}, {Maxted}, {Queloz}, {Triaud}, {West},
  {Bentley}, {Enoch}, {Pepe}, {Pollacco}, {Segransan}, {Smith}, {Southworth},
  {Udry}, {Wheatley}, {Wood}, \& {Bento}}]{Smalley2010}
{Smalley}, B., {Anderson}, D.~R., {Collier Cameron}, A., {et~al.} 2010, \aap,
  520, A56

\bibitem[{{Souami} \& {Souchay}(2012)}]{Souami2012}
{Souami}, D. \& {Souchay}, J. 2012, \aap, 543, A133

\bibitem[{{Southworth}(2011)}]{Southworth2011}
{Southworth}, J. 2011, \mnras, 417, 2166

\bibitem[{{Southworth} {et~al.}(2011){Southworth}, {Dominik}, {J{\o}rgensen},
  {Rahvar}, {Snodgrass}, {Alsubai}, {Bozza}, {Browne}, {Burgdorf}, {Calchi
  Novati}, {Dodds}, {Dreizler}, {Finet}, {Gerner}, {Hardis}, {Harps{\o}e},
  {Hellier}, {Hinse}, {Hundertmark}, {Kains}, {Kerins}, {Liebig}, {Mancini},
  {Mathiasen}, {Penny}, {Proft}, {Ricci}, {Sahu}, {Scarpetta}, {Sch{\"a}fer},
  {Sch{\"o}nebeck}, \& {Surdej}}]{2011A&A...527A...8S}
{Southworth}, J., {Dominik}, M., {J{\o}rgensen}, U.~G., {et~al.} 2011, \aap,
  527, A8

\bibitem[{{Steiner} {et~al.}(2023){Steiner}, {Attia}, {Ehrenreich}, {Lendl},
  {Bourrier}, {Lovis}, {Seidel}, {Sousa}, {Mounzer}, {Astudillo-Defru},
  {Bonfils}, {Bonvin}, {Dethier}, {Heng}, {Lavie}, {Melo}, {Ottoni}, {Pepe},
  {S{\'e}gransan}, \& {Wyttenbach}}]{2023A&A...672A.134S}
{Steiner}, M., {Attia}, O., {Ehrenreich}, D., {et~al.} 2023, \aap, 672, A134

\bibitem[{{Szentgyorgyi} {et~al.}(2005){Szentgyorgyi}, {Geary}, {Latham},
  {Groner}, {Amato}, {Bennett}, {Falco}, {Peters}, {Ordway}, \&
  {Fata}}]{Szentgyorgyi2005}
{Szentgyorgyi}, A.~H., {Geary}, J.~G., {Latham}, D.~W., {et~al.} 2005, in
  American Astronomical Society Meeting Abstracts, Vol. 207, American
  Astronomical Society Meeting Abstracts, 110.10

\bibitem[{{Tejada Arevalo} {et~al.}(2021){Tejada Arevalo}, {Winn}, \&
  {Anderson}}]{Tejada2021}
{Tejada Arevalo}, R.~A., {Winn}, J.~N., \& {Anderson}, K.~R. 2021, \apj, 919,
  138

\bibitem[{{Telting} {et~al.}(2014){Telting}, {Avila}, {Buchhave}, {Frandsen},
  {Gandolfi}, {Lindberg}, {Stempels}, {Prins}, \& {NOT staff}}]{Telting2014}
{Telting}, J.~H., {Avila}, G., {Buchhave}, L., {et~al.} 2014, Astronomische
  Nachrichten, 335, 41

\bibitem[{{Temple} {et~al.}(2017){Temple}, {Hellier}, {Albrow}, {Anderson},
  {Bayliss}, {Beatty}, {Bieryla}, {Brown}, {Cargile}, {Collier Cameron},
  {Collins}, {Col{\'o}n}, {Curtis}, {D'Ago}, {Delrez}, {Eastman}, {Gaudi},
  {Gillon}, {Gregorio}, {James}, {Jehin}, {Joner}, {Kielkopf}, {Kuhn},
  {Labadie-Bartz}, {Latham}, {Lendl}, {Lund}, {Malpas}, {Maxted}, {Myers},
  {Oberst}, {Pepe}, {Pepper}, {Pollacco}, {Queloz}, {Rodriguez},
  {S{\'e}gransan}, {Siverd}, {Smalley}, {Stassun}, {Stevens}, {Stockdale},
  {Tan}, {Triaud}, {Udry}, {Villanueva}, {West}, \&
  {Zhou}}]{2017MNRAS.471.2743T}
{Temple}, L.~Y., {Hellier}, C., {Albrow}, M.~D., {et~al.} 2017, \mnras, 471,
  2743

\bibitem[{{Valsecchi} \& {Rasio}(2014)}]{Valsecchi2014}
{Valsecchi}, F. \& {Rasio}, F.~A. 2014, \apj, 786, 102

\bibitem[{{Vanderburg} {et~al.}(2017){Vanderburg}, {Becker}, {Buchhave},
  {Mortier}, {Lopez}, {Malavolta}, {Haywood}, {Latham}, {Charbonneau},
  {L{\'o}pez-Morales}, {Adams}, {Bonomo}, {Bouchy}, {Collier Cameron},
  {Cosentino}, {Di Fabrizio}, {Dumusque}, {Fiorenzano}, {Harutyunyan},
  {Johnson}, {Lorenzi}, {Lovis}, {Mayor}, {Micela}, {Molinari}, {Pedani},
  {Pepe}, {Piotto}, {Phillips}, {Rice}, {Sasselov}, {S{\'e}gransan},
  {Sozzetti}, {Udry}, \& {Watson}}]{2017AJ....154..237V}
{Vanderburg}, A., {Becker}, J.~C., {Buchhave}, L.~A., {et~al.} 2017, \aj, 154,
  237

\bibitem[{Virtanen {et~al.}(2020)Virtanen, Gommers, Oliphant, Haberland, Reddy,
  Cournapeau, Burovski, Peterson, Weckesser, Bright, {van der Walt}, Brett,
  Wilson, Millman, Mayorov, Nelson, Jones, Kern, Larson, Carey, Polat, Feng,
  Moore, {VanderPlas}, Laxalde, Perktold, Cimrman, Henriksen, Quintero, Harris,
  Archibald, Ribeiro, Pedregosa, {van Mulbregt}, \& {SciPy 1.0
  Contributors}}]{scipy}
Virtanen, P., Gommers, R., Oliphant, T.~E., {et~al.} 2020, Nature Methods, 17,
  261

\bibitem[{{Vogt} {et~al.}(1994){Vogt}, {Allen}, {Bigelow}, {Bresee}, {Brown},
  {Cantrall}, {Conrad}, {Couture}, {Delaney}, {Epps}, {Hilyard}, {Hilyard},
  {Horn}, {Jern}, {Kanto}, {Keane}, {Kibrick}, {Lewis}, {Osborne},
  {Pardeilhan}, {Pfister}, {Ricketts}, {Robinson}, {Stover}, {Tucker}, {Ward},
  \& {Wei}}]{Vogt1994}
{Vogt}, S.~S., {Allen}, S.~L., {Bigelow}, B.~C., {et~al.} 1994, in Society of
  Photo-Optical Instrumentation Engineers (SPIE) Conference Series, Vol. 2198,
  Instrumentation in Astronomy VIII, ed. D.~L. {Crawford} \& E.~R. {Craine},
  362

\bibitem[{{von Braun} {et~al.}(2011){von Braun}, {Boyajian}, {ten Brummelaar},
  {Kane}, {van Belle}, {Ciardi}, {Raymond}, {L{\'o}pez-Morales}, {McAlister},
  {Schaefer}, {Ridgway}, {Sturmann}, {Sturmann}, {White}, {Turner},
  {Farrington}, \& {Goldfinger}}]{2011ApJ...740...49V}
{von Braun}, K., {Boyajian}, T.~S., {ten Brummelaar}, T.~A., {et~al.} 2011,
  \apj, 740, 49

\bibitem[{{Wang} {et~al.}(2022){Wang}, {Rice}, {Wang}, {Pu}, {Stef{\'a}nsson},
  {Mahadevan}, {Radzom}, {Giacalone}, {Wu}, {Esposito}, {Dalba}, {Avsar},
  {Holden}, {Skiff}, {Polakis}, {Voeller}, {Logsdon}, {Klusmeyer}, {Schweiker},
  {Wu}, {Beard}, {Dai}, {Lubin}, {Weiss}, {Bender}, {Blake}, {Dressing},
  {Halverson}, {Hearty}, {Howard}, {Huber}, {Isaacson}, {Jackman}, {Llama},
  {McElwain}, {Rajagopal}, {Roy}, {Robertson}, {Schwab}, {Shkolnik}, {Wright},
  \& {Laughlin}}]{Wang2022}
{Wang}, X.-Y., {Rice}, M., {Wang}, S., {et~al.} 2022, \apjl, 926, L8

\bibitem[{{Wang} {et~al.}(2017){Wang}, {Wang}, {Liu}, {Hinse}, {Laughlin},
  {Wu}, {Zhang}, {Zhou}, {Wu}, {Zhou}, {Wittenmyer}, {Eastman}, {Zhang},
  {Hori}, {Narita}, {Chen}, {Ma}, {Peng}, {Zhang}, {Zou}, {Nie}, \&
  {Zhou}}]{2017AJ....154...49W}
{Wang}, Y.-H., {Wang}, S., {Liu}, H.-G., {et~al.} 2017, \aj, 154, 49

\bibitem[{{Ward}(1981)}]{Ward1981}
{Ward}, W.~R. 1981, \icarus, 47, 234

\bibitem[{Wiecki {et~al.}(2023)Wiecki, Salvatier, Vieira, Kochurov, Patil,
  Osthege, Willard, Engels, Carroll, Martin, Seyboldt, Rochford, Paz,
  rpgoldman, Meyer, Coyle, Gorelli, Abril-Pla, Kumar, Lao, Andreani, Yoshioka,
  Ho, Kluyver, Beauchamp, \& Andorra}]{pymc}
Wiecki, T., Salvatier, J., Vieira, R., {et~al.} 2023, pymc-devs/pymc: v5.8.2

\bibitem[{{Winn} {et~al.}(2010{\natexlab{a}}){Winn}, {Fabrycky}, {Albrecht}, \&
  {Johnson}}]{Winn2010}
{Winn}, J.~N., {Fabrycky}, D., {Albrecht}, S., \& {Johnson}, J.~A.
  2010{\natexlab{a}}, \apjl, 718, L145

\bibitem[{{Winn} {et~al.}(2010{\natexlab{b}}){Winn}, {Johnson}, {Howard},
  {Marcy}, {Isaacson}, {Shporer}, {Bakos}, {Hartman}, \&
  {Albrecht}}]{2010ApJ...723L.223W}
{Winn}, J.~N., {Johnson}, J.~A., {Howard}, A.~W., {et~al.} 2010{\natexlab{b}},
  \apjl, 723, L223

\bibitem[{{Winn} {et~al.}(2005){Winn}, {Noyes}, {Holman}, {Charbonneau},
  {Ohta}, {Taruya}, {Suto}, {Narita}, {Turner}, {Johnson}, {Marcy}, {Butler},
  \& {Vogt}}]{Winn+2005}
{Winn}, J.~N., {Noyes}, R.~W., {Holman}, M.~J., {et~al.} 2005, \apj, 631, 1215

\bibitem[{{Winters} {et~al.}(2022){Winters}, {Cloutier}, {Medina}, {Irwin},
  {Charbonneau}, {Astudillo-Defru}, {Bonfils}, {Howard}, {Isaacson}, {Bean},
  {Seifahrt}, {Teske}, {Eastman}, {Twicken}, {Collins}, {Jensen}, {Quinn},
  {Payne}, {Kristiansen}, {Spencer}, {Vanderburg}, {Zechmeister}, {Weiss},
  {Wang}, {Wang}, {Udry}, {Terentev}, {St{\"u}rmer}, {Stef{\'a}nsson},
  {Shporer}, {Shectman}, {Sefako}, {Schwengeler}, {Schwarz}, {Scarsdale},
  {Rubenzahl}, {Roy}, {Rosenthal}, {Robertson}, {Petigura}, {Pepe},
  {Omohundro}, {Murphy}, {Murgas}, {Mo{\v{c}}nik}, {Montet}, {Mennickent},
  {Mayo}, {Massey}, {Lubin}, {Lovis}, {Lewin}, {Kasper}, {Kane}, {Jenkins},
  {Huber}, {Horne}, {Hill}, {Gorrini}, {Giacalone}, {Fulton}, {Forveille},
  {Figueira}, {Fetherolf}, {Dressing}, {D{\'\i}az}, {Delfosse}, {Dalba}, {Dai},
  {Cort{\'e}s}, {Crossfield}, {Crane}, {Conti}, {Collins}, {Chontos}, {Butler},
  {Brown}, {Brady}, {Behmard}, {Beard}, {Batalha}, \& {Almenara}}]{Winters2022}
{Winters}, J.~G., {Cloutier}, R., {Medina}, A.~A., {et~al.} 2022, \aj, 163, 168

\bibitem[{{Winters} {et~al.}(2019){Winters}, {Medina}, {Irwin}, {Charbonneau},
  {Astudillo-Defru}, {Horch}, {Eastman}, {Vrijmoet}, {Henry}, {Diamond-Lowe},
  {Winston}, {Barclay}, {Bonfils}, {Ricker}, {Vanderspek}, {Latham}, {Seager},
  {Winn}, {Jenkins}, {Udry}, {Twicken}, {Teske}, {Tenenbaum}, {Pepe}, {Murgas},
  {Muirhead}, {Mink}, {Lovis}, {Levine}, {L{\'e}pine}, {Jao}, {Henze},
  {Fur{\'e}sz}, {Forveille}, {Figueira}, {Esquerdo}, {Dressing}, {D{\'\i}az},
  {Delfosse}, {Burke}, {Bouchy}, {Berlind}, \& {Almenara}}]{Winters2019}
{Winters}, J.~G., {Medina}, A.~A., {Irwin}, J.~M., {et~al.} 2019, \aj, 158, 152

\bibitem[{{Wright} {et~al.}(2023){Wright}, {Rice}, {Wang}, {Hixenbaugh}, \&
  {Wang}}]{2023AJ....166..217W}
{Wright}, J., {Rice}, M., {Wang}, X.-Y., {Hixenbaugh}, K., \& {Wang}, S. 2023,
  \aj, 166, 217

\bibitem[{{Yee} {et~al.}(2023){Yee}, {Winn}, {Hartman}, {Bouma}, {Zhou},
  {Quinn}, {Latham}, {Bieryla}, {Rodriguez}, {Collins}, {Alfaro}, {Barkaoui},
  {Beard}, {Belinski}, {Benkhaldoun}, {Benni}, {Bernacki}, {Boyle}, {Butler},
  {Caldwell}, {Chontos}, {Christiansen}, {Ciardi}, {Collins}, {Conti}, {Crane},
  {Daylan}, {Dressing}, {Eastman}, {Essack}, {Evans}, {Everett},
  {Fajardo-Acosta}, {For{\'e}s-Toribio}, {Furlan}, {Ghachoui}, {Gillon},
  {Hellier}, {Helm}, {Howard}, {Howell}, {Isaacson}, {Jehin}, {Jenkins},
  {Jensen}, {Kielkopf}, {Laloum}, {Leonhardes-Barboza}, {Lewin}, {Logsdon},
  {Lubin}, {Lund}, {MacDougall}, {Mann}, {Maslennikova}, {Massey}, {McLeod},
  {Mu{\~n}oz}, {Newman}, {Orlov}, {Plavchan}, {Popowicz}, {Pozuelos},
  {Pritchard}, {Radford}, {Reefe}, {Ricker}, {Rudat}, {Safonov}, {Schwarz},
  {Schweiker}, {Scott}, {Seager}, {Shectman}, {Stockdale}, {Tan}, {Teske},
  {Thomas}, {Timmermans}, {Vanderspek}, {Vermilion}, {Watanabe}, {Weiss},
  {West}, {Van Zandt}, {Zejmo}, \& {Ziegler}}]{2023ApJS..265....1Y}
{Yee}, S.~W., {Winn}, J.~N., {Hartman}, J.~D., {et~al.} 2023, \apjs, 265, 1

\bibitem[{{Zahn}(1977)}]{Zahn1977}
{Zahn}, J.~P. 1977, \aap, 57, 383

\bibitem[{{Zanazzi} \& {Chiang}(2024)}]{ZanazziChiang2014}
{Zanazzi}, J.~J. \& {Chiang}, E. 2024, \mnras, 527, 7203

\bibitem[{{Zanazzi} {et~al.}(2024){Zanazzi}, {Dewberry}, \&
  {Chiang}}]{Zanazzi+2024}
{Zanazzi}, J.~J., {Dewberry}, J., \& {Chiang}, E. 2024, arXiv e-prints,
  arXiv:2403.05616

\bibitem[{{Zanazzi} \& {Lai}(2018)}]{Zanazzi2018}
{Zanazzi}, J.~J. \& {Lai}, D. 2018, \mnras, 477, 5207

\bibitem[{{Zechmeister} {et~al.}(2018){Zechmeister}, {Reiners}, {Amado},
  {Azzaro}, {Bauer}, {B{\'e}jar}, {Caballero}, {Guenther}, {Hagen}, {Jeffers},
  {Kaminski}, {K{\"u}rster}, {Launhardt}, {Montes}, {Morales}, {Quirrenbach},
  {Reffert}, {Ribas}, {Seifert}, {Tal-Or}, \& {Wolthoff}}]{Zechmeister2018}
{Zechmeister}, M., {Reiners}, A., {Amado}, P.~J., {et~al.} 2018, \aap, 609, A12

\bibitem[{{Zhao} {et~al.}(2023){Zhao}, {Kunovac}, {Brewer}, {Llama},
  {Millholland}, {Hedges}, {Szymkowiak}, {Roettenbacher}, {Cabot}, {Weiss}, \&
  {Fischer}}]{2023NatAs...7..198Z}
{Zhao}, L.~L., {Kunovac}, V., {Brewer}, J.~M., {et~al.} 2023, Nature Astronomy,
  7, 198

\bibitem[{{Zhou} {et~al.}(2019){Zhou}, {Bakos}, {Bayliss}, {Bento}, {Bhatti},
  {Brahm}, {Csubry}, {Espinoza}, {Hartman}, {Henning}, {Jord{\'a}n}, {Mancini},
  {Penev}, {Rabus}, {Sarkis}, {Suc}, {de Val-Borro}, {Rodriguez}, {Osip},
  {Kedziora-Chudczer}, {Bailey}, {Tinney}, {Durkan}, {L{\'a}z{\'a}r}, {Papp},
  \& {S{\'a}ri}}]{Zhou2019}
{Zhou}, G., {Bakos}, G.~{\'A}., {Bayliss}, D., {et~al.} 2019, \aj, 157, 31

\bibitem[{{Zhou} {et~al.}(2018){Zhou}, {Rodriguez}, {Vanderburg}, {Quinn},
  {Irwin}, {Huang}, {Latham}, {Bieryla}, {Esquerdo}, {Berlind}, \&
  {Calkins}}]{2018AJ....156...93Z}
{Zhou}, G., {Rodriguez}, J.~E., {Vanderburg}, A., {et~al.} 2018, \aj, 156, 93

\end{thebibliography}


\appendix
\section{Additional figures}


\begin{figure*}[t!]
    \centering
    \begin{tabular}{c c}
        \includegraphics[width=0.45\textwidth]{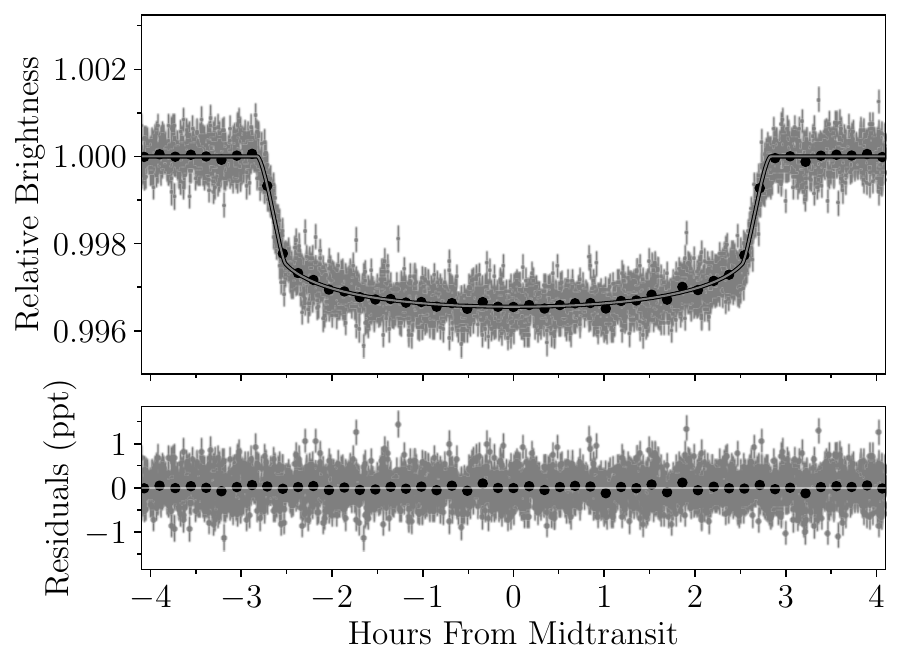} & \includegraphics[width=0.45\textwidth]{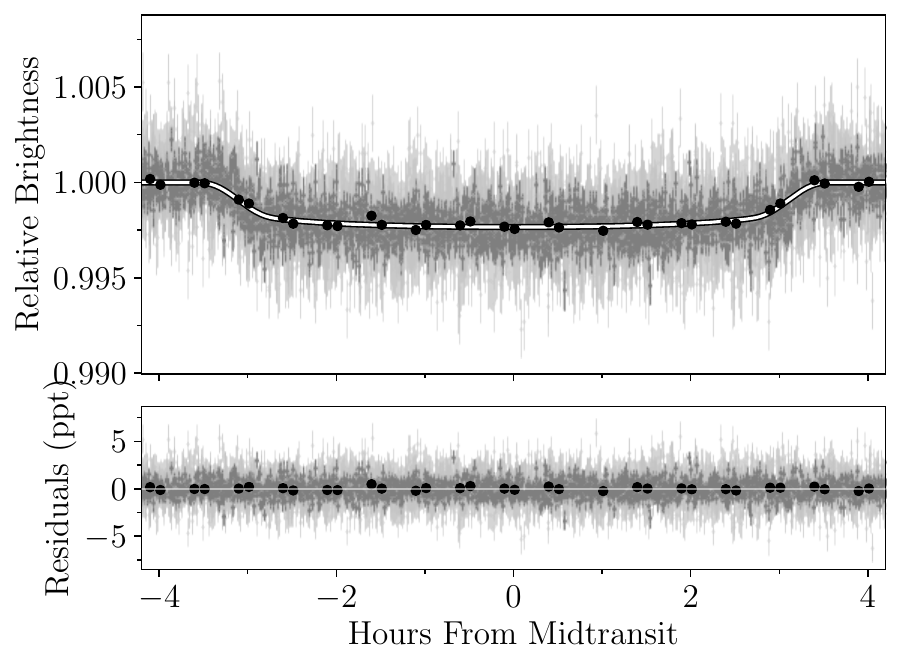} \\
         (a) HD~118203~b & (b) HD~149193~b \\
        \includegraphics[width=0.45\textwidth]{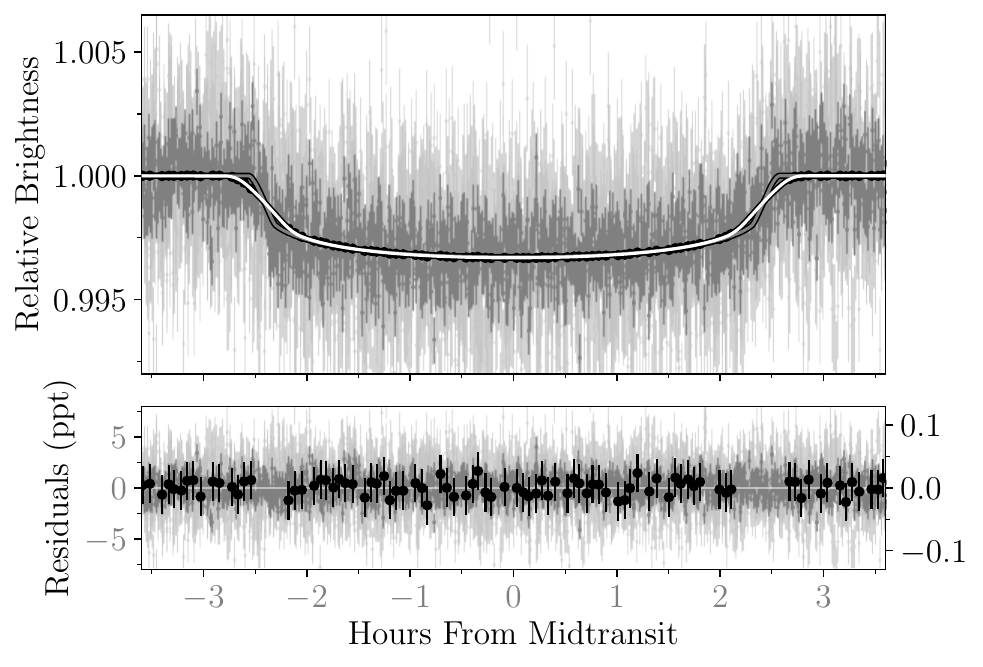} & \includegraphics[width=0.45\textwidth]{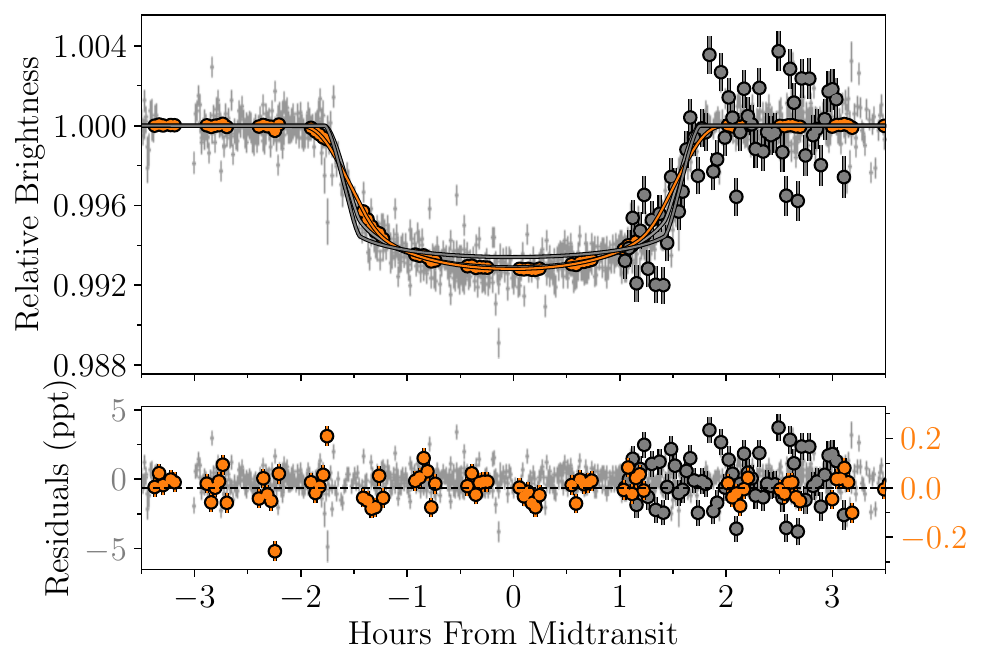} \\
         (c) K2-261~b & K2-287~b (d) \\
        \includegraphics[width=0.45\textwidth]{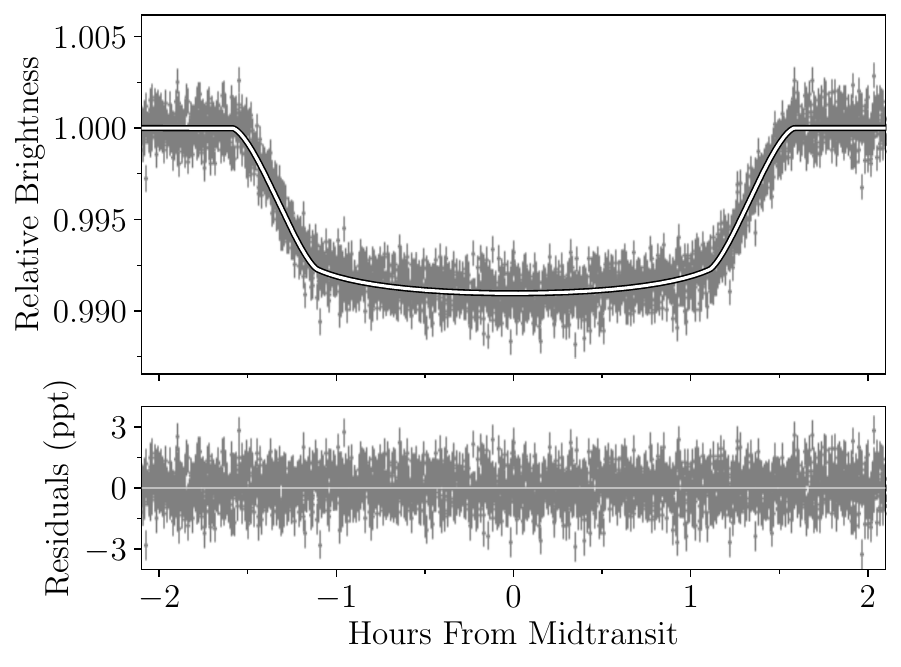} & \includegraphics[width=0.45\textwidth]{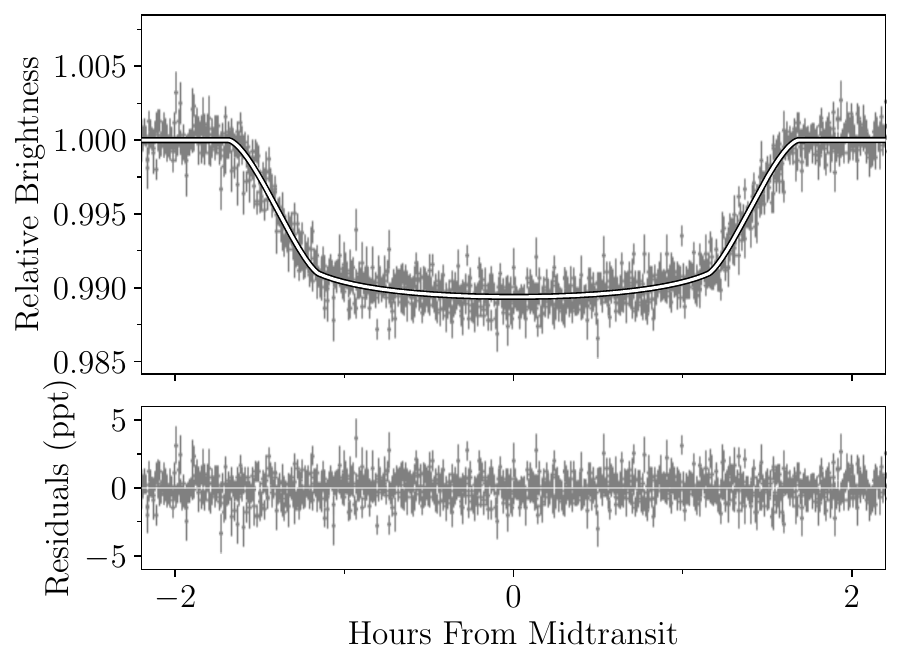} \\
         (e) KELT-3~b & (f) KELT-4Ab \\
    \end{tabular} 
    \caption{{\bf Phasefolded light curves.} The best-fitting models are shown as the white line in the top panel and the residuals are given in the bottom panel. (a) HD~118203~b: \tess 2~min. cadence data shown as gray points with error bars. Black markers are binned data in $\sim$10~min. intervals. (b) HD~149193~b: \tess light curves 20~sec., 2~min., and 30~min. cadence data are shown in light gray, gray, and black, respectively. (c) K2-261~b: K2 light curve in black, \tess 20~sec. and 2~min. in light gray and gray, respectively. A light curve for each passband is shown. (d) K2-287~b: K2 30~min cadence shown in orange with CHEOPS 1~min. and ground-based 2~min. observations in light gray and gray, respectively. A light curve for each passband is also shown with a matching color. (e) KELT-3~b: \tess 2~min. cadence data shown in gray. (f) KELT-4Ab: \tess 2~min. cadence data shown in gray.}
    \label{fig:gridone}
\end{figure*}

\begin{figure*}[t!]
    \centering
    \begin{tabular}{c c}
        \includegraphics[width=0.45\textwidth]{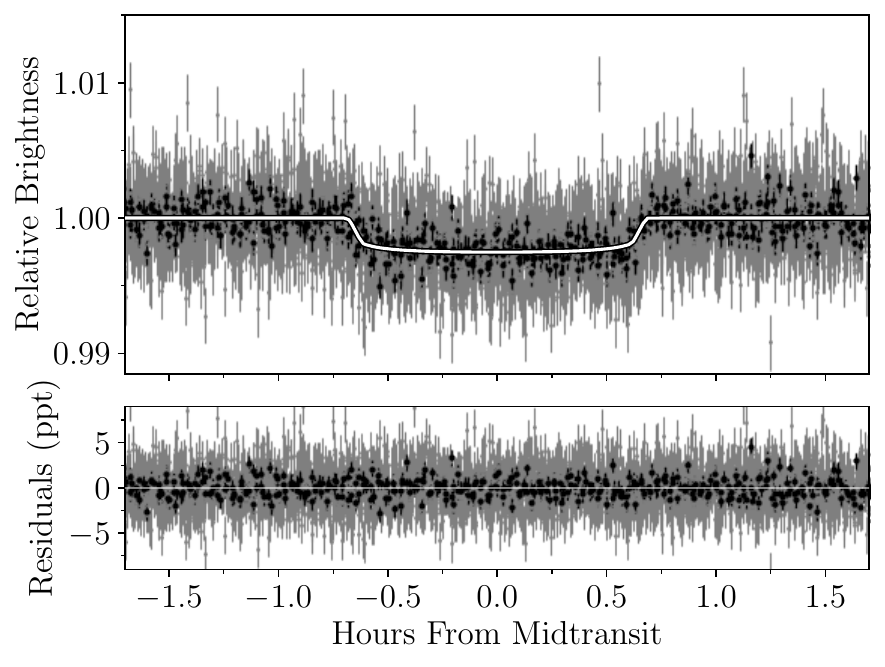} & \includegraphics[width=0.45\textwidth]{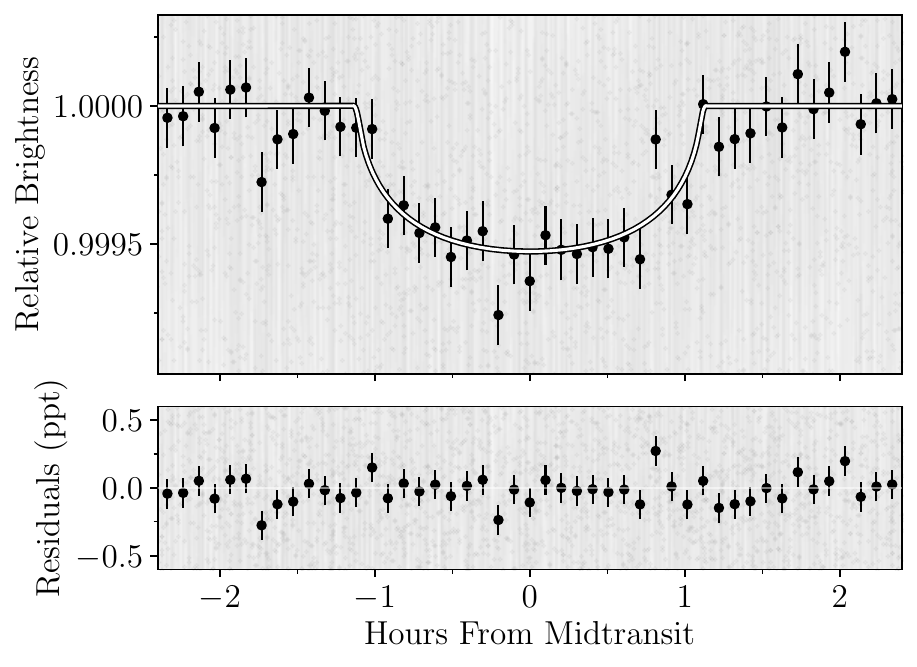} \\
         (a) LTT~1445Ab & (b) TOI-451Ab \\
        \includegraphics[width=0.45\textwidth]{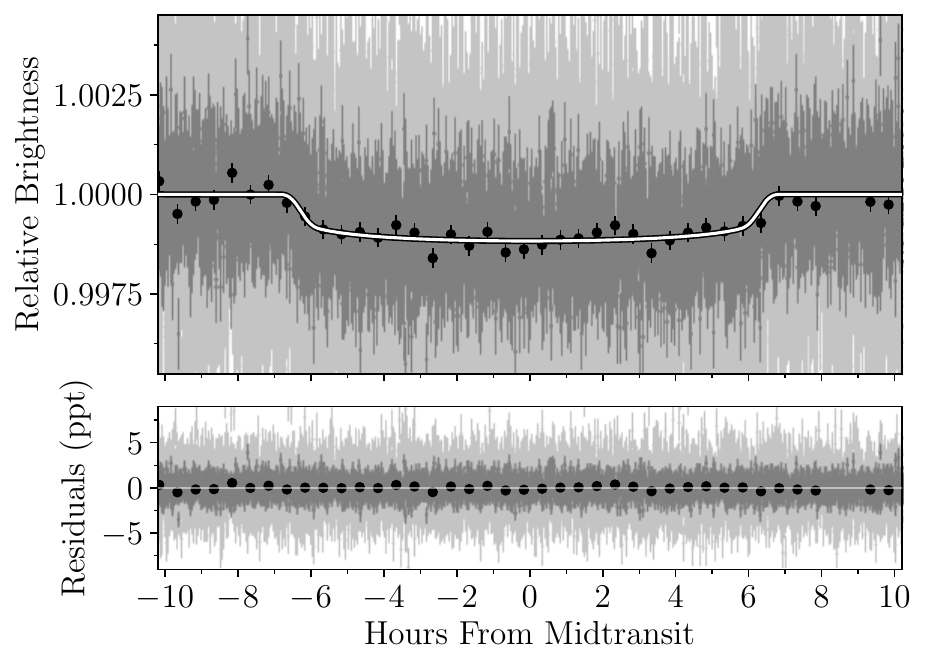} & \includegraphics[width=0.45\textwidth]{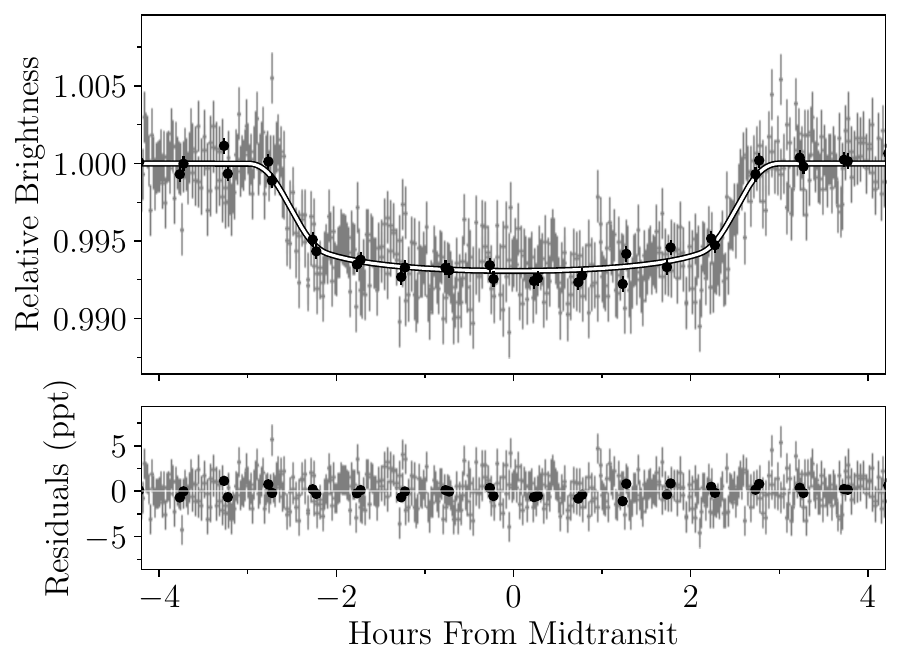} \\
         (c) TOI-813~b & (d) TOI-892~b \\
        \includegraphics[width=0.45\textwidth]{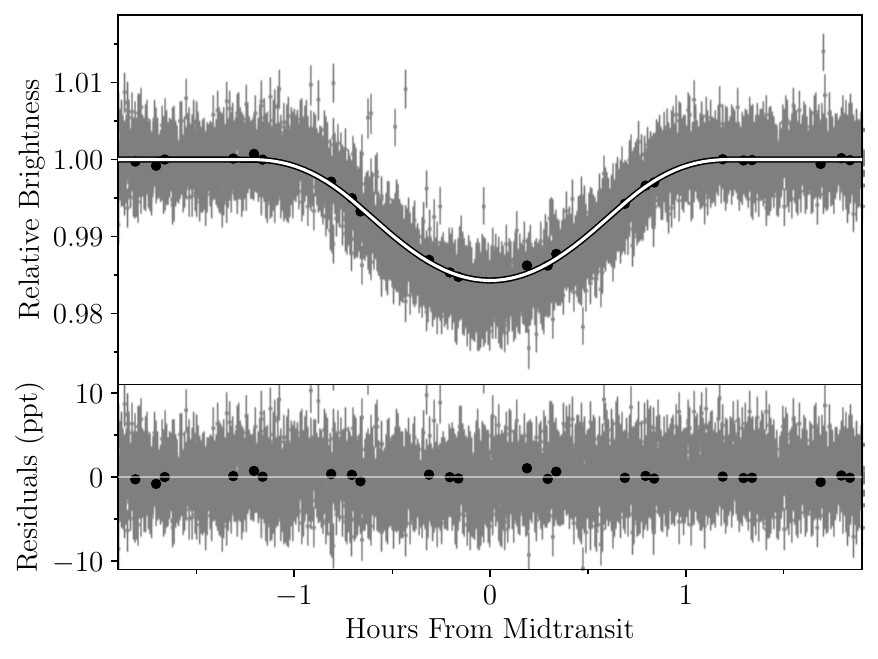} & \includegraphics[width=0.45\textwidth]{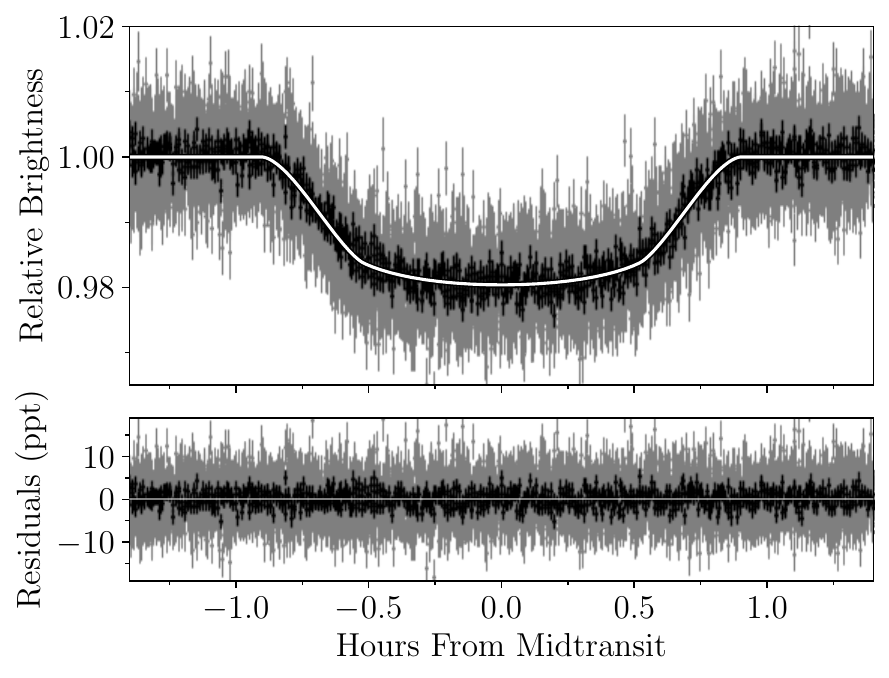} \\
         (e) TOI-1130~c & (f) WASP-50~b \\
    \end{tabular} 
    \caption{{\bf Phasefolded light curves.} Same as in \figref{fig:gridone}. (a) LTT~1445Ab: Black \tess 2~min. and gray \tess 20~sec. (b) TOI-451Ab: Unbinned 2~min. cadence data are shown in gray and in black the binned ($\sim$6~min.) data are shown. (c) TOI-813~b: \tess 20~sec., 2~min., and 30~min. cadence data are shown in light gray, gray, and black, respectively. (d) TOI-892~b: \tess 2~min. and 30~min. cadence data are shown in gray and black, respectively. (e) TOI-1130~c: \tess 20~sec. and 30~min. cadence data shown in gray and black, respectively. (f) WASP-50~b: \tess 20~sec. and 2~min cadence data shown in gray and black, respectively.}
    \label{fig:gridtwo}
\end{figure*}

\begin{figure*}[t!]
    \centering
    \begin{tabular}{c c}
        \includegraphics[width=0.45\textwidth]{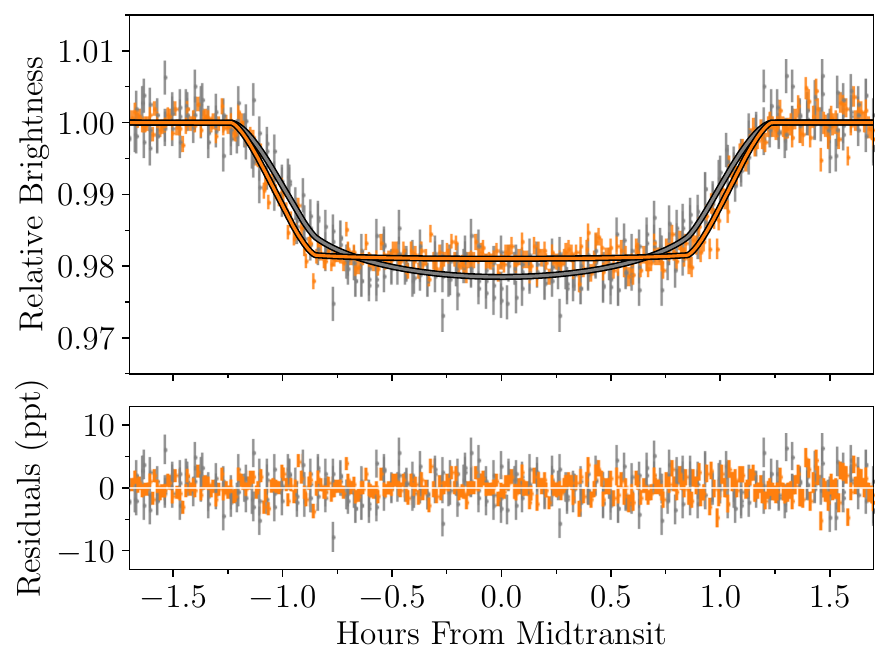} & \includegraphics[width=0.45\textwidth]{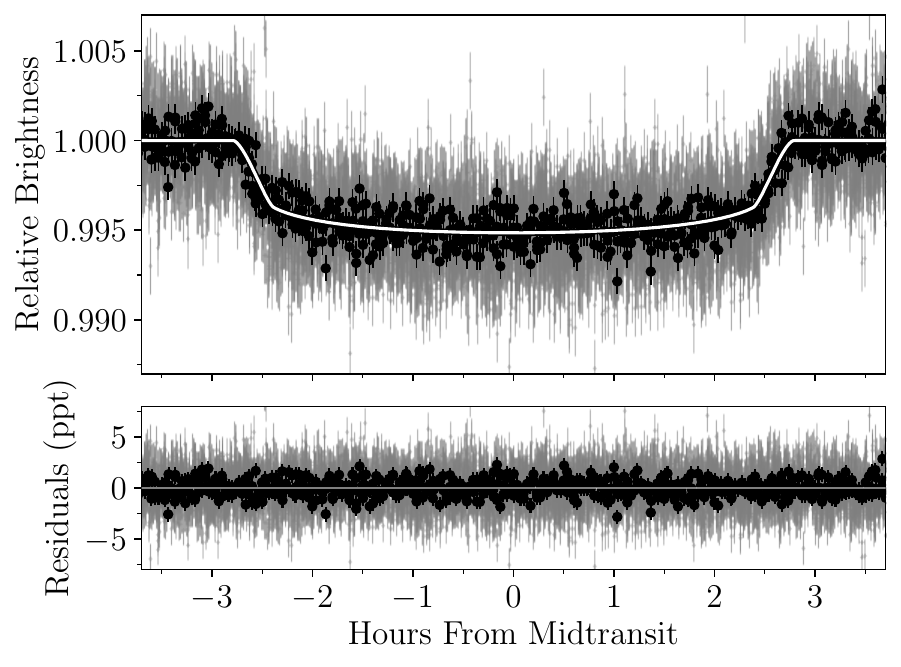} \\
         (a) WASP-59~b & (b) WASP-136~b \\
        \includegraphics[width=0.45\textwidth]{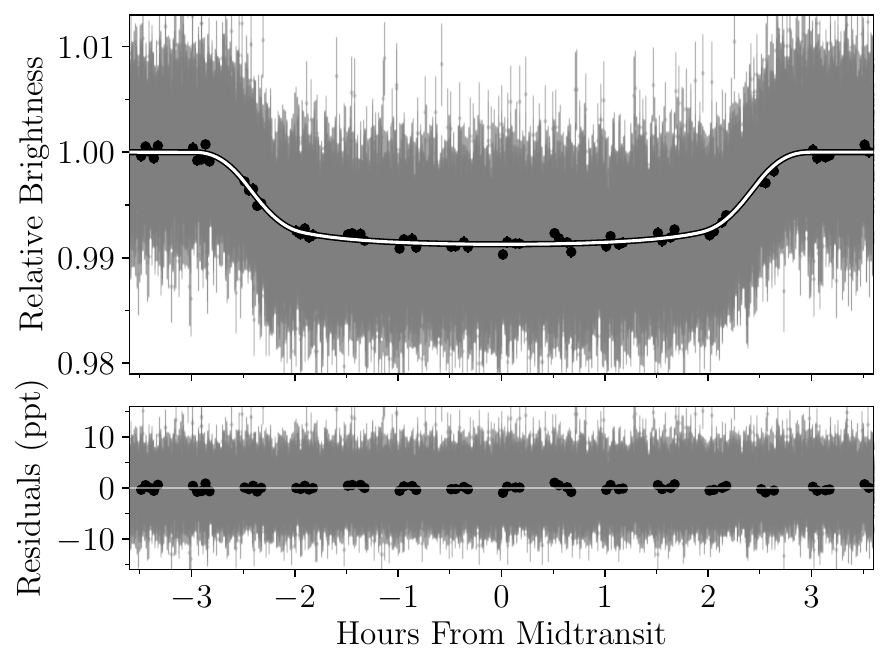} & \includegraphics[width=0.45\textwidth]{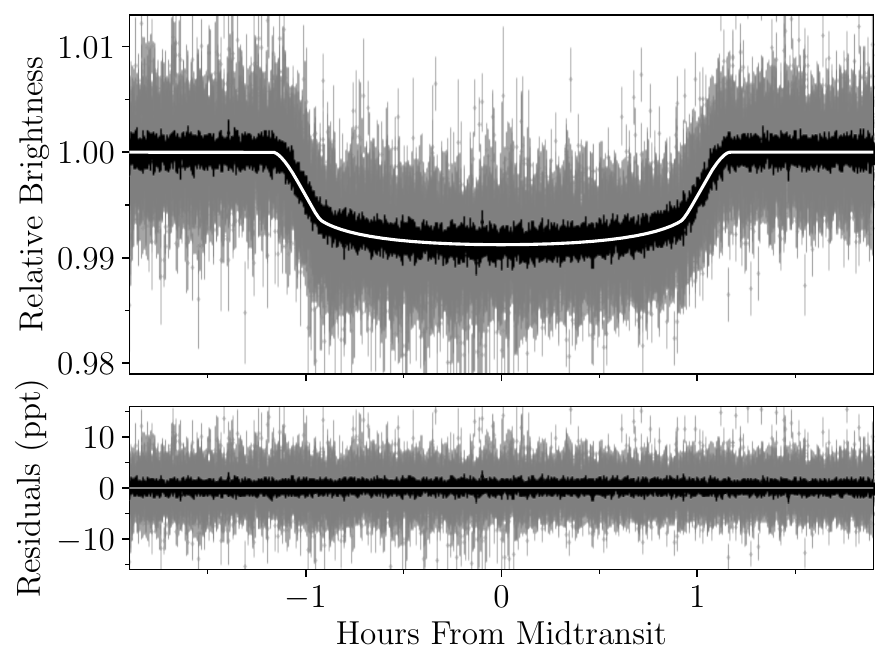} \\
         (c) WASP-172~b & (d) WASP-173Ab \\
        \includegraphics[width=0.45\textwidth]{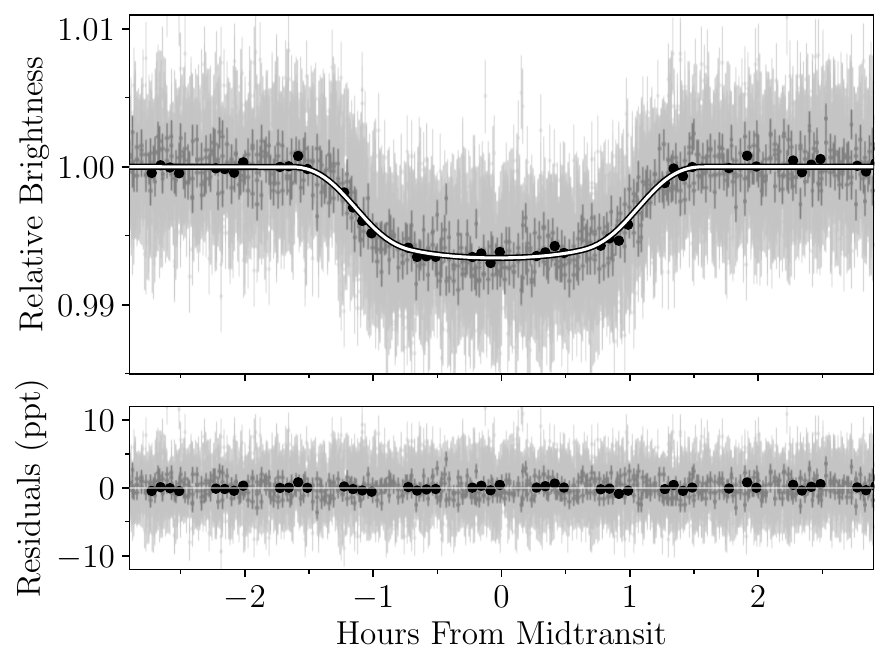} & \includegraphics[width=0.45\textwidth]{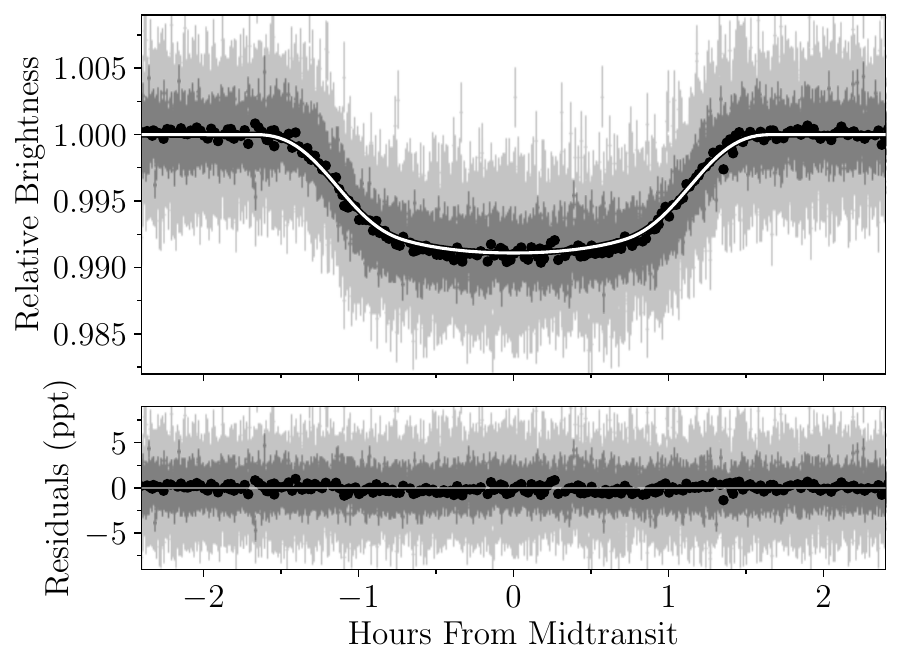} \\
         (e) WASP-186~b & (f) XO-7~b \\
    \end{tabular} 
    \caption{{\bf Phasefolded light curves.} Same as in \figref{fig:gridone}. (a) WASP-59~b: \tess 2~min. cadence data shown in gray and ground-based 2~min. KeplerCam photometry in orange. (b) WASP-136~b: \tess 20~sec. and 2~min. cadence data shown in gray and black, respectively. (c) WASP-172~b: \tess 20~sec. and 30~min. cadence data shown in gray and black, respectively. (d) WASP-173Ab: \tess 20~sec. and 2~min. cadence data shown in gray and black, respectively. (e) WASP-186~b: \tess 20~sec., 2~min., and 30~min. cadence data are shown in light gray, gray, and black, respectively. (f) XO-7~b: \tess 20~sec., 2~min., and 30~min. cadence data are shown in light gray, gray, and black, respectively. }
    \label{fig:gridthree}
\end{figure*}


\begin{figure*}
    \centering
    \includegraphics[width=\textwidth]{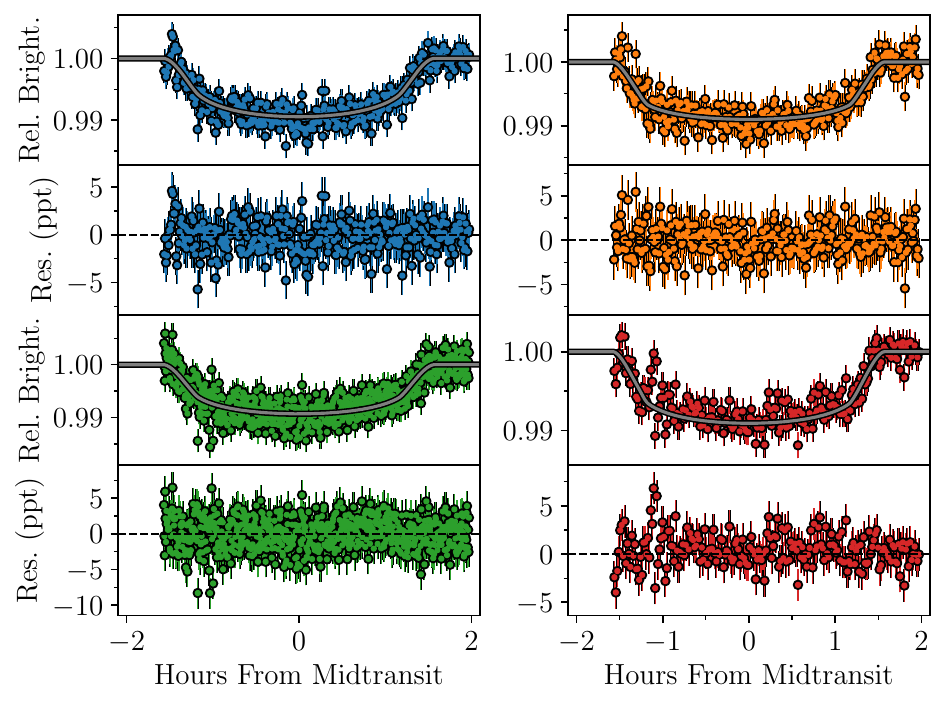}
    \caption{{\bf Phasefolded light curves of WASP-148~b.} The MuSCAT-2 photometry obtained simultaneous with our spectroscopic transit observations. The observations in the different $griz$ filters are shown as blue, orange, green, and red markers, respectively.}
    \label{fig:lc_wasp148}
\end{figure*}



\begin{figure}[h]
    \centering
    \includegraphics[width=\columnwidth]{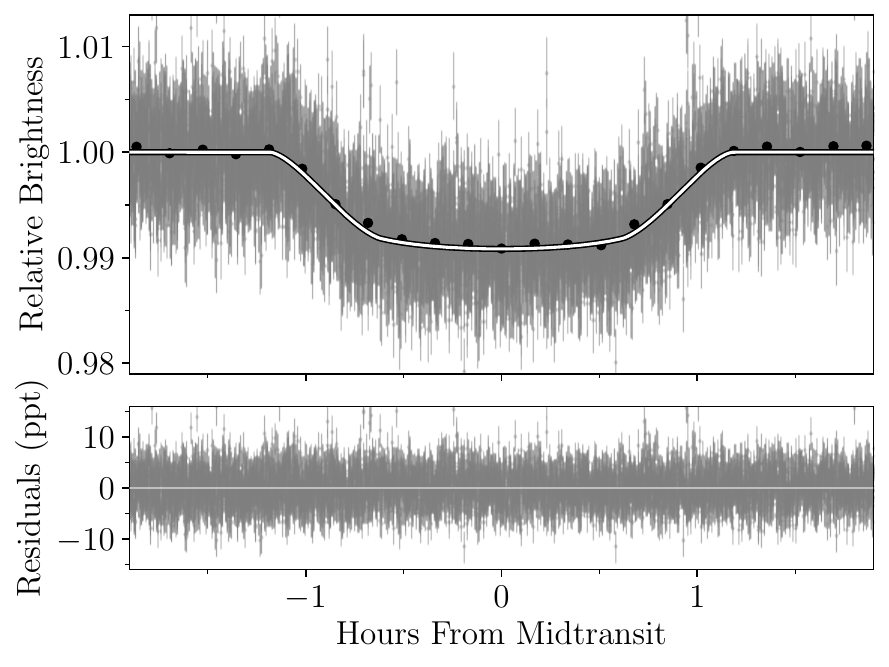}
    \caption{{\bf Phasefolded light curve of WASP-26~b.} Unbinned 20~sec. cadence data are shown in gray and in black the binned ($\sim$6~min.) data are shown.}
    \label{fig:lc_wasp26}
\end{figure}



\clearpage

\section{Additional tables}



\setlength{\tabcolsep}{6pt}
\begin{table}
\begin{threeparttable}
\caption{MCMC posteriors.}
\label{tab:poster}
\begin{tabular}{l c c c c c c}
\toprule
& $a/R_\star$ & $R_p/R_\star$ & $\cos i$ & $K$ & $\gamma$ & $\sigma$ \\
 &  &  &  & (m~s$^{-1}$) & (m~s$^{-1}$) & (m~s$^{-1}$) \\ 
\midrule
HD~118203~b & $7.23^{+0.16}_{-0.18}$ & $0.0546^{+0.0005}_{-0.0004}$ & $0.027^{+0.010}_{-0.026}$ & $196^{+17}_{-19}$ & $-28881^{+6}_{-7}$ & $11^{+2}_{-3}$\\
HD~148193 & $23.1^{+1.7}_{-0.8}$  & $0.0455^{+0.0006}_{-0.0005}$  & $0.016_{-0.007}^{+0.008}$  & $42^{+21}_{-33}$  & $-50340.6^{+1.3}_{-1.2}$  & $1.2^{+0.5}_{-0.5}$ \\
K2-261~b & $14\pm2$ & $0.0524^{+0.0005}_{-0.0006}$ & $0.025^{+0.011}_{-0.025}$ & $18^{+8}_{-11}$ & $3338^{+2}_{-3}$ & $0.8^{+0.3}_{-0.8}$\\  
K2-287~b & $25.4^{+1.1}_{-1.2}$ & $0.0783\pm0.0007$ & $0.026\pm0.004$ & $52^{+13}_{-12}$ & $32872\pm6$ & $1.3^{+0.3}_{-0.4}$\\ 
KELT-3~b & $5.64^{+0.05}_{-0.06}$ & $0.0932\pm0.0002$ & $0.120\pm0.003$ & $282\pm13$ & $28333\pm2$ &  \ldots \\  
KELT-4Ab & $6.02^{+0.09}_{-0.10}$ & $0.1011^{+0.0006}_{-0.0009}$ & $0.110^{+0.005}_{-0.004}$ & $53^{+28}_{-23}$ & $-23145^{+5}_{-4}$ & $8^{+3}_{-7}$\\  
LTT~1445Ab & $29.9^{+1.5}_{-1.3}$ & $0.0487^{+0.0017}_{-0.0018}$ & $0.009^{+0.004}_{-0.009}$ & $-4^{+7}_{-16}$ & $-4088^{+8}_{-10}$ & $6.7^{+1.4}_{-2.1}$\\  
TOI-451Ab & $6.2^{+0.8}_{-0.6}$ & $0.0203^{+0.0010}_{-0.0011}$ & $0.06^{+0.03}_{-0.06}$ & $5^{+3}_{-4}$ & $19755.7\pm0.8$ & $1.9\pm0.7$\\  
TOI-813~b & $47^{+4}_{-2}$ & $0.0324\pm0.0007$ & $0.008^{+0.004}_{-0.005}$ & $95^{+51}_{-95}$ & $1199.6^{+1.7}_{-1.3}$ & $0.5^{+0.2}_{-0.5}$\\  
TOI-892~b & $15.1^{+0.4}_{-0.3}$ & $0.0790^{+0.0007}_{-0.0008}$ & $0.021^{+0.005}_{-0.006}$ & $54^{+23}_{-54}$ & $41907^{+30}_{-49}$ &  \ldots \\  
TOI-1130~b & $22.1\pm0.4$ & $0.176^{+0.013}_{-0.021}$ & $0.0432^{+0.0016}_{-0.0024}$ & $152\pm15$ & $-8033.6\pm1.6$ & $1.6\pm0.8$\\  
WASP-50~b & $7.47^{+0.08}_{-0.09}$ & $0.1371\pm0.0006$ & $0.093\pm0.002$ & $253.5^{+1.5}_{-1.4}$ & $25421.1\pm0.3$ & $0.6^{+0.3}_{-0.5}$\\  
WASP-59~b & $25.4^{+1.1}_{-1.0}$ & $0.136\pm0.002$ & $0.019^{+0.003}_{-0.004}$ & $186^{+20}_{-19}$ & $-0.6^{+1.1}_{-1.0}$ & $2.7\pm0.9$\\
WASP-136~b & $7.4\pm0.2$ & $0.0672\pm0.0005$ & $0.041^{+0.016}_{-0.013}$ & $146^{+52}_{-59}$ & $-3^{+7}_{-8}$ & $12^{+7}_{-9}$\\  
WASP-148~b & $14.4\pm0.3$ & $0.09082\pm0.00020$ &  \ldots  & $11^{+6}_{-11}$ & $-5507\pm3$ & $1.6^{+0.8}_{-1.6}$\\  
WASP-172~b & $7.58\pm0.17$ & $0.0893\pm0.0005$ & $0.061^{+0.007}_{-0.006}$ & $53^{+25}_{-43}$ & $-20410\pm5$ & $11\pm3$\\  
WASP-173Ab & $5.12^{+0.03}_{-0.02}$ & $0.1176\pm0.0006$ & $0.012^{+0.006}_{-0.012}$ & $625^{+5}_{-4}$ & $-7922\pm3$ & $9.6^{+1.5}_{-1.8}$\\  
WASP-186~b & $9.9\pm0.3$ & $0.0812^{+0.0007}_{-0.0006}$ & $0.090\pm0.005$ & $529^{+132}_{-154}$ & $-240^{+43}_{-42}$ & $16^{+7}_{-16}$\\  
XO-7~b & $6.54^{+0.07}_{-0.06}$ & $0.0923\pm0.0003$ & $0.107\pm0.002$ & $128^{+48}_{-59}$ & $-13050^{+10}_{-11}$ & $7^{+4}_{-5}$\\ 
WASP-26~b & $6.7\pm0.3$ & $0.097^{+0.013}_{-0.009}$ & $0.122^{+0.009}_{-0.008}$ & $142\pm15$ & $-16\pm3$ & $10.2^{+1.5}_{-1.7}$ \\
\bottomrule
\end{tabular}
\begin{tablenotes}
    \item For WASP-148~b we were stepping in $i$, where we got $i=86.57_{-0.16}^{+0.16 \circ}$.    
\end{tablenotes}
\end{threeparttable}
\end{table}

\onecolumn
{\small\addtolength{\tabcolsep}{-1.5pt}
\begin{center}
\begin{ThreePartTable}
\begin{TableNotes}
    {\catcode`\&=12 
    \item Values for the new literature systems are drawn from the following references (with bold denoting the source for $\lambda$): 55~Cnc~e: 1 \citet{2021ApJS..255....8R}, \textbf{2} \citet{2023NatAs...7..198Z}. CoRoT-36~b: \textbf{1} \citet{2022MNRAS.516..636S}. HAT-P-11~b: 1 \citet{2011ApJ...743...61S},  \textbf{2} \citet{2010ApJ...723L.223W}. HAT-P-3~b: 1 { \citet{2017A&A...602A.107B} }, 2 {\citet{2023A&A...669A..63B}}, \textbf{3} {\citet{2018A&A...613A..41M}}. HAT-P-30~b: 1 {\citet{2017A&A...602A.107B}},  \textbf{2} {\citet{2023A&A...674A.174C}}. HAT-P-32~b: 1 {\citet{2017A&A...602A.107B}},  2 {\citet{2023A&A...677L..12C}},  \textbf{3} \citet{2012ApJ...757...18A}. HAT-P-33b: 1 \citet{2017AJ....154...49W},  \textbf{2} {\citet{2023A&A...669A..63B}}. HAT-P-49~b: 1 \citet{2014AJ....147...84B},  \textbf{2} {\citet{2023A&A...669A..63B}}. HD 3167~c: \textbf{1} {\citet{2021A&A...654A.152B}}. HD 106315~c: 1 \citet{2018AJ....156...93Z},  \textbf{2} {\citet{2023A&A...669A..63B}}. HD 189733~b: 1 {\citet{2017A&A...602A.107B}},  \textbf{2} {\citet{2024A&A...682A..28C}}. HIP 41378~d: 1 \citet{2019AJ....158..248L},  2 \citet{2019arXiv191107355S},  \textbf{3} {\citet{2022A&A...668A.172G}}. K2-29~b: \textbf{1} \citet{2016ApJ...824...55S}. K2-105~b: 1 \citet{2017PASJ...69...29N},  2 \citet{2022MNRAS.509.1075C},  \textbf{3} {\citet{2023A&A...669A..63B}}. KELT-10~b: 1 \citet{2016MNRAS.459.4281K},  \textbf{2} {\citet{2023A&A...672A.134S}}. KELT-11~b: 1 \citet{2017AJ....154...25B}, \textbf{2} \citet{2022A&A...668A...1M}. KELT-19~b: 1 \citet{2024MNRAS.528..270K},  \textbf{2} \citet{2018AJ....155...35S}. Kepler-63~b: \textbf{1} \citet{2013ApJ...775...54S}. MASCARA-2~b: \textbf{1} \citet{arXiv:2311.03264}. MASCARA-4~b: 1 \citet{2020ApJ...888...63A}, \textbf{2} \citet{2020A&A...635A..60D}. Qatar-6~b: \textbf{1} \citet{2023AJ....165...65R}. TOI-640~b: \textbf{1} \citet{2023A&A...671A.164K}. TOI-677~b: \textbf{1} \citet{2023AJ....166..130S}. TOI-1670~c: \textbf{1} \citet{2023ApJ...959L...5L}. TOI-858Bb: \textbf{1} \citet{2023A&A...679A..70H}. TOI-1136~d: \textbf{1} \citet{2023AJ....165...33D}. TOI-1478~b: 1 \citet{2021AJ....161..194R},  \textbf{2} \citet{2022AJ....164..104R}. TOI-1842~b: \textbf{1} \citet{2023ApJ...949L..35H}. TOI-1937~b: \textbf{1} \citet{2023ApJS..265....1Y}. TOI-2025~b: \textbf{1} \citet{2022A&A...667A..22K}. TOI-2076~b: 1 \citet{2022A&A...664A.156O}, \textbf{2} \citet{2023ApJ...944L..41F}. TOI-2202~b: \textbf{1} \citet{2023AJ....166..266R}. TOI-3884~b: \textbf{1} \citet{2023AJ....165..249L}. WASP-7~b: 1 \citet{2017A&A...602A.107B},  \textbf{2} \citet{2012ApJ...744..189A}. WASP-12~b: 1 \citet{2017A&A...602A.107B},  \textbf{2} \citet{2012ApJ...757...18A}. WASP-33~b: \textbf{1} \citet{2015ApJ...810L..23J}. WASP-47~b: 1 \citet{2023A&A...669A..63B},  \textbf{2} \citet{2015ApJ...812L..11S}. WASP-52~b: 1 \citet{2017A&A...602A.107B},  \textbf{2} \citet{2023A&A...674A.174C}. WASP-62~b: 1 \citet{2020AJ....159..137G},  \textbf{2} \citet{2017MNRAS.464..810B}. WASP-76~b: \textbf{1} \citet{2020Natur.580..597E}. WASP-85~b: \textbf{1} \citet{2016AJ....151..150M}. WASP-94Ab: \textbf{1} \citet{2014A&A...572A..49N}. WASP-106~b: \textbf{1} \citet{2023AJ....166..217W}. WASP-107~b: 1 \citet{2021AJ....161...70P},  \textbf{2} \citet{2023A&A...669A..63B}. WASP-121~b: \textbf{1} \citet{2020A&A...635A.205B}. WASP-131~b: \textbf{1} \citet{2023MNRAS.522.4499D}. WASP-156~b: 1 \citet{2018A&A...610A..63D},  2 \citet{2021AJ....162...18S},  \textbf{3} \citet{2023A&A...669A..63B}. WASP-166~b: \textbf{1} \citet{2023A&A...669A..63B}. WASP-167~b: \textbf{1} \citet{2017MNRAS.471.2743T}. XO-6~b: \textbf{1} \citet{2017AJ....153...94C}.
    }
\end{TableNotes}

\begin{longtable}{l c c c c c c c}
    \caption{Extension to Table A1 of \citet{Albrecht2022}.}\label{tab:lamlit} \\

    \hline Planet & $a/R_\star$ & $M_{\rm p}$ & $R_{\rm p}$ & $e$  & $\lambda$ & $i_\star$ & $\psi$ \\
       &   & (M$_{\rm J}$) & (R$_{\rm J}$) &   & ($^\circ$) & ($^\circ$) & ($^\circ$) \\ \hline 
\endfirsthead

\multicolumn{8}{c}%
{{\bfseries \tablename\ \thetable{} -- continued from previous page}} \\
\hline Planet & $a/R_\star$ & $M_{\rm p}$ & $R_{\rm p}$ & $e$  & $\lambda$ & $i_\star$ & $\psi$ \\
       &   & (M$_{\rm J}$) & (R$_{\rm J}$) &   & ($^\circ$) & ($^\circ$) & ($^\circ$) \\  \hline 
\endhead

\hline \multicolumn{8}{r}{{Continued on next page}} \\ \hline
\endfoot


\hline
\insertTableNotes  
\endlastfoot
    
55~Cnc~e & $3.52\pm0.04$ & $0.0251\pm0.0010$ & $0.165\pm0.002$ & $0.04\pm0.03$ & $10^{+17}_{-20}$ & $75^{+11}_{-17}$ & $23^{+15}_{-23}$ \\ 
CoRoT-36~b & $9.3\pm1.0$ & $<0.7$ & $1.41 \pm 0.14$ & \ldots & $276 \pm 11$ & \ldots  & \ldots \\ 
HAT-P-11~b & $16.6 \pm 0.3$ & $0.087 \pm 0.010$ & $0.389 \pm 0.005$ & $0.2644 \pm 0.0006$ & $103^{+26}_{-10}$ & $67^{+2}_{-4}$  & $97^{+8}_{-4}$ \\ 
HAT-P-3~b & $9.8 \pm 0.3$ & $0.59 \pm 0.02$ & $0.91 \pm 0.03$ & $0.000 \pm 0.010$ & $21 \pm 9$ & $16^{+6}_{-7}$  & $76 \pm 8$ \\ 
HAT-P-30~b & $7.4 \pm 0.3$ & $0.71 \pm 0.03$ & $1.34 \pm 0.07$ & $0.000 \pm 0.016$ & $70 \pm 3$ & \ldots  & \ldots \\ 
HAT-P-32~b & $6.05 \pm 0.12$ & $0.80 \pm 0.14$ & $1.81 \pm 0.03$ & $0.00 \pm 0.04$ & $85.0 \pm 1.5$ & \ldots  & $84.9 \pm 1.5$ \\ 
HAT-P-33~b & $5.7 \pm 0.6$ & $0.72^{+0.13}_{-0.12}$ & $1.9^{+0.3}_{-0.2}$ & $0.18^{+0.11}_{-0.10}$ & $-6 \pm 4$ & \ldots  & \ldots \\ 
HAT-P-49~b & $5.13^{+0.19}_{-0.30}$ & $1.7 \pm 0.2$ & $1.41^{+0.13}_{-0.08}$ & \ldots & $-97.7 \pm 1.8$ & \ldots  & \ldots \\ 
HD 3167~c & $43.9^{+0.8}_{-0.9}$ & $0.031 \pm 0.004$ & $0.27^{+0.04}_{-0.03}$ & $0.0 \pm 0.3$ & $109^{+6}_{-75}$ & \ldots  & $108 \pm 5$ \\ 
HD 106315~c & $24.8 \pm 0.4$ & $0.038 \pm 0.012$ & $0.391 \pm 0.008$ & $0.22 \pm 0.15$ & $-3 \pm 3$ & \ldots  & \ldots \\ 
HD 189733~b & $9.0 \pm 0.3$ & $1.15 \pm 0.04$ & $1.15 \pm 0.04$ & $0.000 \pm 0.004$ & $-1.0 \pm 0.2$ & \ldots  & $14 \pm 7$ \\ 
HIP 41378~d & $190 \pm 20$ & $0.0145$ & $0.318 \pm 0.005$ & $0.06 \pm 0.06$ & $46^{+28}_{-37}$ & \ldots  & \ldots \\ 
K2-29~b & $10.54 \pm 0.14$ & $0.73 \pm 0.04$ & $1.19 \pm 0.02$ & $0.07^{+0.02}_{-0.07}$ & $2 \pm 9$ & \ldots  & \ldots \\ 
K2-105~b & $17.39 \pm 0.19$ & $0.09 \pm 0.06$ & $0.37^{+0.04}_{-0.03}$ & \ldots & $-81^{+50}_{-47}$ & \ldots  & \ldots \\ 
KELT-10~b & $9.0^{+0.3}_{-0.2}$ & $0.68 \pm 0.02$ & $1.41 \pm 0.05$ & \ldots & $-5 \pm 3$ & \ldots  & \ldots \\ 
KELT-11~b & $5.02\pm0.07$ & $0.22\pm0.02$ & $1.51\pm0.09$ & $0.0020^{+0.0005}_{-0.0014}$ & $-77\pm2$ & \ldots & \ldots \\ 
KELT-19~b & $7.5 \pm 0.5$ & $0 \pm 4$ & $1.91 \pm 0.11$ & $0.0 \pm 1.0$ & $-180 \pm 4$ & \ldots  & $155^{+17}_{-21}$ \\ 
Kepler-63~b & $19.1 \pm 0.7$ & $0.0 \pm 0.4$ & $0.544 \pm 0.018$ & $0.0 \pm 0.4$ & $-110^{+22}_{-14}$ & $138 \pm 7$  & $104^{+9}_{-14}$ \\ 
MASCARA-2~b & $7.4^{+0.3}_{-0.4}$ & $<3.5$ & $1.74^{+0.07}_{-0.08}$ & \ldots & $3.9\pm1.1$ & $89^{+18}_{-20}$ & $5.0\pm1.1$ \\ 
MASCARA-4~b & $5.3 \pm 0.5$ & $3.1 \pm 0.9$ & $1.53^{+0.07}_{-0.04}$ & $0.0 \pm 1.0$ & $247.5^{+1.5}_{-1.7}$ & $-63^{+10}_{-7}$  & $104^{+7}_{-13}$ \\ 
Qatar-6~b & $12.4 \pm 0.3$ & $0.68 \pm 0.05$ & $1.16 \pm 0.06$ & $0.05 \pm 0.03$ & $0 \pm 3$ & $67^{+10}_{-23}$  & $22^{+9}_{-18}$ \\ 
TOI-640~b & $6.33^{+0.07}_{-0.06}$ & $0.57 \pm 0.02$ & $1.72 \pm 0.05$ & \ldots & $184 \pm 3$ & $23^{+3}_{-2}$  & $104 \pm 2$ \\ 
TOI-677~b & $15.9^{+1.6}_{-1.3}$ & $1.24 \pm 0.07$ & $1.17 \pm 0.03$ & $0.44 \pm 0.02$ & $0.3 \pm 1.3$ & \ldots  & \ldots \\ 
TOI-1670~c & $40.66 \pm 0.06$ & $0.58^{+0.06}_{-0.55}$ & $0.97 \pm 0.02$ & $0.067^{+0.019}_{-0.018}$ & $-0 \pm 2$ & \ldots  & \ldots \\ 
TOI-858Bb & $7.3 \pm 0.3$ & $1.10^{+0.08}_{-0.07}$ & $1.25 \pm 0.04$ & \ldots & $99 \pm 4$ & $35^{+4}_{-5}$  & $94 \pm 3$ \\ 
TOI-1136~d & $23.5^{+0.6}_{-0.5}$ & $8.0^{+2.4}_{-1.9}$ & $4.63^{+0.08}_{-0.07}$ & $0.016^{+0.013}_{-0.010}$ & $5 \pm 5$ & \ldots  & \ldots \\ 
TOI-1478~b & $18.5^{+0.7}_{-0.6}$ & $0.88^{+0.11}_{-0.12}$ & $1.07^{+0.16}_{-0.10}$ & $0.024^{+0.032}_{-0.017}$ & $6 \pm 6$ & \ldots  & \ldots \\ 
TOI-1842~b & $12 \pm 3$ & $0.19^{+0.06}_{-0.04}$ & $1.06^{+0.07}_{-0.06}$ & $0.13^{+0.16}_{-0.09}$ & $-68^{+21}_{-15}$ & $46^{+12}_{-10}$  & $73^{+16}_{-13}$ \\ 
TOI-1937~b & $3.85^{+0.09}_{-0.10}$ & $2.01^{+0.17}_{-0.16}$ & $1.25 \pm 0.06$ & \ldots & $4 \pm 4$ & \ldots  & \ldots \\ 
TOI-2025~b & $12.7^{+0.5}_{-0.4}$ & $4.4 \pm 0.3$ & $1.117 \pm 0.009$ & $0.41 \pm 0.02$ & $9^{+33}_{-31}$ & \ldots  & \ldots \\ 
TOI-2076~b & $25.0\pm0.3$ & \ldots & $0.00792\pm0.00011$ & \ldots & $-3^{+16}_{-15}$ & $79^{+8}_{-11}$ & $18^{+10}_{-9}$ \\ 
TOI-2202~b & $26.0 \pm 0.3$ & $0.90^{+0.09}_{-0.10}$ & $0.977 \pm 0.016$ & $0.022^{+0.022}_{-0.015}$ & $26^{+12}_{-15}$ & $90 \pm 17$  & $31^{+13}_{-11}$ \\ 
TOI-3884~b & $25.9^{+1.0}_{-0.7}$ & $0.10 \pm 0.02$ & $0.574 \pm 0.018$ & $0.06^{+0.06}_{-0.04}$ & $75 \pm 10$ & $25 \pm 5$  & \ldots \\ 
WASP-7~b & $9.1 \pm 0.6$ & $0.98 \pm 0.13$ & $1.37 \pm 0.09$ & $0.00 \pm 0.05$ & $86 \pm 6$ & \ldots  & \ldots \\ 
WASP-12~b & $3.04^{+0.11}_{-0.10}$ & $1.47^{+0.08}_{-0.07}$ & $1.90^{+0.06}_{-0.04}$ & $0.00 \pm 0.02$ & $59^{+15}_{-20}$ & \ldots  & \ldots \\ 
WASP-33~b & $3.69^{+0.05}_{-0.10}$ & $2.2 \pm 0.2$ & $1.679^{+0.019}_{-0.030}$ & $0.0 \pm 1.0$ & $-112.9 \pm 0.2$ & \ldots  & \ldots \\ 
WASP-47~b & $9.67 \pm 0.15$ & $1.14 \pm 0.02$ & $1.123 \pm 0.013$ & $0.028^{+0.004}_{-0.002}$ & $0 \pm 24$ & $70^{+11}_{-9}$  & $29^{+11}_{-13}$ \\ 
WASP-52~b & $7.2 \pm 0.2$ & $0.43 \pm 0.02$ & $1.25 \pm 0.03$ & $0.00 \pm 0.09$ & $0.6 \pm 0.9$ & \ldots  & \ldots \\ 
WASP-62~b & $9.5 \pm 0.4$ & $0.57 \pm 0.04$ & $1.39 \pm 0.06$ & $0.0061 \pm 0.0006$ & $19 \pm 5$ & \ldots  & \ldots \\ 
WASP-76~b & $4.02 \pm 0.16$ & $0.894^{+0.019}_{-0.013}$ & $1.85 \pm 0.08$ & $0.00 \pm 0.05$ & $61^{+8}_{-5}$ & \ldots  & \ldots \\ 
WASP-85~b & $9.0 \pm 0.3$ & $1.26 \pm 0.07$ & $1.24 \pm 0.03$ & $0.0 \pm 1.0$ & $0 \pm 14$ & \ldots  & \ldots \\ 
WASP-94Ab & $7.3^{+0.3}_{-0.2}$ & $0.45^{+0.04}_{-0.03}$ & $1.72^{+0.06}_{-0.05}$ & $0.00 \pm 0.13$ & $151^{+16}_{-23}$ & \ldots  & \ldots \\ 
WASP-106~b & $13.2^{+0.3}_{-0.4}$ & $1.93 \pm 0.15$ & $1.080^{+0.016}_{-0.017}$ & $0.023^{+0.027}_{-0.016}$ & $6^{+17}_{-16}$ & $90 \pm 25$  & $26^{+12}_{-17}$ \\ 
WASP-107~b & $17.7 \pm 0.7$ & $0.096 \pm 0.005$ & $0.92 \pm 0.02$ & $0.06 \pm 0.04$ & $-158^{+15}_{-18}$ & $15.10 \pm 0.04$  & $103.5^{+1.7}_{-1.8}$ \\ 
WASP-121~b & $3.80 \pm 0.11$ & $1.18 \pm 0.06$ & $1.86 \pm 0.04$ & $0.00 \pm 0.07$ & $87.2 \pm 0.4$ & \ldots  & $88.1 \pm 0.2$ \\ 
WASP-131~b & $8.37 \pm 0.15$ & $0.273 \pm 0.019$ & $1.23 \pm 0.04$ & \ldots & $162.4^{+1.3}_{-1.2}$ & $41^{+13}_{-8}$  & $124^{+13}_{-8}$ \\ 
WASP-156~b & $12.75 \pm 0.03$ & $0.128^{+0.010}_{-0.009}$ & $0.51 \pm 0.02$ & \ldots & $106 \pm 14$ & \ldots  & \ldots \\ 
WASP-166~b & $11.3 \pm 0.6$ & $0.102 \pm 0.004$ & $0.63 \pm 0.03$ & $0.00 \pm 0.07$ & $-0.7 \pm 1.6$ & $1^{+23}_{-19}$  & $0 \pm 56$ \\ 
WASP-167~b & $4.38 \pm 0.14$ & $0 \pm 8$ & $1.56 \pm 0.05$ & $0.0 \pm 1.0$ & $-165 \pm 5$ & \ldots  & \ldots \\ 
XO-6~b & $8.1 \pm 1.0$ & $2.0 \pm 0.7$ & $2.08 \pm 0.18$ & $0.0 \pm 1.0$ & $-21 \pm 2$ & \ldots  & \ldots \\
HD~118203~b & $7.23^{+0.16}_{-0.18}$ & $2.17^{+0.07}_{-0.08}$ & $1.14\pm0.03$  & $0.314 \pm 0.017$ & \hdg  & $17\pm2$ & $75^{+3}_{-5}$ \\
HD~148193~b & $23.1^{+1.7}_{-0.8}$ & $0.092\pm0.015$ & $0.764^{+0.018}_{-0.017}$ & $0.13_{-0.09}^{+0.12}$ & \hdtwos & \ldots & \ldots \\
K2-261~b & $14\pm2$  & $0.22\pm0.03$ & $0.85^{+0.03}_{-0.02}$ & $0.42 \pm 0.03$ & \ksixoneg & \ldots & \ldots \\
K2-287~b & $25.4^{+1.1}_{-1.2}$ & $0.32\pm0.03$ & $0.833\pm0.013$ & $0.48\pm0.03$ & \keightseveng & \ldots & \ldots \\
KELT-3~b & $5.64^{+0.05}_{-0.06}$ & $1.47\pm0.07$ & $1.35 \pm 0.07$ & 0 & \keltthreeu & \ldots & \ldots \\
KELT-4Ab & $6.02^{+0.09}_{-0.10}$ & $0.90\pm0.06$ & $1.70\pm0.05$ & $0.03^{+0.03}_{-0.02}$ & \keltfourg  & \ldots &  \ldots \\
LTT~1445Ab & $29.9^{+1.5}_{-1.3}$ & $0.0090\pm0.0008$ & $0.116^{+0.006}_{-0.005}$ & $<0.110$ & \lttg & \ldots & \ldots \\
TOI-451Ab & $6.93^{+0.11}_{-0.16}$ & \ldots & $0.170\pm0.011$ & $0$ & \toifourfiveg & $69^{+11}_{-8}$ & $6.2^{+0.8}_{-0.6}$  \\
TOI-813~b  & $47.2^{+4}_{-2}$ & \ldots & $0.60\pm0.03$ & \ldots & \toieightthirteeng & \ldots & \ldots \\
TOI-892~b & $15.1^{+0.4}_{-0.3}$ & $0.95\pm0.07$ & $1.07\pm0.02$ & $<0.125$ & \toieightnines & \ldots & \ldots  \\
TOI-1130~c & $22.1\pm0.4$ & $0.97\pm0.04$ & $1.5^{+0.3}_{-0.2}$ & $0.047^{+0.040}_{-0.027}$ & \toithirtyg & \ldots & \ldots \\
WASP-50~b & $7.47^{+0.08}_{-0.09}$ & $1.47\pm0.09$ & $1.15\pm0.05$ & $0.009^{+0.011}_{-0.006}$ & \waspfiftyu & \ldots & \ldots \\
WASP-59~b & $25.4^{+1.1}_{-1.0}$ & $0.86\pm0.05$ & $0.78\pm0.07$ & $0$ & \waspfiftynineu & \ldots & \ldots \\
WASP-136~b & $7.4\pm0.2$ & $1.51\pm0.08$ & $1.38 \pm 0.16$ & $0$ & \waspthirtyeightg & $38^{+7}_{-6}$ & $60\pm8$ \\
WASP-148~b & $14.4 \pm 0.3$ & $0.29\pm0.03$ & $0.72\pm0.06$ & $0.220 \pm 0.063$ & \waspfortyeightg  & $68^{+8}_{-21}$ & $21^{+9}_{-17}$ \\
WASP-172~b & $7.58\pm0.17$  & $0.47\pm0.10$ & $1.57\pm0.10$ & $0$ & \waspseventwog & $73^{+9}_{-14}$ & $112\pm6$ \\
WASP-173Ab & $5.12^{+0.03}_{-0.02}$ & $3.69 \pm 0.18$ & $1.20 \pm 0.06$ & $0$ & \waspseventhreeg & $71^{+10}_{-13}$ & $30\pm14$  \\
WASP-186~b & $9.9\pm0.3$  & $4.22\pm0.18$ & $1.11\pm0.03$ & $0.33\pm0.01$ & \waspeightsixg  & $75^{+7}_{-15}$ & $22^{+11}_{-14}$ \\
XO-7~b & $6.54^{+0.07}_{-0.06}$   & $0.71\pm0.03$ & $1.37\pm0.03$ & $0.04 \pm 0.03$ & \xoseveng & $14.5^{+1.6}_{-1.4}$ & $70\pm1.7$  \\ 
WASP-26~b & $6.5^{+0.3}_{-0.2}$ & $1.02\pm0.03$ & $1.32\pm0.08$ & $0$ & $-16^{+14}_{-10}$ &  \ldots & \ldots  \\
\end{longtable}
\end{ThreePartTable} 
\end{center}
}
\clearpage
\twocolumn



\onecolumn
\begin{center}
\begin{ThreePartTable}
    \begin{TableNotes}
    {\catcode`\&=12 
    \item Values for the new literature systems are drawn from the following references: 55~Cnc: 1 \citet{2011ApJ...740...49V}, 2 \citet{2023NatAs...7..198Z}. CoRoT-36: 1 \citet{2022MNRAS.516..636S}. HAT-P-11: 1 \citet{2010ApJ...710.1724B}, 2 \citet{2010ApJ...723L.223W}. HAT-P-3: 1 \citet{2023A&A...669A..63B}, 2 \citet{2018A&A...613A..41M}. HAT-P-30: 1 \citet{2011ApJ...735...24J}, 2 \citet{2023A&A...674A.174C}. HAT-P-32: 1 \citet{2017A&A...602A.107B}, 2 \citet{2012ApJ...757...18A}. HAT-P-33: 1 \citet{2017AJ....154...49W}, 2 \citet{2023A&A...669A..63B}. HAT-P-49: 1 \citet{2014AJ....147...84B}, 2 \citet{2023A&A...669A..63B}. HD 3167: 1 \citet{2017AJ....154..122C}, 2 \citet{2019A&A...631A..28D}, 3 \citet{2021A&A...654A.152B}. HD 106315: 1 \citet{2017AJ....153..255C}, 2 \citet{2023A&A...669A..63B}. HD 189733: 1 \citet{2016A&A...588A.127C}. HIP 41378: 1 \citet{2019AJ....158..248L}, 2 \citet{2022A&A...668A.172G}. K2-29b: 1 \citet{2016ApJ...824...55S}. K2-105: 1 \citet{2022MNRAS.509.1075C}, 2 \citet{2023A&A...669A..63B}. KELT-10: 1 \citet{2016MNRAS.459.4281K}, 2 \citet{2023A&A...672A.134S}. KELT-11: 1 \citet{2017AJ....154...25B}, 2 \citet{2022A&A...668A...1M}. KELT-19: 1 \citet{2018AJ....155...35S}. Kepler-63: 1 \citet{2013ApJ...775...54S}. MASCARA-2: 1 \citet{2017AJ....154..194L}, 2 \citet{arXiv:2311.03264}. MASCARA-4: 1 \citet{2020A&A...635A..60D}. Qatar-6: 1 \citet{2018AJ....155...52A}, 2 \citet{2023AJ....165...65R}. TOI-640: 1 \citet{2021AJ....161..194R}, 2 \citet{2023A&A...671A.164K}. TOI-677: 1 \citet{2023AJ....166..130S}. TOI-1670: 1 \citet{2023ApJ...959L...5L}. TOI-858B: 1 \citet{2023A&A...679A..70H}. TOI-1136: 1 \citet{2023AJ....165...33D}. TOI-1478: 1 \citet{2021AJ....161..194R}, 2 \citet{2022AJ....164..104R}. TOI-1842: 1 \citet{2023ApJ...949L..35H}. TOI-1937: 1 \citet{2023ApJS..265....1Y}. TOI-2025: 1 \citet{2022A&A...667A..22K}. TOI-2076: 1 \citet{2023ApJ...944L..41F}. TOI-2202: 1 \citet{2023AJ....166..266R}. TOI-3884: 1 \citet{2023AJ....165..249L}. WASP-7: 1 \citet{2011A&A...527A...8S}, 2 \citet{2012ApJ...744..189A}. WASP-12: 1 \citet{2017A&A...602A.107B}, 2 \citet{2012ApJ...757...18A}. WASP-33: 1 \citet{2011A&A...535A.110M}, 2 \citet{2015ApJ...810L..23J}.  WASP-47: 1 \citet{2017AJ....154..237V}, 2 \citet{2015ApJ...812L..11S}. WASP-52: 1 \citet{2017A&A...602A.107B}, 2 \citet{2023A&A...674A.174C}. WASP-62: 1 \citet{2017MNRAS.464..810B}. WASP-76: 1 \citet{2020Natur.580..597E}. WASP-85: 1 \citet{2014arXiv1412.7761B}, 2 \citet{2016AJ....151..150M}. WASP-94A: 1 \citet{2017A&A...602A.107B}, 2 \citet{2014A&A...572A..49N}. WASP-106: 1 \citet{2023AJ....166..217W}. WASP-107: 1 \citet{2021AJ....161...70P}, 2 \citet{2023A&A...669A..63B}. WASP-121: 1 \citet{2020A&A...635A.205B}. WASP-131: 1 \citet{2023MNRAS.522.4499D}. WASP-156: 1 \citet{2018A&A...610A..63D}, 2 \citet{2023A&A...669A..63B}. WASP-166: 1 \citet{2019MNRAS.488.3067H}, 2 \citet{2023A&A...669A..63B}. WASP-167: 1 \citet{2017MNRAS.471.2743T}. XO-6: 1 \citet{2017AJ....153...94C}.
    }
    \end{TableNotes}

\begin{longtable}{l c c c c c}
    \caption{Extension to Table A2 of \citet{Albrecht2022}.}\label{tab:syslit} \\

    \hline System & $T_{\rm eff}$ & $M_\star$ & $R_\star$ & Age  & \vsini \\
       & (K) & (M$_\odot$) & (R$_\odot$) & (Gyr) & (km~s$^{-1}$) \\ \hline 
\endfirsthead

\multicolumn{6}{c}%
{{\bfseries \tablename\ \thetable{} -- continued from previous page}} \\
\hline System & $T_{\rm eff}$ & $M_\star$ & $R_\star$ & Age  & \vsini \\
       & (K) & (M$_\odot$) & (R$_\odot$) & (Gyr) & (km~s$^{-1}$) \\  \hline 
\endhead

\hline \multicolumn{6}{r}{{Continued on next page}} \\ \hline
\endfoot


\hline
\insertTableNotes  
\endlastfoot

55~Cnc & $5272\pm24$ & $0.943\pm0.010$ & $0.905\pm0.015$ & \ldots & $2.0^{+0.43}_{-0.47}$ \\ 
CoRoT-36 & $6730 \pm 140$ & $1.32 \pm 0.09$ & $1.52^{+0.20}_{-0.10}$ & \ldots & $25.6 \pm 0.3$ \\ 
HAT-P-11 & $4780 \pm 50$ & $0.80 \pm 0.03$ & $0.683 \pm 0.009$ & $6^{+6}_{-4}$ & $1.0^{+0.9}_{-0.6}$ \\ 
HAT-P-3 & $5190 \pm 80$ & $0.93 \pm 0.04$ & $0.85 \pm 0.02$ & $2.6 \pm 0.6$ & $0.5 \pm 0.2$ \\ 
HAT-P-30 & $6338 \pm 42$ & $1.24 \pm 0.04$ & $1.22 \pm 0.05$ & $1.0^{+0.8}_{-0.5}$ & $3.63 \pm 0.07$ \\ 
HAT-P-32 & $6269 \pm 64$ & $1.18 \pm 0.05$ & $1.225 \pm 0.017$ & $2.7 \pm 0.8$ & $20.6 \pm 1.5$ \\ 
HAT-P-33 & $6460^{+300}_{-290}$ & $1.42^{+0.16}_{-0.15}$ & $1.9^{+0.3}_{-0.2}$ & \ldots & $15.6 \pm 0.3$ \\ 
HAT-P-49 & $6820 \pm 52$ & $1.54 \pm 0.05$ & $1.83^{+0.14}_{-0.08}$ & \ldots & $10.7 \pm 0.5$ \\ 
HD 3167 & $5261 \pm 60$ & $0.84^{+0.05}_{-0.04}$ & $0.880^{+0.012}_{-0.013}$ & $8 \pm 4$ & $2.1 \pm 0.4$ \\ 
HD 106315 & $6364 \pm 87$ & $1.15 \pm 0.04$ & $1.27 \pm 0.02$ & $4.0 \pm 1.0$ & $9.7^{+0.6}_{-0.7}$ \\ 
HD 189733 & $5050\pm50$ & $0.84\pm0.04$ & $0.752\pm0.025$ & $6.2\pm3.4$ & $3.25\pm0.02$ \\ 
HIP 41378 & $6290 \pm 77$ & $1.22^{+0.03}_{-0.02}$ & $1.300 \pm 0.009$ & \ldots & $3.8 \pm 1.0$ \\ 
K2-29 & $5358 \pm 38$ & $0.94 \pm 0.02$ & $0.860 \pm 0.010$ & $2.6^{+1.2}_{-2.3}$ & $3.7 \pm 0.5$ \\ 
K2-105 & $5636^{+49}_{-52}$ & $1.05 \pm 0.02$ & $0.970 \pm 0.010$ & \ldots & $2.1^{+1.0}_{-0.9}$ \\ 
KELT-10 & $5948 \pm 74$ & $1.11 \pm 0.06$ & $1.21^{+0.05}_{-0.04}$ & \ldots & $2.58 \pm 0.12$ \\ 
KELT-11 & $5375\pm25$ & $1.80\pm0.07$ & $2.69\pm0.04$ & \ldots & $1.99^{+0.06}_{-0.07}$ \\ 
KELT-19 & $7500 \pm 110$ & $1.6 \pm 0.2$ & $1.83 \pm 0.10$ & $1.10 \pm 0.10$ & $84 \pm 2$ \\ 
Kepler-63 & $5576 \pm 50$ & $0.98 \pm 0.04$ & $0.90^{+0.03}_{-0.02}$ & $0.21 \pm 0.05$ & $5.6 \pm 0.8$ \\ 
MASCARA-2 & $8730^{+250}_{-260}$ & $1.76^{+0.14}_{-0.20}$ & $1.561^{+0.058}_{-0.064}$ & $0.20^{+0.10}_{-0.05}$ & $116\pm1$ \\ 
MASCARA-4 & $7800 \pm 200$ & $1.75 \pm 0.05$ & $1.92 \pm 0.11$ & $0.7 \pm 0.2$ & $46.5 \pm 1.0$ \\ 
Qatar-6 & $5052 \pm 66$ & $0.82 \pm 0.02$ & $0.72 \pm 0.02$ & \ldots & $2.9^{+0.9}_{-0.7}$ \\ 
TOI-640 & $6460^{+130}_{-150}$ & $1.54^{+0.07}_{-0.08}$ & $2.08 \pm 0.06$ & \ldots & $5.9 \pm 0.4$ \\ 
TOI-677 & $6295 \pm 80$ & $1.16 \pm 0.03$ & $1.281 \pm 0.012$ & \ldots & $7.4 \pm 0.5$ \\ 
TOI-1670 & $6330^{+68}_{-70}$ & $1.22^{+0.06}_{-0.07}$ & $1.31 \pm 0.03$ & \ldots & $8.9 \pm 0.5$ \\ 
TOI-858B & $5842^{+84}_{-79}$ & $1.08^{+0.08}_{-0.07}$ & $1.31 \pm 0.04$ & \ldots & $6.4 \pm 0.2$ \\ 
TOI-1136 & $5770 \pm 50$ & $1.02 \pm 0.03$ & $0.97 \pm 0.04$ & \ldots & $6.7 \pm 0.6$ \\ 
TOI-1478 & $5595 \pm 83$ & $0.95^{+0.06}_{-0.04}$ & $1.05 \pm 0.03$ & \ldots & $1.24 \pm 0.16$ \\ 
TOI-1842 & $6033^{+95}_{-93}$ & $1.45^{+0.07}_{-0.14}$ & $2.03 \pm 0.07$ & \ldots & $6.0 \pm 0.9$ \\ 
TOI-1937 & $5814^{+91}_{-93}$ & $1.07 \pm 0.06$ & $1.08^{+0.03}_{-0.02}$ & \ldots & $6.0 \pm 0.9$ \\ 
TOI-2025 & $5880 \pm 53$ & $1.32 \pm 0.14$ & $1.56 \pm 0.03$ & \ldots & $6.0 \pm 0.3$ \\ 
TOI-2076 & $5180 \pm 110$ & $0.883 \pm 0.017$ & $0.772^{+0.015}_{-0.016}$ & \ldots & $5.3 \pm 0.2$ \\ 
TOI-2202 & $5169^{+80}_{-78}$ & $0.84 \pm 0.03$ & $0.81 \pm 0.02$ & \ldots & $2.1^{+0.3}_{-0.2}$ \\ 
TOI-3884 & $3180 \pm 88$ & $0.298 \pm 0.018$ & $0.302 \pm 0.012$ & \ldots & $3.6 \pm 0.9$ \\ 
WASP-7 & $6520\pm70$ & $1.317\pm0.072$ & $1.48\pm0.09$ & $2.4\pm1.0$ & $14\pm2$ \\ 
WASP-12 & $6313 \pm 52$ & $1.43^{+0.11}_{-0.09}$ & $1.66^{+0.05}_{-0.04}$ & $2.0^{+0.7}_{-2.0}$ & $1.6^{+0.8}_{-0.4}$ \\ 
WASP-33 & $7430\pm100$ & $1.56^{+0.05}_{-0.08}$ & $1.51^{+0.02}_{-0.03}$ & $0.1^{+0.4}_{-0.09}$ & $86.6^{+0.3}_{-04}$ \\ 
WASP-47 & $5576 \pm 67$ & $1.04 \pm 0.03$ & $1.137 \pm 0.013$ & $6.7^{+1.5}_{-1.1}$ & $1.80^{+0.24}_{-0.16}$ \\ 
WASP-52 & $5000 \pm 100$ & $0.80 \pm 0.05$ & $0.786 \pm 0.016$ & $10.7^{+1.9}_{-4.5}$ & $2.06 \pm 0.04$ \\ 
WASP-62 & $6230 \pm 80$ & $1.25 \pm 0.05$ & $1.28 \pm 0.05$ & $0.8 \pm 0.6$ & $9.3 \pm 0.2$ \\ 
WASP-76 & $6329 \pm 65$ & $1.46 \pm 0.02$ & $1.76 \pm 0.07$ & $1.8 \pm 0.3$ & $1.5 \pm 0.3$ \\ 
WASP-85 & $5685 \pm 65$ & $1.09 \pm 0.08$ & $0.94 \pm 0.02$ & $0.50^{+0.30}_{-0.10}$ & $3.4 \pm 0.9$ \\ 
WASP-94A & $6170 \pm 80$ & $1.45 \pm 0.09$ & $1.62^{+0.05}_{-0.04}$ & $2.7 \pm 0.6$ & $4.2 \pm 0.5$ \\ 
WASP-106 & $6002 \pm 164$ & $1.18^{+0.08}_{-0.07}$ & $1.470^{+0.016}_{-0.017}$ & \ldots & $7.0^{+1.1}_{-1.0}$ \\ 
WASP-107 & $4425 \pm 70$ & $0.683^{+0.017}_{-0.016}$ & $0.67 \pm 0.02$ & $7^{+4}_{-3}$ & $0.51^{+0.07}_{-0.09}$ \\ 
WASP-121 & $6586 \pm 59$ & $1.38 \pm 0.02$ & $1.44 \pm 0.03$ & $1.5 \pm 1.0$ & $13.6 \pm 0.7$ \\ 
WASP-131 & $5990 \pm 50$ & $1.06 \pm 0.06$ & $1.56 \pm 0.04$ & \ldots & $3.0 \pm 0.9$ \\ 
WASP-156 & $4910 \pm 61$ & $0.84 \pm 0.05$ & $0.76 \pm 0.03$ & \ldots & $3.2^{+0.7}_{-0.8}$ \\ 
WASP-166 & $6050\pm50$ & $1.19\pm0.06$ & $1.22\pm0.06$ & $2.1\pm0.9$ & $5.40\pm0.14$ \\ 
WASP-167 & $7043^{+89}_{-68}$ & $1.59 \pm 0.08$ & $1.79 \pm 0.05$ & $1.5 \pm 0.4$ & $49.94 \pm 0.04$ \\ 
XO-6 & $6720 \pm 100$ & $1.47 \pm 0.06$ & $1.93 \pm 0.18$ & $1.9^{+0.9}_{-0.2}$ & $48 \pm 3$ \\ 
HD~118203 &  $5683 \pm 85$ & $1.13^{+0.05}_{-0.06}$ & $2.10\pm0.05$ & $5.32^{+0.96}_{-0.73}$ & \hdvsinig \\
HD~148193 & $6198 \pm 100$ & $1.23^{+0.02}_{-0.05}$ & $1.63^{+0.03}_{-0.02}$ & $3.5^{+1.3}_{-0.5}$ & \hdtwovsinig \\
K2-261  &  $5537\pm71$ & $1.105 \pm 0.019$ & $1.669 \pm 0.022$ & $8.8^{+0.4}_{-0.3}$ & \ksixonevsinig \\
K2-287 & $5695 \pm 58$ & $1.06\pm0.02$ & $1.070\pm0.010$ & $4.5\pm1$ &  \keightsevenvsinig \\
KELT-3 & $6306 \pm 50$ & $1.28\pm0.06$ & $1.47\pm0.07$ & $3\pm0.2$ & \keltthreevsinig \\
KELT-4A & $6207\pm75$ & $1.20^{+0.07}_{-0.06}$ & $1.60\pm0.04$ & $4.44^{+0.78}_{-0.89}$ & \keltfourvsinig \\
LTT~1445A & $3340\pm150$ & $0.257\pm0.014$ & $0.265^{+0.011}_{-0.010}$ & \ldots & \lttvsinig \\
TOI-451A & $5550\pm56$ & $0.95\pm0.02$ & $0.88\pm0.03$ & $0.125\pm0.008$ & \toifourfivevsinig \\
TOI-813 & $5907 \pm 150$ & $1.32\pm0.06$ & $1.94 \pm 0.10$ & $3.73\pm0.62$ & \toieightthirteenvsinig \\
TOI-892 & $6261\pm80$ & $1.28\pm0.03$ & $1.39\pm0.02$ & $2.2\pm0.5$ & \toieightninevsinis \\
TOI-1130 & $4250\pm67$ & $0.684^{+0.016}_{-0.017}$ & $0.687\pm0.015$ & $8.2^{+3.8}_{-4.9}$ & \toithirtyvsinig \\
WASP-50 & $5400 \pm 100$ & $0.89^{+0.08}_{-0.07}$ & $0.84\pm0.03$ & $7\pm3.5$ &  \waspfiftyvsinig  \\
WASP-159 & $4650 \pm 150$ & $0.72\pm0.04$ & $0.61\pm0.04$ & $0.5^{+0.7}_{-0.4}$ & \waspfiftyninevsiniu \\
WASP-136 & $6250 \pm 100$  & $1.41\pm0.07$ & $2.2\pm0.2$ & $3.62\pm0.70$ & \waspthirtyeightvsinig \\
WASP-148 & $5460\pm130$ & $1.00\pm0.08$ & $1.03\pm0.20$ & \ldots & \waspfortyeightvsinig \\
WASP-172 & $6900 \pm 150$ & $1.49\pm0.07$ & $1.91\pm0.10$ & $1.79\pm0.28$ & \waspseventwovsinig \\
WASP-173A & $5700 \pm 150$ & $1.05 \pm 0.08$ & $1.11\pm0.05$ & $6.78\pm2.93$ & \waspseventhreevsinig \\
WASP-186 & $6300\pm100$ & $1.21^{+0.07}_{-0.08}$ & $1.46\pm0.02$ & $3.1^{+1.0}_{-0.8}$ & \waspeightsixvsinig \\
XO-7 & $6250 \pm 100$ & $1.405 \pm 0.059$ & $1.480\pm0.022$ & $1.18^{+0.98}_{-0.71}$ & \xosevenvsinig \\ 
WASP-26 & $5950\pm100$ & $1.12\pm0.03$ & $1.34\pm0.06$ & $6\pm2$ & $2.9\pm0.8$ \\
\end{longtable}
\end{ThreePartTable}
\end{center}
\clearpage
\twocolumn


{\setlength{\tabcolsep}{6pt}
\begin{table*}
    \centering
    \begin{threeparttable}
    \caption{Hyperparameters for the beta-distribution.}
    \label{tab:beta}
    \begin{tabular}{c c c c c c c c}
    \toprule
         Population & $N$ & $\mu_0$ & $\kappa_0$ & $w_0$ & $\mu_1$ & $\kappa_1$ & $w_1$ \\
        \midrule
        all systems$^{a)}$ & 151 & $0.43 \pm 0.09$ & $4 \pm 6$ & $0.28 \pm 0.09$ & $0.98 \pm 0.02$ & $14 \pm 27$ & $0.72\pm0.09$ \\
        all systems & 205 & $0.47\pm0.06$ & $5\pm4$ & $0.31\pm0.07$ & $0.981 \pm 0.016$ & $33 \pm 43$ & $0.69\pm0.07$ \\
        not applying info $i_\star$ & 81 & $0.42\pm0.05$ & $8\pm6$ & $0.31\pm0.06$ & $0.986\pm0.018$ & $31\pm35$ & $0.69\pm0.06$ \\
        applying info on $i_\star$ & 81 & $0.45\pm0.03$ & $11 \pm 3$ & $0.35 \pm 0.05$ & $0.990\pm0.006$ & $30\pm17$ & $0.65\pm0.05$ \\
        $T_{\rm eff} \geq 6250$~K, $a/R_\star<7$ & 41 & $0.46\pm0.05$ & $13\pm11$ & $0.49\pm0.11$ & $0.95\pm0.05$ & $20\pm24$ & $0.51\pm0.11$ \\
        $M_{\rm p}<0.2$~M$_{\rm J}$ & 17 & $0.40\pm0.09$ & $23\pm28$ & $0.52\pm0.16$ & $0.92\pm0.07$ & $21\pm30$ & $0.48\pm0.16$ \\
        \hdashline
        $T_{\rm eff} \geq 6250$~K, $a/R_\star<7$ & 41 & $0.43\pm0.15$ & $130\pm400$ & $0.3\pm0.2$ & $0.85\pm0.10$ & $80\pm300$ & $0.7\pm0.2$ \\
        $M_{\rm p}<0.2$~M$_{\rm J}$ & 17 & $0.39\pm0.13$ & $1300\pm4000$ & $0.4\pm0.2$ & $0.82\pm0.11$ & $400\pm2000$ & $0.6\pm0.2$ \\
         \bottomrule
    \end{tabular}
    \begin{tablenotes}
        \item Subscript 0 denotes the parameters for the misaligned component, while subscript 1 is for the aligned component. The true mean of $\cos \psi$ is given as $2\mu - 1$. $^{a)}$Results from \citet{Dong2023}. A Gaussian prior was applied to $\log \kappa$ \citep[$\mathcal{N}(0,3)$ as in][]{Dong2023} for the runs above the dashed line, whereas a uniform prior of $\mathcal{U}(-4,10)$ for the runs below.
    \end{tablenotes}
    \end{threeparttable}
\end{table*}
}


{\setlength{\tabcolsep}{8pt}
\begin{table}
    \centering
    \begin{threeparttable}
    \caption{Radial velocity measurements for all targets. }
    \label{tab:rvs}
    \begin{tabular}{c c c c c}
    \toprule
    Target & Instrument & BJD & RV & $\sigma_{\rm RV}$ \\
      &   & (TDB) & (m~s$^{-1}$) & (m~s$^{-1}$) \\
    \midrule
        HD 118203 & FIES & 2458933.38776 & -28917.17361 & 3.14228 \\ 
        HD 118203 & FIES & 2458933.40158 & -28906.96003 & 2.81819 \\ 
        HD 118203 & FIES & 2458933.41544 & -28895.14674 & 4.15899 \\ 
        \vdots & \vdots & \vdots & \vdots & \vdots \\ 
        XO-7 & FIES+ & 2459727.62884 & -13079.66931 & 5.87951 \\ 
        XO-7 & FIES+ & 2459727.64049 & -13085.85461 & 6.76278 \\ 
        XO-7 & FIES+ & 2459727.65148 & -13070.58831 & 7.04130 \\  
    \bottomrule
    \end{tabular}
    \begin{tablenotes}
        \item This table is available online in its entirety in machine-readable form.
    \end{tablenotes}
    \end{threeparttable}
\end{table}
}


\end{document}